\documentclass[12pt,oneside,a4paper,openright]{report}
\usepackage{times}                                 %for pgelayout command and options(http://en.wikibooks.org/wiki/LaTeX/Page_Layout)
\usepackage{amsmath,amssymb,amsfonts}
\usepackage{ifpdf}
\usepackage{graphicx,lineno}
\usepackage{graphics}   % Include figure files
\usepackage{dcolumn,enumerate}    % Align table columns on decimal point
\usepackage{bm,bbm}         % bold math
\usepackage{mathrsfs}
\usepackage{mathtools}
\usepackage{xcolor}
\usepackage[T1]{fontenc}
\usepackage{rotating}
\usepackage{upgreek}
\usepackage{float}
\usepackage[caption = false]{subfig}
\newcommand{\proj}[1]{\outpr{#1}{#1}}
\usepackage[utf8]{inputenc}
\usepackage{titlesec}
\usepackage{txfonts}
\usepackage{url}
\numberwithin{equation}{chapter}       % includes chapter number within equation
\usepackage{geometry}                   %   for setting layout
\usepackage{fancyhdr}                   % extensive use of setting of header and footer options
\usepackage[small,bf]{caption}
\usepackage{setspace}
\usepackage[bookmarks,
                  bookmarksopen = true,
                  bookmarksnumbered = true,
                  breaklinks = true,
                  linktocpage,
                  colorlinks = true,
                  linkcolor = blue,
                  urlcolor  = blue,
                  citecolor = blue ,
                  anchorcolor = green,
                  hyperindex = true,
                  hyperfigures
                   ]{hyperref} 

\ifpdf
\else 
\fi
\DeclareMathVersion{normal2}
\renewcommand{\headrulewidth}{0.5pt}

\fancypagestyle{plain}{
  \fancyhead{}
  \renewcommand{\headrulewidth}{0pt}
}
\definecolor{lightgray}{rgb}{0.98, 0.81, 0.69}
\makeatletter
\newcommand\longleftrightarrowfill@{%
  \arrowfill@\leftarrow\relbar\rightarrow}
\makeatother

\titleformat*{\section}{\large}{\cal}

\newcommand{\tr}[1]{\mathrm{Tr}[{#1}]}
\newcommand{\ket}[1]{\vert{#1}\rangle}
\newcommand{\bra}[1]{\langle{#1}\vert}
\newcommand{\outpr}[2]{\vert{#1}\rangle\langle{#2}\vert}

\newcommand{\expv}[1]{\langle{#1}\rangle}
\newcommand{\modulous}[1]{\vert{#1}\vert}

\newcommand{\blankpage}{\newpage\thispagestyle{empty}\mbox{}\newpage}
\begin{document}
% % % % % title page
\begin{titlepage}
\begin{center}
\vspace{0.5cm}
{\Large \bfseries Ancilla Assisted Quantum Information Processing:
General protocols and NMR implementations}\\[2.5cm]

A thesis\\
Submitted in partial fulfillment of the requirements\\
Of the degree of\\
\vspace{0.8cm}
{\normalsize \bfseries \color{blue}{\large DOCTOR OF PHILOSOPHY}}\\[1cm]
By\\[1.2cm]
{\large Abhishek Shukla}\\
20093040 \\[1.8cm]

\begin{figure}[h]
\begin{center}
\includegraphics[trim = 2cm 6cm 2cm 6cm, clip=true, width=4cm]{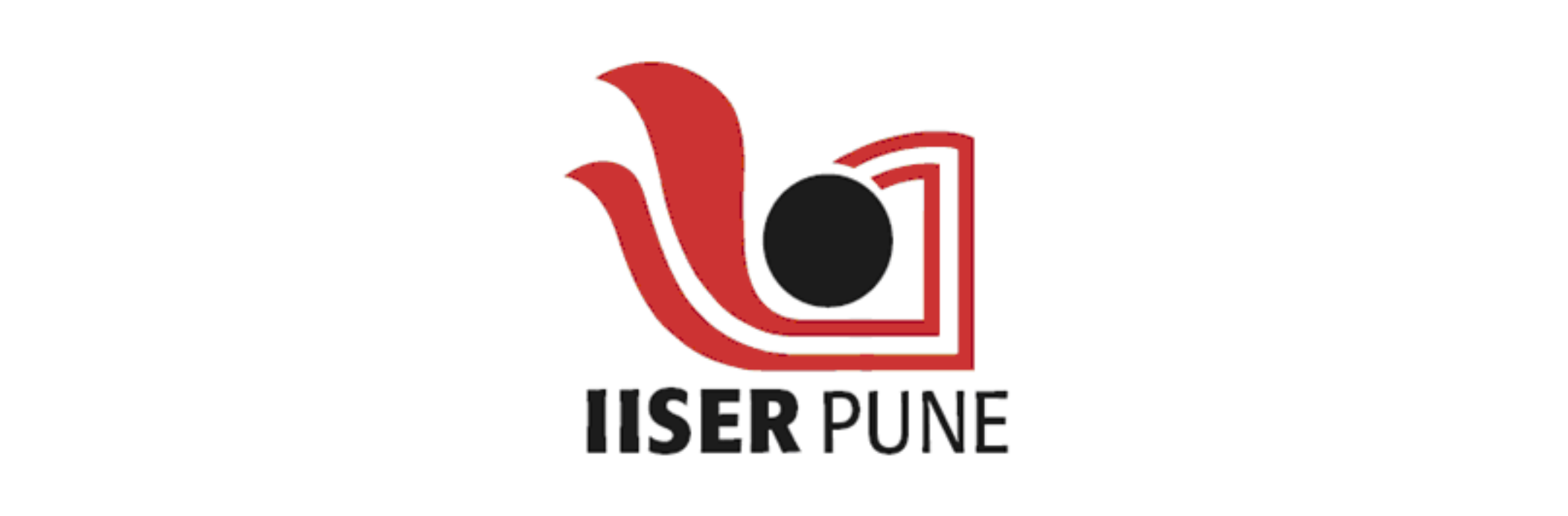}
\end{center}
\end{figure}

%\vspace{-2cm}
{INDIAN INSTITUTE OF SCIENCE EDUCATION AND RESEARCH, PUNE}\\[1cm]

{December, 2015}

\end{center}
\pagenumbering{roman}
\end{titlepage}
\blankpage
% % % % % % % % % % % % Parents

\thispagestyle{plain}
\begin{center}
\topskip0pt
\vspace*{\fill}
\Large{\bfseries\LARGE\textit{{Dedicated to my parents}}}
\vspace*{\fill}
\end{center}
\cleardoublepage
\thispagestyle{plain}
\blankpage
% % % % % % % % % % % % Declearation

\chapter*{Declaration}\addcontentsline{toc}{section}{\normalsize
\textbf{Declaration}}
I declare that this written submission represents my ideas in my own words and where others' ideas have been included, I have adequately cited and referenced the original sources. I also declare that I have adhered to all principles of academic honesty and integrity and have not misrepresented or fabricated or falsified any idea/data/fact/source in my submission. I understand that violation of the above will be cause for disciplinary action by the Institute and can also evoke penal action from the sources which have thus not been properly cited or from whom proper permission has not been taken when needed.\\[3cm]

Date: \hspace{10.8cm} \textbf{Abhishek Shukla}\\
\vspace{2cm}
\hspace{12.4cm} Roll No.- 20093040
\cleardoublepage
\blankpage
% % % % % % % % % % certificate
\chapter*{Certificate}\addcontentsline{toc}{section}{\normalsize \textbf{Certificate}}
Certified that the work incorporated in the thesis entitled \textit{\textcolor{blue} {"Ancilla Assisted quantum Information Processing: General protocols and NMR implementations"}}, submitted by \textit{Abhishek Shukla} was carried out by the candidate, under my supervision. The work presented here or any part of it has not been included in any other thesis submitted previously for the award of any degree or diploma from any other University or institution.\\[3cm]
\begin{flushright}
Dr. T. S. Mahesh\\
\end{flushright}
Date:
\begin{flushright}
(Supervisor)
\end{flushright}
\cleardoublepage
\blankpage
% % % % % %

% % % % % % % % % % % Acknow

\addcontentsline{toc}{section}{\normalsize
\textbf{Acknowledgement}}
%\chapter*{Acknowledgement}\addcontentsline{toc}{section}
\thispagestyle{plain}   
\begin{center}
\textbf{\huge Acknowledgement}
\end{center}
\vspace{1.5cm}

I must admit that this thesis would have not been possible without the help and support of many
persons whom I met while pursuing PhD at IISER Pune.  I have been very privileged
to have so many wonderful friends and collaborators.

First of all, I thank to my research supervisor Dr. T. S. Mahesh for his guidance and
encouragement.  I am grateful to him for teaching me nuts and bolts of Quantum Information Processing and NMR.  His deep understanding about the subject and generous behaviour played the key role behind my research. 
I consider myself fortunate to have a supervisor like him.  I pay my gratitude to him for his support and
affection.  

I am thankful to Prof. Anil Kumar for various insightful discussions and his concern about
my research.  His immense knowledge of the field and positive attitude has always been a source of
inspiration for me.  I thank Prof. A. K. Rajagopal, Prof. Ushadevi, and H. S. Karthik for our collaborative work on 
moment inversion, which enhanced my curiosity for foundational issues in quantum physics. It was nice to work
with Hemant and Manvendra as collaborator.  I learned some good things from them.  I also thanks to my other
collaborators Sharad and Gaurav.

I thank my RAC members- Dr. Arijit Bhattacharyay, Dr. T. G. Ajitkumar, and Dr. K.
Gopalakrishnan for their support through their comments and suggestions during annual evaluation of progress in my research.  I also thank Dr. R. G. Bhat, Dr. V. G. Anand, and Dr.
H. N. Gopi, for their help and affection.  I am thankful to our Director Prof. K. N. Ganesh for
providing all the necessary experimental facilities in the lab.  I would like to thank
IISER-Pune for the graduate scholarships that I received during my PhD.  I also acknowledge administrative staff, security service people, and housekeeping persons of IISER-Pune.  My academic life may have not taken shape in this way without 
my M.Sc. teacher Dr. Aanand Kumar.  I am thankful to him for inciting my interest in Quantum Physics and also for his moral 
support.  The list is incomplete without thanking my all other teachers.

I thank all the members of NMR Research Center -past and present- with whom I
have worked.  It was a pleasure working with my senior Soumya, his always
positive attitude was really amazing.  I can't forget relay of arguments between me and Swathi on various topics.
 I also thank her for giving me company at tea time for number of times in Innovation park canteen.
It was really nice to work with Hemant and Koteswar (Sir Ji), Hemant's great attention for his work was always admirable.
 It was nice to listen Koteswar for basics of NMR.
 A person is always enjoys benefit if he has such a perfection loving friend.  Optimizing the diffusion parameters with Manvendra
 was really enjoyable.  I thank Pooja didi for teaching me how to operate NMR spectrometers.
 It was wonderful time in NMR lab at Innovation park with all group members.  I enjoyed discussions 
 with Ravi Shanker, Bhargava, Nitesh, Sudheer, along with the most young members Deepali, Deepak, Anjusha, and other project students.  At last, I would like to pay a special thank to Sachin Kate for his easy availability and help on spectrometer issues.
 
 Outside the lab, the life was enjoyable because of many friends and I thank all of them.  First of all I thank all my friends in Physics department, especially friends from 2008 batch
 for their support and company.  Anyone will feel lucky to have friends like them, I will always remember Mayur for his unconditional
 helps to everyone, Arun for his keen analysis, and Murthy for trying to tease me with his famous 'because of you' sentence. I always
 enjoyed Arthur's 'wo kya' approach.  Resmi's keen observation and Kanika's optimistic behavior was always inspiring.  I thank Vimal and Kajari for giving me company while studying in their offices.  I thank all my friends in Chemistry
 department for providing me chemicals.  
 It was awesome to roam, to fight, and to laugh with unforgettable friends Sumit and Abhigyan.  I would like 
 to thank to Ramya and Padma to tighten the knot as my sisters.
 At last I thanks my M.Sc. friends along with my long time friends Praveen, Manu, Rajesh, Neeraj, Anil, Vivek, Sarvesh, and Suchita for their support. 
 
 My research career would have not been possible without the great support of my family. I thank my mother, father, and sister for their love, support and encouragement.  I also like to thank my cousin sisters and brothers for replacing my unavailability to support my parents in odd times.  I found myself lucky to have a friend like Rajeev, I thank him for always 
standing with me in any circumstances.  I also thank to all my near and dear for their support and encouragement.
 
\vspace{1cm}
\hspace{10cm}
{\Large\textit{Abhishek Shukla}}
\thispagestyle{plain}   
\blankpage                          % (looks ok)

% % % % % % % % % % % % LOPub
\fancyhf{}
\thispagestyle{plain}
\addcontentsline{toc}{section}{\normalsize
\textbf{List of Publications}}
\vspace{2.5cm}
\textbf{\huge List of Publications}
\vspace{0.2cm}
\begin{enumerate}
\item
Soumya Singha Roy, \textbf{Abhishek Shukla}, and T. S. Mahesh,
\textit{NMR implementation of Quantum Delayed-Choice Experiment,}
Phys. Rev. A 85, 022109 (2012).

\item
Hemant Katiyar, \textbf{Abhishek Shukla}, Rama Koteswara Rao, and T. S. Mahesh,
\textit{Violation of Entropic Leggett-Garg Inequality in Nuclear Spin Ensembles,}
Phys. Rev. A 87, 052102 (2013).

\item
H. S. Karthik, Hemant Katiyar, \textbf{Abhishek Shukla}, T. S. Mahesh, A. R. Usha Devi, A. K. Rajagopal,
\textit{Inversion of moments to retrieve joint probabilities in quantum sequential measurements,}
Phys. Rev. A 87, 052118 (2013).

\item
\textbf{Abhishek Shukla}, K. Rama Koteswara Rao, and T. S. Mahesh,
\textit{Ancilla Assisted Quantum State Tomography in Many-Qubit Registers,}
Phys. Rev. A 87, 062317 (2013).

\item
\textbf{Abhishek Shukla}, Manvendra Sharma, and T. S. Mahesh,
\textit{NOON states in star-topology spin-systems: Applications in diffusion studies and RF inhomogeneity mapping}
Chem. Phys. Lett. 592, 227 (2014).

\item
Sharad Joshi, \textbf{Abhishek Shukla}, Hemant Katiyar, Anirban Hazra, and T. S. Mahesh,
\textit{Estimating Franck-Condon factors using an NMR quantum processor,}
Phys. Rev. A 90, 022303 (2014).

\item
\textbf{Abhishek Shukla} and T. S. Mahesh,
\textit{Single-Scan Quantum Process Tomography,}
Phys. Rev. A 90, 052301(2014).

\item
\textbf{Abhishek Shukla} and T. S. Mahesh,
\textit{Dynamical Decoupling of Spin-Clusters using Solid State NMR,}
arXiv: quant-ph:1110.1473.

\item
Gaurav Bhole, \textbf{Abhishek Shukla}, T. S. Mahesh, 
\textit{Benford distributions in NMR,}
arXiv:1406.7077 [physics.data-an].

\item
Gaurav Bhole, \textbf{Abhishek Shukla}, T. S. Mahesh, 
\textit{Benford Analysis: A useful paradigm for spectroscopic analysis,} 
Chem. Phys. Lett. 639, 36 (2015).

\item
T. S. Mahesh, \textbf{Abhishek Shukla}, Swathi S. Hegde, C. S. Sudheer Kumar, Hemant Katiyar, Sharad Joshi, K. R. Koteswara Rao,
\textit{Ancilla assisted measurements on quantum ensembles: General protocols and applications in NMR quantum information processing,
} Current Science 109, 1987 (2015).
\end{enumerate}

\pagenumbering{roman}
\thispagestyle{plain}
\blankpage
% % % % % % % % % % % % TOC

\fancyhf{}
\fancyhead[RO]{\nouppercase{\emph{\rightmark}}}
\fancyhead[LE]{\nouppercase{\emph{\leftmark}}}
\fancyfoot[C]{\thepage}
\setlength{\headheight}{0.4in}
\pagestyle{fancy}
\tableofcontents{\normalsize}
\cleardoublepage
% % % % % %
\blankpage
%% % % % % % % % % % % % % LOF LOT
%%\fancyhf{}
%\thispagestyle{plain}
%\addcontentsline{toc}{section}{\normalsize \textbf{List of Figures}}
%\listoffigures    %(I have to see how to add list of figures)
%\nopagebreak
%\addcontentsline{toc}{section}{\normalsize \textbf{List of Tables}}\
%\listoftables
%\clearpage{\pagestyle{empty}\cleardoublepage} 
%% % % % % %
%\blankpage

%\addcontentsline{toc}{section}{\normalsize \textbf{Abstract}}
\addcontentsline{toc}{section}{\normalsize
\textbf{Abstract}}
\thispagestyle{plain}
\pagenumbering{roman}
\nopagebreak
\begin{center}
\textbf{\Huge Abstract}
\end{center}
\vspace{1.5cm}

While a \textit{bit} is the fundamental unit of binary classical information, a \textit{qubit} is a fundamental unit of quantum information.  In quantum information processing (QIP), it is customary to call the qubits under study as system qubits, and the additional qubits as ancillary qubits.  In this thesis, I describe various schemes to exploit the ancillary qubits to efficiently perform many QIP tasks and their experimental demonstrations in nuclear magnetic resonancee (NMR) systems.  Particularly, we have showed that, in the presence of sufficient ancillary qubits, it is possible to completely characterize a general quantum state as well as a general quantum dynamics in a single measurement.  In addition, it is also possible to exploit ancillary qubits for realizing noninvasive quantum measurements required for several experiments related to quantum physics.  Finally, I will also illustrate some interesting applications of ancillary qubits in spectroscopy.  The abstracts of individual chapters are given below.

Chapter \ref{chp1} is the introduction to this thesis.  Here I describe about classical/quantum information, quantum information processing (QIP), nuclear magnetic resonance (NMR), NMR-QIP, and finally ancilla-assisted QIP.

The standard method of Quantum State Tomography (QST) relies on the
measurement of a set of noncommuting observables, realized in 
a series of independent experiments. Ancilla Assisted QST (AAQST) 
greatly reduces the number of independent
measurements by exploiting an ancilla register in a known initial state.
In suitable conditions AAQST allows mapping out 
density matrix of an input register in a single experiment.
 In \textbf{chapter} \ref{chp2}, I describe methods for explicit construction of 
AAQST experiments in multi-qubit registers.  I also report NMR implementations of
AAQST on certain qubit-systems and the experimental results confirm the
effectiveness of AAQST in such many-qubit registers.

In \textbf{chapter} \ref{chp3}, I present a procedure to characterize a general quantum
process in a single ensemble measurement.  The standard procedure for quantum process tomography (QPT) requires
a series of experiments.  Each experiment involves initialization of the system to a particular
basis state, applying the quantum process $\varepsilon$ on the system, and finally characterizing
the output state by quantum state tomography (QST).  The output states
collected for a complete set of basis states enable us to 
calculate the $\chi$ matrix characterizing the process $\varepsilon$.
The standard procedure for QST itself requires independent experiments 
each involving measurement of a set of commuting observables.
 Thus QPT procedure demands a number of independent measurements, and moreover, 
this number increases rapidly with the size of the system.
However in ensemble systems, the total number of independent measurements can be greatly reduced 
with the availability of ancilla qubits.
 Here we combine AAPT with AAQST to realize a `single-scan QPT' (SSPT), a procedure to characterize a general quantum
process in a single ensemble measurement.
 We demonstrate experimental SSPT by characterizing several single-qubit processes using
a three-qubit NMR quantum register. Furthermore, using the SSPT procedure
we experimentally characterize the twirling process and compare the results
with theory.

The measurement as described in quantum mechanics is in general invasive.  An invasive measurement may affect subsequent dynamics of the quantum system.  In \textbf{chapter} \ref{chp4}, I report use of ancilla assisted noninvasive measurement to study following two problems.  In section \ref{41}, I describe violation of entropic Leggett-Garg inequality in nuclear spin ensembles.
 Entropic Leggett-Garg inequality (ELGI) places a bound on the statistical measurement outcomes of dynamical observables describing a macrorealistic system \cite{elgiUshadevi}.  Such a bound is not necessarily obeyed by quantum systems and therefore provides an important way to distinguish quantumness from classical behaviour.  We studied ELGI using a two-qubit NMR system and the experimental results showed a clear violation of ELGI by over four standard deviations.
 In section \ref{42}, I describe our experiments on retrieving joint probabilities by inversion of moments. 
  Further, we studied sequential measurements of a single quantum system and investigated their moments and joint probabilities \cite{MI} and demonstrated that the moments and the probabilities are inconsistent with each other. 

The NOON state is a special multiple-quantum coherence that can
be prepared easily using a star-topology spin-system. 
In \textbf{chapter} \ref{chp5}, I describe two important application of such
systems: (i) measuring translational diffusion constants in liquids 
and (ii) quantitative characterization of radio-frequency (RF)
inhomogeneity of NMR probes.
 When compared with the standard single quantum method,
the NOON state method requires shorter diffusion delays or weaker
pulsed-field-gradients.  Similarly, Torrey oscillations with NOON states decay at a faster rate than that of single
quantum coherences and allow accurate characterization of RF inhomogeneity at higher RF powers.  

\textbf{chapter} \ref{chp6}, contains an experimental study of the efficiency of various dynamical decoupling sequences for suppressing decoherence of single as well as multiple quantum coherences (MQC) on large spin-clusters.   The system involved crystallites of a powdered sample containing a large number of molecular protons interacting via long-range inter molecular dipole-dipole interaction.  We invoked single as well as MQC using this interaction followed by an application of various DD sequences namely CPMG, UDD, and RUDD.  The experiments reveal superior performance of RUDD sequences in suppressing decoherence.  We have also analysed performances of CPMG, UDD, and RUDD sequences used in our experimental study via filter function analysis.  The analysis confirms superior performance of RUDD and hence supports our experimental results.

% % % % % % % % % % % % % % Fancypagestyle chapters

\fancyhead[LE,RO]{\nouppercase {\color{black} \bfseries \rightmark}}
\fancyhead[LO,RE]{\nouppercase {\color{black} \bfseries \chaptername{ }\thechapter}}
\fancypagestyle{plain}{\fancyhead{}
\renewcommand{\headrulewidth}{0pt}}
\clearpage{\pagestyle{empty}\cleardoublepage} 
\thispagestyle{plain}
\pagenumbering{arabic}

% % % % % % % % % % % % % % %
\blankpage
% % % % % % chaptertitle formating
\titlespacing*{\chapter}{0pt}{-50pt}{20pt}
\titleformat{\chapter}[display]{\normalfont\Large\bfseries}{\chaptertitlename\ \thechapter}{20pt}{\Large}
 \chapter{Introduction \label{chp1}}
 %\vspace{1cm}
 The term information refers to the amount of knowledge contained either in a message or observation.  Depending on the nature of states used for encoding and manipulating the information it can be classified as classical information and quantum information.

 \section{\textbf{Classical Information}}
  The digital information processing relies on encoding information by a set of discrete values.  The smallest unit of information in the binary system is known as a $bit$,
  which is a mathematical object encoding two states of a computational device.  Although information can be quantified in various other units such as byte, nat, trit, decimal etc, the bit being the simplest one, is the most popular unit. 
   Claude Shannon, the founder of information theory, in his seminal paper has proposed a measure for the amount of information contained in a message \cite{shannon2001mathematical}.  Consider a message as a string of random variables, say $(X_{i})$.  We may calculate the amount of information contained in such a message by calculating the uncertainty in values of random variables before we measure them or in other words the amount of information we gain after we know the value of $(X_{i})$ \cite{chuangbook}.  Entropy of this string of random variables is a function of the probabilities of different possible values the random variable takes.  In his classic paper \cite{shannon2001mathematical} Shannon showed that if $p_{1}, p_{2}, \cdots, p_{n}$ are the probabilities of values of random variables $X_1, X_2, \cdots, X_{n}$ consecutively, then entropy (and hence information) associated to such probability distribution must be of the form 
  \begin{equation}
  H(X)= -\sum_{i} p_{i}\log_2{p_{i}}.
  \end{equation}
  For example: Information revealed in single flip of an unbiased coin is one bit, i.e., $H(X) = 1$, whereas in two coin flips it is $H(X) = 2$.
  \subsection{Conditional Entropy and Mutual Information}
%  \textbf{Conditional Entropy and Mutual Information: }
  Consider two random variables X and Y.  The total uncertainty corresponding to the simultaneous values of both variables, is known as the joint entropy $H(X,Y)$ of variables $X$ and $Y$ \cite{shannon2001mathematical}.  It depends on joint probability distribution $p(x,y)$ of outcomes of variable $X$ and $Y$ and can be calculated by the formula 
  \begin{equation}
  H(X,Y) = - \sum_{x,y} p(x,y)\log p(x,y).
  \end{equation}
  This definition of entropy can be extended to any set of random variables.  The uncertainty associated to the values of one random variable say $X$, while knowing the value of the other variable say $Y$, can be calculated by the conditional entropy $H(X \vert Y)$ \cite{shannon2001mathematical}.  Suppose entropy of the known variable is $H(Y)$, then conditional entropy
   \begin{equation} 
   H(X \vert Y) =  H(X,Y) - H(Y).  
  \label{CE}
  \end{equation}
  Mutual information stored in variables $X$ and $Y$ can be calculated by subtracting joint entropy $H(X,Y)$ from the sum of individual entropies $H(X)$ and $H(Y)$ \cite{cover2012elements}, i.e.,
  \begin{equation}
  H(X:Y) = H(X)+ H(Y)- H(X,Y).
 \label{MI1}
 \end{equation}
Using \ref{CE} and \ref{MI1}, the expression for mutual information can be rewritten as
  \begin{equation}
   H(X:Y) = H(X)-  H(X \vert Y).
  \label{MI2}
  \end{equation}  
  
  \section{\textbf{Quantum Information and its processing}}
  Quantum Information Processing (QIP)
  is the branch of information processing in which the resources used are quantum mechanical systems \cite{chuangbook}.  In 1973, A. Holevo
  has proposed an inequality which puts an upper bound over the classical mutual information \cite{holevo1973bounds}.  It infers that the encoding of $n$ bits of classical information requires at least $n$ bits of classical resources \cite{chuangbook}.  However because of the superposition principle, quantum mechanical systems are supposed to have a better encoding efficiency than their classical counterparts.  This sets the motivation to use quantum resources for information processing.  
  
  Here I describe the chronological development of QIP.  In 1973, C. H. Bennett showed that computation can be made logically reversible \cite{bennett1973logical}.  In 1975, R. P. Poplavaskii in his thermodynamic model of information processing" showed the computational infeasibility of simulating quantum systems" \cite{Poplavskii1975}.  In 1981, Paul Beinoff proposed a model for a "non dissipative Turing machine" using quantum mechanical resources \cite{benioff1980quantum,benioff1982quantum}.  In 1976 Polish mathematician Roman Stanislaw Ingarden, in his seminal work proposed a generalised concepts of using quantum systems for information processing \cite{ingarden1976quantum}.  In 1981, Yuri Manin had first proposed the idea of quantum computing \cite{manin} but it was R. Feynmann who actually set the platform.  He observed that simulating quantum systems using classical computers is inefficient.  He presented a model of quantum computers for simulating quantum systems \cite{feynman1982simulating}.  In 1984 C. Bennett and G. Brassard presented {\it cryptographic key distribution} model using Weisner's conjugate coding \cite{bennett1983quantum}. This work opened a new perspective of QIP namely Quantum Communication.  In 1991, A. Eckart invented an entanglement based protocol for secure quantum communication \cite{ekert1991quantum}.  Year 1994 witnessed a milestone in development of QIP when Peter Shor from Bell labs developed an algorithm for efficient solution of (i) factorization problem  (ii) descrete log problems \cite{shorpolynomial}.  Empowered with Shor's algorithm, quantum Computation become able to break many of the present days encryption codes.  In 1994, P. Shor and A. Steane presented first scheme for error correction code \cite{shor1995RDscheme,steane1996error}.  In 1996 Lov Grover from Bell labs came up with a search algorithm \cite{grover1996fast}.  Unlike Shor's algorithm, quantum simulation, which provides an exponential speed up to computation, Grover's algorithm provides only quadratic speed up over existing classical algorithm.  However, the algorithm can be used for a variety of problems, including database search \cite{grover1996fast,grover1997quantum}.  In 1997, D.Simon had invented an oracle problem for which a quantum computer was found exponentially faster than their classical counterparts \cite{simon1997power}.  As an important step in development of QIP, P. Divincenzo proposed a list of minimal requirements for any physical architecture to be able to realize QIP tasks \cite{divincenzoCrieteria97}.  
  
  Divincenzo's criteria includes the following:
\begin{itemize}
\item[1.]
Physical states to realize individually adressible qubits. 
\item[2.]
 Ability to initialize system to any quantum state.
\item[3.]
 Universal set of quantum gates.
\item[4.]
Qubit-specific measurement.
\item[5.]
 Sufficiently long coherence time(Depending on gate time).
\item[~]
     Other than these five points Divincenzo has also proposed two more points which are essential for quantum communication.  These criteria are
\item[6.] Interconvertibility of stationary and flying qubits
\item[7.] Transmit flying qubits between distant locations
\end{itemize}
  Execution of a QIP task can be divided into three parts, namely \textbf{initialization}, \textbf{evolution}, and \textbf{measurement}. 
  These points are described below.
  
 \subsection{Quantum bit}
Analogous to a classical bit, a quantum bit (\textit{qubit}) is a two-dimensional
 mathematical object encoded by a two-level quantum system.  Similar to the two states of a bit, usually represented by $0$ and $1$, the states of a qubit are represented by $\ket{0}$ or  $\ket{1}$.  In contrast, however, any quantum superposition of $\ket{0}$ and $\ket{1}$ is also
 a valid state of the qubit.  Here $\ket{0}$ and $\ket{1}$ forms an orthonormal basis for state space of a qubit and can be used as \textit{computational basis} in QIP.  In this basis, the state of a qubit is in general represented as
 \begin{eqnarray} 
 \ket{\psi} =  \alpha \ket{0} + \beta \ket{1},
 \end{eqnarray} 
 where $\vert \alpha \vert^2 + \vert \beta \vert^2 = 1$. 
%  Here quantities $\vert \alpha \vert^2$ and $\vert \beta \vert^2$ are probabilities of system being in state $\ket{0}$ and $\ket{1}$  respectively. 
 Here quantities $\alpha$ and $\beta$ are the complex numbers and forms a two dimensional complex vector space. In this space
 state, a qubit can be represented by a vector
 $ \left[
 \begin{array}{c}
 \alpha\\
 \beta\\
 \end{array}
 \right]$.
 Consider the situation where we are interested in knowing the state of a classical bit.  The information about the state of a classical bit is inherited in it, measurement just reveals it thus one can determine state with certainty. 
 On the contrary, when a quantum mechanical system, say qubit, initially in state $\ket{\psi}$, is subjected to the measurement process, we get an outcome $\ket{0}$ with probability  $ \vert \alpha \vert^2$ and an outcome $\ket{1}$ with probability  $\vert \beta \vert^2$.  This probabilistic outcome reveals the fact that the state of a system can not be measured by using a single copy in one experiment.  To compute probabilities, we need either simultaneous measurement of a large number of copies of the qubit or a large number of measurements of a single qubit with repeated state preparation.  Above is the description of state of a single quantum mechanical system.  Below we describe the state of a multi-qubit quantum register.
 
 Consider a register of $n$ qubits. The most general state of such a register is given by
 \begin{eqnarray}
\ket{\psi} = \sum_j \alpha_j \ket{\psi_{1j}} \otimes \ket{\psi_{2j}} \otimes \cdots \otimes \ket{\psi_{nj}},
\end{eqnarray}
where $\ket{\psi_{ij}}$ refers to $i$th qubit in $j$th term of the superposition, and $\alpha_j$
are the complex coefficients which together normalize to unity.
We may choose $\ket{\psi_{ij}} \in \{\ket{0},\ket{1}\}$, so that we can represent the combined state $\ket{\psi}$ by a set (product basis) of $2^n$ basis elements.

For example in the case of a two qubit system, the product basis contains four elements.  These four basis elements are  \\
\begin{center}
$
\begin{array}{cccc}
\ket{\phi_1} = \left[
\begin{array}{c}
1 \\
0 \\
\end{array}
\right] \otimes  \left[
\begin{array}{c}
1\\
0\\
\end{array}
\right] = \left[
\begin{array}{c}
1\\
0\\
0\\
0\\
\end{array}
\right];
&
&
&
\ket{\phi_2} = \left[
 \begin{array}{c}
 1 \\
 0 \\
 \end{array}
 \right] \otimes  \left[
 \begin{array}{c}
 0\\
 1\\
 \end{array}
 \right]=\left[
 \begin{array}{c}
 0\\
 1\\
 0\\
 0\\
 \end{array}
 \right];
\\
\\
  \ket{\phi_3} = \left[
 \begin{array}{c}
 0 \\
 1 \\
 \end{array}
 \right] \otimes  \left[
 \begin{array}{c}
 1\\
 0\\
 \end{array}
 \right] = \left[
 \begin{array}{c}
 0\\
 0\\
 1\\
 0\\
 \end{array}
 \right];
 &
 &
 &
\ket{\phi_4} = \left[
  \begin{array}{c}
  0 \\
  1 \\
  \end{array}
  \right] \otimes  \left[
  \begin{array}{c}
  0\\
  1\\
  \end{array}
  \right]=\left[
  \begin{array}{c}
  0\\
  0\\
  0\\
  1\\
  \end{array}
  \right].
  \end{array}
  $
\end{center}

 States of a qubit can be represented geometrically in terms of polar coordinates.  In this representation all pure states lie on the surface of a sphere of unit radius known as Bloch sphere. The most general state of a single qubit can then be written as
 \begin{equation}
 \ket{\psi} = \cos{\frac{\theta}{2}}\ket{0} + e^{i\phi} \sin{\frac{\theta}{2}}\ket{1}
 \end{equation}   
 where $\theta$ and $\phi$ are spherical polar co-ordinates.  Bloch sphere representation is very helpful in visualizing the effects of quantum operations on the qubit.
 \begin{figure}[h]
 \begin{center}
 \includegraphics[clip=true,width=10cm,trim= 2cm 3cm 8cm 6cm]{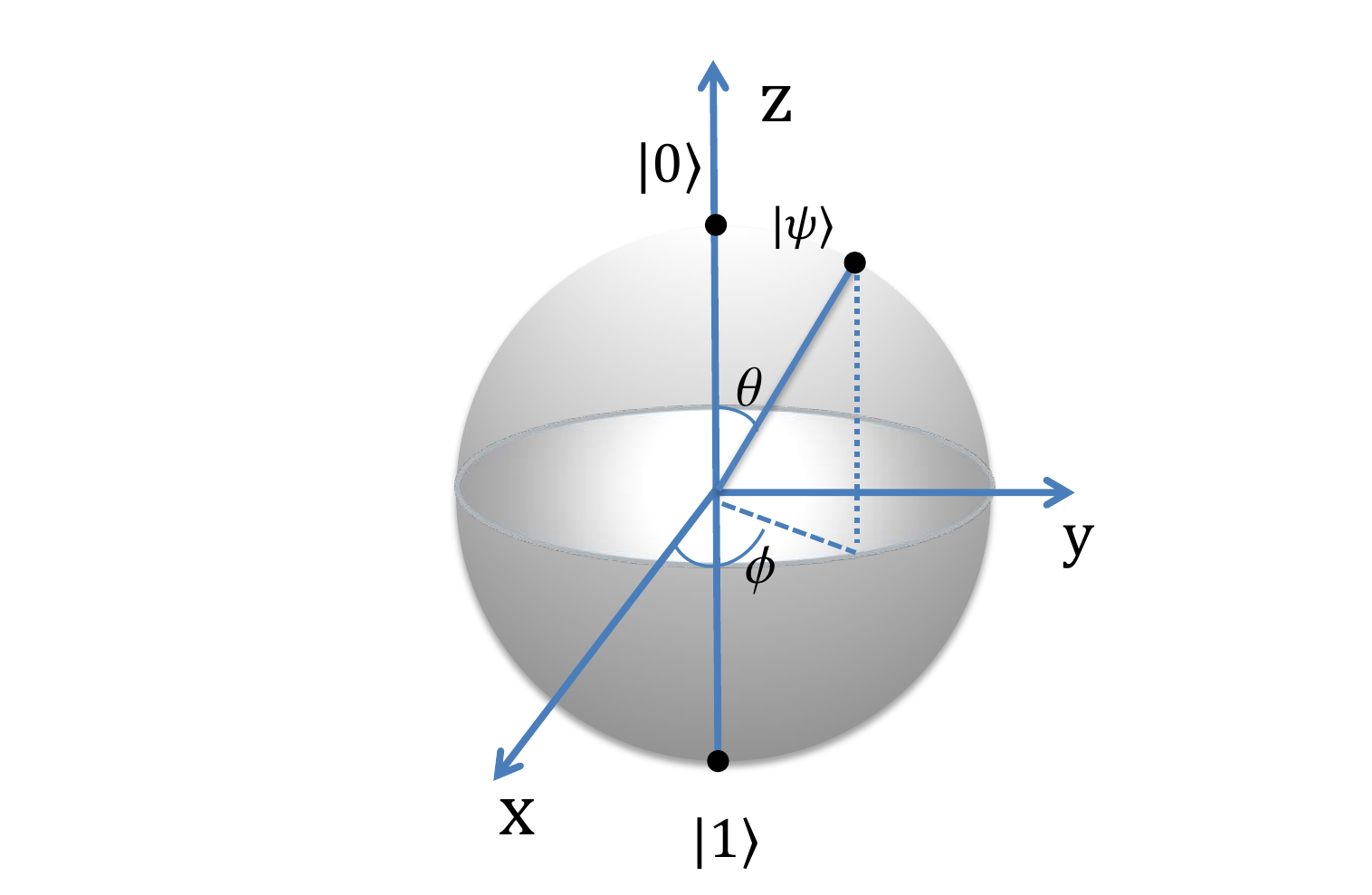}
 \caption{Bloch sphere representation of a single qubit states}
 \end{center}
 \end{figure}

 \subsection{Density Matrix Formulation}  
 \label{dmf}
 A quantum state can also be represented by an operator known as density operator.  
The density matrix formulation is very useful in describing the state of an ensemble quantum system such as in nuclear magnetic resonance (NMR).
 The diagonal terms of the density matrix correspond to the populations of the eigenstates.  The off-diagonal terms represent coherences.  
 
 For an ensemble of states $\ket{\psi_{i}}$ each with probability $p_{i}$ the density operator is $\rho = \sum_{i} p_{i}\outpr{\psi_{i}}{\psi_{i}}$ (here $\sum_{i} p_{i} = 1$).  If all the members of the ensemble are in the same state, we call it a \textit{pure state}.  The density operator corresponding to a pure state $\ket{\psi} = \sum_{j} c_{j}\ket{j}$ is 
 \begin{equation}
  \rho_{pure} = \ket{\psi}\bra{\psi} = \sum_{j} \sum_{k} c_{j}c_{k}^*\ket{j}\bra{k}.
 \end{equation} 
  where $c_{j}$'s are the probability amplitudes in basis ${\ket{j}}$.   
%   Necessary condition for a density operator to be pure is that $trace(\rho^2)$ should be $1$.
  In the case of a \textit{mixed state}, there exists a distribution of sates $\ket{\psi_{i}}$ with respective probabilities $p_{i}$.  In this case the density operator can be written as
 \begin{eqnarray}
 \rho_{mixed} &=& \sum_{i} p_{i} \ket{\psi_{i}}\bra{\psi_{i}} = \sum_{j} \sum_{k} \overline {c_{j}c_{k}^*} \ket{j}\bra{k} \nonumber \\
             &=& \sum_{i} p_{i} \rho_{i},
 \end{eqnarray}
where $\rho_{i}$ is the state of an individual quantum system of ensemble with probabilities $p_{i}$.  A density operator $\rho$ satisfies three important properties
\begin{enumerate}
\item $\rho$ is Hermitian, i.e., $\rho^\dagger = \rho$.
\item $\rho$ is a positive operator, i.e., eigenvalues are all non-negative.
\item $\tr{\rho} = 1$.\\
Other than the above properties, density matrix also satisfies the following properties.  
\item
For a pure state, density operator $\rho$ is idempotent, i.e., $\rho^2 = \rho$ so that
$\tr{\rho^2} = \tr{\rho} = 1$, while for a mixed state, $\tr{\rho^2} < 1$.
\end{enumerate}

In terms of Bloch sphere, the most general state of a single qubit can be written as
\begin{equation}
\rho = \frac{\mathbb I + \vec{r} \cdot \vec{\sigma}}{2}
\end{equation}
where $\vec{\sigma} = \sum_i \vec{\sigma}_{i}$ is the Pauli vector operator and $\vec{r}$ is a three dimensional Bloch vector s.t. $\vert \vert \vec{r} \vert \vert \le 1$.

\subsection{Entangled state}
Like various other non classical features, entanglement is also a completely quantum mechanical phenomenon.  An entangled state of a composite quantum system is that which cannot be expressed  in terms of states of its components.  This criteria is known as separability criteria.  Consider an $n$-component composite system. If $\ket{\psi_{1}}$, $\ket{\psi_{2}}$, $\cdots,\ket{\psi_{n}}$ are the states of  the components $A_{1}, A_{2}$, $\cdots, A_{n}$, then an entangled state ${\ket{\psi_{EN}}}$ can not be expressed as $\ket{\psi_{1}} \otimes \ket{\psi_{2}} \otimes \cdots \otimes \ket{\psi_{n}}$.  

Entanglement plays a very crucial role in various protocols of QIP. Non separability criteria of entanglement lies in the very heart of various quantum communication protocols such as quantum cryptography, superdense coding, and quantum teleportation.  An entangled state is called a maximally entangled state if on tracing out one subsystem from the composite system, the rest of the system falls into a maximally mixed state.

Simple examples for maximally entangled states include Bell states of the form
\begin{eqnarray}
\ket{\phi_\pm}= \frac{\ket{00} \pm \ket{11}}{\sqrt{2}}.
\end{eqnarray}
Suppose we measure the state of first qubit, and the outcome is $0$.  This can happen with a probability $\frac{1}{2}$.  Then this measurement outcome infers that the state of the second qubit must also be $0$.  Similarly, if the measurement outcome for the first qubit is $1$, then the outcome of the second qubit must also be $1$.
The other two Bell states are of the form
\begin{eqnarray}
\ket{\psi_\pm}= \frac{\ket{01} \pm \ket{10}}{\sqrt{2}}.
\end{eqnarray}

\subsection{Quantum gates}
In a classical computer, classical information can be manipulated using physical tools known as classical logic gates e.g. NOT, OR, NOR, etc.  Similarly in quantum computation, quantum information can be manipulated using quantum gates which can be realized by unitary operators $U$ ($UU^{\dagger} = \mathbb{I}$).   Thus quantum logic gates are reversible in nature.  An efficient implementation of such a quantum operation can be achieved by decomposing it into one qubit gates (local gates) and two qubit gates (non-local gates).  Local gates together with C-NOT gates (explained below) form universal gates.

Below we describe some important one qubit and two qubit gates. 
\subsubsection{Single qubit gates} 
\textbf{$X$ gates}:  Similar to the classical NOT gate, a quantum NOT gate transforms the state $\ket{0}$ into $\ket{1}$ and vice-versa.  Matrix form of this gate is same as that of the Pauli operator $\sigma_{x}$:
\begin{center}
$ X = \begin{bmatrix}
      0 & & 1      \\    
      1 & & 0          
\end{bmatrix}.$
\end{center}    
Operation of NOT gate can be written as 
\begin{center}
$  \begin{bmatrix} 0 & & 1 \\ 1 & & 0 \end{bmatrix} 
  \begin{bmatrix} 0 \\ 1 \end{bmatrix}
  =
   \begin{bmatrix} 1 \\ 0 \end{bmatrix}.$
\end{center}

\textbf{Hadamard gates}:  The Hadamard gate transforms state $\ket{0}$ into state $\ket{+} = \frac{\ket{0}+\ket{1}}{\sqrt{2}}$ and $\ket{1}$ into state $\ket{-} = \frac{\ket{0}-\ket{1}}{\sqrt{2}}$.  The Hadamard gate is represented by $H$ and the matrix form is 
\begin{center}
$ H = \frac{1}{\sqrt{2}}
\begin{bmatrix}
      1 & & 1          \\
      1 & & -1        
\end{bmatrix}.$
\end{center} 
Since $H^{2} = \textbf{I}$, two consecutive applications of $H$ operator does not change the initial state.   
\textbf{$Z$ gates}:  The $Z$ gate introduces a relative phase of $\pi$ to the state $\ket{1}$.  Matrix form of Z-gate is
\begin{center}
$ Z = \begin{bmatrix}
      1 & & 0          \\
      0 & & -1         
\end{bmatrix}$.
\end{center} 

\textbf{Phase gate}:  The phase gate introduces a relative phase factor 'i' corresponding to a phase of $\frac{\pi}{2}$ to the state $\ket{1}$.  Matrix form of phase gate (denoted by $S$) is
\begin{center}
$ S = \begin{bmatrix}
      1 & & 0          \\
      0 & & i          
\end{bmatrix}$.
\end{center} 
 \subsubsection{Multi qubit gates}
\begin{figure} 
\hspace{-1.2cm}
\centering 
\includegraphics[trim=0.5cm 1cm 3.2cm 4cm, clip=true,width=8cm]{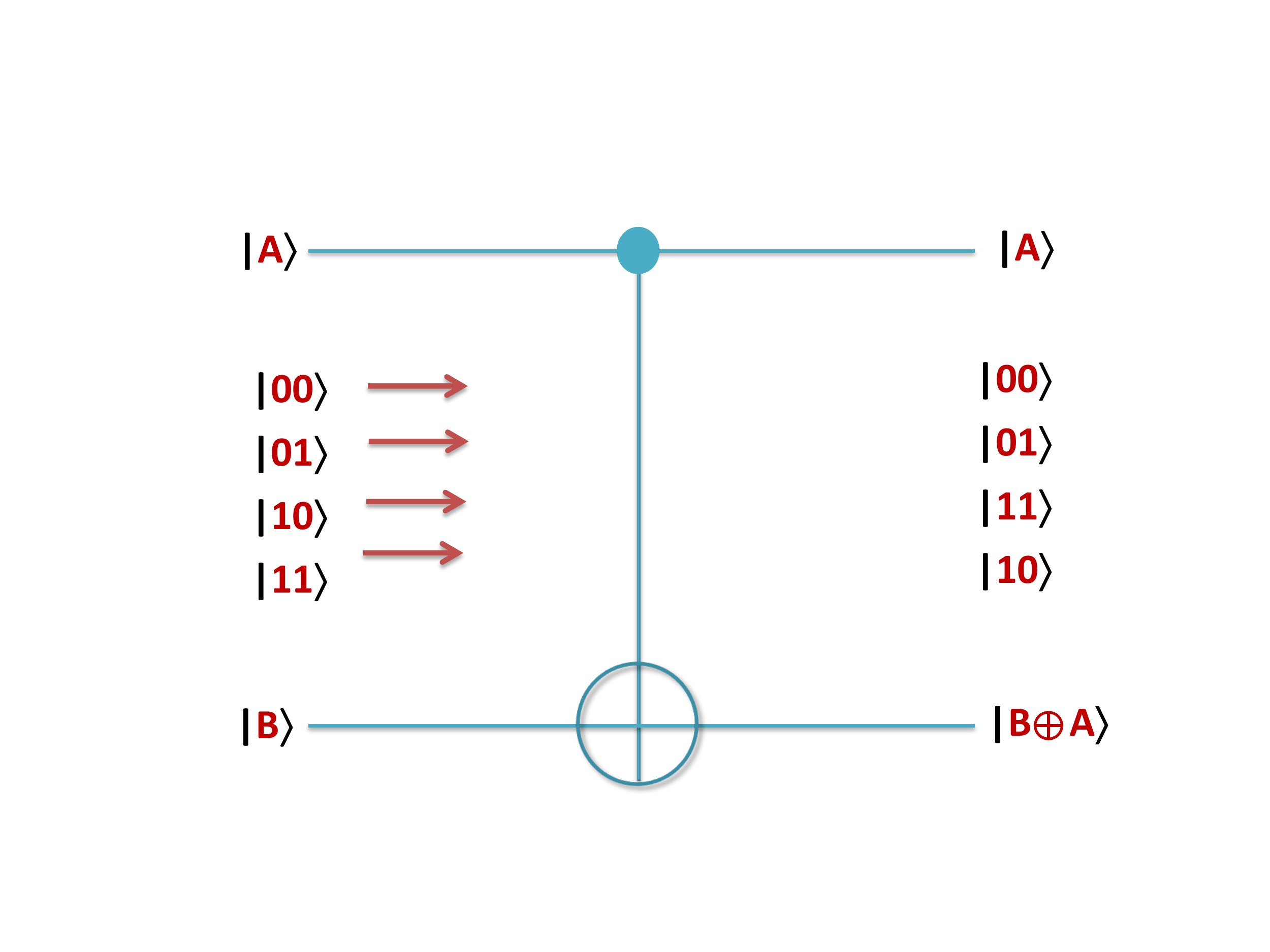}  
\caption{Circuit for the C-NOT gate.} 
\label{CNOT} 
\end{figure}

\textbf{Control-NOT (C-NOT) gates}:  C-NOT is a two qubit gate, in which one qubit works as control while other qubit works as target. Application of the C-NOT gate leads to selective inversion of target qubit w.r.t. $\ket{1}$ state of control qubit.  Quantum circuit and truth table for C-NOT gate are shown in Fig. \ref{CNOT}.
%\begin{centre}

Matrix form of C-NOT gate is the following:
\begin{center}
$ $C-NOT$ = \begin{bmatrix}
      1 & & 0 & & 0 & & 0    \\
      0 & & 1 & & 0 & & 0    \\
      0 & & 0 & & 0 & & 1    \\
      0 & & 0 & & 1 & & 0    
\end{bmatrix}$.
\end{center}   
% Action of $C-NOT$ gate is similar to that of classical $XOR$ gate, so $C-NOT$ can be assumed as generalisation of $XOR$ gate.  Analogous to the classical NAND gate in classical computation, $C-NOT$ and single qubit gate together forms a set of universal quantum gate.
 
 \subsection{Quantum measurement}
  Some important measurement schemes used in quantum information and quantum computation include  (i) projective measurement, (ii) positive operator valued measure (POVM), (iii) weak measurement, and (iv) protective measurement. I shall describe the first two schemes in the following.
 
 Consider a quantum system in the state $\ket{\psi} =\sum_{m} c_{m}\ket{m}$ 
 being projectively measured by an observable $M$ having a spectral decomposition s.t. $M = \sum_{m} m M_m$.  Here $m$ are the eigenvalues and $M_m$ are the corresponding projectors which form a
 complete basis s.t.  $\sum_m M_m = \mathbb{I}$.
 According the measurement postulate of quantum theory \cite{sakurai2011modern,chuangbook,preskill1998lecture},
 the post-measurement state of the system is given by,
 \begin{eqnarray}
 \ket{\psi_{m}} =  \frac{M_{m}\ket{\psi}}{p_{m}}.
 \end{eqnarray}  
Here $p_{m} = \bra{\psi}M^{\dagger}_{m}M_{m}\ket{\psi}$ is the probability of getting the outcome $m$. 

   Often, when knowledge of the post measurement state is of less importance, it is convenient to use the POVM formalism.  
%   Projective measurements on a larger system, act on a subsystem in ways that cannot be described by a Projective valued Measure on the subsystem alone. 
 Here one considers a set of measurement operators $\{M_{m}\}$ which are not necessarily orthogonal. If the measurement is performed on a system with state $\ket{\psi}$, the probability of outcome $m$ is $p_{m} = \bra{\psi}M^{\dagger}_{m}M_{m}\ket{\psi}$.  We define the POVM elements $E_{m} = M^{\dagger}_{m}M_{m}$. Then $\sum_{m}E_{m} = I$ and the probability $p_{m}  = \bra{\psi} E_{m} \ket{\psi}$.  Set $\{E_{m}\}$ is known as POVM \cite{chuangbook}.  POVM has been utilized in studying many of the foundational problems of quantum mechanics.  Furthermore it has also been applied to quantum state tomography \cite{helstrom1976quantum} and quantum cryptography \cite{helstrom1976quantum}.
  
   \subsection{Experimental Architectures}
   Even after a lot of development in theoretical QIP, so far a physical architecture for a universal quantum computer is still a mirage. As described earlier, any physical device must satisfy Divincenzo criteria in order to qualify as a general quantum processor \cite{divincenzoCrieteria97}.  
   Till date there exist no architecture which can full fill all these criteria in one set up. Nevertheless several architecture are being explored for the efficient implementation of various QIP tasks.  Some of these techniques are
  
  \begin{enumerate}
\item    NMR
\item    Nitrogen vacancy centers 
\item    Quantum dots 
\item    Superconducting quantum interface devices
\item    Ion/atom trap 
\item    Linear optics  
  \end{enumerate}
  
 A comparison of the merits and the demerits of various QIP architectures is given below \cite{roy2012nuclear,ExCNotIonTrap}.
%
% % % % % % % % % % % % % % % % % % % % % % %
%\centering
\begin{center}
\begin{table}
\scriptsize
\vspace{0cm}
\begin{tabular}{|c|c|c|c|c|c|c|c|}
%\begin{array}{c|c|c|c|c|c|c|}
\hline
\textbf{Architecture}            &    \textbf{NMR}     &    \textbf{SQUIDS}      &  \textbf{Linear optics}    &  \textbf{NV centre}  &   \textbf{quantum dots} &   \textbf{Trapped Ions} \\
\hline
\textbf{System type}          &   Nuclear spins     &    Flux/charge    &    Photons   &   Defect centres  &  Semiconductor  & Atoms \\
                              &                     &    qubits         &              &                   &                 &        \\ 
\hline
% \textbf{Maximum qubits for QIP} & 12(Entangled state)/SSC spin state(64) &  Fabricated(512)/ Entangled(3)    &   Entangled(10) &  Realized qubits(3)& %Realized qubits(1) & stored(10-10^{3})/Entangled(14) \\

 \textbf{Maximum} & Entangled (12) &  Fabricated (512)    &   Entangled (10) &  Realized (3)& Realized (1) & Stored(10-10$^{3}$) \\
  \textbf{qubits for QIP}         &    Correlated   &  Entangled (3) & & & & Entangled (14)\\
                                  &    in solids  (64) & & & & &                            \\     
%   \textbf{Maximum qubits for QIP} & 12(a)\\b(64) &  c(512)/ d(3)    &   e &  f& g & h/ k(14) \\
 \hline
 \textbf{Control}    &     RF pulses     &     MW pulses,   &    Optical  &  RF, optical &    RF, optical &  MW, Optical,  \\
 
\textbf{technique's} & & Current, Voltage & &  Electrical &  Electrical & RF, Electrical\\
\hline
\textbf{Coherence time}         &     $\sim$  (1-3) sec &    $\sim$  10 $\mu$ s   &     $\sim$ 100 $\mu$ s  &  (1-10) ms  &    (1-10) $\mu$ s  & $\ge$ 1s   \\
                                &     $\sim$ ms & & & & & \\  
\hline
 \textbf{Gate fidelity}          &    (0.99-0.999)      &     $\sim$ (0.90-0.95)   &    $\ge$ 0.95  &       $\sim$ 0.9  &    $\sim$ 0.9  &  (0.99-0.999) \\
\hline
\textbf{Measurement}             &     Bulk  &  flux & Optical    &  Electrical and &  Electrical and & Fluorescence   \\
& Magnetization & Charge & responses& Optical  & Optical  &  Optical  \\
\hline
\end{tabular}
\caption{Comparison of the merits and the demerits of various architectures}. 
\scriptsize
\label{architecture}
\end{table}
\end{center}
\fontsize{12}{15}

% % % % % % % % % % % % % % % % %
\section{\textbf{Nuclear Magnetic Resonance}}
 Story of nuclear magnetism begins from 1922, with one of the most famous experiment of quantum mechanics, The Stern-Gerlach experiment.  In this experiment Stern and Gerlach demonstrated the existence of energy levels in presence of inhomogeneous magnetic field \cite{gerlach1922magnetische}.  Further Fresch, Estermann, and Stern in between 1933-1937 succeeded in measuring proton magnetism.  On the basis of these experiments, understanding of nuclear magnetism is as follows.  Nuclei of most isotopes posses an intrinsic physical property called \textit{spin}.  
% These spins behave like tiny bar magnets \cite{levitt2008spin}.  
All spins have spin angular momentum and hence an associated magnetic moment.  When a spin having a net magnetic moment is placed in an external magnetic field it aligns in the direction of this magnetic field.  This magnetic field lifts degeneracy and splits energy level into multiple levels depending on spin angular momentum number.  For example, In case of a spin $\frac{1}{2}$ nucleus, external magnetic field of  strength $B_{0}$    splits energy level into two sub levels say $\ket{1/2}$ and $\ket{-1/2}$ with energies $-\frac{\gamma B_{0}\hbar}{2}$ and $\frac{\gamma B_{0}\hbar}{2}$.  Such a system is irradiated with an RF radiation.  When the frequency of the RF becomes equal to the frequency corresponding to the energy difference between the two levels, resonance occurs. 
%This magnetic moment due to spins can be tilted by RF field, of intensity less than the Zeeman field, by an order of magnitude. 
This phenomenon is known as nuclear magnetic resonance (NMR) \cite{abragam1998principles}.
% which leads to an absorption spectrum .  
 The coherent oscillations between two levels were first observed by Rabi in 1937 \cite{rabi1939molecular}.  These oscillations lead to an absorption spectrum typically known as a NMR spectrum.  In 1945 Purcell and his co-workers \cite{purcell1945}, and Bloch and co-workers \cite{bloch1946nuclear}, independently observed the first NMR signal from bulk matter.
 
 NMR techniques depend on the state of the sample, and broadly classified into liquid state NMR, liquid crystal NMR, and solid state NMR.  In liquid state NMR, the solute containing the system of interest, is dissolved in a sui solvent to prepare an isotropic solution.  In such samples, molecular tumbling motions average out anisotropic and dipole-dipole interactions and lead to a simple spectrum but at the cost of certain information on spin interactions.  In solid state NMR, because of the closely spaced and spatially oriented spin-systems, strong anisotropic spin interactions arise including the dipolar interactions.  Although in this case one obtains a broad spectrum, by suitably modulating the anisotropic interactions, it is possible to extract  average spatial information about the spin system.   Various techniques for modulating the interactions have been developed over the years \cite{haberlen}. 
  
 A spin system partially oriented in a liquid crystal has restricted motion and hence possesses reduced anisotropic interactions, while still leading to highly resolved spectral lines.  Availability of dipolar coupling and high resolution makes such partially oriented spin-systems an attractive candidate for studying quantum information.  Firstly, such systems display only orientational order, but no spatial order, and hence form well-defined qubit-systems with only intramolecular interactions.  Secondly, the longer range of the intramolecular dipolar couplings may allow realizing larger number of qubits.  Thirdly, the residual dipolar interactions are much stronger than the scalar interactions (which are the only spin-spin interactions surviving in liquids) and allow synthesis of faster quantum gates. 
\subsection{NMR Interactions} 
 The NMR interactions in general can be described by second order tensors. In the following I shall describe different types of interactions.

 \subsubsection{Zeeman interaction}
 Consider a spin$-I$ ($I$ is the spin-quantum number which can take integral or half-integral values) nucleus with an associated magnetic moment $\boldsymbol{\mu} = \gamma \hbar \mathbf{I}$, where $\gamma$ is the gyromagnetic ratio, $\hbar$ is the reduced Planck constant ($h/2\pi$), and $\mathbf{I}$ is the spin operator.  
 If the spin is placed in an external magnetic field $B_{0}$ applied along the $\hat{z}$ direction, the Zeeman interaction is described by the Hamiltonian
 \begin{eqnarray}
  H_{\mathrm{Z}} = -\gamma \hbar B_{0} I_{z},
 \end{eqnarray}
 where $I_z$ is the $z$ component of the spin operator $\mathbf{I}$ \cite{cavanagh}.
% here $\omega_{\_{0}} = -\gamma B_{\mathrm{0}}$ is known as Larmor frequency. 
 The $(2I+1)$ eigenvalues of the Zeeman Hamiltonian are $E_{\mathrm m} = -\gamma \hbar B_{\mathrm{0}}m $, where $m = -I,-I+1,\cdots,I-1,I$.
 The energy gap between the successive levels is $\Delta E = \hbar \omega_0$, where $\omega_0 = -\gamma B_0$ is known as the Larmor frequency. \\

 \noindent \textit{Thermal Equilibrium State:} \\
 \indent At thermal equilibrium the populations of energy levels are governed by Boltzmann statistics.  The state of the system at temperature $T$ and magnetic field $B_{\mathrm {0}}$ is described 
 by the density operator
 \begin{eqnarray}
 \rho = \frac{{e}^{-H_{\mathrm{Z}}/k_{\mathrm {B}}T}}{Z}
 \end{eqnarray}
 where $Z$ is the partition function, $k_{\mathrm {B}}$ is the Boltzmann constant, and $T$ is the absolute temperature.  For an ensemble of spin-$1/2$ nuclei, the ratio of populations
 $p_0$ and $p_1$ respectively corresponding to states $\ket{0} \equiv \ket{1/2}$ and $\ket{1} \equiv \ket{-1/2}$ is
 \begin{eqnarray}
 \frac{p_{0}}{p_{1}} = 1 + \frac{\hbar \omega_{0}}{k_{\mathrm {B}}T}
 \end{eqnarray}
under high temperature and low field approximation.
 For a Zeeman-field of 10 T at room temperature quantity $\frac{\hbar \omega_{0}}{k_{\mathrm {B}}T} \sim 10^{-5}$.  Thus population of the ground state is slightly more than the excited state.  This slight imbalance in populations leads to a net magnetization in $\hat{z}$ direction.  The nuclear magnetization for an ensemble of $n$ spin-1/2 nuclei at thermal equilibrium
 is given by
 \begin{eqnarray}
 M_{\mathrm {0}} = \frac{\mu_0  \gamma^2 \hbar^2 B_{\mathrm {0}}}{4k_{\mathrm {B}}T}
 \end{eqnarray} 
where $\mu_0$ is the permeability of free-space \cite{levitt2001spin}.

% \hspace{3cm}
 \begin{table}[h]
 \begin{center}
 $
 \begin{array}{|c|c|c|c|c|}
%% \begin{array}{c|c|c|c|c|c|c|}
 \hline
  \mathrm{Nucleus} &  \mathrm{Abundance} & \mathrm{Spin} & \gamma ~(10^6 ~\mathrm{rad/T/s}) & 
  \mathrm{Absolute~sensitivity} \\ 
  \hline
   ^{1}H     &     \sim 100\%   &     1/2   &   267.522   &    1   \\
  \hline
   ^{13}C    &     \sim 1.1\%   &     1/2  &   67.283   &     1.76^{*}10^{-4} \\
  \hline
  ^{15}N    &     \sim 0.4\%   &     1/2  &   -27.12  &      3.85^{*}10^{-6} \\
  \hline
  ^{19}F    &     \sim 100\%   &     1/2  &   251.815  &     0.83      \\
  \hline
  ^{31}P    &     \sim 100\%   &     1/2  &  10.394   &      6.63^{*}10^{-2} \\
  \hline
  \end{array}
  $
  \caption{Comparison of properties of some spin-1/2 nuclei}
 \end{center} 
  \end{table}

 \subsubsection{Chemical shift interaction}
 For a nucleus in a molecule, the local magnetic field is different from the external magnetic field due to the induced field by the nearby diamagnetic electrons.  Since the density of the electron cloud depends on the chemical environment, so is the local field.  
 This local field is given by  
 \begin{eqnarray}
 B_{\mathrm{loc}} =  B_{\mathrm{0}}(1-\sigma_{\mathrm{0}}),
 \end{eqnarray}
  \cite{cavanagh}, where $\sigma_{\mathrm{0}}$ is known as the \textit{chemical shift tensor}. The induced field is generally small compared to $B_0$, but is sufficient to create a measurable shift in Larmor frequency.  This shift, called \textit{Chemical Shift}, is an important tool in structural analysis.  
 
 \subsubsection{Dipole-dipole interaction} 
A pair of nearby nuclei can exhibit dipole-dipole (DD) interaction by inducing local fields at the site of each other.  Such a through-space interaction leads to what is known as the \textit{direct dipolar coupling}.  The Hamiltonian for the DD interaction between two nuclei $j$
and $k$ is given by \cite{haberlen}
 \begin{eqnarray}
 H_{\mathrm {DD}}^{jk} = \mathbf{I}^{j} \cdot \mathbf{D}^{jk} \cdot \mathbf{I}^{k}.
 \end{eqnarray}
 Here $\mathbf{D}^{jk}$ is a 2nd rank tensor and $\mathbf{I}^{j}$, $\mathbf{I}^{k}$ are 1st rank spin tensors.  On expressing $\mathbf{I}^{j}$ and $\mathbf{I}^{k}$ in terms of their z components ($\boldmath{I^{j}_{z}}$, $\boldmath{I^{k}_{z}}$), raising spin operators ($\boldmath{I^{j}_{+}}$, $\boldmath{I^{k}_{+}}$), and lowering spin operators ($\boldmath{I^{j}_{-}}$, $\boldmath{I^{k}_{-}}$), the dipolar Hamiltonian can be written as \cite{slitcher}.
 \begin{eqnarray}
 H^{jk}_{\mathrm {DD}}  =  b^{jk}(A + B + C + D + E + F),
 \end{eqnarray}
 where,
 \begin{eqnarray}
 A &=& I^{j}_{z}I^{k}_{z}(1-3\cos^{2}{\theta})\\ \nonumber
 B &=& -\frac{1}{4}(I^{j}_{+}I^{k}_{-}+I^{j}_{-}I^{k}_{+})(1-3\cos^{2}{\theta})\\ \nonumber
 C &=& -\frac{3}{2}(I^{j}_{+}I^{k}_{z}+I^{j}_{z}I^{k}_{+})\sin{\theta}\cos{\theta}e^{-i \phi}\\ \nonumber
 D &=& -\frac{3}{2}(I^{j}_{-}I^{k}_{z}+I^{j}_{z}I^{k}_{-})\sin{\theta}\cos{\theta}e^{i \phi}\\  \nonumber
 E &=& -\frac{3}{4}(I^{j}_{+}I^{k}_{+})\sin^{2}{\theta}e^{-2i\phi}\\  \nonumber
 F &=& -\frac{3}{4}(I^{j}_{-}I^{k}_{-})\sin^{2}{\theta}e^{2i\phi},
 \end{eqnarray}
with $\theta$ being the angle between $\vec{r}^{jk}$ (radius vector from spin $j$ to $k$) 
and $\hat{z}$, and
 $b^{jk} = -\frac{\mu_{0} \gamma^{j}\gamma^{k} \hbar}{4\pi({r^{jk}})^{3}}$.  Under the secular approximation, the terms $C$, $D$, $E$, and $F$ can be discarded and the truncated Hamiltonian becomes
  \begin{eqnarray}
 H^{jk}_{\mathrm {DD}} = d^{jk}(3I^{j}_{z}I^{k}_{z}-  \mathbf{I}^{j} \cdot \mathbf{I}^{k}),
 \end{eqnarray}
where $d^{jk} = \frac{b^{jk}}{2}(3 \cos^{2}{\theta} - 1)$. 
In weakly-dipolar-coupled systems (wherein the dipolar coupling is much smaller than the corresponding chemical shift difference) or in heteronuclear spin systems, the dipolar coupling Hamiltonian can be further approximated to
 \begin{eqnarray}
 H^{jk}_{\mathrm {DD}} = 2d^{jk}I^{j}_{z}I^{k}_{z}
 \end{eqnarray}
 \cite{slitcher}.  
 
 In an \textit{isotropic liquid}, molecules execute fast tumbling motion thus averaging the intramolecular dipolar coupling to zero \cite{levitt2001spin}, i.e.,
 \begin{eqnarray}
 \int\limits_{0}^{\pi} (3\cos^{2}\theta - 1) \sin{\theta} d \theta = 0.
 \end{eqnarray}
 In an \textit{anisotropic liquid}, partial alignment of molecules leads to an incomplete averaging of dipolar coupling.  Thus in this case the residual dipolar coupling constant amounts to $d_{jk} = b_{jk}S$, where the average quantity $S = \overline{(3 \cos^{2}\theta - 1)/2}$ is the order parameter of the oriented spin-systems.

 \subsubsection{Quadrupolar interaction} 
 Nuclei with spin $s > 1/2$ posses an asymmetric charge distribution due to nucleons.  Such nuclei posses quadrupolar charge distribution and hence an associated electric quadrupolar moment.  The electric quadrupolar moment interacts strongly with surrounding electric field gradients.  If $\theta$ is the orientation of atomic framework w.r.t. external magnetic field direction $\hat{z}$, the quadrupolar Hamiltonian for corresponding nucleus is \cite{levitt2001spin} 
 \begin{eqnarray}
 H_{Q}(\theta) = \frac{eQ}{2I (2I-1)} \mathbf{I} \cdot \mathbf{V}(\theta) \cdot \mathbf{I}.
 \end{eqnarray}
 Here $Q$ is the quadrupolar moment, $I$ is the spin quantum number, $\mathbf{I}$ is the nuclear spin tensor, and $\mathbf{V}(\theta)$ is the electric field gradient tensor at the cite of nucleus.  Usually strength of the quadrupolar interaction varies from a few kHz to hundreds of MHz.  If the interaction is weak compared to the Zeeman interaction, we may discard higher order terms, but more often quadrupolar interaction is strong enough.  So we need to keep higher order terms, mostly till second order.  The full Hamiltonian is 
 \begin{eqnarray}
  H_{\mathrm Q} =  H^{(1)}_{\mathrm Q} + H^{(2)}_{\mathrm Q}.
  \end{eqnarray}
  Here first order coupling constant $H^{(1)}_{\mathrm Q}$ is
  \begin{eqnarray}
 H^{(1)}_{\mathrm Q} = \omega^{(1)}_{\mathrm Q} \frac{1}{6}\left\{3I^2_z - I(I+1)  \mathbb{I}\right\}, 
 \end{eqnarray}
where $\omega^{(1)}_{\mathrm Q}$ = $\frac{3e Q \overline{V}_{zz}}{2I(2I-1)\hbar}$.
 
 The second order term $H^{(2)}_{\mathrm Q}$ is more complex, however, the second order coupling is such that $\omega^{(2)}_{\mathrm Q} \sim  \left\vert \frac{\omega^{(1)}_Q}{\omega_{0}} \right\vert$,
 where the denominator is the Larmor frequency. 
 
 \subsubsection{Radio Frequency (RF) interaction}
In NMR, the spin manipulations are realized with the help of RF pulses. Consider
the magnetic component of an RF field, $2B_{1} \cos(\omega_{\mathrm{RF}}t + \phi) \hat{x}$.  The corresponding 
Hamiltonian is
 \begin{eqnarray}
 H_{\mathrm{RF}} = -\boldsymbol{\mu} \cdot 2{B_{1}}(t)\hat{x} = 2 \omega_{1} \hbar I_{x} \cos(\omega_{\mathrm{RF}}t + \phi).
 \end{eqnarray}
 Here $\omega_{1} = - \gamma B_{1}$ is the effective RF amplitude, $\omega_{\mathrm{RF}}$ is the RF frequency, and $\phi$ is the RF phase.  When $\omega_{\mathrm{RF}}$ = $\omega_{0}$ we achieve resonance condition.  Difference between the two frequencies is known as offset.  
  
 \textit{Rotating frame}:
The linearly polarized RF field $2B_{1} \cos(\omega_{\mathrm{RF}}t + \phi) \hat{x}$ can be decomposed into right ($\mathbf{B_{r}}$) and left ($\mathbf{B_{l}}$) circularly polarized RF fields 
 \begin{eqnarray}
\mathbf{B_{r}} &=& B_{1}\left\{\cos(\omega_{\mathrm{RF}}t+ \phi)\hat{x} + \sin(\omega_{\mathrm{RF}}t+ \phi)\hat{y}\right\}, ~~ \mathrm{ and} \nonumber \\
\mathbf{B_{l}} &=& B_{1}\left\{\cos(\omega_{\mathrm{RF}}t+ \phi)\hat{x} - \sin(\omega_{\mathrm{RF}}t+ \phi)\hat{y}\right\}. 
 \end{eqnarray}
If we choose $\omega_{\mathrm{RF}} = \omega_0$,
the first component ($\mathbf{B_{r}}$) precesses about the $\hat{z}$ axis
in the same sense as the nuclear spin, realizing resonance condition.  
The other RF component precesses in opposite direction (equivalent to frequency $-\omega_{\mathrm{RF}}$) and hence will be out of resonance.  The latter component
is ignored under high-field conditions, but causes Bloch-Siegert shift 
as RF strength becomes comparable to the Larmor frequency \cite{bloch1940}.
Now we choose a frame, also rotating about $\hat{z}$ axis with frequency $\omega_{\mathrm{RF}}$.  In this frame the RF component $\mathbf{B_{r}}$ appears
static, 
\begin{eqnarray}
\mathbf{B_{r}}' = B_{1}\left\{\cos\phi\hat{x} + \sin\phi\hat{y}\right\}.
\end{eqnarray}
In the off-resonant case, with offset $\Omega = \omega_0 - \omega_{\mathrm{RF}}$,
the effective field in the rotating frame is the vector sum of the RF in the transverse
direction and the offset in the longitudinal direction,
\begin{eqnarray}
\mathbf{B_{\mathrm{eff}}} = -\frac{\Omega}{\gamma}\hat{z}
- \frac{\omega_1}{\gamma}\left\{\cos\phi\hat{x} + \sin\phi\hat{y}\right\}.
\end{eqnarray}
The RF Hamiltonian in the rotating frame thus becomes,
\begin{eqnarray}
H_{\mathrm{eff}} & = & -\boldsymbol{\mu} \cdot \mathbf{B_{\mathrm{eff}}}  \nonumber \\
& = & \Omega \hbar I_z + \omega_1 \hbar \left\{I_x\cos\phi  +  I_y \sin\phi\right\}.
\end{eqnarray}

The evolution of the quantum ensemble under the effective field in the rotating frame is described by the time-dependent density operator,
 \begin{eqnarray}
  \rho_{\mathrm r}(t) = e^{-iH_{\mathrm{eff}}t}\rho_{\mathrm r}(0)e^{iH_{\mathrm{eff}}t},
 \end{eqnarray}
 where $\rho_{\mathrm r}(0)$ describes the initial condition.

If $\Omega = 0$ (on-resonance) and $\phi=0$, the effective field is along $\hat{x}$ direction.
Then, in the rotating frame, the magnetization nutates about $\hat{x}$.  Choosing the duration
of RF $\tau$ such that $\omega_1 \tau = \theta$, it is possible to obtain a $\theta_x$ pulse.
In general, the effective nutation frequency about the effective field is $\omega_{\mathrm{eff}} = \sqrt{\Omega^2 + \omega_1^2}$.

 \subsection{Relaxation of nuclear spins}
    Relaxation is the process which brings nuclear spins back to thermal equilibrium.  Nuclear spin relaxation is a consequence of interaction of the spin system to the surrounding environment or lattice.  The lattice can introduce a random time dependent magnetic field fluctuations at the site of a spin.  Frequency components of this fluctuating local field can be decomposed in two parts, parallel and perpendicular to the static field.  Depending On the direction of frequency components of fluctuating local field at the site of spin system, NMR relaxation can be classified in following two ways \cite{cavanagh}. 

  \subsubsection{Longitudinal relaxation}
  The transverse frequency components of the local fluctuating fields are responsible for non-adiabatic contributions to relaxation.  Random magnetic fields with a magnitude equal to the Larmor frequency induce a transition of spin system population from higher energy state to lower energy state and corresponding lattice population from lower energy state to higher energy state via an energy conserving process.  This loss of energy of spin system to the lattice brings spin system in thermal equilibrium.  Furthermore being a non-adiabatic transition, it also introduces some uncertainty in the energies of eigenstates. This uncertainty leads to fluctuations in Larmor frequency of spin system, resulting in the decay of phase coherence in time.  Consequently non-adiabatic transitions lead to both longitudinal as well as transverse relaxation.
  
  \subsubsection{Transverse relaxation}   
  On the other hand the longitudinal frequency components which are responsible for adiabatic transition cause a fluctuation in Larmor frequency leading to the loss of phase coherence.  As a result the off diagonal terms in the density matrix undergo decay.  Since adiabatic transitions do not contribute to any exchange of energy between the spin system and lattice, the populations of the spin states remain unaltered.  So adiabatic transitions are responsible only for transverse relaxation.

\section{\textbf{NMR quantum information processing}}
NMR has already provided a number of applications in spectroscopy, imaging, and many other fields of science and technology.  The experimental realization of QIP is another recent application of NMR.  In 1997, Cory et al and Chuang et al have independently proposed that NMR can be used as an experimental architecture for QIP \cite{corypnas1997,GershenfeldLogical}.  They proposed that the difficulty of achieving pure states in NMR can be overcome by preparing a special mixed state, called pseudopure state.  This approach provided a realistic way for NMR-QIP in ensemble systems.  After that, till date a number of QIP tasks have been implemented using NMR.  Availability of numerous well developed techniques in conventional NMR and favourable properties of spin systems helped in establishing NMR as a good candidate for QIP.  Some of the details about NMR architecture are mentioned below. 

  The developments in control of spin dynamics via specially designed RF pulses have facilitated realization of quantum gates with high fidelities.
 One utilizes internal Hamiltonian (Zeeman, Spin-Spin interactions) along with an external control Hamiltonian (RF), to realize universal quantum gates.  Also, the NMR spin systems, being weakly coupled to the environment, exhibit long coherence times compared to many other architectures.  Even with all these merits, liquid state NMR-QIP is limited by the scalability criterion.  Thus in the present scenario, implementation of a large scale quantum computation using NMR is difficult.  Till now, the largest register used for computation involves a dozen-qubit \cite{maheshbenchmark2006}.   Hope is still alive with solid-state NMR spin systems where one may achieve  larger registers and with shorter gate times.
  A comparative study of various architectures has already been given by \ref{architecture}.  NMR-QIP protocols can be broadly divided into three parts: (a) initialization (b) realization of a quantum circuit via robust quantum gates, and (c) measurements or characterization of the output states.   Below we will explain these in detail along with a brief description of NMR systems.
            	 
  \subsection{NMR qubits}  
  The natural choice to mimic a qubit is a spin-1/2 nucleus.  A multispin system with spin-spin interactions can be used to realize a multiqubit register. 
  
  \subsection{Initialization}  
  A quantum algorithm usually begins with a definitely known initial state or a pure state.
  Achieving a pure state in conventional NMR is not practical.  One way to overcome this problem is by preparing a pseudopure state which is isomorphic to a pure state.
  
\begin{figure}[h]
  \begin{center}
%  \vspace*{5cm}
  \includegraphics[trim=0cm 5cm 0cm 1cm clip=true, width=12cm]{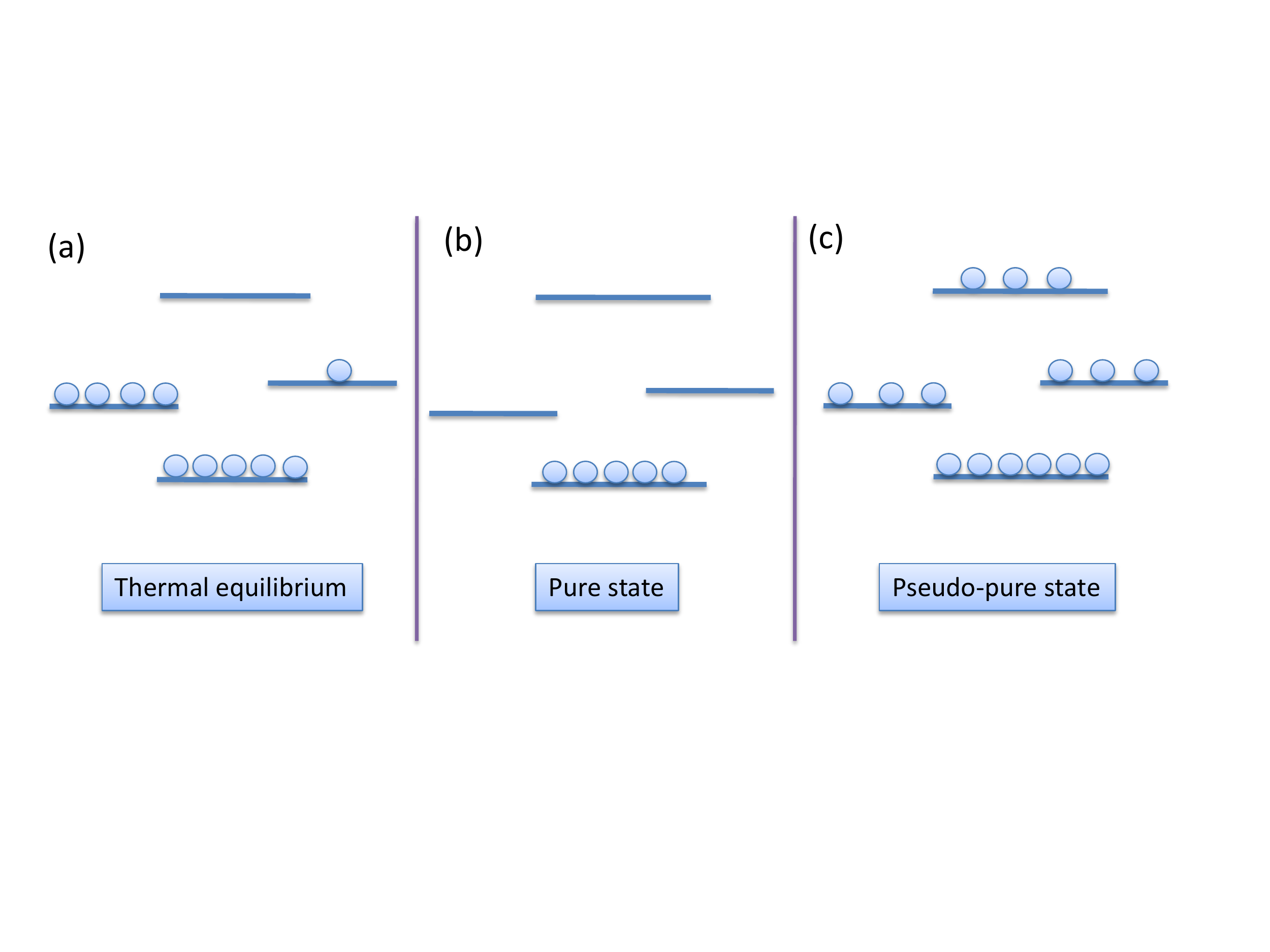}
  \caption{Distribution of population for a heteronuclear two-spin system (a) for thermal equilibrium state, (b) pure state, and (c) pseudopure state}
  \label{pps}
  \end{center}
  \end{figure}

  \textit{Pseudopure state}:  An ensemble is said to be in a pure state $\ket{\psi}$, when all its members are in the state $\ket{\psi}$.  The density operator for the pure state is 
  given by $\rho_{\mathrm{pure}} = \outpr{\psi}{\psi}$.  An ensemble is said to be in a
  pseudopure state if it is described by the density operator of the type 
  $\rho_{\mathrm{pps}} = (1-\epsilon) \mathbb{I}/N + \epsilon \rho_{\mathrm{pure}}$.
  An example is illustrated in Figure \ref{pps}.

  \subsubsection{Spatial averaging}
  This method of preparing pseudopure states involves RF pulses and pulsed-field-gradients (PFGs) \cite{corypnas1997}.  In the following, we describe this method by considering 
%  a $^1$H-$^{13}$C spin system,
  a homonuclear spin-pair using product operator formalism:
%\begin{eqnarray}
%  \rho_{z} = I^{1}_{z} + \frac{1}{4}I^{2}_{z}
%  \end{eqnarray}
%  %\vspace{-0.4cm}
%  \begin{eqnarray}
%  \rho_{00} &=& \frac{1}{4} (\mathbb{I_{\mathrm {n}}}+\epsilon \ket{00}\bra{00})\\  
%          &=& I^{1}_{z} + I^{2}_{z} + 2I^{1}_{z}I^{2}_{z}
%  \end{eqnarray}

  Below we show preparation of $\rho_{00}$ pseudopure state in a two spin system using product operator formalism.
  \begin{eqnarray}
  I^{1}_{z} + I^{2}_{z}
  & \xrightarrow{\hspace{0.5cm}(\frac{\pi}{3})^{2}_{x} \hspace{0.5cm}} & I^{1}_{z} + \frac{1}{2}I^{2}_{z} + \frac{\sqrt{3}}{2}I^{2}_{y} \nonumber \\
  & \xrightarrow{\hspace{0.5cm} G_{z} \hspace{0.5cm}} & I^{1}_{z} + \frac{1}{2}I^{2}_{z}  \nonumber \\
  & \xrightarrow{\hspace{0.5cm}(\frac{\pi}{4})^{1}_{x} \hspace{0.5cm}} & \frac{1}{\sqrt{2}}I^{1}_{z} - \frac{1}{\sqrt{2}}I^{1}_{y} + \frac{1}{2}I^{2}_{z} \nonumber \\ 
  & \xrightarrow{\hspace{0.5cm} \frac{1}{2J_{12}}\hspace{0.5cm}}& \frac{1}{\sqrt{2}}I^{1}_{z} +\frac{1}{\sqrt{2}}2I^{1}_{x}I^{2}_{z} + \frac{1}{2}I^{2}_{z} \nonumber \\   
  & \xrightarrow{\hspace{0.5cm} (\frac{\pi}{4})^{1}_{-y} \hspace{0.5cm}}& \frac{1}{2}I^{1}_{z} - \frac{1}{2}I^{1}_{x} + \frac{1}{2}2I^{1}_{x}I^{2}_{z} + \frac{1}{2}2I^{1}_{z}I^{2}_{z}+ \frac{1}{2}I^{2}_{z} \nonumber \\
  &  \xrightarrow{\hspace{0.5cm} G_{z} \hspace{0.5cm}} &\frac{1}{2}(I^{1}_{z}+I^{2}_{z}+2I^{1}_{z}I^{2}_{z})  \nonumber \\
  && \equiv \frac{1-\epsilon}{4}\mathbb{I_{\mathrm {n}}}+\epsilon \ket{00}\bra{00}.
  \end{eqnarray}
  In the above set of expressions, $\frac{1}{2J_{12}}$ represents an evolution under spin-spin interaction ($J$-coupling) for the duration $\frac{1}{2J_{12}}$, and $G_{z}$ represents PFG applied to destroy transverse magnetization.
  	    
  \subsubsection{Logical labelling}
\begin{figure}[h]
  \begin{center}
  \vspace*{1.5cm}
  \includegraphics[trim=1cm 4cm 2cm 4cm clip=true, width=11cm]{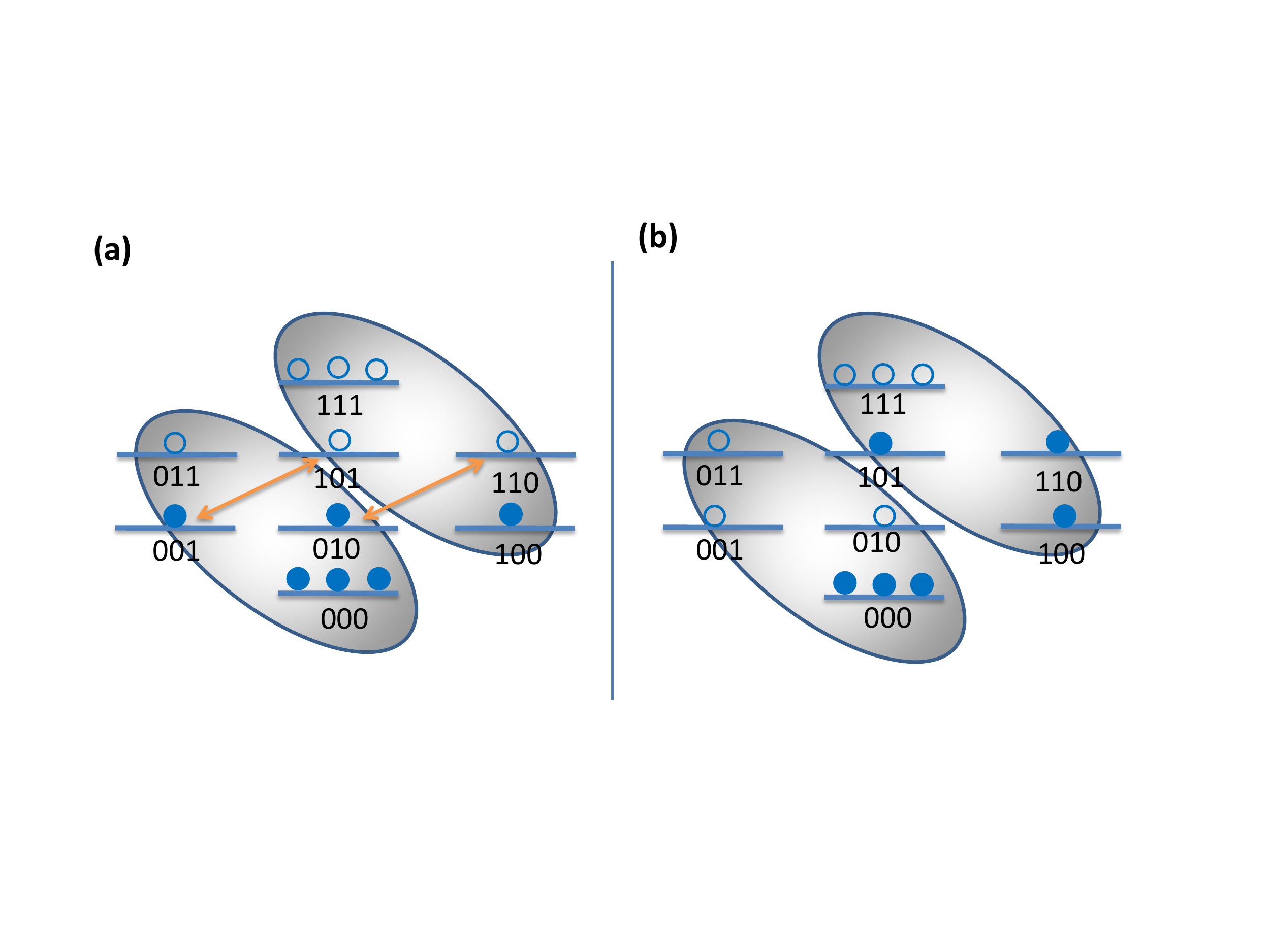}
 \caption{Populations (a) of density matrix for a homonuclear $3$ spin system at thermal equilibrium and (b) of deviation density matrices
 $\Delta\rho^{0}$ and $\Delta \rho^{1}$}
 \label{logicallabelling}
 \end{center}
 \end{figure}

  In 1997 Gershenfield and chuang proposed a method to prepare subsystem in pseudopure state \cite{GershenfeldLogical}.  For example: in the case of a homonuclear system of three spin-1/2 nuclei at thermal equilibrium, the deviation-populations (proportional to diagonal terms of deviation density matrix) are $\Delta\rho = I^{1}_{z}+I^{2}_{z}+I^{3}_{z}$ are $\{3,1,1,1,-1,-1,-1,-3\}$.  To prepare pseudopure state for $2^{nd}$ and $3^{rd}$ qubit subsystem, the populations between states $\ket{001}$ and $\ket{101}$, $\ket{010}$ and $\ket{110}$ are inverted by application of transition selective pulses as  shown in Fig. {\ref{logicallabelling}}. 
   The new  populations corresponding to subsystems $ \{ \ket{000}, \ket{001}, \ket{010}, \ket{011}\}$ and $\{ \ket{100}, \ket{101}, \ket{110}, \ket{111}\}$ are respectively, $\{3,-1,-1,-1\}$, and $\{1,1,1,-3\}$ respectively.  The deviation part of pseudopure state density matrix, corresponding to $0$ and $1$ subspace of $1^{st}$ qubit are $\Delta\rho^{0} = 4\outpr{00}{00} - \mathbb{I}$ and $\Delta \rho^{1} = \mathbb{I} - 4\outpr{11}{11}$.

 \subsection{NMR quantum gates}
 In the following we give examples of some important single qubit gates and their NMR realization.  We will describe single and multiple qubit gates.
 
 \subsubsection{Single qubit gates} 
 The most important single qubit quantum gates are the NOT gate, Hadamard gate and the phase gate.  In principle, one can always achieve any desired local rotation using at most three Euler angle rotations.  
 
 \subsubsection{NOT-gate}
   It flips spin state $\ket{0}$ to $\ket{1}$ and vice-versa.  In NMR, it can be realized by a $\pi_{x}$ pulse  \cite{jonesjmr98}
 since 
 \begin{eqnarray}
 \pi_{x} = e^{-i \pi I_{x}} &=&  
 -i \begin{bmatrix}
            0 & & 1          \\[0.3em]
            1 & & 0          \\[0.3em]
           \end{bmatrix} .
 \end{eqnarray}
Here the undetectable global phase can be ignored.

 \subsubsection{Hadamard gate}
It prepares a qubit in state $\ket{0}$ into equal superposition of $\ket{0}$ and $\ket{1}$.  In NMR it can be realized using the pulses $\left(\frac{\pi}{2}\right)_{y}\pi_{x}$, since
 \begin{eqnarray}
 \left(\frac{\pi}{2}\right)_{y}\pi_{x} \equiv e^{-i \pi I_{x}} e^{-i \frac{\pi}{2} I_{y}}  =  \frac{1}{\sqrt{2}}\begin{bmatrix}
                                         1 & & 1          \\[0.3em]
                                         1 & & -1          \\[0.3em]
                                        \end{bmatrix},
 \end{eqnarray}
 up to a global phase.
 
 \subsubsection{Phase gate}
 A phase gate implements a rotation around $\hat{z}$-axis.  For example the application of a phase gate $\phi$ on a qubit state prepared in equal superposition of two of its states introduces a phase factor $e^{i \phi}$ to $\ket{1}$.  In NMR, RF pulses can be applied only in the transverse direction, nevertheless a $\hat{z}$- rotation can be realized by a composite pulse using pulses of phases $x$ and $y$ (or by using chemical shift)
 \begin{eqnarray}
 \phi_{z} = \pi_{x}\phi_{y}\pi_{-x} \equiv e^{-i \frac{\pi}{2} I_{x}}e^{-i \phi I_{y}}e^{i \frac{\pi}{2} I_{x}} = e^{-i \phi I_{z}}=  e^{-i \frac{\phi}{2}}\begin{bmatrix}
                                         0 & & 0          \\[0.3em]
                                         0 & & e^{i \phi}          \\[0.3em]
                                        \end{bmatrix}.
 \end{eqnarray}
 
 \subsubsection{Multiqubit gates}  
 A general multiqubit gate can be realized with an appropriate combination of qubit selective RF pulses (local gates) and coupling interactions (nonlocal gates).  An NMR pulse sequence to realize $C$- NOT gate is given below \cite{jonesjmr98}.

% \begin{figure}[h]
% \begin{center}
% \includegraphics[trim= 8.5cm 6cm 11.5cm 4.5cm clip=true, width=4.5cm]{fig/figIntroduction/pscnot1.pdf}
% \caption{NMR pulse sequence for the C-NOT gate.}
% \end{center}
% \end{figure} 

 \begin{figure}[h]
 \begin{center}
 \includegraphics[trim= 5cm 6.5cm 6cm 0.5cm clip=true, width=11cm]{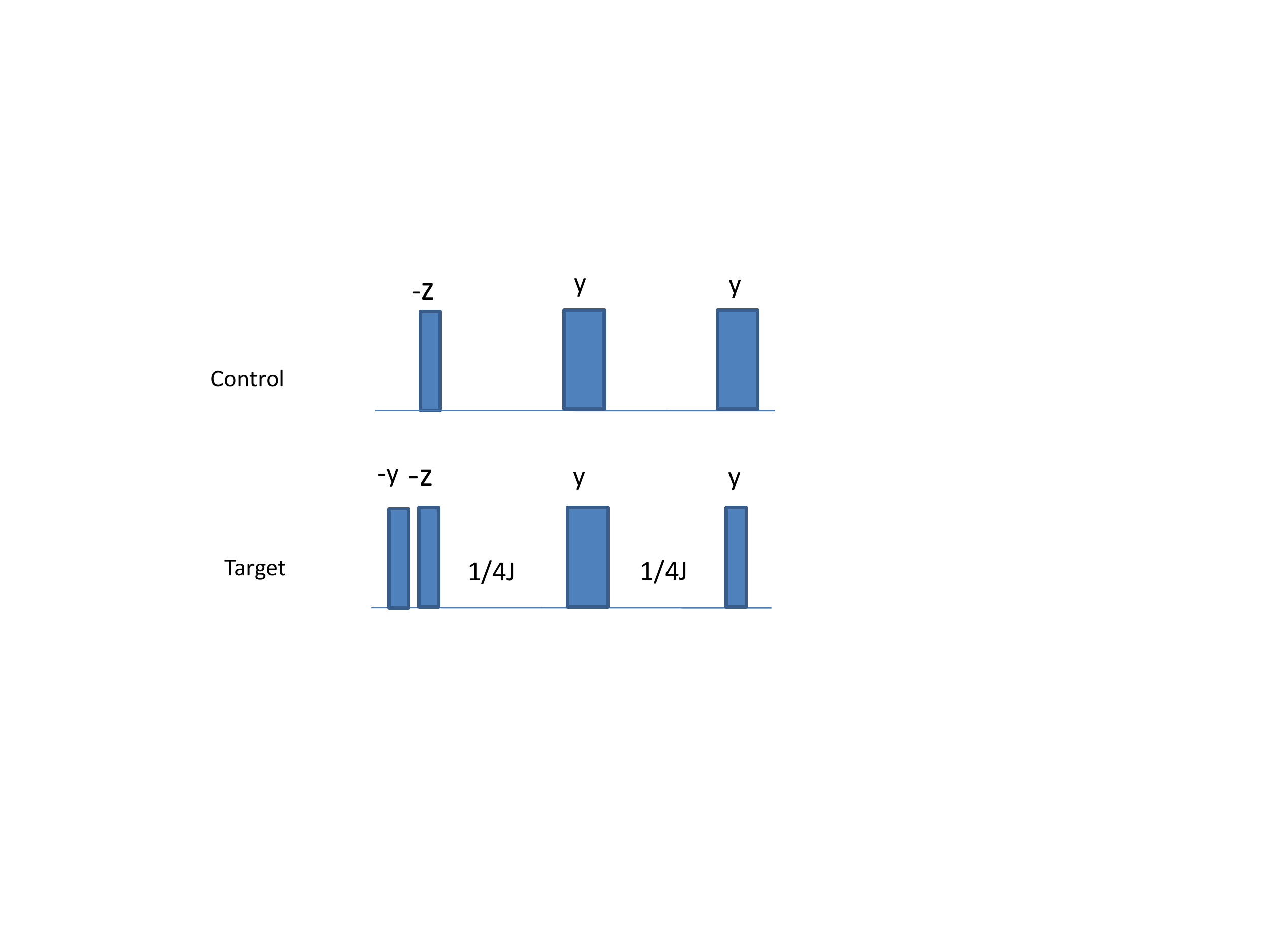}
 \caption{NMR pulse sequence for the C-NOT gate.  Thin pulses are $\frac{\pi}{2}$ pulses and
  broad pulses are $\pi$ pulses.  Delay $\frac{1}{4J}$ represent free evolution under secular part of indirect spin-spin interaction.}
 \end{center}
 \end{figure}
 
 \subsection{Numerically optimized pulses}
NMR quantum gates can be realized either by a combination of RF pulses and delays where evolution under internal Hamiltonian can take place. In a heteronuclear spin system, due to the large dispersion of Larmor frequencies and due to the availability of multi-channel RF coils, it is easy to introduce spin-selective rotations. On the other hand, in the case of homonuclear systems, realizing spin-selective rotations is nontrivial, and often requires specially designed RF modulations \cite{coryrfijcp}.  Recently many numerical methods are available for this purpose.  These methods not only take into account of complete Hamiltonian of the spin-system, but also attempt to realize gates that are robust against the experimental errors such as RF inhomogeneity (RFI) \cite{coryrfijcp,khaneja2005optimal}.

\begin{figure}
  \begin{center}
  \includegraphics[trim= 4cm 8cm 6cm 14cm clip=true, width=7cm]{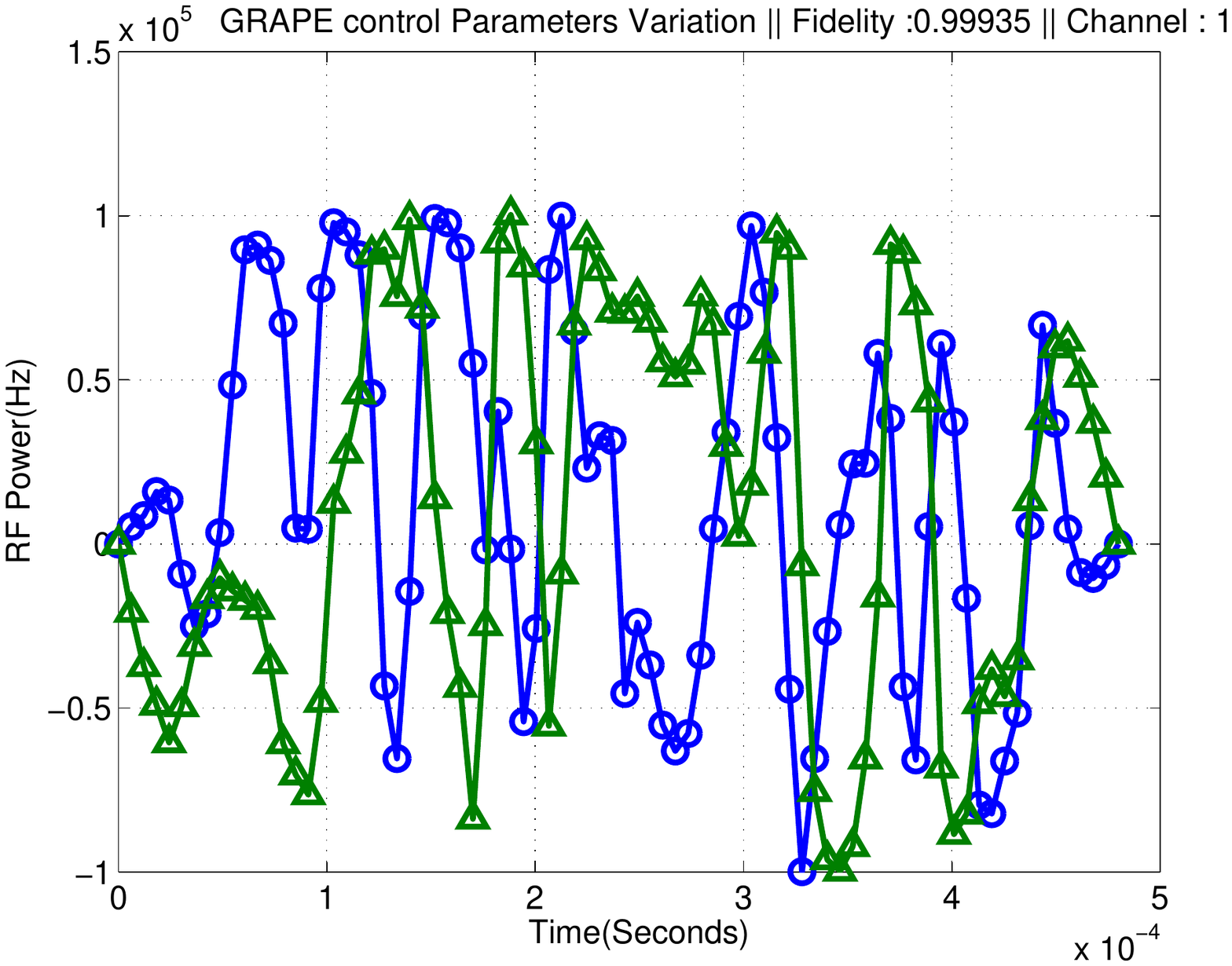}
  \includegraphics[trim=0cm 8cm 1.8cm 6.5cm, clip=true, width=11.5cm]{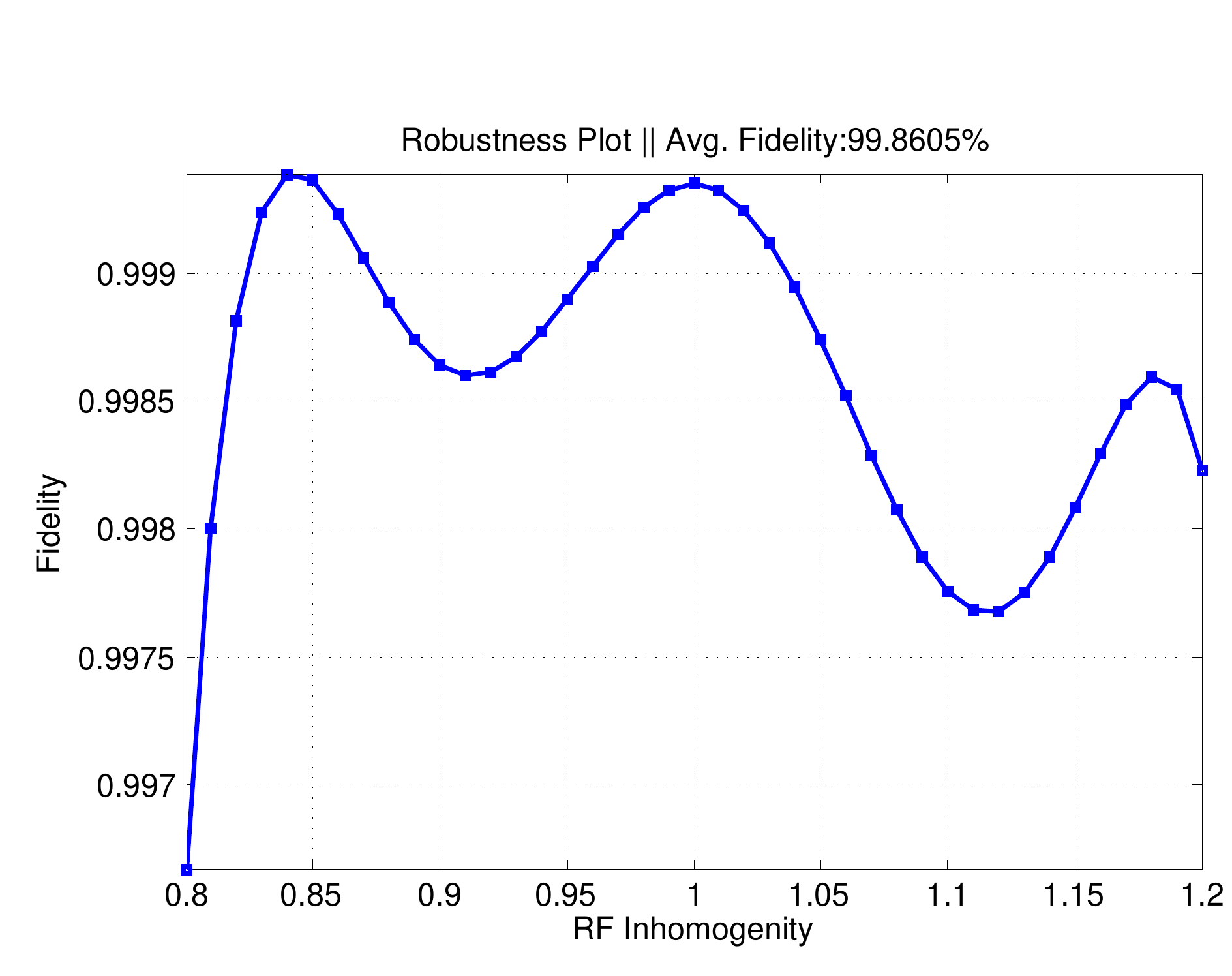}
  \caption{GRAPE generated RF profiles of $\pi_{x}$ pulse on $^{19}$F spins of iodotrifluoroethylene: X- component is shown in blue, Y- component is shown in green (top), and robustness profile against RFI (bottom).}
  \label{grape}
  \end{center}
  \end{figure}

 Fig.\ref{grape} shows a GRAPE modulation for a $\pi_{x}$ pulse on a FFF spin system realized in NMR sample of iodotrifluoroethylene (C$_{2}$F$_{3}$I)  and it's performance against RFI.  The fidelity of the numerically generated pulse profile with respect to ideal pulse has been calculated using the expression \ref{fidelity}
 \begin{equation}
F =  \frac{\vert \tr{U_{targ}^\dagger U_{sim}} \vert}{N}  \label{fidelity}.
 \end{equation}
 Here $N = 2^{n}$ is the dimension of  unitary operators.
% $({\mathrm{C}_{2}}{\mathrm{F}_{3}}{\mathrm{I})$  
 
 \section{\textbf{Ancilla assisted quantum information processing (AAQIP)}}
  \subsection{System qubits and Ancilla qubits}
  In a QIP task those qubits which are under study are known as system qubits. The additional qubits which assist in this study but may or may not be measured at the end of the computation are often referred as ancilla qubits.  Such ancilla qubits may be very useful in efficient implementation of certain QIP protocol.
  
 \subsection{Applications of ancilla qubits}
 Availability of ancilla qubits can be exploited in various ways.  In the following I discuss three of them.
 
\begin{enumerate}
\item[(i)]
\textbf{Extended work space:}\\
Consider a qubit system of size $n$ in the state $\rho$, along with additional ancillary qubits $n_{a}$ all prepared in the state $\rho_{a}$.  Thus total number of the qubits in combined system and ancillary register are $\tilde{n} = n + n_{a}$.  Now the state of the combined system can be written as $\tilde{\rho} = \rho \otimes \rho_{a}$.  This extended space, can be exploited for efficient implementation of various QIP tasks.  Below are the two cases where extended space is useful.  
 
 \textit{Characterization of quantum state}:  The extended Hilbert space allows a larger set
 of observables which can be simultaneously measured, and hence allow more efficient quantum
 state tomography of system qubits.  This aspect is discussed in chapter \ref{chp2}.
 
 \textit{Characterization of quantum process}: The extended Hilbert space can also simultaneously encode different input states of the system qubits and hence allows more
 efficient quantum process tomography.  This aspect is discussed in chapter \ref{chp3}.

\item[(ii)]
\textbf{Non-invasive measurements: }\\
   While a truly noninvasive quantum measurement is an 
  idealized process, it is often possible to extract certain information about the quantum state via indirect measurement.  One method involves letting the system qubits interact with ancilla qubits, and the measurement of ancilla qubits at a later stage, thus indirectly extracting the information about the system qubits.  This aspect is discussed in chapter \ref{chp4}.
  
\item[(iii)]  
\textbf{Spectroscopy: }\\
We show in chapter \ref{chp5} that the availability of ancilla qubits can be utilized for
sensitive encoding of relative phase leading to certain interesting applications in spectroscopy.  For example efficient measurement of translational diffusion constant and RF inhomogeneity.
\end{enumerate}  
  
\begin{figure}
 \begin{center}
 \includegraphics[trim= 4cm 5cm 4cm 0cm clip=true, width=9cm]{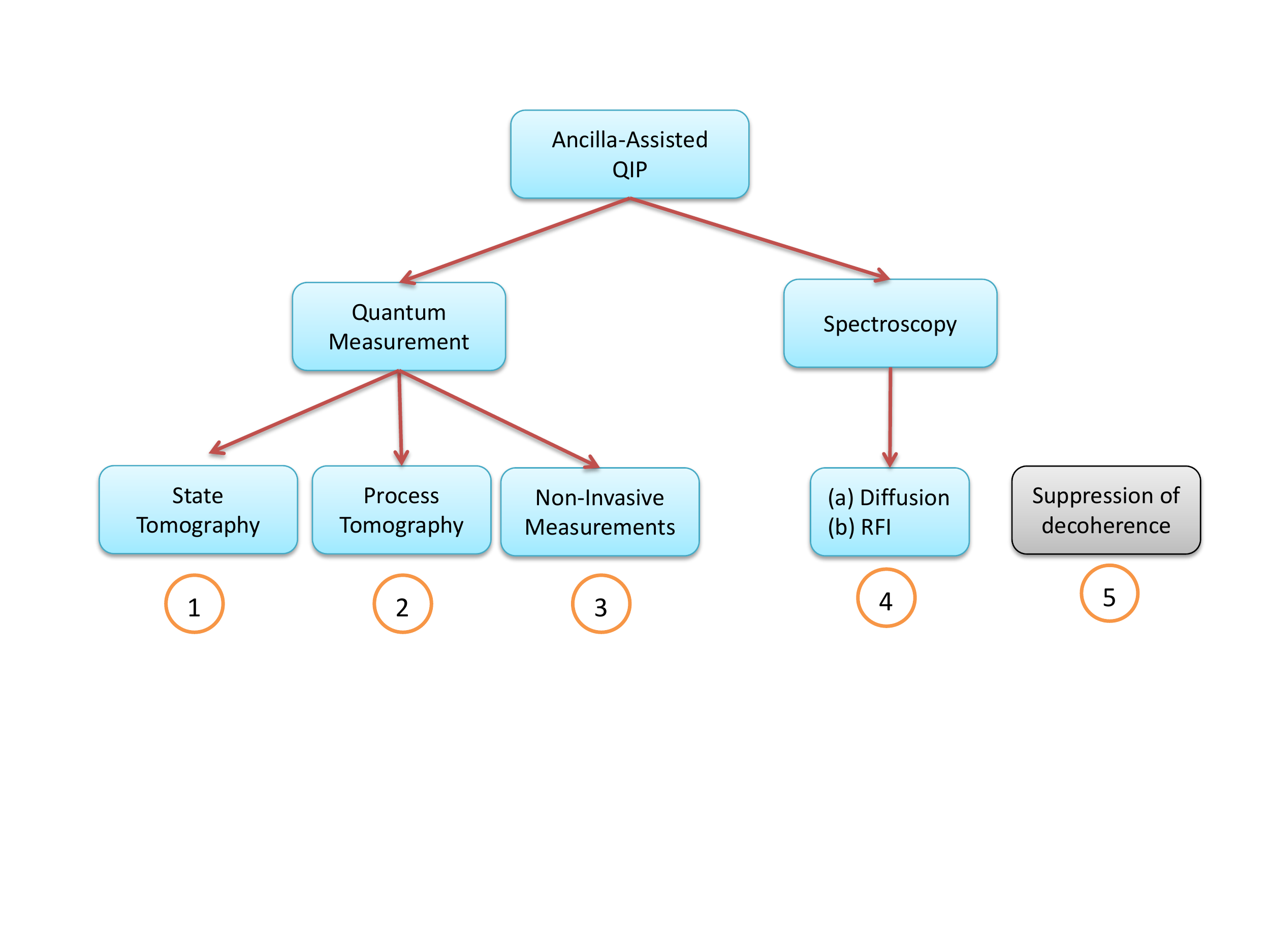}
 \caption{Topics studied in this thesis summarized as different chapters.\label{chps}}
 \end{center}
 \end{figure}

  Figure \ref{chps} summarizes various QIP and spectroscopy related topics studied in this thesis.    
 
\thispagestyle{empty}

% % % % % % % % ch2
\titlespacing*{\chapter}{0pt}{-50pt}{20pt}
\titleformat{\chapter}[display]{\normalfont\Large\bfseries}{\chaptertitlename\ \thechapter}{20pt}{\Large}
\chapter{Ancilla assisted quantum state tomography \label{chp2}}

\section{\textbf{Introduction}\label{21}}
Quantum computers have the potential to carry-out certain computational tasks 
with an efficiency that is beyond the reach of their classical
counterparts \cite{chuangbook}.  In practice however, harnessing the computational 
power of a quantum system has been an enormously challenging task
\cite{exptqipbookeveritt2005}.
The difficulties include imperfect control on the quantum dynamics
and omnipresent interactions between the quantum system and its
environment leading to an irreversible loss of quantum coherence.
In order to optimize the control fields and to understand the effects
of environmental noise, it is often necessary to completely characterize the quantum state.
In experimental quantum information studies,
Quantum State Tomography (QST) is an important tool that is routinely used
to characterize an instantaneous quantum state \cite{chuangbook}.  

QST on an initial state is usually carried out to confirm the efficiency of
initialization process.  Though QST of the final state is usually not  
part of a quantum algorithm, it allows one
to measure the fidelity of the output state.  
QSTs in intermediate stages often help experimentalists
to tune-up the control fields better.  

QST can be performed by a series of measurements of
noncommuting observables which together enables one to reconstruct 
the complete complex density matrix.  In the standard method,
the required number of independent experiments grows
exponentially with the number of input qubits \cite{ChuangPRSL98,ChuangPRA99}.
Anil Kumar and co-workers have illustrated QST using a single
two-dimensional NMR spectrum \cite{aniltomo}.  They showed that a two-dimensional
NMR experiment consisting of a series of identical measurements with systematic
increments in evolution time, can be used to quantitatively estimate all
the elements of the density matrix.  
Later Nieuwenhuizen and co-workers have shown that 
it is possible to reduce the number of independent experiments
in the presence of an ancilla register initialized to a known
state \cite{Nieuwenhuizen}.
They pointed out that in suitable situations, it is possible to 
carry-out QST with a single measurement of a set of factorized
observables.  We refer to this method as Ancilla Assisted QST (AAQST).
This method was experimentally illustrated by Suter
and co-workers using a single input qubit and a single ancilla 
qubit \cite{sutertomo}.  Recently Peng and coworkers have studied the effectiveness
of the method for qutrit-like systems using numerical simulations 
\cite{pengtomo}.  
Single shot mapping of density matrix by AAQST method not only
reduces the experimental time, but also alleviates the need to 
prepare the target state several times.  Often slow variations
in system Hamiltonian may result in systematic errors in 
repeating the state preparation.  Further, environmental noises
lead to random errors in multiple preparations.  
These errors play important roles in the quality of the 
reconstruction of the target state. Therefore AAQST has the 
potential to provide a more reliable way of tomography.  

In this chapter, I first report our revisit of the theory of QST and AAQST and also provide
methods for explicit construction of the constraint matrices,
which will allow extending the tomography procedure for large registers.
An important feature of the method described here is that
it requires only global rotations and short evolutions under the collective 
internal Hamiltonian.  
I also describe our NMR demonstrations 
of AAQST on two different types of systems:
(i) a two-qubit input register using a one-qubit ancilla in an isotropic liquid-state system 
and (ii) a three-qubit input register using a two-qubit ancilla register
in a partially oriented system. 

In the following \S I briefly describe the theory of QST and AAQST.
In \S \ref{23} I describe experimental demonstrations  and finally I conclude in \S \ref{24}.
\section{\textbf{Theory}\label{22}}
\subsection{Quantum State Tomography}
We consider an $n$-qubit register 
formed by a system of $n$ mutually interacting spin-1/2 nuclei
with distinct resonance frequencies $\omega_i$ and mutual
interaction frequencies $2\pi J_{ij}$.
The Hamiltonian under weak-interaction limit ($2 \pi J_{ij}\ll \vert \omega_i-\omega_j \vert$) 
consists of the Zeeman part and spin-spin interaction part, i.e.,
\begin{eqnarray}
{\cal H} = -\sum\limits_{i=1}^{n}\omega_i \sigma_z^i /2 + 
\sum\limits_{i=1}^{n}\sum\limits_{j=i+1}^{n} 2\pi J_{ij} \sigma_z^i \sigma_z^j /4
\label{ham}
\end{eqnarray}
respectively, where $\sigma_{z}^i$ and $\sigma_{z}^j$ are the $z$-components of Pauli operators
of $i$th and $j$th qubits \cite{cavanagh}.  
The set of $N=2^n$ eigenvectors $\{ \vert m_1 m_2 \cdots m_n \rangle \}$ 
of the Zeeman Hamiltonian form a complete orthonormal computational basis.  
We can order the eigenvectors based on the decimal value $m$
of the binary string $(m_1 \cdots m_n)$, i.e., $m = m_1 2^{n-1}+\cdots+m_n 2^0$.

The general density matrix can be decomposed as
$\mathbb{I}/N+\epsilon \rho$ where the identity part is known as the background,
the trace-less part $\rho$ is known as the \textit{\large{deviation density matrix}},
and the dimensionless constant $\epsilon$ is the purity factor 
\cite{corypnas1997}.
In this context, QST refers to complete characterization of the deviation density
matrix, which can be expanded in terms of $N^2-1$ real unknowns:
\begin{eqnarray}
\rho &=& 
\sum\limits_{m=0}^{N-2} \rho_{mm}(\vert m \rangle \langle m \vert -\vert N-1 \rangle \langle N-1 \vert)  \nonumber \\
&&+ \sum_{m=0}^{N-2}\sum_{m'=m+1}^{N-1} \{
R_{mm'}(\vert m \rangle \langle m' \vert+\vert m' \rangle \langle m \vert)
+ iS_{mm'}(\vert m \rangle \langle m' \vert-\vert m' \rangle \langle m \vert)
\}.
\label{dmm}
\end{eqnarray}
Here first part consists of $N-1$ diagonal unknowns $\rho_{mm}$ with
the last diagonal element $\rho_{N-1,N-1}$ being constrained by the trace-less condition.  
$R$ and $S$
each consisting of $(N^2-N)/2$ unknowns correspond to real and imaginary 
parts of the off-diagonal elements respectively.
Thus a total of $N^2-1$ real unknowns needs to be determined.

Usually an experimental technique allows a particular set of observables 
to be measured directly.
To explain the NMR case, we introduce $n$-bit binary strings,\\
$j_\nu = \nu_1 \nu_2 \cdots \nu_{j-1}
 0 \nu_{j} \cdots \nu_{n-1}$ and
$j'_{\nu} = \nu_1 \nu_2 \cdots \nu_{j-1} 1 \nu_{j} \cdots \nu_{n-1}$
differed only by the flip of the $j$th bit.
Here $\nu = \nu_1 2^{n-2} + \nu_2 2^{n-3} + \cdots + \nu_{n-1} 2^0$ is the 
value of the $n-1$ bit binary string $(\nu_1,\nu_2,\cdots,\nu_{n-1})$
and $\nu$ can take a value between $0$ and $\gamma = N/2-1$.
The real and imaginary parts of an NMR signal recorded in a 
quadrature mode corresponds to the expectation values of 
transverse magnetization observables $\sum\limits_{j=1}^{n}\sigma_{jx}$ and 
$\sum\limits_{j=1}^{n}\sigma_{jy}$ respectively \cite{cavanagh}.
The background part of the density matrix neither evolves under
unitaries nor gives raise to any signal, and therefore we ignore it.
Under suitable conditions (when all the transitions are resolved), 
a single spectrum directly yields $nN$
matrix elements $\{R_{j_\nu,j_\nu'},S_{j_\nu,j_\nu'}\}$
as complex intensities of spectral lines.  These matrix elements
are often referred to as single quantum elements since they
connect eigenvectors related by the flip of a single qubit.  
We refer the single-quantum terms $R_{j_\nu,j_\nu'}$
and $S_{j_\nu,j_\nu'}$ respectively
as the real and imaginary parts of $\nu$th spectral line of $j$th qubit.
%Since each qubit can have $N/2$ spectral lines,
%a total of $n N/2$ complex matrix elements of $\rho$ can be measured
%directly through the single spectrum.
Thus a single spectrum of an $n$-qubit system in an
arbitrary density matrix can yield $n N$ real unknowns. 
In order to quantify the remaining elements,
one relies on multiple experiments all starting from the same initial state
$\rho$.  The $k$th experiment consists of applying a unitary $U_k$
to the state $\rho$, leading to $\rho^{(k)} = U_k \rho U_k^\dagger$, and measuring 
the single-quantum spectrum $\{R_{j_\nu,j_\nu'}^{(k)},S_{j_\nu,j_\nu'}^{(k)}\}$. 
From eqn. (\ref{dmm}) we obtain
\begin{eqnarray}
&&R^{(k)}_{j_\nu,j_\nu'} 
 = \sum\limits_{m} a_{j\nu}^{(k)}(m) \rho_{mm} + 
 \sum\limits_{m,m'>m}c_{j\nu}^{(k)}(m,m') R_{mm'} + e_{j\nu}^{(k)}(m,m') S_{mm'}, 
\nonumber \\
&&S^{(k)}_{j_\nu,j_\nu'} 
 = \sum\limits_{m} b_{j\nu}^{(k)}(m) \rho_{mm} + 
\sum\limits_{m,m'>m} d_{j\nu}^{(k)}(m,m') R_{mm'} + f_{j\nu}^{(k)}(m,m') S_{mm'}, 
\label{leq}
\end{eqnarray}
in terms of the unknowns $\rho_{mm'}$ and the known real constants $\{a, \cdots, f\}$:
\begin{eqnarray}
a_{j\nu}^{(k)}(m,m) + ib_{j\nu}^{(k)}(m,m) &=& 
\langle j_\nu \vert U_k \vert m \rangle \langle m \vert U_k^\dagger \vert j'_\nu\rangle-
\langle j_\nu \vert U_k \vert N-1 \rangle \langle N-1 \vert U_k^\dagger \vert j'_\nu\rangle, 
\nonumber \\
c_{j\nu}^{(k)}(m,m') +i d_{j\nu}^{(k)}(m,m')&=& 
\langle j_\nu \vert U_k \vert m \rangle \langle m' \vert U_k^\dagger \vert j'_\nu\rangle + 
\langle j_\nu \vert U_k \vert m' \rangle \langle m \vert U_k^\dagger \vert j'_\nu\rangle, 
\nonumber \\
e_{j\nu}^{(k)}(m,m')+if_{j\nu}^{(k)}(m,m') &=& 
i\langle j_\nu \vert U_k \vert m \rangle \langle m' \vert U_k^\dagger \vert j'_\nu\rangle - 
i\langle j_\nu \vert U_k \vert m' \rangle \langle m \vert U_k^\dagger \vert j'_\nu\rangle
\end{eqnarray}
\cite{maheshtomo}.
After $K$ experiments, we can setup the matrix equation

%%\baselinestretch{1}
%\vspace{-5.75cm}
\newpage
\begin{eqnarray}
%\vspace{-5.5cm}
\hspace{-0.2cm}
M
\hspace{0.3cm}
\left[
\begin{array}{c}
\rho_{0,0} \\
\cdots \\
\rho_{N-2,N-2} \\
-------- \\
R_{0,1} \\
\cdots \\
R_{0,N-1} \\
\cdots \\
R_{m,m'>m} \\
\cdots \\
R_{N-2,N-1} \\
-------- \\
S_{0,1} \\
\cdots \\
S_{0,N-1} \\
\cdots \\
S_{m,m'>m} \\
\cdots \\
S_{N-2,N-1} \\
\end{array}
\right]
\hspace{1.1cm}
 = 
 \hspace{1.1cm}
 \left[
 \begin{array}{c}
R^{(1)}_{1_0,1_0'}  \\
 \cdots  \\
R^{(1)}_{1_\gamma,1_\gamma'}  \\
R^{(1)}_{2_0,2_0'}  \\
 \cdots  \\
 \cdots  \\
R^{(K)}_{n_\gamma,n_\gamma'} \\
-------\\
S^{(1)}_{1_0,1_0'}  \\
 \cdots  \\
S^{(1)}_{1_\gamma,1_\gamma'}  \\
S^{(1)}_{2_0,2_0'}  \\
 \cdots  \\
 \cdots  \\
S^{(K)}_{n_\gamma,n_\gamma'} \\
 \end{array}
 \right].
\label{meq}
\end{eqnarray}

%%
%\baselinestretch{1.5}
Here the left column vector is formed by the $N^2-1$ unknowns of $\rho$: diagonal elements in the top, real off-diagonals in the middle,
and imaginary off-diagonals in the bottom.  The right column vector is formed by $KnN$ numbers - the real and imaginary parts of the experimentally obtained spectral intensities ordered according to the value of the binary string $\nu$, the qubit number $j$, and the experiment number $k$.  The $KnN\times(N^2-1)$ dimensional constraint matrix is of the form
\begin{eqnarray}
&&M = \nonumber \\
&&\left[
\begin{array}{c c|c c|c c}
a_{1,0}^{(1)}(0,0) & \cdots & c_{1,0}^{(1)}(m,m') & \cdots  & e_{1,0}^{(1)}(m,m') & \cdots \\
\cdots &\cdots &\cdots &\cdots &\cdots &\cdots   \\
a_{1,\gamma}^{(1)}(0,0) & \cdots & c_{1,\gamma}^{(1)}(m,m') & \cdots  & e_{1,\gamma}^{(1)}(m,m') & \cdots \\
\cdots &\cdots &\cdots &\cdots &\cdots &\cdots   \\
a_{n,0}^{(1)}(0,0) & \cdots & c_{n,0}^{(1)}(m,m') & \cdots  & e_{n,0}^{(1)}(m,m') & \cdots \\
\cdots &\cdots &\cdots &\cdots &\cdots &\cdots   \\
\cdots &\cdots &\cdots &\cdots &\cdots &\cdots   \\
a_{n\gamma}^{(K)}(0,0) & \cdots & c_{n\gamma}^{(K)}(m,m') & \cdots  & e_{n\gamma}^{(K)}(m,m') & \cdots \\
\hline
b_{1,0}^{(1)}(0,0) & \cdots & d_{1,0}^{(1)}(m,m') & \cdots  & f_{1,0}^{(1)}(m,m') & \cdots \\
\cdots &\cdots &\cdots &\cdots &\cdots &\cdots   \\
b_{1,\gamma}^{(1)}(0,0) & \cdots & d_{1,\gamma}^{(1)}(m,m') & \cdots  & f_{1,\gamma}^{(1)}(m,m') & \cdots \\
\cdots &\cdots &\cdots &\cdots &\cdots &\cdots   \\
b_{n,0}^{(1)}(0,0) & \cdots & d_{n,0}^{(1)}(m,m') & \cdots  & f_{n,0}^{(1)}(m,m') & \cdots \\
\cdots &\cdots &\cdots &\cdots &\cdots &\cdots   \\
\cdots &\cdots &\cdots &\cdots &\cdots &\cdots   \\
b_{n\gamma}^{(K)}(0,0) & \cdots & d_{n\gamma}^{(K)}(m,m') & \cdots  & f_{n\gamma}^{(K)}(m,m') & \cdots \\
\end{array}
\right]\nonumber .\\
\label{mmat}
\end{eqnarray}
Note that each column of the constraint matrix corresponds to contribution of a particular unknown element of $\rho$  to the various spectral intensities.  By choosing the unitaries $\{U_k\}$ such that $\mathrm{rank} (M) \ge N^2-1$ (the number of unknowns), eqn.  (\ref{meq}) can be solved either by singular value decomposition or by Gaussian elimination method
\cite{maheshtomo}. 
Fig. \ref{exptscaling} illustrates the minimum number ($K$) 
of experiments required for QST. As anticipated, $K$ increases 
rapidly as $O(N/n)$ with the number of input qubits.
In the following we describe how it is possible to speed-up QST,
in the presence of an ancilla register, with fewer experiments.

\subsection{Ancilla Assisted QST (AAQST)}

\begin{figure}[h]
\begin{center}
%\hspace{2.8cm}
\includegraphics[width=10cm]{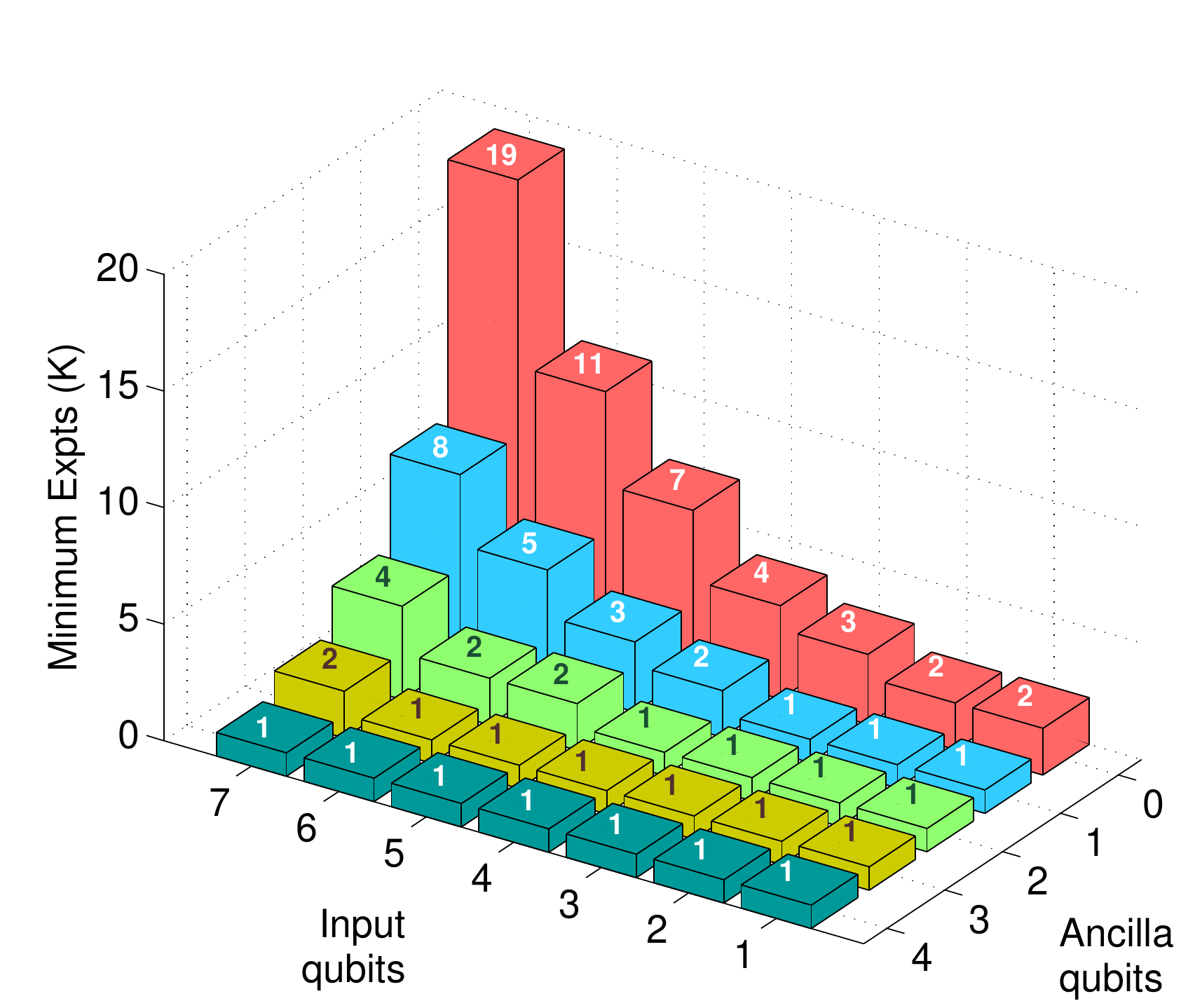} 
\caption{Minimum number of independent experiments required in QST
(without ancilla) and AAQST (with different number of ancilla register).  
%The cases with experimental demonstrations are indicated with circles.
}
\label{exptscaling}
\end{center} 
%\noindent\makebox[\linewidth]{\rule{15 cm}{1pt}}
\vspace*{0.5cm}
\end{figure}

Suppose the input register of $n$-qubits is associated with an ancilla register consisting of $\hat{n}$ qubits. 
The dimension of the combined system of $\tilde{n} = n+\hat{n}$ qubits is $\tilde{N} = N\hat{N}$, where $\hat{N} = 2^{\hat{n}}$.  For simplicity we assume that each qubit interacts sufficiently with all other qubits so as to obtain a completely resolved spectrum yielding $\tilde{n}\tilde{N}$ real parameters.
Following method is applicable even if there are spectral overlaps, 
albeit with lower efficiency (i.e., with higher number $(K)$ of minimum
experiments).  Further for simplicity, 
we assume that the ancilla register
begins with the maximally mixed initial state, with no contribution
to the spectral lines from it.  
Otherwise, we need to add the contribution of the ancilla to
the final spectrum and the eqn. (\ref{meq}) will become inhomogeneous.
As explained later in the experimental
section, initialization of maximally mixed state can be achieved with high precision.
Thus the deviation density matrix of the combined system is $\tilde{\rho} = \rho \otimes \mathbb{I}/\hat{N}$.
Now applying only local unitaries neither leads to ancilla coherences
nor transfers any of the input coherences to ancilla.  
Therefore we consider applying a non-local unitary 
exploiting the input-ancilla interaction,
\begin{eqnarray}
\tilde{U}_k = V \sum\limits_{a=0}^{\hat{N}-1} U_{ka} \otimes \vert a \rangle \langle a \vert,
\end{eqnarray}
where $U_{ka}$ is the $k$th unitary on the input register dependent 
on the ancilla state $\vert a \rangle$ and $V$ is the local 
unitary on the ancilla.
The combined state evolves to
\begin{eqnarray}
&& \tilde{\rho}^{(k)} = \tilde{U}_k \tilde{\rho} \tilde{U}_k^\dagger \nonumber \\
&& = \frac{1}{\hat{N}}\sum\limits_{m,m',a} \rho_{mm'} U_{ka} \vert m \rangle \langle m' \vert U_{ka}^\dagger
\otimes V \vert a \rangle \langle a \vert V^\dagger.
\end{eqnarray}
We now record the spectrum of the combined system corresponding
to the observable $\sum\limits_{j=1}^{\tilde{n}} \sigma_{jx}+i\sigma_{jy}$.
Each spectral line can again be expressed in terms of the unknown
elements of the ancilla matrix in the form given in eqn. (\ref{leq}).
The spectrum of the combined system yields $\tilde{n}\tilde{N}$ linear
equations.  The minimum number of independent experiments
needed is now $O(N^2/(\tilde{n}\tilde{N}))$.
Since we can choose $\tilde{N} \gg N$, AAQST needs 
fewer than O($N/n$) experiments required in the standard QST.
In particular, when $\tilde{n}\tilde{N} \ge N^2$, a single
optimized unitary suffices for QST.
Fig. \ref{exptscaling} illustrates the minimum number ($K$) 
of experiments required for various sizes of input and ancilla 
registers. As illustrated, QST can be achieved with only one
experiment, if an ancilla of sufficient size is provided along with.

\subsection{Building the constraint matrix}
The major numerical procedure in AAQST is obtaining the constraint matrix $M$.
For calculating the constraint coefficients $c_{rj}^{(k)}$,
one may utilize an elaborate decomposition of $U_k$
using numerical or analytical methods.
Alternatively, as described below, we can use a simple algorithmic approach to 
construct the constraint matrix.

First imagine a diagonal state $\rho$ for the ancilla register
(eqn. (\ref{dmm})) with $\rho_{00}=1$ and $\rho_{mm}=0$ for all
other $1 \le m \le N-2$, $R_{mm'}=S_{mm'}=0$.  
Applying the unitary $U_k$ on the composite deviation density matrix 
$\tilde{\rho} = \rho \otimes \mathbb{I}/\hat{N}$, we obtain all the
spectral intensities (using eqn. (\ref{leq}))
\begin{eqnarray}
a^{k}_{j\nu}(0,0) = R^{(k)}_{j\nu,j\nu'}, ~
b^{k}_{j\nu}(0,0) = S^{(k)}_{j\nu,j\nu'}.
\end{eqnarray}
Thus the spectral lines indicate the contributions only from $\rho_{00}$
(and $\rho_{N-1,N-1}$).
Repeating
the process with all the unitaries $\{U_k\}$ yields
the first column in $M$ matrix (eqn. (\ref{mmat})) corresponding to the unknown $\rho_{00}$.
Same procedure can be used for all the diagonal elements $\rho_{mm}$ with
$0 \le m \le N-2$.  
To determine $M$ matrix column corresponding to a
real off-diagonal unknown $R_{mm'}$, 
we start with an input-register density matrix 
$R_{mm'} = 1$ and all other elements  
set to zero.  Again by applying the unitary 
$U_k$ on the composite density matrix, and
using eqn. (\ref{leq}) we obtain
\begin{eqnarray}
c^{k}_{j\nu}(m,m') = R^{(k)}_{j\nu,j\nu'}, ~
d^{k}_{j\nu}(m,m') = S^{(k)}_{j\nu,j\nu'}.
\end{eqnarray}
Repeating the process with all unitaries $\{U_k\}$ 
determines the column of $M$ corresponding to the unknown $R_{mm'}$. 

To determine $M$ matrix column corresponding to 
an imaginary off-diagonal unknown $S_{mm'}$, 
we set $S_{mm'} = 1$ and all other elements to zero, 
and apply $U_k$ on the composite state to obtain
\begin{eqnarray}
e^{k}_{j\nu}(m,m') = R^{(k)}_{j\nu,j\nu'}, ~
f^{k}_{j\nu}(m,m') = S^{(k)}_{j\nu,j\nu'}.
\end{eqnarray}
Proceeding this way, by selectively setting the unknowns
one by one, the complete constraint matrix can be built easily.
\subsection{Optimization of Unitaries}
Solving the matrix equation (\ref{meq}) requires that
$\mathrm{rank}(M) \ge N^2-1$, the number of unknowns.
But having the correct rank is not sufficient.
The matrix $M$ must be well conditioned in order to ensure that small
errors in the observed intensities $\{R^{(k)}_{j\nu,j\nu'},S^{(k)}_{j\nu,j\nu'}\}$ 
do not contribute to large errors in the values of the elements $\rho_{mm'}$.  
The quality of
the constraint matrix can be measured by a scalar quantity called condition number
$C(M)$  defined as the ratio of the largest singular
value of $M$ to the smallest \cite{NLalgebra}.  Smaller the value of $C(M)$, better conditioned
is the constraint matrix $M$ for solving the unknowns.  Thus the condition 
number provides a convenient scalar quantity to optimize the set $\{U_k\}$
of unitaries to be selected for QST.  As explained in the experimental
section, we used a simple unitary model $U_1(\tau_1,\tau_2)$ as an initial
guess and used genetic algorithm to minimize the condition number and 
optimize the parameters $(\tau_1,\tau_2)$.
The necessary number ($K$) of independent experiments
is decided by the rank of the constraint matrix and the desired precision.
The rank condition requires that $KnN \ge N^2-1$. Introducing
additional experiments renders the problem over-determined, thus 
reducing the condition number and increasing the precision.  
In the following section
we describe the experimental results of AAQST for registers with (i) $n=2,\hat{n}=1,\tilde{n}=3$ and
(ii) $n=3,\hat{n}=2,\tilde{n}=5$ respectively.

\section{\textbf{Experiments}\label{23}}

We report experimental demonstrations of AAQST on two 
spin-systems of different sizes and environments.  
In each case, we have chosen two density matrices for tomography.
All the experiments described below are carried out on a Bruker 500 MHz
spectrometer at an ambient temperature of 300 K 
using high-resolution nuclear magnetic resonance techniques.  
  In the following, we describe experimental implementation of AAQST on two different spin systems.

%\vspace*{-0.5cm}

\begin{figure}[h]
\begin{center}
%\begin{array}{rr}
\includegraphics[trim= 0cm 0cm 0cm 0cm,clip=true,width= 11cm]{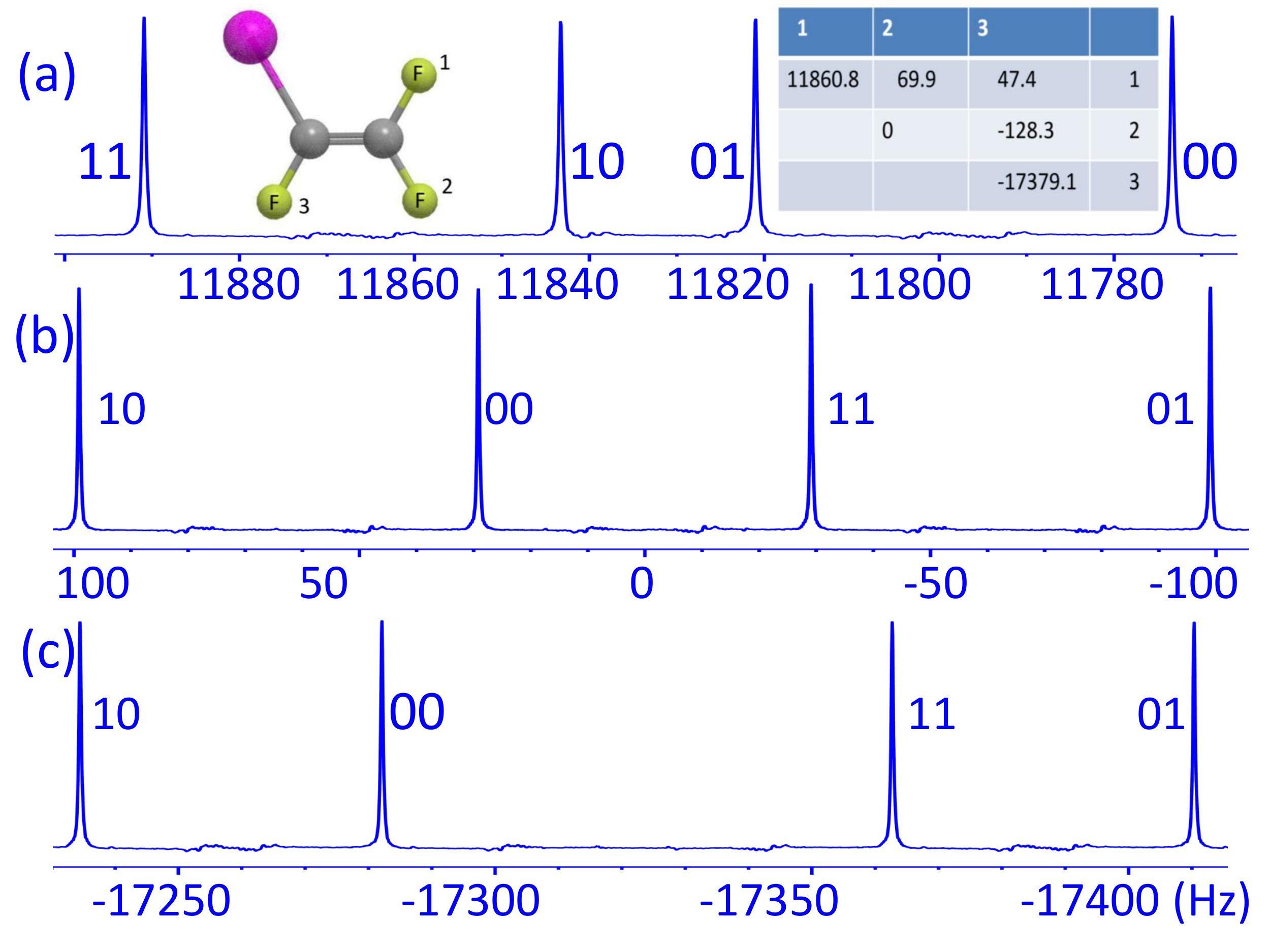}
%\includegraphics[trim= 0cm 0cm 0cm 0cm,clip=true,width=11cm]{tfi1.pdf}
%\end{array}$
\end{center}
\caption{Molecular structure of iodotrifluoroethylene and
the table of Hamiltonian parameters in Hz:
chemical shifts (diagonal elements) and J-coupling constants
(off-diagonal elements) are given in upper trace.  Single quantum transitions of 
(a) F$_1$, (b) F$_2$, and  (c) F$_3$ spins, labelled by states of other spins are also shown.}
\label{tfi1}
\end{figure}

\begin{figure}[h]
	%\noindent\makebox[\linewidth]{\rule{15 cm}{1pt}}
	\begin{center}
		\includegraphics[trim= 5cm 6.5cm 5cm 3cm, clip=true,width=10cm]{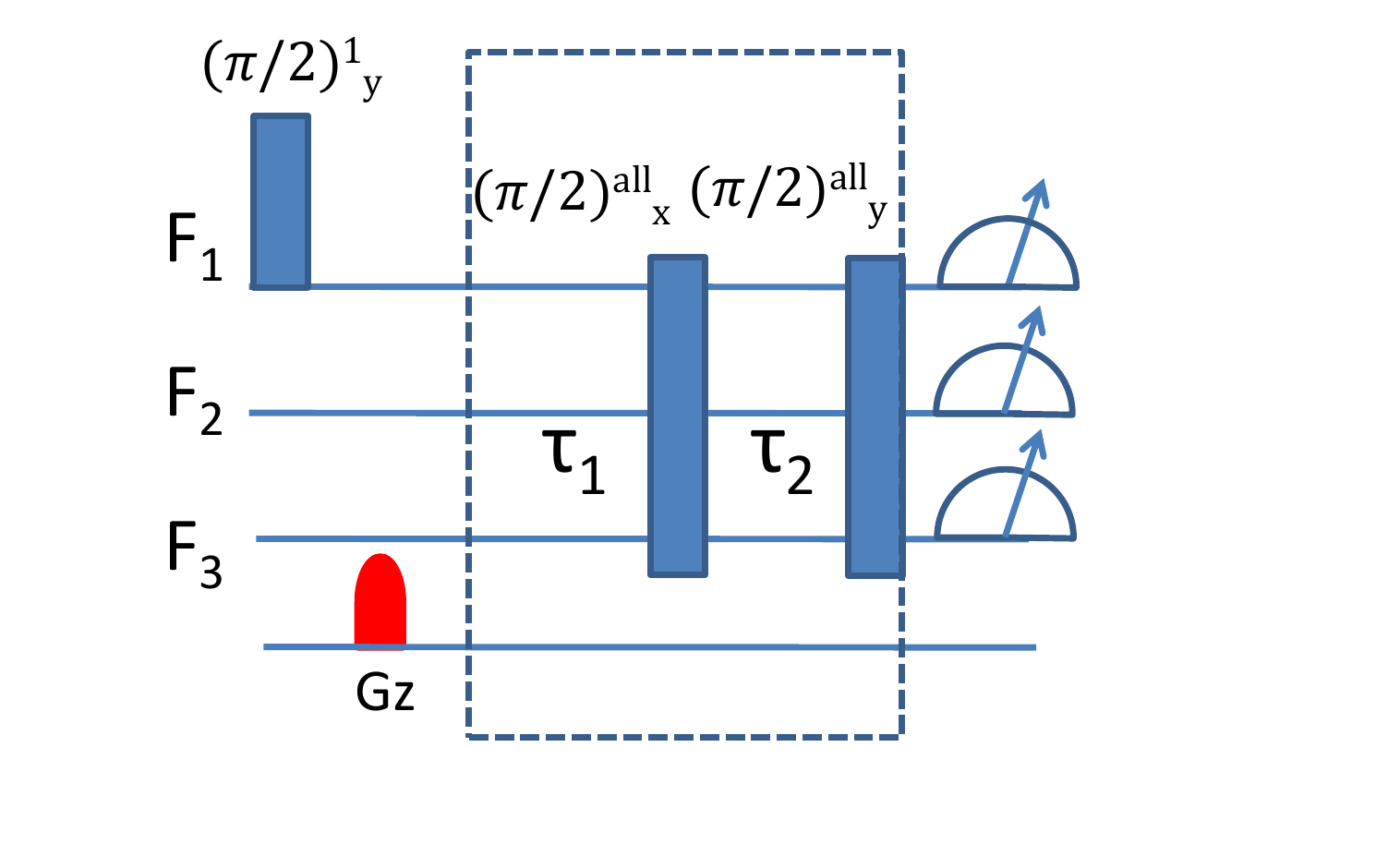} 
		\caption{
			The pulse sequence for two qubit AAQST.  A $\frac{\pi}{2}$ pulse followed by a gradient prepares first spin into maximally mixed state.
			The pulse sequence corresponding to unitary $U_{1}$ is shown inside the dotted block.  Unitaries $U_{int}(\tau_1)$ and $U_{int}(\tau_2)$ are realized by delays $\tau_{1}$ and $\tau_{2}$.  The ${\frac{\pi}{2}}^{all}$ are shown by solid boxes.}
		\label{aaqstpul} 
	\end{center}
	%\noindent\makebox[\linewidth]{\rule{15 cm}{1pt}}
\end{figure}

\begin{figure}[h]
\begin{center}
%\noindent\makebox[\linewidth]{\rule{15 cm}{1pt}}
\includegraphics[trim=0.6cm 0cm 0.5cm 0.2cm, clip=true,width=14.8cm]{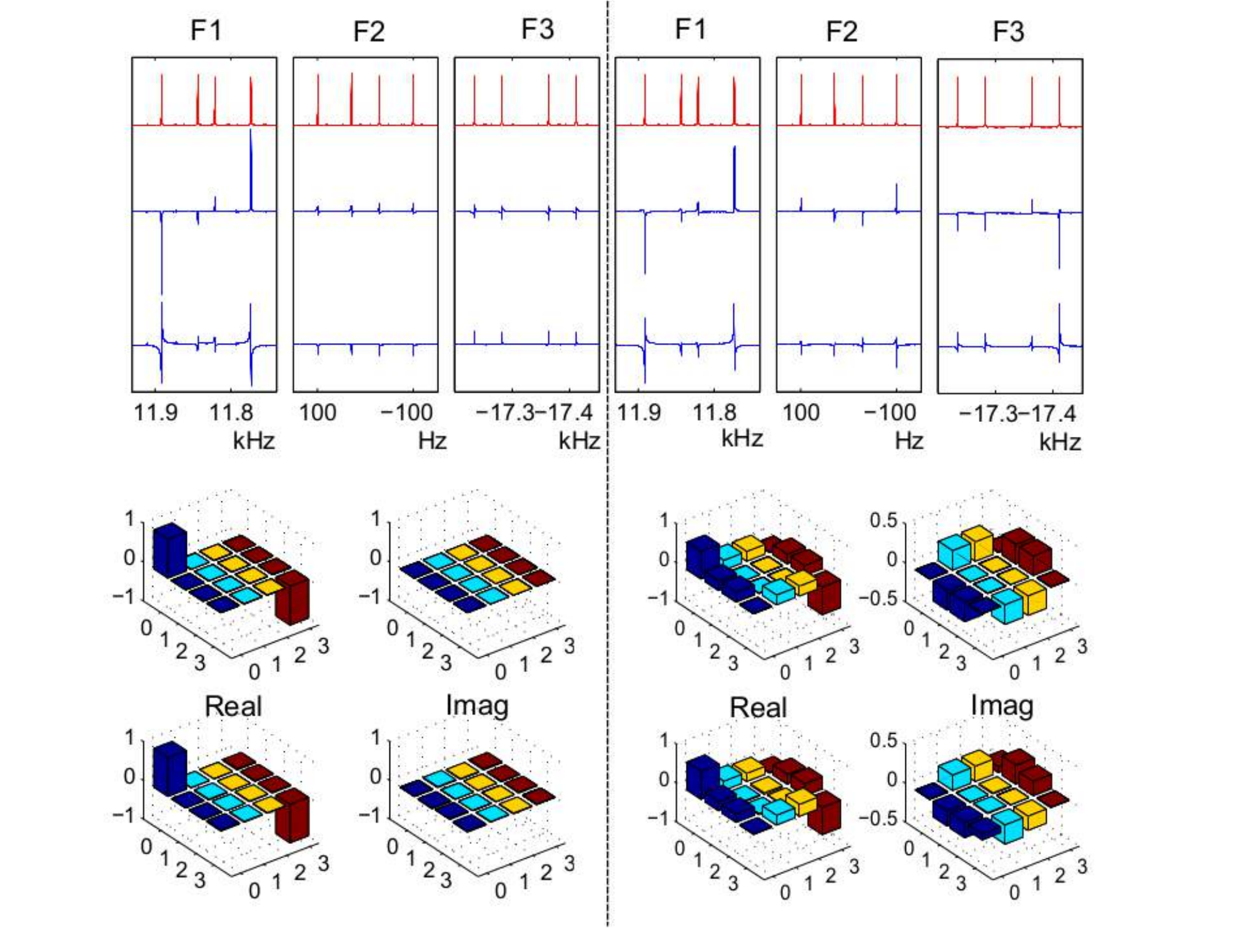} 
\caption{
AAQST results for
thermal equilibrium state $\rho_1$ (left column), and that of state $\rho_2$
(right column), described in the text.  
The reference spectra is in the top trace.  
 The spectra
corresponding to the real part ($R_{j\nu,j\nu'}^{(1)}$, middle trace)
and the imaginary part ($S_{j\nu,j\nu'}^{(1)}$, bottom trace) 
of the $^{19}$F signal are obtained in a single shot AAQST experiment.
The bar plots correspond to theoretically expected
states (top row) and those obtained from AAQST 
experiments (bottom row).
Fidelities of the states are 0.997 and 0.99 respectively
for the two density matrices.
}
\label{fffres} 
%\noindent\makebox[\linewidth]{\rule{15 cm}{1pt}}
\end{center}
\end{figure}

\subsection{Two-qubit input, One-qubit ancilla} \label{tqoa}
Here we use three spin-1/2 $^{19}$F nuclei of 
iodotrifluoroethylene (C$_2$F$_3$I) dissolved in acetone-D$_6$
as a 3-qubit system.
The molecular structure and the Hamiltonian parameters are given
in upper trace of Fig. \ref{tfi1}.  Single quantum transitions of 
each spin, labelled by other spin states are also shown in Fig. \ref{tfi1}(a, b, c).
As can be seen in Fig.\ref{fffres}, all the 12 transitions 
of this system are clearly resolved.

The pulse sequence for the AAQST experiment is shown in Fig. \ref{aaqstpul}.
We have chosen $F_1$ as the ancilla
qubit and $F_2$ and $F_3$ as the input qubits.
QST was performed for two different density matrices
(i) thermal equilibrium state, i.e., 
$\rho_1 = \frac{1}{2} \left( \sigma_z^2+\sigma_z^3 \right)$,
and 
(ii) state after a $(\pi/4)_{\pi/4}$ pulse applied
to the thermal equilibrium state, i.e.,
$
\rho_2 =
\frac{1}{2}\left(\sigma_x^2+\sigma_x^3\right)
-\frac{1}{2}\left(\sigma_y^2+\sigma_y^3\right)
+\frac{1}{\sqrt{2}}\left(\sigma_z^2+\sigma_z^3\right)
$.
In both the cases, the first qubit was initialized
into a maximally mixed state by applying a selective $(\pi/2)_y$
pulse on $F_1$ and followed by a strong pulsed-field-gradient (PFG)
in the $z$-direction.
The selective pulse was realized by GRAPE technique \cite{khaneja2005optimal}.

AAQST of each of the above density matrices required just one 
unitary evolution followed by the measurement of complex NMR signal. 
We modelled the AAQST unitary as follows: 
$U_1 = \left(\frac{\pi}{2}\right)_y U_\mathrm{int}(\tau_2) 
\left(\frac{\pi}{2}\right)_x U_\mathrm{int}(\tau_1)$,
where $U_\mathrm{int}(\tau) = \exp\left(-i{\cal H}\tau\right)$ is
the unitary operator for evolution under the internal Hamiltonian
${\cal H}$ (see eqn. (\ref{ham})) for a time $\tau$, and $\left(\frac{\pi}{2}\right)$ rotations
are realized by non selective radio frequency pulses applied to all the spins
along the directions indicated by the subscripts.  
The constraint matrix $M$ had 15 columns corresponding
to the unknowns and 24 rows corresponding to the real and imaginary
parts of the 12 spectral lines.
Only the durations
$\left\{\tau_1,\tau_2 \right\}$ needed to be optimized to minimize the condition
number $C(M)$.  We used a genetic algorithm for the optimization and obtained
$C(M) = 17.3$ for $\tau_1 =  6.7783$ ms and $\tau_2 = 8.0182$ ms.
The real and imaginary parts of the single shot experimental AAQST spectrum,
along with the reference spectrum,
are shown in the top part of Fig. \ref{fffres}.  The intensities
$\{R_{j\nu,j\nu'}^{(1)},S_{j\nu,j\nu'}^{(1)}\}$
were obtained by simple curve-fit routines, and the matrix eqn. (\ref{meq})
was solved to obtain all the unknowns.  The reconstructed density matrices
along with the theoretically expected ones are shown below the spectra in 
Fig. \ref{fffres}. 
The fidelities of experimental states with
the theoretically expected states ($\rho_1$ and $\rho_2$)
are respectively 0.998 and 0.990.  The high fidelities indicated successful
AAQST of the prepared states.

\begin{figure}[h]
\begin{center}
%\begin{array}{rr}
\includegraphics[trim= 0cm 0cm 0cm 0cm,clip=true,width= 13cm]{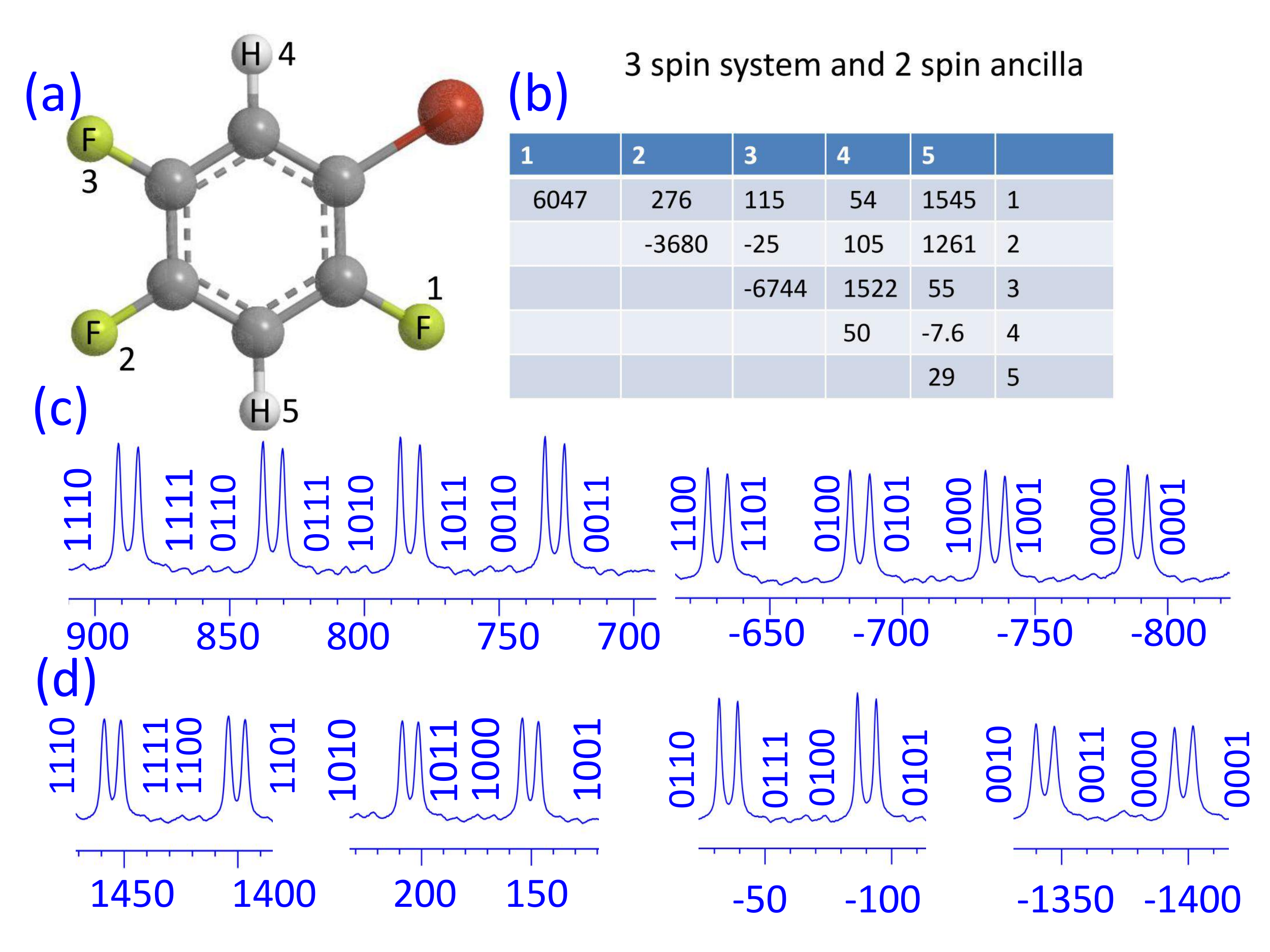}
%\includegraphics[trim= 0cm 0cm 0cm 0cm,clip=true,width=11cm]{labelbtfbzF2.pdf}
%\end{array}$
\caption{(a)  Molecular structure and
(b) the table of Hamiltonian parameters of 1-bromo 2,4,5- trifluorobenzene in Hz:
chemical shifts (diagonal elements) and J-coupling constants (off-diagonal elements). Single quantum transitions for (c) H$_4$, (d) H$_5$ spins, labelled by states of other spins are also shown.}
\label{btfbz1}
\end{center}
\end{figure}

\begin{figure}[h]
\begin{center}
\includegraphics[trim= 0cm 2cm 1cm 2cm,clip=true,width=13cm]{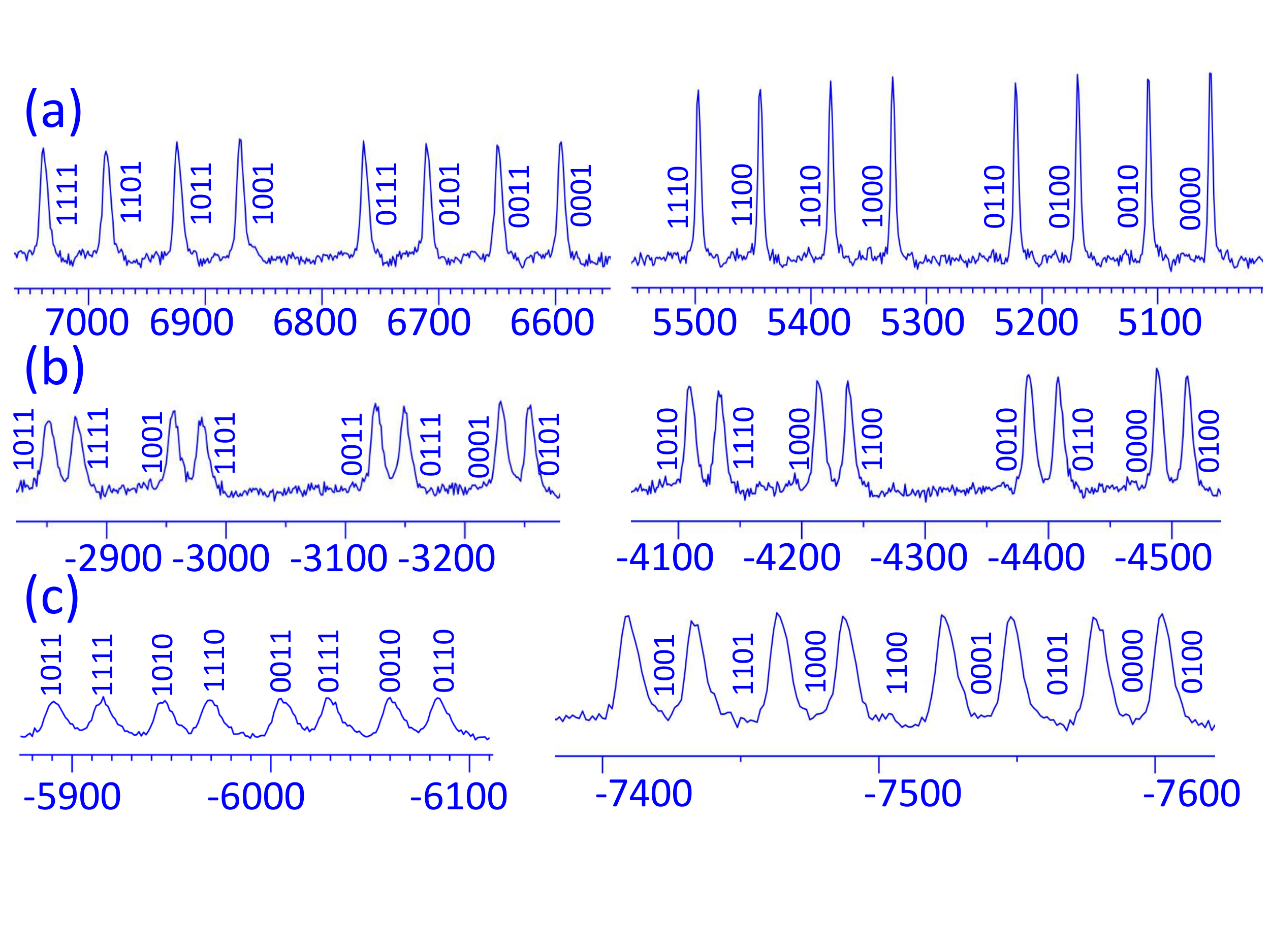}
\caption{Single quantum transitions of (a) F$_1$, (b) F$_2$, and (c) F$_3$ spins in 1-bromo-2,4,5-trifluorobenzene are shown.  Transitions are labelled by states of other spins.}
\label{btfbz2}
\end{center}
\end{figure}

\subsection{Three-qubit input, Two-qubit ancilla} \label{tqta}
We use three $^{19}$F nuclei and two $^1$H nuclei
of 1-bromo-2,4,5-trifluorobenzene partially oriented
in a liquid crystal namely, N-(4-methoxybenzaldehyde)-4-
butylaniline (MBBA). Due to the partial orientational
order, the direct spin-spin interaction (dipolar interaction) 
does not get fully averaged out, but gets scaled down
by the order parameter \cite{dongbook1997}.  The molecular structure,
 the chemical shifts, the 
strengths of the effective couplings, and $^{1}$H and $^{19}$F spectra
 of the above sample are shown in Fig. \ref{btfbz1} and \ref{btfbz2}.
Single quantum transitions corresponding to each spins as shown in 
Fig. \ref{btfbz1}(c, d, e) and Fig. \ref{btfbz2}a, b, c are labelled 
by states of other spins.are shown in
Fig. \ref{btfbz1}a and Fig. \ref{btfbz1}b.    As is evident, the partially oriented
system can display stronger and longer-range coupling
network leading to a larger register.  Here we choose 
the three $^{19}$F nuclei forming the input register
and two $^1$H nuclei forming the ancilla register.
The Hamiltonian for the heteronuclear dipolar interaction 
(between $^1$H and $^{19}$F)
has an identical form as that of J-interaction \cite{dongbook1997}.
The homonuclear dipolar couplings 
(among $^{19}$F, as well as among $^{1}$H nuclei) were small
compared to their chemical shift differences enabling us to approximate
the Hamiltonian in the form of eqn. (\ref{ham}).

\begin{figure}[h]
%\noindent\makebox[\linewidth]{\rule{15 cm}{1pt}}
\begin{center}
%\vspace*{0.5cm}
%\hspace{2cm}
\includegraphics[trim=1cm 2cm 0cm 0cm, clip=true,width= 9.8cm]{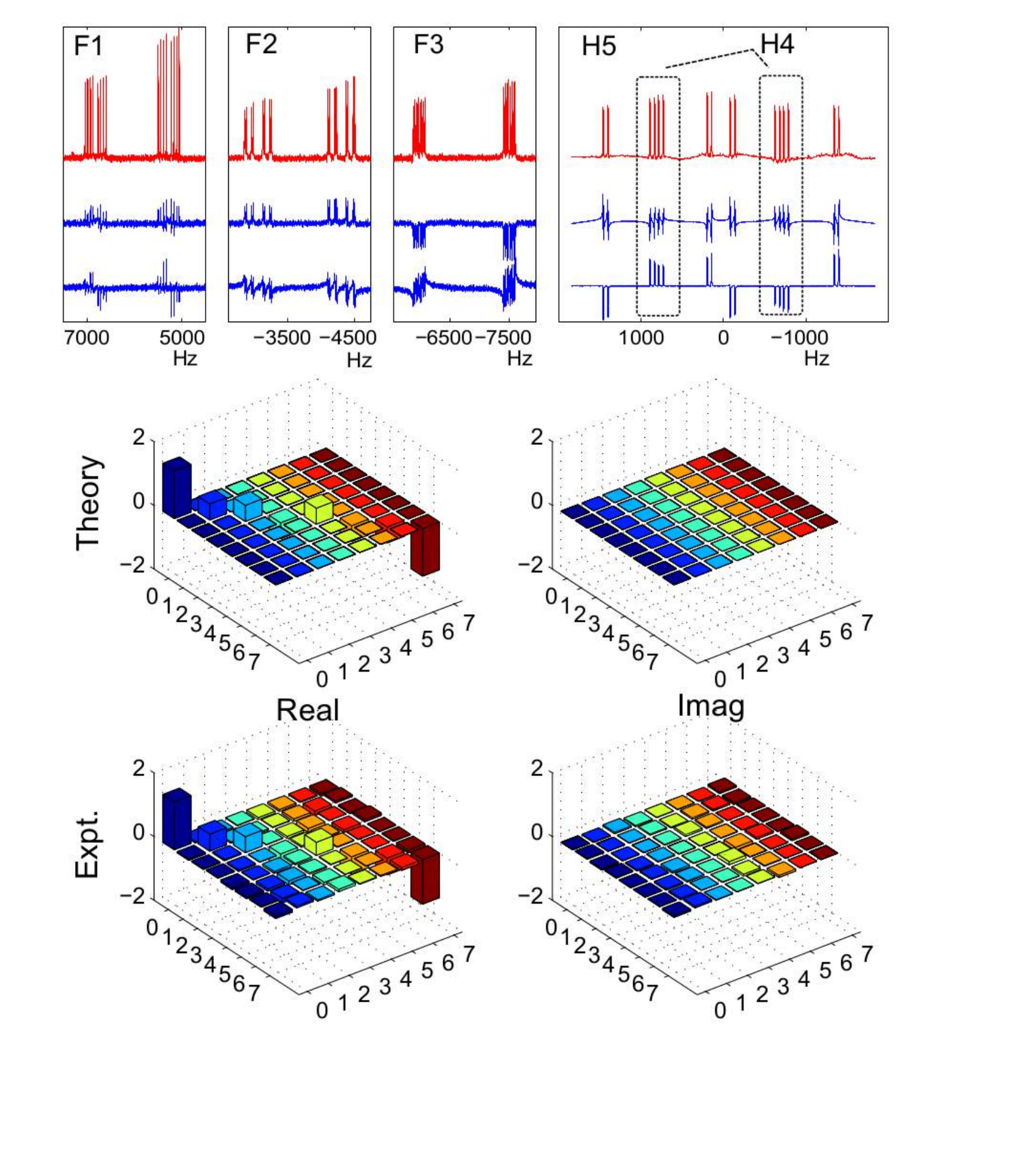} 
\caption{
AAQST results for
thermal equilibrium state, i.e., $(\sigma_z^1+\sigma_z^2+\sigma_z^3)/2$.
The reference spectrum is in the top trace.  The spectra
corresponding to the real part ($R_{j\nu,j\nu'}^{(1)}$, middle trace)
and the imaginary part ($R_{j\nu,j\nu'}^{(1)}$, bottom trace) 
of the $^{19}$F signal are obtained in a single shot AAQST experiment.
The bar plots correspond to theoretically expected
states (top row) and those obtained from AAQST 
experiments (bottom row). Fidelity of the AAQST state is 0.98.
}
\label{btfbzres1} 
\end{center}
%\noindent\makebox[\linewidth]{\rule{15 cm}{1pt}}
%\vspace*{0.5cm}
\end{figure}

\begin{figure}[h]
%\noindent\makebox[\linewidth]{\rule{15 cm}{1pt}}
\begin{center}
\includegraphics[trim=0.7cm 1.8cm 1.2cm 0cm, clip=true,width=10cm]{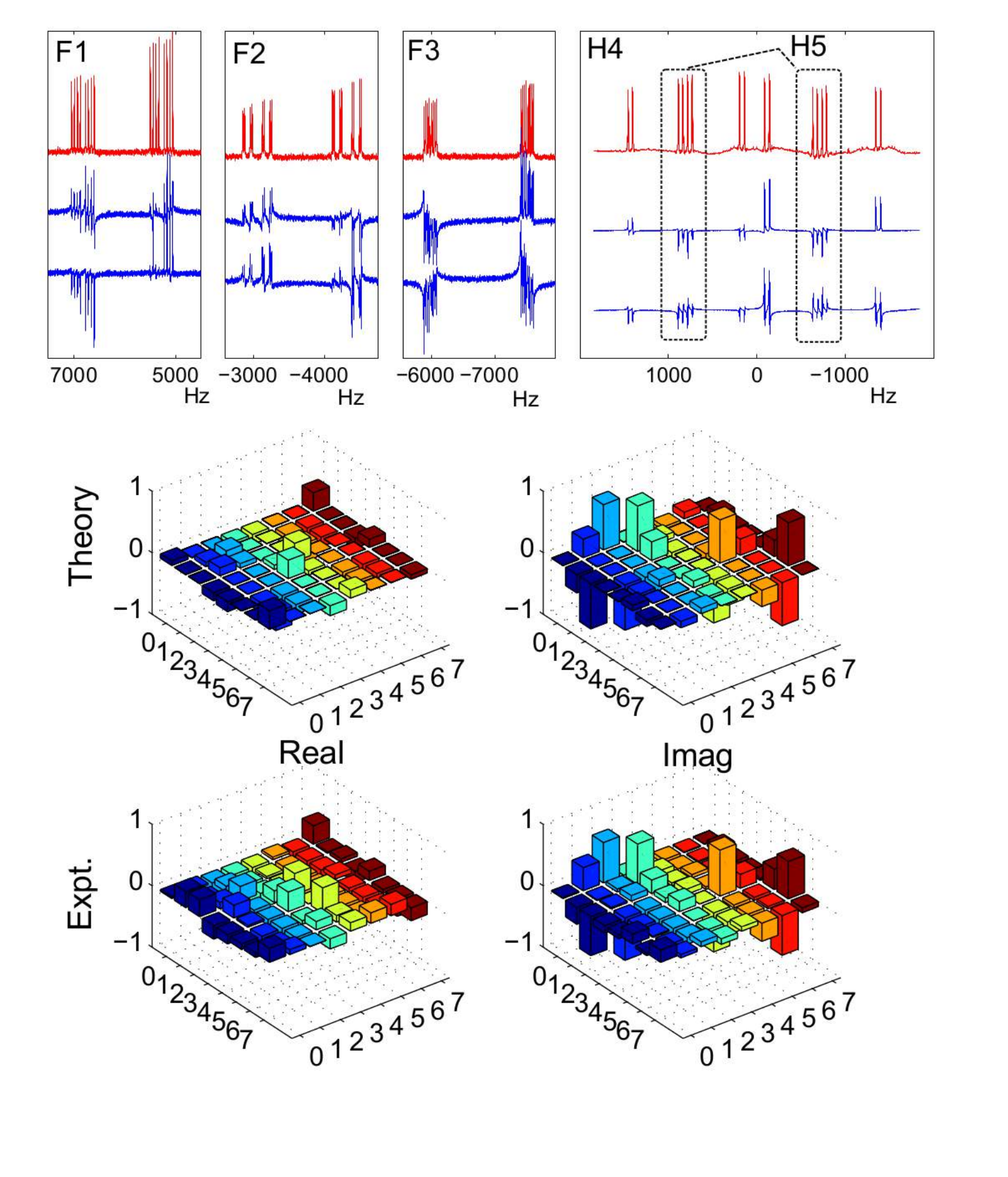} 
\caption{
AAQST results for the state $\rho_2$ described in the text.
The reference spectrum is in the top trace.  The
real (middle trace) and the imaginary spectra (bottom trace) 
are obtained in a single shot AAQST experiment.
The bar plots correspond to theoretically expected
states (top row) and those obtained from AAQST 
experiments (bottom row). Fidelity of the AAQST state is 0.95.
}
\label{btfbzres2} 
\end{center}
%\noindent\makebox[\linewidth]{\rule{15 cm}{1pt}}
\end{figure}

The partially oriented spin-system yields all the 80 transitions
sufficiently resolved.  
Again we use just one experiment for the complete AAQST of the 3-qubit
input register.
We modelled the AAQST unitary in a similar way as before: 
$U_1 = \left(\frac{\pi}{2}\right)_x U_\mathrm{int}(\tau_2) 
\left(\frac{\pi}{2}\right)_x U_\mathrm{int}(\tau_1)$
where $U_\mathrm{int}(\tau) = \exp\left(-i{\cal H}\tau\right)$ is
the unitary operator for evolution under the internal Hamiltonian
${\cal H}$ (see eqn. (\ref{ham})) for a time $\tau$, and $\left(\frac{\pi}{2}\right)_x$ are 
global x-rotations.  The constraint matrix $M$ had 63 columns corresponding
to the unknowns and 160 rows corresponding to the real and imaginary
parts of 80 spectral lines.
After optimizing the durations by minimizing the condition number
using a genetic algorithm, we obtained $C(M) = 14.6$
for $\tau_1 = 431.2 \upmu$s and $\tau_2 = 511.5\upmu$s.
Again we study AAQST on two states:
(i) Thermal equilibrium of the $^{19}$F spins: $\rho_1 = (\sigma_z^1+\sigma_z^2+\sigma_z^3)/2$, 
and 
(ii) a random density matrix $\rho_2$ obtained by applying unitary
$
U_0 = \left(\frac{\pi}{2}\right)_x^{F} \tau_0 (\pi)_x^{H} \tau_0 \left(\frac{\pi}{2}\right)_y^{F_1},
$ with $\tau_0 = 2.5$ ms, on thermal equilibrium state, i.e., $\rho_2 = U_0 \rho_1 U_0^\dagger$.
In both the cases, we initialize the ancilla i.e., the $^1$H qubits on to
a maximally mixed state by first applying a $(\pi/2)^{H}$ pulse followed
by a strong PFG in the $z$-direction.
\begin{figure}[h]
%\noindent\makebox[\linewidth]{\rule{15 cm}{1pt}}
\begin{center}
\includegraphics[trim=0cm 0cm 0cm 0cm, clip=true,width=10cm]{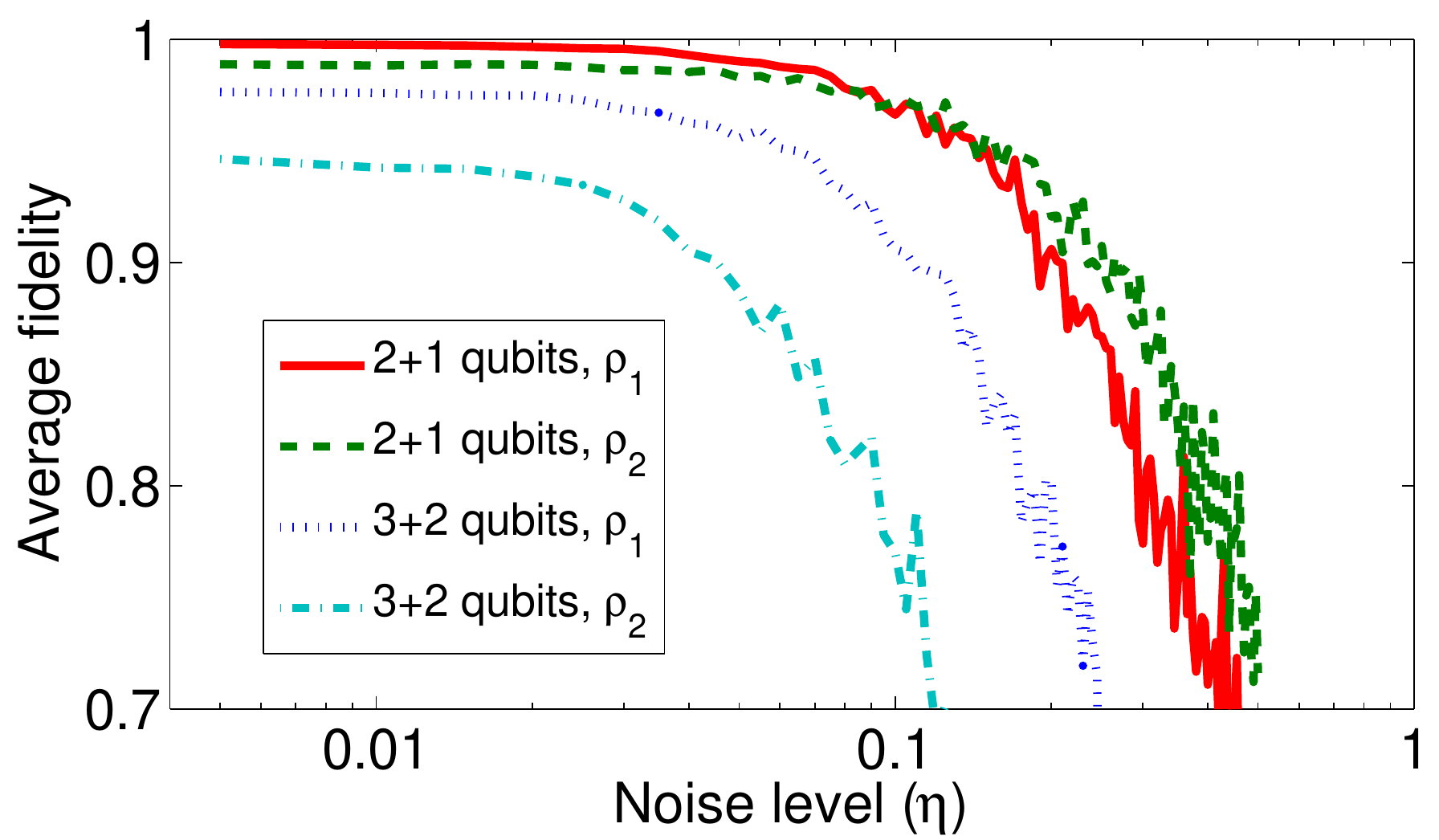} 
\caption{
Average fidelity of the characterised state for simulated noise level. The simulation procedure is described in the text.  We have carried out this study for all four experimentally characterised density matrices  as shown in the legend.  The expressions for density matrices $\rho_1$ and $\rho_2$ are mentioned in \S\ref{tqoa} and \S \ref{tqta}.  Results show robustness of the protocal}.
\label{robust} 
\end{center}
%\noindent\makebox[\linewidth]{\rule{15 cm}{1pt}}
\end{figure}
The real and imaginary parts of the single shot AAQST spectra, 
%needed for the complete tomography of the 3-qubit input register
%which is prepared in the two target density matrices ($\rho_1$ and $\rho_2$), 
along with the reference spectra, are shown in
Figs. \ref{btfbzres1} and \ref{btfbzres2} respectively.  
Again the line intensities
$\{R^{(1)}_{j\nu,j\nu'},S^{(1)}_{j\nu,j\nu'}\}$ are obtained by curve-fitting, and all the 63
unknowns of the 3-qubit deviation density matrix are obtained by solving the 
matrix eqn. (\ref{meq}). The reconstructed density matrices 
along with the theoretically expected states ($\rho_1$ and $\rho_2$)
are shown below the spectra 
in Figs. \ref{btfbzres1} and \ref{btfbzres2}. 
The fidelities of experimental states with
the theoretically expected states ($\rho_1$ and $\rho_2$)
are respectively 0.98 and 0.95.  The lower fidelity in the latter case
is mainly due the imperfections in the preparation of the target state
$\rho_2$.  The overall poorer performance in the liquid crystal system
is due to the lower fidelities of the QST pulses, spatial and temporal 
variations of solute order-parameter, and stronger decoherence rates
compared to the isotropic case.  In spite of these difficulties, 
the three-qubit density matrix with 63 unknowns could be 
estimated quantitatively through a single NMR experiment.
\section{Robustness}
We have also simulated the robustness of the AAQST protocal against simulated noise and the results are shown in Fig. \ref{robust}. To realize a noisy experimental output we have we have externally added the noise (a random-number array) into the measured output signals before reconstructing the  density matrix. This procedure has been repeated for the various noise levels $\eta$. The average fidelity drop against the corresponding noise level for all four cases namely two-qubit input, single-qubit ancilla ($\rho_1$ and $\rho_2$) and three-qubit input, two-qubit ancilla ($\rho_1$ and $\rho_2$) are shown.    

\section{\textbf{Conclusions}\label{24}}
Quantum state tomography is an important part of experimental 
studies in quantum information processing.  The standard method 
involves a large number of independent measurements to reconstruct
a density matrix.  The ancilla-assisted quantum state tomography
introduced by Nieuwenhuizen and co-workers allows complete reconstruction
of complex density matrix with fewer experiments by letting the 
unknown state of the input register to interact with an ancilla
register initialized in a known state.  
Ancilla registers are essential in many of the quantum algorithms.
Usually, at the end of the quantum algorithms,
ancilla is brought to a state which is separable with the input 
register.  The same ancilla register which is used for computation
can be utilized for tomography after the computation. 
The ancilla register can be prepared into a maximally mixed state
by dephasing all the coherences and equalizing the populations.
We provided methods for
explicit construction of tomography matrices in large registers.  
We also discussed the optimization of tomography 
experiments based on minimization of the condition number of the 
constraint matrix.  Finally, we demonstrated the experimental 
ancilla-assisted quantum state tomography in two systems:  
(i) a system with two input qubits and one ancilla qubit in an isotropic medium
and (ii) a system with three input qubits and two ancilla qubits
in a partially oriented medium.  In both the cases, we 
successfully reconstructed the target density matrices 
with a single quadrature detection of transverse magnetization.  
The methods introduced in this work should be useful for extending 
the range of quantum state tomography to larger registers.
%\section{\textbf{Appendix}}

%\begin{figure}[h]
%\begin{center}
%%\hspace{2.8cm}
%\includegraphics[width=10cm]{labelbtfbzF2.pdf} 
%\caption{Minimum number of independent experiments required in QST
%(without ancilla) and AAQST (with different number of ancilla register).  
%%The cases with experimental demonstrations are indicated with circles.
%}
%\label{exptscaling}
%\end{center} 
%%\noindent\makebox[\linewidth]{\rule{15 cm}{1pt}}
%\vspace*{0.5cm}
%\end{figure}

%\begin{figure}[h]
%\begin{center}
%%\hspace{2.8cm}
%\includegraphics[width=12cm]{labelbtfbzH.pdf} 
%\caption{Minimum number of independent experiments required in QST
%(without ancilla) and AAQST (with different number of ancilla register).  
%%The cases with experimental demonstrations are indicated with circles.
%}
%\label{exptscaling}
%\end{center} 
%%\noindent\makebox[\linewidth]{\rule{15 cm}{1pt}}
%\vspace*{0.5cm}
%\end{figure}

%\bibliographystyle{apsrev4-1}
%\bibliography{biblio}
%\end{document}
\thispagestyle{empty}

%  % % % % % %ch3
\titlespacing*{\chapter}{0pt}{-50pt}{20pt}
\titleformat{\chapter}[display]{\normalfont\Large\bfseries}{\chaptertitlename\ \thechapter}{20pt}{\Large}
\chapter{Single-scan quantum process tomography \label{chp3}}

\section{\textbf{Introduction}\label{31}}
An open quantum system may undergo an evolution
due to intentional control fields as well as due to
unintentional interactions with stray fields caused by environmental
fluctuations.  In practice, even a carefully designed control field may
be imperfect to the extent that one might need to characterize
the overall process acting on the quantum system.  Such a
characterization, achieved by a procedure called quantum
process tomography (QPT), is crucial in the physical realization of a
fault-tolerant quantum processor  \cite{chuang97,zollar97}. 
QPT is realized by considering the quantum process as a map from
a complete set of initial states to final states, and experimentally
characterizing each of the final states using quantum state tomography (QST) \cite{ChuangPRSL98}.
Since the spectral decomposition of a density matrix may involve
noncommuting observables,
Heisenberg's uncertainty principle demands multiple experiments to 
characterize the quantum state.
Thus QST by itself involves the measurement of a series of observables 
after identical preparations of the system in the quantum state. 
Hence, QPT in general requires a number of independent 
experiments, each involving initialization of the quantum system, 
applying the process to be characterized, and finally QST.
Furthermore, the total number of independent measurements required for QPT increases 
exponentially with the size of the system undergoing the process.

The physical realization of QPT has been demonstrated on various 
experimental setups such as NMR \cite{ChuangPRA2001,QPTofQFT}, linear optics \cite{EAPT1Exp,altepeter,obrian,bellstatefilter},
ion traps \cite{qptITprl2006,QPT_IonTrapNature2010}, superconducting qubits
\cite{QPT_SQUID,SQUID2009,MartiniSQUID2010,QPT2spSQUID,QPT_SQUIDChow2011,SQUID_Dewes2012}, and
NV center qubit \cite{suterprotectedgate}.  
Several developments in the methodology of QPT have also been
reported \cite{simplifiedQPT1,simplifiedQPT2}.  In particular, it has been shown that ancilla assisted 
process tomography (AAPT) can characterize a process with a single QST
\cite{mazzei2003pauli,EAPT1Exp,PRLAriano2003,altepeter}.
However, it still requires multiple
measurements each taken over a set of commuting observables. 
On the other hand, if sufficient ancilla qubits are available, QST can be
carried out with a single ensemble measurement (i.e., a single scan)
over the entire system-ancilla space.
This procedure, known as ancilla assisted quantum state tomography
(AAQST), has been studied both theoretically and experimentally
\cite{Nieuwenhuizen,sutertomo,pengtomo,abhishekQST2013,QPT_SQUID}.
Here we combine AAPT with AAQST and realize a `single-scan quantum process tomography' 
(SSPT), which can characterize a general process in a single ensemble measurement 
of the system-ancilla state.

In the next section, after briefly revising QPT and AAPT, I describe SSPT procedure.  In
 \S \ref{33}, I illustrate our SSPT procedure using a three-qubit NMR 
quantum register.  I also present our characterization of certain unitary processes corresponding to 
standard quantum gates and a nonunitary process, namely twirling
operation.  Finally I conclude in \S \ref{34}.
\vspace{-0.5 cm}
\begin{figure}
\vspace{0.5 cm}
%\noindent\makebox[\linewidth]{\rule{20 cm}{1 pt}}
\begin{center}
\hspace*{0.3 cm}
\includegraphics[trim=0cm 0cm 0cm 0cm,clip=true,width=13cm,height= 15 cm]{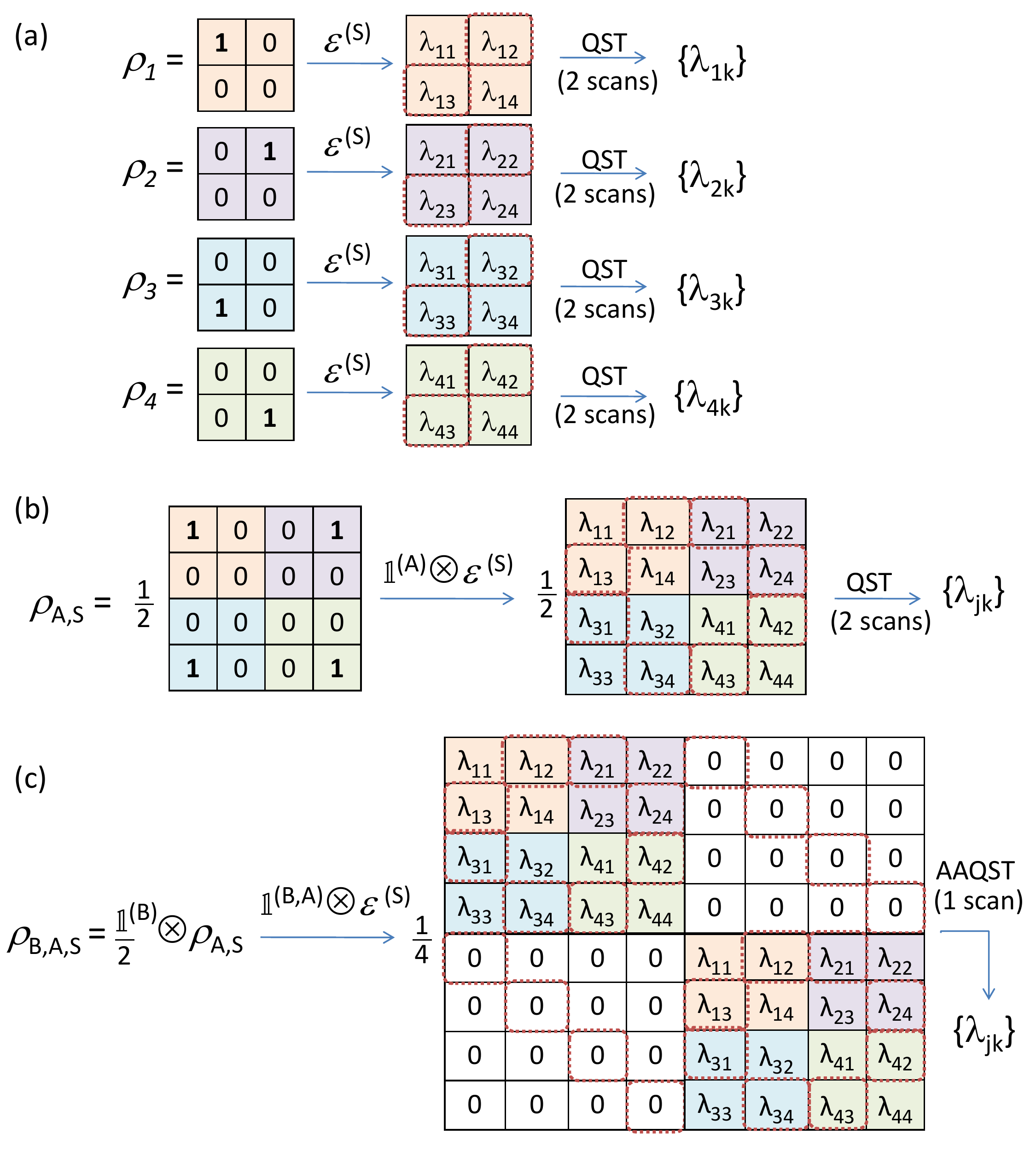} 
\caption{Illustrating (a) single-qubit QPT 
requiring a total of 8 NMR measurements, (b) AAPT requiring 2 NMR measurements, and (c) SSPT
requiring a single NMR measurement. 
}
\label{sum} 
\end{center}
\vspace{0.5 cm}
%\noindent\makebox[\linewidth]{\rule{20 cm}{1 pt}}
\end{figure}

\section{\textbf{Theory}\label{32}}
\subsection{Quantum Process Tomography (QPT)}
A process $\varepsilon$ maps a quantum state $\rho$ 
to another state $\varepsilon(\rho)$.  Here we consider an $n$-qubit system with
$N^2( = 2^{2n})$-dimensional Liouville space $S$.
In order to characterize $\varepsilon$,
we let the process act on each linearly independent element of a complete basis set $\{\rho_1, \rho_2, \cdots, \rho_{N^2}\}$. 
Expressing each output state in the complete basis we obtain 
\begin{eqnarray}
\varepsilon(\rho_j) = \sum_k \lambda_{jk}\rho_k,
\label{lb}
\end{eqnarray}
where the complex coefficients $\lambda_{jk}$ can be extracted after QST.

The outcome of a trace-preserving quantum process $\varepsilon$ also has an operator-sum representation 
\begin{eqnarray}
\varepsilon(\rho) = \sum_i E_i \rho E_i^\dagger,
\end{eqnarray}
where the \textit{Kraus operators} $E_i$ satisfy the completeness relation $\sum_i E_i^\dagger E_i = \mathbb{I}$.
To assist experimental characterization of the process, we utilize 
a fixed set of basis operators $\{\tilde{E}_m\}$, and express 
$E_i = \sum_m e_{im} \tilde{E}_m$.
The process is now described by
\begin{eqnarray}
\varepsilon(\rho) = \sum_{mn}\tilde{E}_m \rho \tilde{E}_n^\dagger \chi_{mn},
\label{tildeE}
\end{eqnarray}
where $\chi_{mn} = \sum_i e_{im} e_{in}^*$ form a complex matrix which completely characterizes the process $\varepsilon$.
Since the set $\{\rho_k\}$ forms a complete basis, it is also possible to express
\begin{eqnarray}
\tilde{E}_m\rho_j\tilde{E}_n^\dagger = \sum_k \beta_{jk}^{mn}\rho_k,
\label{beta}
\end{eqnarray}
where $\beta_{jk}^{mn}$ can be calculated theoretically. Eqns. \ref{lb}, \ref{tildeE}, and \ref{beta}
lead to
\begin{eqnarray}
\varepsilon(\rho_j) =  \sum_k \lambda_{jk}\rho_k  = \sum_k \sum_{mn} \beta_{jk}^{mn} \chi_{mn} \rho_k.
\end{eqnarray}
Exploiting the linear independence of $\{\rho_k\}$, one obtains the matrix equation
\begin{eqnarray}
\beta \chi = \lambda,
\label{chi}
\end{eqnarray}
from which $\chi$-matrix can be extracted by standard methods in linear algebra.

For example, in the case of a single qubit, one can choose the linearly independent basis
$\{\outpr{0}{0},\outpr{0}{1},\outpr{1}{0},\outpr{1}{1}\}$ (see Fig. \ref{sum}a). 
While the middle-two elements are non-Hermitian, they
can be realized as a linear combination of Hermitian density operators \cite{chuangbook}.
A fixed set of operators $\{I,X,-iY,Z\}$ can be used to express the
$\chi$ matrix.  Thus the standard single-qubit QPT procedure requires four QST experiments.

QPT on an $N$-dimensional system requires $N^2$-QST experiments, where
a single QST involves several quantum measurements each taken jointly over a set of
commuting observables.  The exact number of measurements required for QST may depend on the
properties of available detectors. 

In NMR, a single-scan experiment allows us to detect all the single-quantum
elements of the density matrix (see Fig. \ref{sum}).  For example,  real and imaginary part of NMR signal of a two qubit system
together consists of eight transitions. Transitions in real spectrum corresponding to the four observables are
Observables, corresponding to the transitions in real part are $\{\sigma_x\otimes\proj{0},\sigma_x\otimes\proj{1},\proj{0}\otimes\sigma_x,\proj{1}\otimes\sigma_x\}$ and 
imaginary part are $\{\sigma_y\otimes\proj{0},\sigma_y\otimes\proj{1},\proj{0}\otimes\sigma_y,\proj{1}\otimes\sigma_y\}$.
Thus a quadrature detected NMR signal directly provides information on four density matrix
elements \cite{cavanagh}.  To measure other elements, one needs to transform the
density matrix by a known unitary, and again record the four transitions.  The intensities
of these transitions are proportional to linear combinations of various elements of the
density matrix.  In principle, two experiments suffice for a 2-qubit QST \cite{abhishekQST2013}. 
In the case of an $n$-qubit NMR system with a well resolved spectrum,
QST requires $\simeq \left\lceil \frac{N}{n} \right\rfloor$
measurements, where $\lceil \rfloor$ rounds the argument to next integer \cite{abhishekQST2013}.
Therefore an $n$-qubit QPT needs a total of $M_\mathrm{QPT} \simeq N^2 \left\lceil \frac{N}{n} \right\rfloor$ measurements. 
Estimates of $M$ for a small number of qubits
shown in the first column of Table 1 illustrate the exponential increase of $M_\mathrm{QPT}$ with $n$.

\fontsize{12}{10}
\begin{center}
\begin{table}
%\vspace{-3 cm}
\hspace{2.5 cm}
$
%\begin{array}{|c|c|cc|cc|}
\begin{array}{|cc|cc|cc|cccc|}
           \hline
           n & M_\mathrm{QPT} & M_\mathrm{AAPT} & (n_{A1})& M_\mathrm{SSPT} &(n_{A1}, n_{A2})\\
           \hline
           \hline
           1      &     8     &      2 & (1)     &     1 &(1,1) \\
           \hline
           2    &      32      &     4 &(2)     &     1 &(2,2) \\
           \hline
           3    &     192      &    11 &(3)     &     1 &(3,3) \\
           \hline
           4    &    1024      &    32 &(4)     &     1 &(4,5) \\
           \hline
           5   &     7168    &     103 &(5)     &     1 &(5,6) \\
           \hline
\end{array}
$
\caption{Comparison of number of independent measurements and 
number of ancilla qubits (in parenthesis) required for 
$n$-qubit QPT, AAPT, and SSPT.}
\label{compsspt}
\end{table}
\end{center}
\fontsize{12}{10}

\subsection{Ancilla-Assisted Process Tomography (AAPT)}
\begin{figure}
%\vspace*{1 cm}
%\noindent\makebox[\linewidth]{\rule{12cm}{1 pt}}
 \begin{center}
% \vspace{-1.5cm}
%\hspace*{-0.5cm}
\includegraphics[trim= 3.2cm 4cm 3cm 1cm, clip=true,width = 12cm]{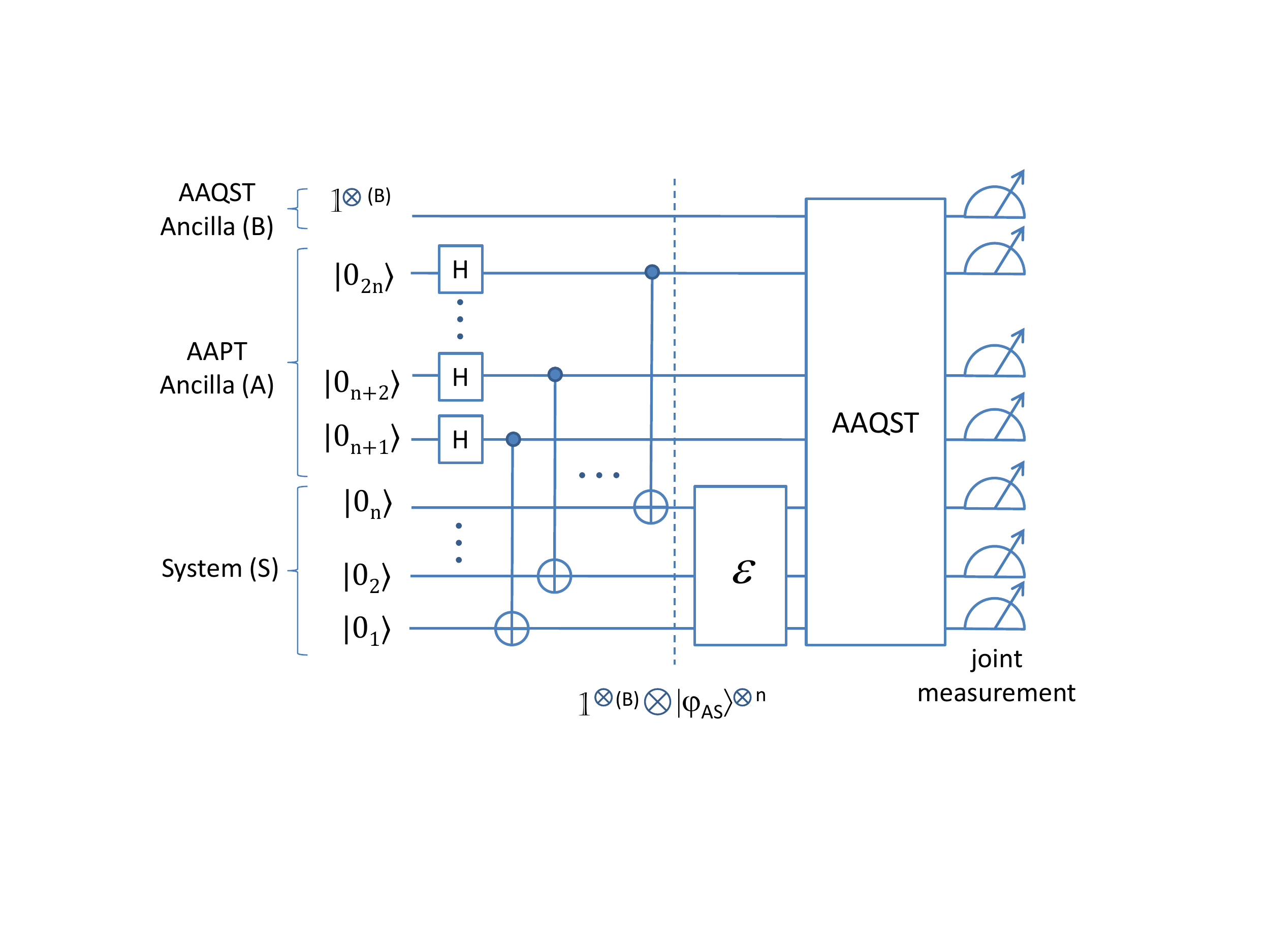} 
\caption{Quantum circuit for SSPT.  Building blocks of the circuit involves initialization of the system and ancilla registers, encoding of the input states into subspaces of system-ancilla register, application of process $\varepsilon$, and finally AAQST.}
\label{qckt} 
\end{center}
%\noindent\makebox[\linewidth]{\rule{12 cm}{1pt}}
%\vspace*{0.5cm}
\end{figure}

If sufficient number of ancillary qubits are available, 
ancilla assisted process tomography (AAPT) can be carried out
by simultaneously encoding all the basis elements onto a higher 
dimensional system-ancilla Liouville space $A \otimes S$ \cite{mazzei2003pauli,EAPT1Exp,PRLAriano2003,altepeter}.
AAPT requires a single final QST, thus greatly reducing the number of independent measurements.
For example, a single-qubit process tomography can be carried out with the
help of an ancillary qubit by preparing the $2-$qubit Bell state 
$\ket{\phi_{AS}} = (\ket{0_A}\ket{0_S}+\ket{1_A}\ket{1_S})/\sqrt{2}$,
applying the process on the system-qubit, and finally carrying out QST of the
two-qubit state (see Fig. \ref{sum}b). While the choice of the
initial state for AAPT is not unique, the above choice provides a simple
way to represent all the four $2\times2$ dimensional basis states directly onto 
different subspaces of the $4\times4$ dimensional density operator (see Figs. \ref{sum}a and \ref{sum}b).
For an $n$-qubit system, all the $N^2$ basis elements
%, each of dimension $N \times N$, 
can be encoded simultaneously in independent subspaces of a single $N^2 \times N^2$ Liouville operator 
belonging to $2n$-qubit space $A \otimes S$. 
A simple choice for the initial state is of the form $\ket{\phi_{AS}}^{\otimes n}$.
The quantum circuit for the preparation of this state is shown in the
first part of Fig. \ref{qckt}.
Thus exactly $n$-ancilla qubits are needed to carry out AAPT
on an $n$-qubit system. 

Although only two independent measurements are needed for a two-qubit QST, this number grows exponentially with the
total number of qubits.
An $n$-qubit AAPT involves a $2n$-qubit QST, and accordingly requires 
$M_{\mathrm{AAPT}}\simeq \left\lceil \frac{N^2}{2n} \right\rfloor$ 
scans \cite{abhishekQST2013}.  The minimum number of scans for a few system-qubits 
are shown in the second column of Table \ref{compsspt}. While AAPT requires significantly
lesser number of measurements compared to QPT, it still scales exponentially with
the number of system-qubits.

%\nopagebreak
\begin{figure}[h]
%\vspace*{1cm}
%\noindent\makebox[\linewidth]{\rule{12cm}{1pt}}
\begin{center}
%\hspace*{-0.5cm}
\includegraphics[trim=0cm 2cm 0cm 1.5cm, clip=true,width= 15cm]{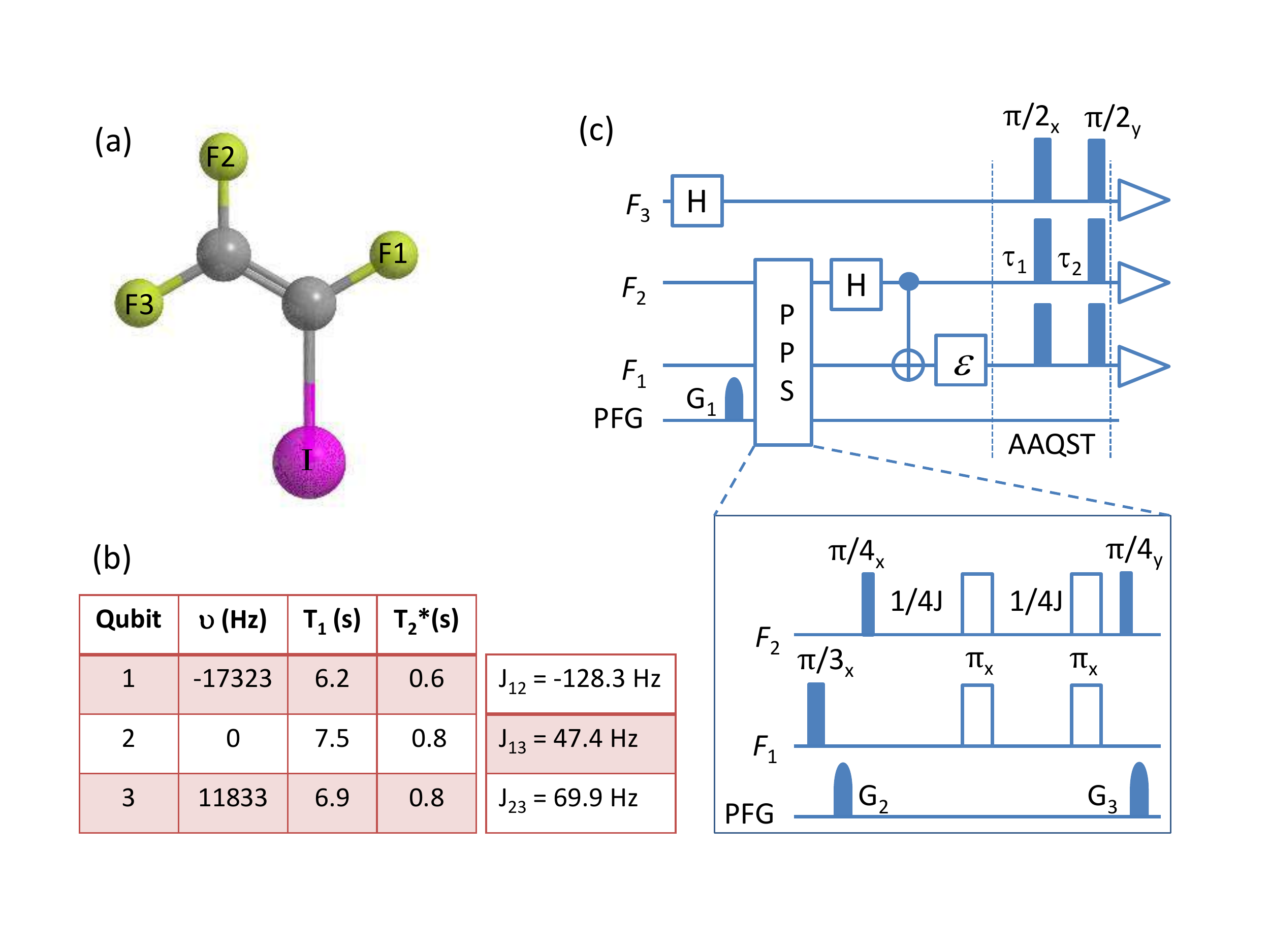} 
\caption{Molecular structure of iodotrifluoroethylene (a), and
the table of Hamiltonian and relaxation parameters (b), NMR pulse-sequence 
to demonstrate SSPT (c). Pulse sequence for preparing $\ket{00}$ pseudopure state
is shown in the inset of (c). \label{fffmol} }
\end{center}
%\vspace*{0.5cm}
%\noindent\makebox[\linewidth]{\rule{20 cm}{1pt}}
\end{figure}

\subsection{Single-Scan Process Tomography (SSPT)}
It had been shown earlier that, if sufficient number 
of ancillary qubits are available, QST of a general density matrix of
arbitrary dimension can be performed with a single-scan  
\cite{sutertomo,pengtomo,abhishekQST2013}.
 This method, known as
ancilla assisted quantum state tomography (AAQST) is based on the redistribution
of all elements of the system density matrix on to a joint density matrix
in the combined system-ancilla Liouville space.
Initially ancilla register for AAQST is prepared in a maximally mixed state thus erasing all information in it
and redistribution of matrix elements is achieved by an optimized joint unitary operator \cite{abhishekQST2013}.
By combining AAPT with AAQST, process tomography can be achieved with
a single-scan measurement of all the qubits (see Fig. \ref{sum}c  and 3rd
column of Table \ref{compsspt}).

If AAQST is carried out with an ancilla space ($B$) of $n_{B}$-qubits, the combined space
$B \otimes A \otimes S$ corresponds to $\tilde{n} = 2n+n_{B}$ qubits.
%AAQST register is  $n_{B}$ and the combined size of 
%A1+S register is $2n$, then the total size of the register is $\tilde{n} = 2n+n_{B}$.
A single-scan measurement suffices if the total number of observables 
is equal to or exceeds the number of real unknowns (i.e., $N^4-1$) in the $2n$-qubit density matrix, i.e.,
if $\tilde{n}{\tilde{N}} \ge  (N^4-1)$,
where $\tilde{N} = 2^{\tilde{n}}$ \cite{abhishekQST2013}.
However, if only pairwise interactions are used between the system and ancilla
of same dimension, then also, a single experiments suffices for AAQST \cite{pengpairwise2014}.
The number of ancillary qubits $n_{A}$ and $n_{B}$ required for SSPT are shown in the third column of
Table \ref{compsspt}.

The complete circuit for SSPT is shown in Fig. \ref{qckt}. It involves two ancilla
registers, one for AAPT and the other for AAQST.  Initially AAQST register is
prepared in a maximally mixed state and the other two registers are set to $\ket{0}^{\otimes{n}}$ states.  
Hadamard gates on the AAPT ancilla followed by C-NOT gates (on system qubits
controlled by ancilla) prepare state $\ket{\phi_{AS}}^{\otimes n}$, which simultaneously 
encodes all the basis elements required for QPT. A single application of the process $\varepsilon$, on the system
qubits, acts simultaneously and independently on all the basis elements $\{\rho_j\}$. 
The final AAQST operation allows estimation of all the elements of the $2n$-qubit
density matrix $\sum_j A^{(j)} \otimes \varepsilon(\rho_j)$, 
where $A^{(j)}$ identifies the $j$th subspace.
The output of each subspace $\varepsilon(\rho_j)$ can now be extracted using a single
scan experiment, and the coefficients $\lambda_{jk} = \mathrm{Tr}[\varepsilon(\rho_j) \rho_k^\dagger]$ can be calculated.

\section{\textbf{Experiments}\label{33}}
We used iodotrifluoroethylene (C$_2$F$_3$I) dissolved in acetone-D$_6$
as a 3-qubit system.  The molecular structure and labelling scheme are shown in Fig. \ref{fffmol}a.  
All the experiments described below are carried out
on a Bruker 500 MHz NMR spectrometer at an ambient temperature of 300 K using high-resolution
NMR techniques.  
The NMR Hamiltonian in this case can be expressed as
\begin{eqnarray}
{\cal H} = -\pi\sum_{i=1}^{3}\nu_i \sigma_z^i  + 
\pi\sum_{i=1,j > i}^{3,3}  J_{ij} \sigma_z^i \sigma_z^j /2
\label{ham}
\end{eqnarray}
where $\sigma_{z}^i$ and $\sigma_{z}^j$ are Pauli $z$-operators
of $i$th and $j$th qubits \cite{cavanagh}. The chemical shifts $\nu_i$,
coupling constants $J_{ij}$, and relaxation parameters (T$_1$ and T$_2^*$)
are shown in Fig.  \ref{fffmol}b.
All the pulses are realized using gradient ascent pulse engineering (GRAPE)
technique \cite{khaneja2005optimal} 
and had average fidelities above 0.99 over 20\% inhomogeneous
RF fields.

We utilize spins F$_1$, F$_2$, and F$_3$ respectively as
the system qubit ($S$), AAPT ancilla ($A$), and AAQST ancilla ($B$). 
The NMR pulse-sequence for SSPT experiments are shown in Fig. \ref{fffmol}c.
It begins with preparing B qubit in the maximally mixed state by
bringing its magnetization into transverse direction using a Hadamard gate,
and subsequently dephasing it using a PFG. 
The remaining qubits are initialized into a pseudopure  $\ket{00}$ state
by applying the standard pulse-sequence shown in the inset of Fig. \ref{fffmol}c \cite{corypnas1997}.
The Bell state $\ket{\phi_{AS}}$ prepared using a
Hadamard-CNOT combination had a fidelity of over 0.99. 
After preparing this state, we applied the process $\varepsilon$ on 
the system qubit. The final AAQST consists of $(\pi/2)_x$ and $(\pi/2)_y$
pulses on all the qubits separated by delays $\tau_1 = 6.7783$ ms and $\tau_2 = 8.0182$ ms
\cite{abhishekQST2013}.
A single-scan measurement of all the qubits now leads to a complex
signal of 12 transitions, from which all the 15 real unknowns of the
2-qubit density matrix $\rho_{AS} = \sum_j A^{(j)} \otimes \varepsilon(\rho_j)$ of $F_1$ and $F_2$
can be estimated \cite{abhishekQST2013} (see Fig. \ref{sum}).  
In our choice of fixed set of operators and basis elements
\begin{eqnarray}
\rho_{AS} = \left[
\begin{array}{cc|cc}
\lambda_{11} & \lambda_{12} & \lambda_{21} & \lambda_{22} \\
\lambda_{13} & \lambda_{14} & \lambda_{23} & \lambda_{24} \\
\hline
\lambda_{31} & \lambda_{32} & \lambda_{41} & \lambda_{42} \\
\lambda_{33} & \lambda_{34} & \lambda_{43} & \lambda_{44}
\end{array}
\right]. 
\label{rhosa1}
\end{eqnarray}
The $\chi$ matrix characterizing the complete process can now be obtained
by solving the eqn. \ref{chi}.

\begin{figure}[t]
%\noindent\makebox[\linewidth]{\rule{20 cm}{1pt}}
\begin{center}
%\vspace*{0.5 cm}
%\hspace*{0.3cm}
\includegraphics[trim=9.5cm 0cm 0cm 1cm, clip=true,width= 13cm]{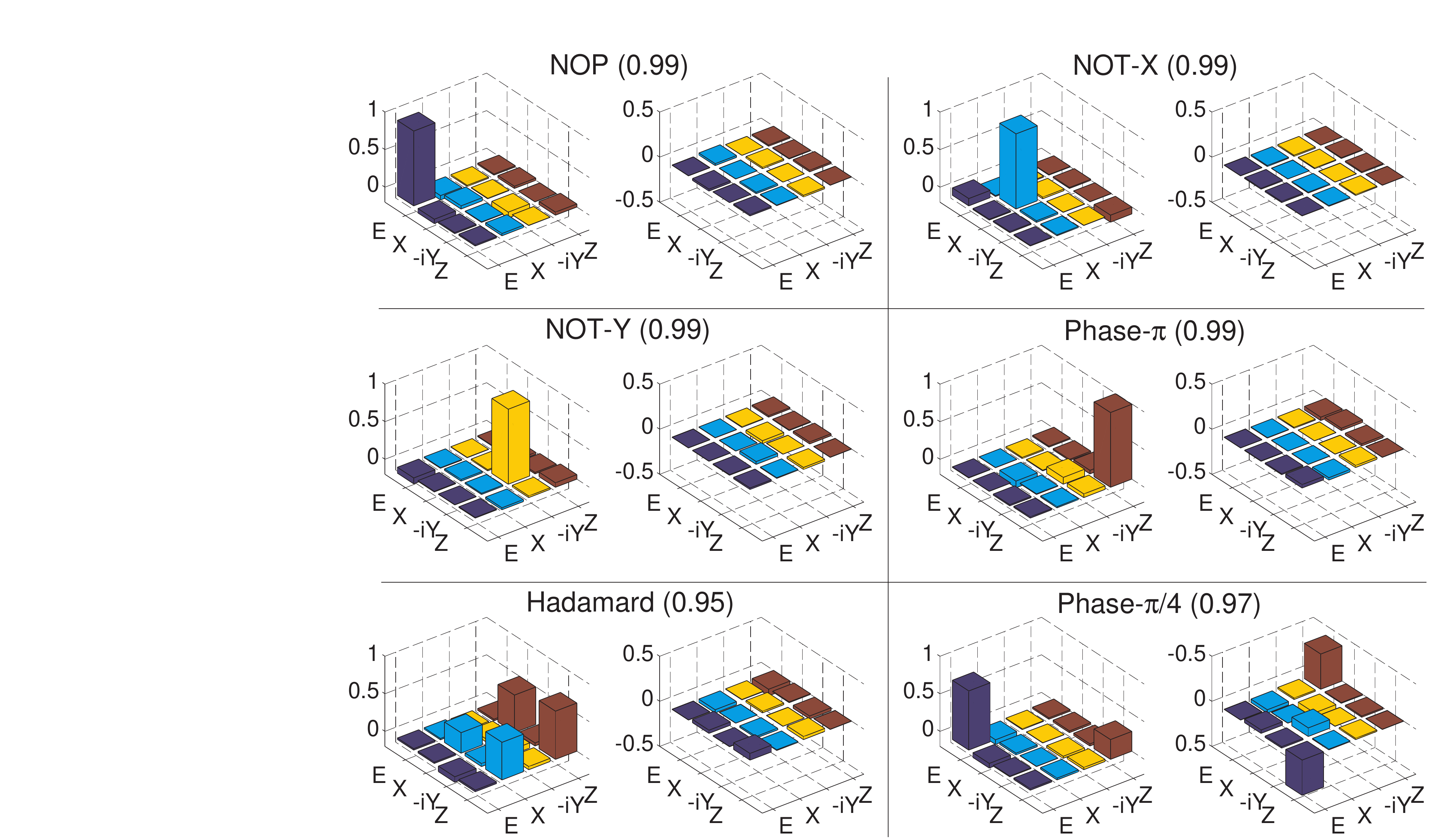} 
\caption{The barplots showing experimental $\chi$-matrices 
for various quantum processes obtained using SSPT.  In each case, the left and right barplots
correspond to the real and imaginary parts respectively, and the fidelities are indicated
in parenthesis.
\label{ssptres}
} 
%\vspace*{0.5 cm}
%\noindent\makebox[\linewidth]{\rule{20 cm}{1pt}}
\end{center}
\end{figure}

\subsection{SSPT of quantum gates}
We now describe experimental characterization of several single-qubit unitary processes.
%The circuit diagram for the experimental SSPT procedure is shown in Fig. \ref{fffmol}c.
The quantum gates to be characterized are introduced as process $\varepsilon$ on
F$_1$ qubit in Fig. \ref{fffmol}c.
The experimental $\chi$-matrices for 
NOP (identity process), NOT-X ($e^{-i \pi X/2}$), NOT-Y ($e^{-i \pi Y/2}$), 
Hadamard, Phase$-\pi$ ($e^{i \pi Z/2}$), and Phase$-\pi/4$ ($e^{i \pi Z/8}$) are
shown in Fig. \ref{ssptres}. 
Starting from thermal equilibrium, each SSPT experiment
characterizing an entire one-qubit process took less than four seconds.
A measure of overlap of the experimental process $\chi_\mathrm{exp}$
with the theoretically expected process $\chi_\mathrm{th}$ is given
by the gate fidelity \cite{suterprotectedgate}
\begin{eqnarray}
F(\chi_\mathrm{exp},\chi_\mathrm{th}) = \frac{\vert Tr[\chi_\mathrm{exp} \chi_\mathrm{th}^\dagger] \vert}
{\sqrt{Tr[\chi_\mathrm{exp}^\dagger \chi_\mathrm{exp}] ~ Tr[\chi_\mathrm{th}^\dagger \chi_\mathrm{th}]}}.
\label{gatefid}
\end{eqnarray}
The gate fidelities for all the six processes are indicated
in Fig. \ref{ssptres}.
Except in the cases of Hadamard and
Phase-$\pi/4$, the gate fidelities were about 0.99.  
The lower fidelities in Hadamard (0.95) and Phase-$\pi/4$ (0.97) are presumed to be
due to RF inhomogeneity and nonlinearities in the pulse implementations.

\begin{figure}[h]
\begin{center}
\includegraphics[trim=0cm 1cm 0cm 0cm, clip=true,width=10cm]{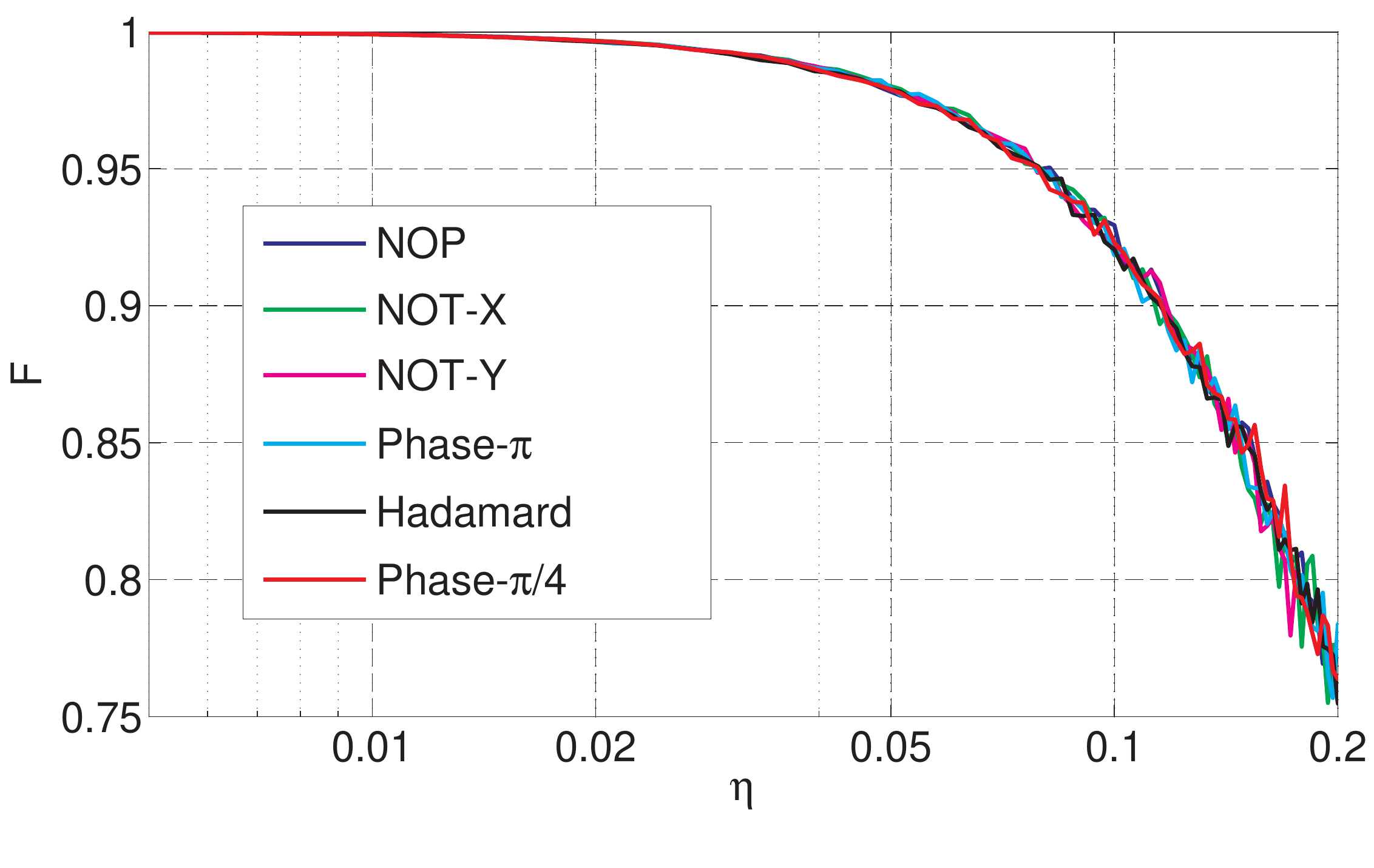} 
\caption{Simulated fidelity of various processes as a function of noise strength $\eta$.
\label{ssptrob} 
}
\end{center}
\end{figure}

In order to study the robustness of SSPT procedure we first considered an ideal
process, simulated the corresponding spectral intensities, and reconstructed
the final density matrix $\rho_{AS}$. 
Using eqn. \ref{rhosa1} we obtained $\lambda_{jk}$ values and calculated
the  matrix $\chi_0$ simulating the noise-free SSPT procedure.
We then introduced noise by adding random numbers in the range $[-\eta,\eta]$ to the spectral intensities 
and used the resulting data for calculating $\chi_\eta$.
The variations of average gate fidelities $F(\chi_0,\chi_\eta)$ for various processes versus noise amplitude
$\eta$ are shown in Fig. \ref{ssptrob}.  Interestingly, the noise has similar effects on
fidelities of all the simulated processes.  We also observe that fidelities remained
above 0.9 for $\eta < 0.1$, indicating that SSPT is fairly robust against the noise
in this range.

\subsection{SSPT of twirling process}
Twirling is essentially a nonunitary process usually realized by an ensemble average
of a set of unitary operations. It was first introduced
by Bennett et al \cite{benett1} for extracting singlet states from 
a mixture of Bell states.  Twirling has been studied in detail 
\cite{benett2,emerson1,emerson2,emerson3,corytwirl}
and various modified twirling protocols have also been suggested \cite{dankert2009,laflammeprl2012}.

In NMR, twirling can be achieved with the help of a pulsed field gradient (PFG), which 
produces a continuous space-dependent spin-rotation, such that the ensemble average
effectively emulates a nonunitary process \cite{anwar}. A $\hat{z}$ PFG produces a 
$z$-dependent unitary $U_\phi(z) = \exp\left(-i \frac{\phi}{2}\sum_{j=1}^n \sigma_{jz}\right)$,
where $j$ is the summation index over all the qubits.  Assuming a linear
gradient introducing a maximum phase $\pm \Phi$ on either ends of 
a sample of length $z_0$, we have $\phi(z) = 2\Phi(z/z_0)$.
When the $\hat{z}$ PFG acts on an initial $n$-qubit density matrix $\rho_{\mathrm{in}} = \sum_{lm} \rho_{lm}\outpr{l}{m}$,
the resultant output density matrix is,

\begin{eqnarray}
\rho_{\mathrm{out}} 
&=& \frac{1}{2\Phi}\int_{-\Phi}^{\Phi} d\phi 
~U_\phi \rho_{\mathrm{in}} U_\phi^\dagger
\nonumber \\
%&=& \sum_{lm} \rho_{lm}\outpr{l}{m} \frac{1}{2\Phi} \int_{-\Phi}^{\Phi} d\phi ~ e^{i q_{lm} \phi} \nonumber \\
&=&  \sum_{lm} \rho_{lm}\outpr{l}{m} ~ \mathrm{sinc}(q_{lm}\Phi).
\label{twirlrho}
\end{eqnarray}
Here $\mathrm{sinc(x)} = \frac{\sin x}{x}$ and $q_{lm} = \frac{1}{2} \sum_j \left[ (-1)^{m_j} - (-1)^{n_j} \right]$ is
the quantum number of the element $\outpr{l_1 l_2 \cdots l_n}{m_1 m_2 \cdots m_n}$, i.e., the
difference in the spin-quantum numbers of the corresponding basis states.
While the diagonal elements $\outpr{l}{l}$ 
and other zero-quantum elements are unaffected by twirling, 
the off-diagonal elements with $q_{lm} \neq 0$ undergo decaying oscillations with increasing $\Phi$ values.

SSPT of twirling process is carried out using the procedure described in
Fig. \ref{fffmol}c after introducing $\delta$-PFG-$\delta$ in place of the process $\varepsilon$,
where $\delta$ is a short delay for switching the gradient.
Applying PFG selectively on the system qubit is not simple, and is also unnecessary.
Since the F$_3$ qubit (AAQST ancilla) is already in a maximally mixed state, twirling has no effect on it.
For the Bell state $\ket{\phi_{AS}}$, applying a strong twirling on either or both spins (F$_1$, F$_2$) has the same
effect, i.e., a strong measurement  reducing the joint-state to a 
maximally mixed state. However, since $\ket{\phi_{AS}}$ corresponds to a two-quantum 
coherence (i.e., $q_{00,11}=2)$, its dephasing is double that of a single-quantum coherence.
Assuming the initial state $\rho_{\mathrm{in}} = \proj{\phi_{AS}}$, and
using expressions \ref{lb} and \ref{twirlrho}, we find that the non-zero elements of $\lambda$ are
\begin{eqnarray}
\lambda_{11} = \lambda_{44} = 1, ~ \mathrm{and,} ~ \lambda_{22} = \lambda_{33} = \mathrm{sinc}(2\Phi).
\end{eqnarray}
Solving expression \ref{chi}, we obtain a real $\chi$ matrix with only nonzero elements 
\begin{eqnarray}
\chi_\mathrm{EE} = \frac{1+\mathrm{sinc}(2\Phi)}{2} ~ \mathrm{and} ~  
\chi_\mathrm{ZZ} =  \frac{1-\mathrm{sinc}(2\Phi)}{2}.
\label{chith}
\end{eqnarray}

\begin{figure}[t]
\begin{center}
\includegraphics[trim=3.5cm 0cm 0cm 1.8cm, clip=true,width= 13.5cm]{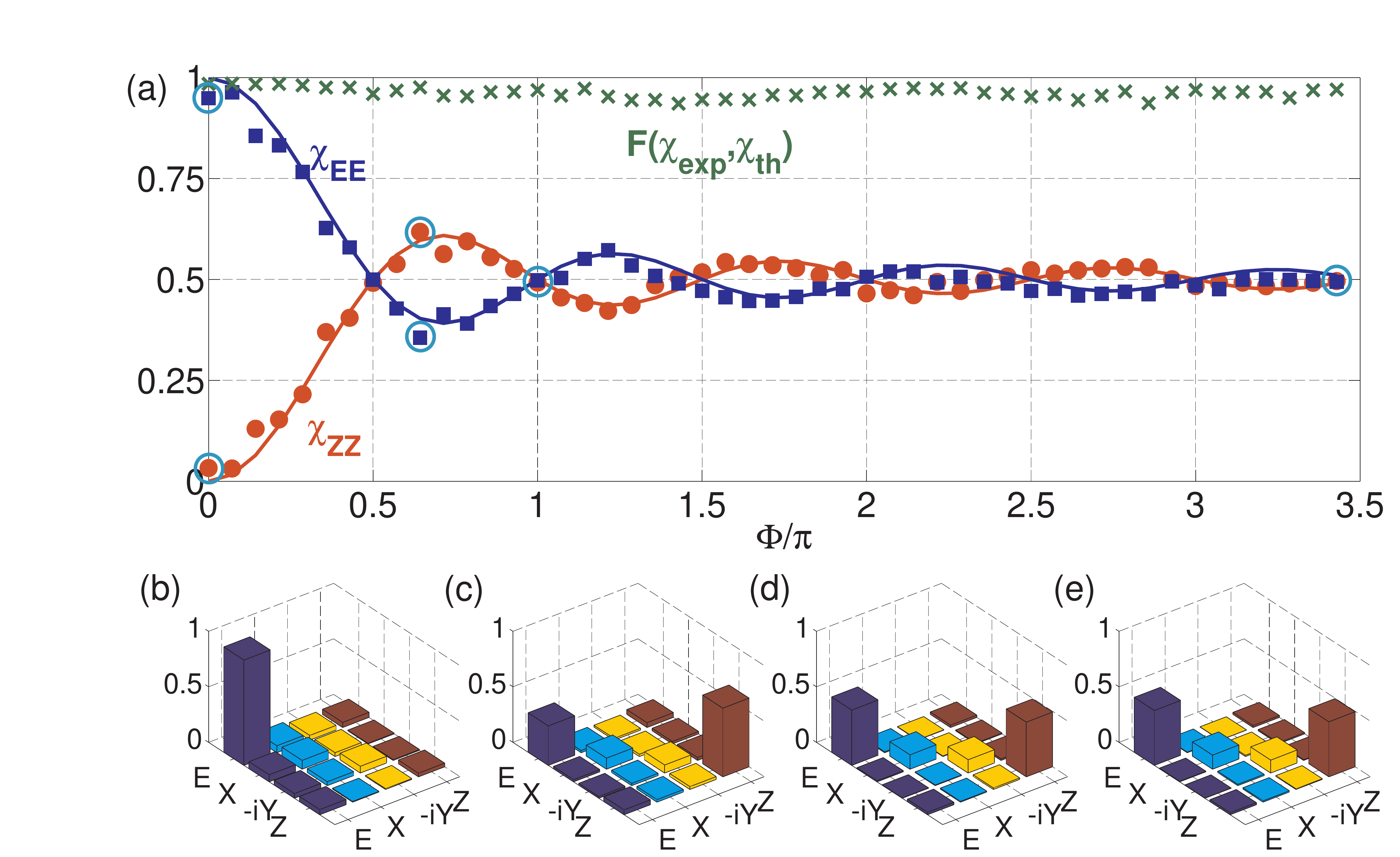} 
\caption{(a) The experimental values of $\vert \chi_{\mathrm{EE}} \vert$ 
($\vert \chi_{\mathrm{ZZ}} \vert$) are shown
by filled squares (filled circles).  
The solid line ($\chi_{\mathrm{EE}}$) in blue and 
($\chi_{\mathrm{ZZ}}$) in red illustrate theory.
The barplots correspond to experimental $\vert \chi \vert$ matrices at 
(b) $\Phi = 0$, (c) $\Phi = 0.64\pi$, (d) $\Phi = \pi$, and (e) $\Phi = 3.43 \pi$.
\label{twirlres} 
}
\end{center}
%\vspace{0.5 cm}
%\noindent\makebox[\linewidth]{\rule{20 cm}{1pt}}
\end{figure}

In our experiments, the duration of PFG and $\delta$ are set to 300 $\upmu$s 
and 52.05 $\upmu$s respectively, 
such that the chemical shift evolutions and J-evolutions are negligible.  The strength of twirling
is slowly varied by increasing the PFG strength from 0 to 2.4 G/cm in steps of 0.05 G/cm.
The results of the experiments are shown in Fig. \ref{twirlres}.
The filled squares (circles) in Fig. \ref{twirlres}a correspond to experimentally
obtained values for $\vert \chi_{\mathrm{EE}}\vert$ ($\vert \chi_{\mathrm{ZZ}}\vert$).
Small imaginary parts observed in experimental $\chi$ matrices are due to minor experimental imperfections.
The smooth lines indicate corresponding theoretical values obtained from eqns. \ref{chith}.
The crosses indicate the gate fidelities $F(\chi_\mathrm{exp},\chi_\mathrm{th})$ 
calculated using eqn. \ref{gatefid}.
The barplots show experimental $\vert \chi \vert$ matrices for 
(b) $\Phi = 0$, (c) $\Phi = 0.64\pi$, (d) $\Phi = \pi$, and (e) $\Phi = 3.43 \pi$, and
$\chi_{\mathrm{EE}}$ and $\chi_{\mathrm{ZZ}}$ values in Fig. \ref{twirlres}a  
corresponding to these $\Phi$ values are circled out.

At zero twirling, the process is essentially a NOP process as is clear
from the bar plot in Fig. \ref{twirlres}b, wherein $\vert \chi_{\mathrm{EE}} \vert \approx 1$ 
and $\vert \chi_{\mathrm{ZZ}} \vert \approx 0$. 
When $\Phi = k \pi/2$ with an integer $k$, the ensemble initially prepared in state $\ket{\psi_{AS}}$ undergoes
an overall phase distribution over $[-k\pi,k\pi]$, and at this stage $\chi_{\mathrm{EE}} =  \chi_{\mathrm{ZZ}} = 0.5$
(eg. Fig. \ref{twirlres}d).  Further increase in $\Phi$ leads to oscillations 
of $ \chi_{\mathrm{EE}} $ and $ \chi_{\mathrm{ZZ}} $ about 0.5, and for large 
$\Phi$ values, both of these elements damp towards 0.5 and all other elements
vanish (eg. Fig. \ref{twirlres}e).
The errors in experimental $\chi_\mathrm{EE}$ and $\chi_\mathrm{ZZ}$ values
were less than 8 \%.  The good agreement of the experimental values with theory
indicates the overall success of SSPT procedure.  The average of the gate fidelities 
was over 0.96. Small deviations of the experimental
values from theory are due to nonlinearities in PFG profile as well as
due to imperfections in RF pulses implementing the SSPT procedure.  

%Characterization of a $$
%As explained in \cite[Nieuwenhuizen], a single operator measurement procedure requires a set of commuting  observables with $N$ number of  total  to be %measured. Characterization of a two qubit density matrix requires knowledge of 15 unknowns. Total no of such observables are n*N/2 ancilla requires %for    

\section{\textbf{Conclusions}\label{34}}
Information processing requires two important physical resources, namely, the size of the register
and the number of operations. Often there
exists an equivalence between these two resources which allows trading  one resource with another.
Likewise, in the present work we show that, if some extra qubits
are available, it is possible to carry out quantum process tomography of the
system qubits with a single-scan ensemble measurement.
We have illustrated this method on a single system qubit and two ancillary
qubits using NMR quantum computing methods.  In particular, 
we extracted the $\chi$ matrices characterizing certain quantum gates and obtained their
gate fidelities with the help of a single ensemble measurement of a three qubit system in each case.
We studied the robustness of SSPT procedure using numerical simulations.
We also characterized twirling operation which is essentially a nonunitary process.

The ensemble nature of NMR systems allows us to determine all the single-quantum observables 
in a single scan experiment. However, a larger ancilla may be required if measurement of
only a commuting set of observables is allowed in a single experiment, as in the case of single-apparatus QST \cite{Nieuwenhuizen}, or if the system-ancilla interactions are constrained,
as in pair-wise interaction case \cite{peng2014pairwise}.
Nevertheless, the overall procedure of SSPT can be generalized to 
apply in other fields such as optical qubits, trapped ions, or superconducting qubits. 

A potential application of single-scan
process tomography could be in high throughput characterization of dynamic processes.
The standard methods require repeated applications of the same process either to
collect independent outputs from all the basis states or to allow quantum state tomography.  
However, the present method requires only one application of the process for the
entire characterization.  

%\bibliographystyle{apsrev4-1}
%\bibliography{ref2}

%\end{document}
\thispagestyle{empty}

%% % % % % % % % ch4
\titlespacing*{\chapter}{0pt}{-50pt}{20pt}
\titleformat{\chapter}[display]{\normalfont\Large\bfseries}{\chaptertitlename\ \thechapter}{20pt}{\Large}
\chapter{Ancilla assisted non-invasive measurements \label{chp4}}

The measurement of a classical object need not affect its subsequent dynamics.  Thus classical measurements are said to be \textit{non-invasive}.  On the other hand,  a strong measurement of a quantum object does affect its subsequent dynamics, and is said to be \textit{invasive}.  The effects of measurements on dynamics of the classical and the quantum objects are illustrated below in Fig. \ref{thepb}.  

\begin{figure}[h]
\begin{center}
\includegraphics[trim= 3cm 1cm 1cm 1cm,clip=true,width=16cm] {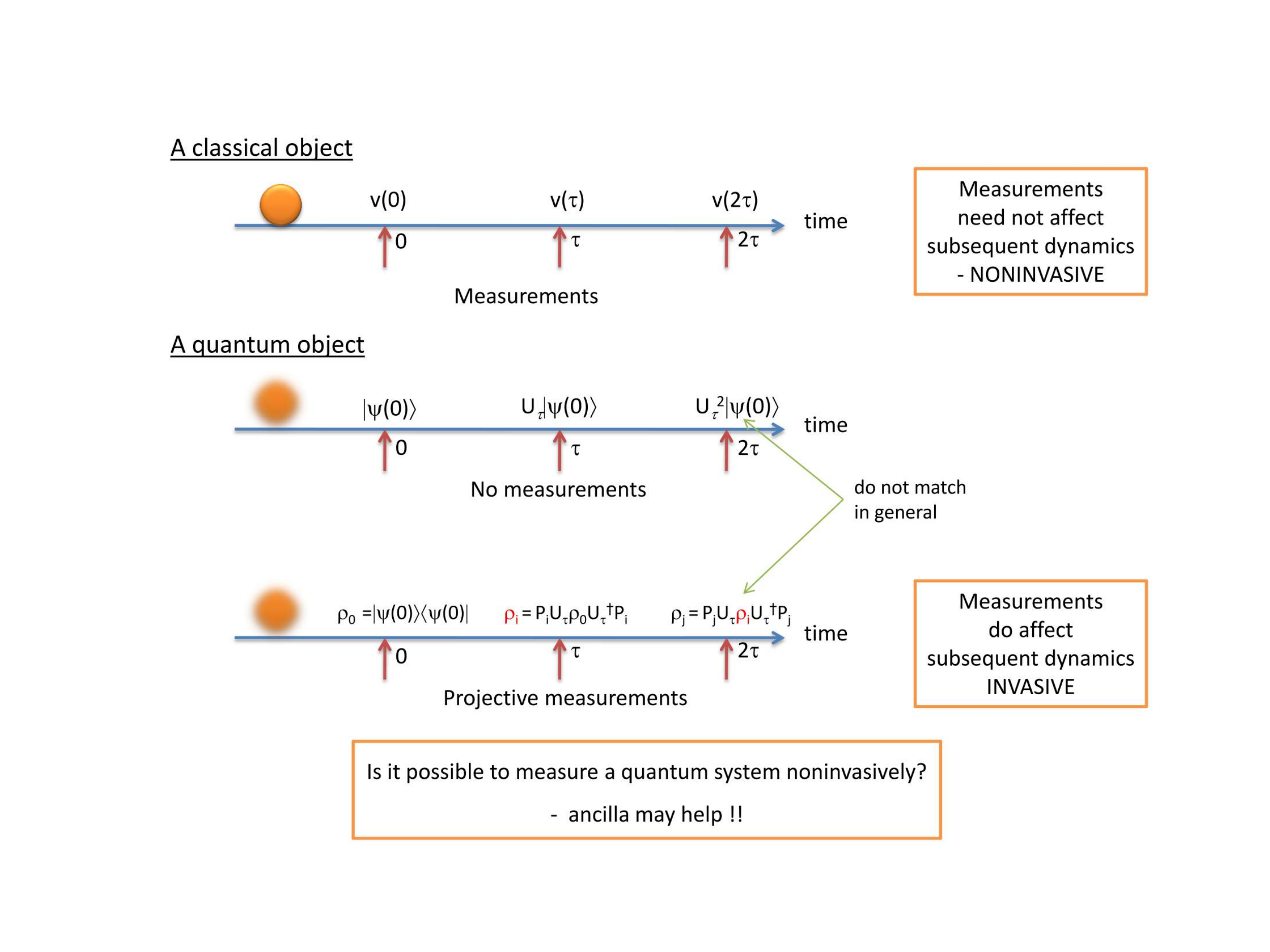}
\caption{Illustrating classical \textit{noninvasive} measurements 
and quantum \textit{invasive} measurements.}
\label{thepb}
\end{center}
\end{figure}

Certain quantum mechanical studies, like Leggett-Garg inequality, are based on the assumption of non-invasive measurability  \cite{palacios2010experimental,goggin2011violation,jordanPRL2011,elgimahesh,ExCNotIonTrap}.  Recently Knee et al \cite{knee2012violation} have proposed a scheme known as \textit{ideal negative result measurement} (INRM), which is more effectively non-invasive than some previous schemes.  Here we describe two interesting quantum physics problems studied using NMR systems, wherein ancilla qubits are utilized for noninvasive measurements.  The problems of interest are:
\begin{enumerate}
\item
Entropic Leggett-Garg Inequality (ELGI) in nuclear spin ensembles and,
\item
Retrieving joint probabilities by inversion of moments in quantum sequential measurements.
\end{enumerate}

\noindent\textbf{Some definitions:} \\
\textit{Joint and Conditional probabilities:}
In probability theory, given at least two random variables $X$ and $Y$ with outcomes $\{x_i\}$ and $\{y_j\}$, the joint probability distribution $p(X, Y)$ gives the probability of combined outcomes $x_i$ and $y_j$ for variable $X$ and $Y$.  Although here I have described it for a two variable case, the concept is general and can be extended to any number of random variables and also for continuous variables.

The Conditional probability distribution $p(Y/X)$ for random variables $X$ and $Y$ gives the probability of getting outcome $y_j$ for given outcome $x_i$.  Given a joint probability distribution $p(X,Y)$ the conditional probability distribution can be calculated as
$p(Y/X) = p(X,Y)/ p(x)$.  This relation is also known as Bayes theorem.

\textit{Marginal and grand probabilities:}
Consider a set of random variable $\{X_i\}$ for $i = 1 \cdots n$ and outcome $\{x_i = \pm 1 \}$. According to the classical probability theory, if the $n$ variable joint probability is $p(x_1,x_2, \cdots x_n)$, then the family of $n-1$ variate probability $p(x_1,x_2, \cdots x_{k-1},x_{k+1} \cdots x_{n-1})$ can be obtained by summing up (marginalizing out) probabilities corresponding to all outcomes variable $x_{k}$.  These probabilities are known as marginals of $p(x_1,x_2, \cdots x_n)$.  By marginalizing different combinations of variables, a complete family of marginal probabilities can be obtained.

\section{\textbf{Entropic Leggett-Garg Inequality in Nuclear Spin Ensembles} \label{41}}

\subsection{Introduction \label{411}}
The behavior of quantum systems is often incomprehensible by classical notions, the
best examples being nonlocality \cite{epr,bell1964einstein} and contextuality \cite{kochen}.
Quantum systems are \textit{nonlocal} since they violate Bell's inequality, which 
assumes that local operations on one of the two space-like separated objects can not disturb
the measurement outcomes of the other \cite{bellRevModPhys}.  
The quantum systems are also \textit{contextual} in the sense that a measurement outcome
depends not only on the system and the property being measured, but also on
the context of the measurement, i.e., on the set of other compatible properties which are
being measured along with.

Another notion imposed on classical objects is macrorealism, which is 
based on two criteria: 
(i) the object remains in one or the other of many possible states at all times, and
(ii) the measurements are noninvasive, i.e., they reveal the state of the object without disturbing 
the object or its future dynamics.
%An important question is how to distinguish the 
%quantum behavior from that of macrorealism. 
%The violation of LGI in quantum systems signify the non-existence of hidden variable models based on macro-realism. 
Quantum systems are incompatible with these criteria and therefore violate
bounds on correlations derived from them.
For instance, Leggett-Garg inequality (LGI) sets up macrorealistic bounds 
on linear combinations of two-time correlations of a dichotomic observable
belonging to a single dynamical system \cite{lgi1985}.
In this sense, LGI is considered as a temporal analogue of Bell's inequality.
Quantum systems do not comply with LGI,
and therefore provide an important way
to distinguish the quantum behavior from macrorealism.
Violations of LGI by quantum systems have been investigated and 
demonstrated experimentally in various systems
\cite{palacios2010experimental,Lambert2011,goggin2011violation,jordanPRL2011,elgimahesh,Oliveira2011timeinequality,etrans,suzuki2012violation,violationQuantumdots,zhou2012experimental,knee2012violation}.
%\cite{exps,noninvasive}

%Most of these previous studies of LGI are based on the measurement of two-time correlations.
For understanding the quantum behavior it is important to investigate
it through different approaches, particularly
from an information theoretical point of view.
For example, an entropic formulation for Bell's inequality has 
been given by Braunstein and Caves \cite{BraunsteinCavesbellInth},
and more recently that for contextuality has been given independently by Rafael and Fritz 
\cite{ChavesFritz2012} and Kurzynski {\it et.al.} \cite{EntropicQContext2012}.
Recently, an entropic formulation of LGI has also been introduced  
by Usha Devi \textit{et al.} \cite{elgiUshadevi}, 
in terms of classical Shannon entropies associated 
with classical correlations.  

Here we report an experimental demonstration of violation of entropic LGI (ELGI) 
in an ensemble of spin $1/2$ nuclei using nuclear magnetic resonance (NMR) techniques.
Although NMR experiments are carried out at a high temperature limit, the nuclear spins
have long coherence times, and their unitary evolutions can be 
controlled in a precise way.
The large parallel computations carried out in an NMR spin ensemble
assists in efficiently extracting the single-event probability (SEP) and joint 
probabilities (JP).
The simplest ELGI study involves three sets of two-time joint measurements of 
a dynamic observable belonging to a `system' qubit at time instants
$(t_1,t_2)$, $(t_2,t_3)$, and $(t_1,t_3)$.  The first measurement in each
case must be `noninvasive' in the sense, it should not influence the outcome
of the second measurement.  These noninvasive measurements (NIM) can be 
performed with the help of an ancilla qubit.

Further, it has been argued in \cite{elgiUshadevi} that the violation of ELGI arises essentially
due to the fact that certain JP are not legitimate in a quantum 
scenario. Here we describe extracting three-time JP
using a three-qubit system, and demonstrate experimentally that it
can not reproduce all the marginal probabilities (MP) and hence is illegitimate. 

 This section is organized as follows. In subec. \ref{412} I briefly revisit the theory of the ELGI \cite{elgiUshadevi}, and then I describe the scheme we designed for the measurement of probabilities in subec. \ref{413}.  Later I detail our experimental study in subsec. \ref{414} and describe the study of the three-time joint probability in subec. \ref{415}. I conclude in subec. \ref{416}.
%%%
\subsection{Theory \label{412}}
Consider a dynamical observable $Q(t_k)=Q_k$ measured at different time instances $t_k$.
Let the measurement outcomes be $ q_k $ with probabilities $ P(q_k) $.
In classical information theory, the amount of information stored 
in the random variable $Q_k$ is given by the Shannon entropy \cite{chuangbook},
\begin{eqnarray}
H(Q_k)=-\sum_{q_k} P(q_k)\log_2{P(q_k)}.
\label{shne} 
\end{eqnarray}
The conditional information stored in $ Q_{k+l} $ at time $ t_{k+l} $,
assuming that the observable $Q_k$ has an outcome $ q_k $,
is 
\begin{eqnarray}
H(Q_{k+l}\vert Q_k=q_k) = 
-\sum_{q_{k+l}} P(q_{k+l}\vert q_k)\log_2P(q_{k+l}\vert q_k) \nonumber,
\end{eqnarray}
where $P(q_{k+l}\vert q_k)$ is the conditional probability.
Then the mean conditional entropy is given by,
\begin{eqnarray}
H(Q_{k+l}\vert Q_k) &=& -\sum_{q_k} P(q_k)H(Q_{k+l}\vert Q_k=q_k).
\end{eqnarray}
Using Bayes' theorem, $ P(q_{k+l}\vert q_k)P(q_k)=P(q_{k+l},q_k)$,
Here $P(q_k)=P(q_{k+l},q_k)$
the mean conditional entropy becomes
\begin{eqnarray}
H(Q_{k+l}\vert Q_k) = H(Q_k,Q_{k+l})-H(Q_k),
 				   \label{condentr}
\end{eqnarray}
where the joint Shannon entropy is given by
\begin{eqnarray}
 H(Q_k,Q_{k+l}) = -\sum_{q_k,q_{k+l}}P(q_{k+l},q_k)
\log_2P(q_{k+l},q_k).
\label{shnjoint}
\end{eqnarray}
These Shannon entropies always follow the inequality \cite{BraunsteinCavesbellInth}
\begin{equation}
H(Q_{k+l} \vert Q_k) \leq H(Q_{k+l}) \leq H(Q_k,Q_{k+l}).
\label{bceqn}
\end{equation}
  The left side of the equation implies
 that removing a constraint never decreases the entropy, and the right side
 implies information stored in two variables is always greater than or equal to 
 that in one \cite{elgiUshadevi}. 
Suppose that three measurements $Q_k$, $Q_{k+l}$, and $Q_{k+m}$, are performed at time instants
$t_k < t_{k+l} < t_{k+m}$. Then,
from equations (\ref{condentr}) and (\ref{bceqn}), the following inequality can 
 be obtained:
\begin{equation}
H(Q_{k+m} \vert Q_k) \leq H(Q_{k+m} \vert Q_{k+l})+H(Q_{k+l} \vert Q_k).
 \end{equation}
For $ n $ measurements $Q_{1},Q_{2}, \dots ,Q_{n}$, at  time instants 
$ t_1 < t_{2} < \dots < t_n $, the above inequality can be 
 generalized to \cite{elgiUshadevi}
\begin{equation}
\sum_{k=2}^nH(Q_{k} \vert Q_{k-1})-H(Q_n \vert Q_1) \geq 0.
\label{elgi}
 \end{equation}
This inequality must be followed by all macro-realistic objects, since 
its satisfaction means the existence of legitimate JP distribution,
which can yield all MP \cite{EnQConRamnathan}.

Usha Devi \textit{et al.} \cite{elgiUshadevi} have shown theoretically that the above 
inequality is violated by a quantum spin-$s$ system, prepared in a 
completely mixed initial state, $\rho_{in} = \mathbb{I}/(2s+1)$.
Consider the $z$-component of the spin evolving under the Hamiltonian ${H}=-\omega S_x$ as 
our dynamical observable, \textit{i.e.} $ Q_t = U_tS_zU_t^\dag $, where
$ U_t=e^{-iHt} $, and $S_x$ and $S_z$ are the components of spin-angular momentum.
Let $n$-measurements occur at regular time instants $\Delta t, ~2 \Delta t, \cdots, n\Delta t$.
Ideally in this case, the conditional entropies $ H(Q_k \vert Q_{k-1}) $ between successive measurements 
are all equal, and can be denoted as $ H [ \theta /(n-1) ] $, where $ \theta/(n-1) = \omega \Delta t $
is the rotation caused by the Hamiltonian in the interval $\Delta t$.
Similarly we can denote $ H(Q_n \vert Q_1) $ as $ H [\theta] $.
The left hand side of inequality (\ref{elgi}) scaled in units of $ \log_2(2s+1) $ is termed as
the information deficit $D$.  For $n$-equidistant measurements, it can be written as
\cite{elgiUshadevi}
\begin{equation}
D_n(\theta)=\frac{(n-1)H[\theta / (n-1)] - H[ \theta ]}{\log_2(2s+1)} \geq 0.
\end{equation}
\subsection{Measurement of Probabilities \label{413}}
\begin{figure}[h]
\begin{center}
 \includegraphics[trim= 2cm 3cm 0cm 2cm,clip=true,width=16cm]{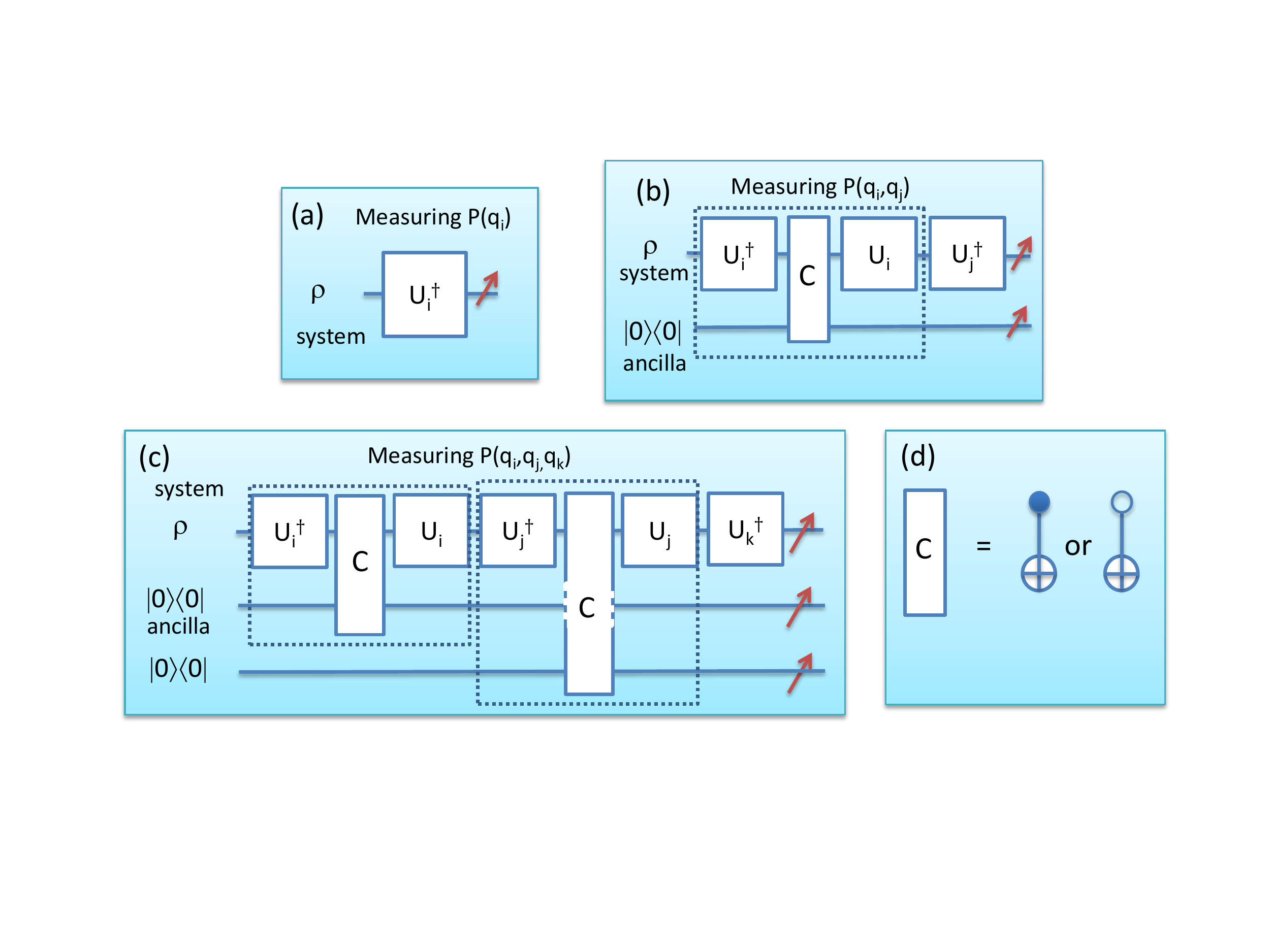}
\caption{Circuits for measuring SEP (a), and two-time JP (b),
and three-time JP (c).
The grouped gates represent measurement in
$\{\Pi_0^t,\Pi_1^t\}$ basis.
The operation $C$ can be either CNOT gate or anti-CNOT gate, as
described in the text.  In (c) the second CNOT gate is applied between
first and third spin.  The pointer at the end in each circuit represents the
measurement of diagonal elements of the density matrix.
(d) Block C represents either CNOT or anti-CNOT gate.}
\label{circuit}
\end{center}
\end{figure}

Consider a spin-1/2 particle as the system qubit.
Using the eigenvectors $\{ \ket{0}, \ket{1}\}$ of $S_z$,  as the computational basis,
the projection operators at time $t=0$ are
$\{\Pi_\alpha = \ket{\alpha} \bra{\alpha}\}_{\alpha = 0,1}$.
For the dynamical observable, the measurement basis is rotating under the 
unitary $ U_t=e^{i\omega S_x t} $, such that
$ \Pi_\alpha^t= U_t\Pi_\alpha U_t^\dag$. 
However, it is convenient to perform the actual measurements in 
the time-independent computational basis.
Since for an instantaneous state $\rho(t)$, 
$\Pi_\alpha^t \rho(t) \Pi_\alpha^t = U_t \Pi_\alpha \left( U_t^\dagger \rho(t) U_t \right) \Pi_\alpha U_t^\dagger$,
measuring in $ \{\Pi_\alpha^t\} $ basis is equivalent to back-evolving the state by 
$U_t^\dagger$,
measuring in computational basis, and lastly forward evolving by
$U_t$. This latter evolution can be omitted if one is interested
only in the probabilities and not in the post measurement state of the
system. For example, in case of multiple-time measurements, the forward
evolution can be omitted after the final measurement.
The method for extracting SEP and JP
involves the quantum circuits shown in Fig. \ref{circuit}.
To measure SEP $P(q_i)$ of system qubit in a
general state $\rho_S$, it is evolved by
$U_i^\dagger = e^{iHt_i}$, and the probabilities $P(q_i)$ are 
obtained using diagonal tomography.
Here a further forward evolution by $U_i$ is not necessary as described earlier.

To measure JP $P(q_i,q_j)$, we utilize an ancilla qubit 
initialized in the state $\ket{0}\bra{0}$
(Fig. \ref{circuit} b).
After back evolution to computational basis, the CNOT gate encodes the probabilities
of the system-qubit $P(q_i)$ on to the ancilla-qubit.  After a further evolution by 
$U_i U_j^\dagger = e^{-i\omega S_x(t_j-t_i)}$, a diagonal tomography of the two
qubit system yields $P(q_i,q_j)$.
A similar scheme, shown in Fig. \ref{circuit} c, 
is employed for extracting three-time JP. These circuits can be generalized for 
higher order JP or for spin-$s>1/2$ systems, 
using appropriate ancilla register.

\begin{figure}[h]
\begin{center}
\includegraphics[trim= 1cm 1.5cm 2cm 0.5cm,clip=true,width=16cm]{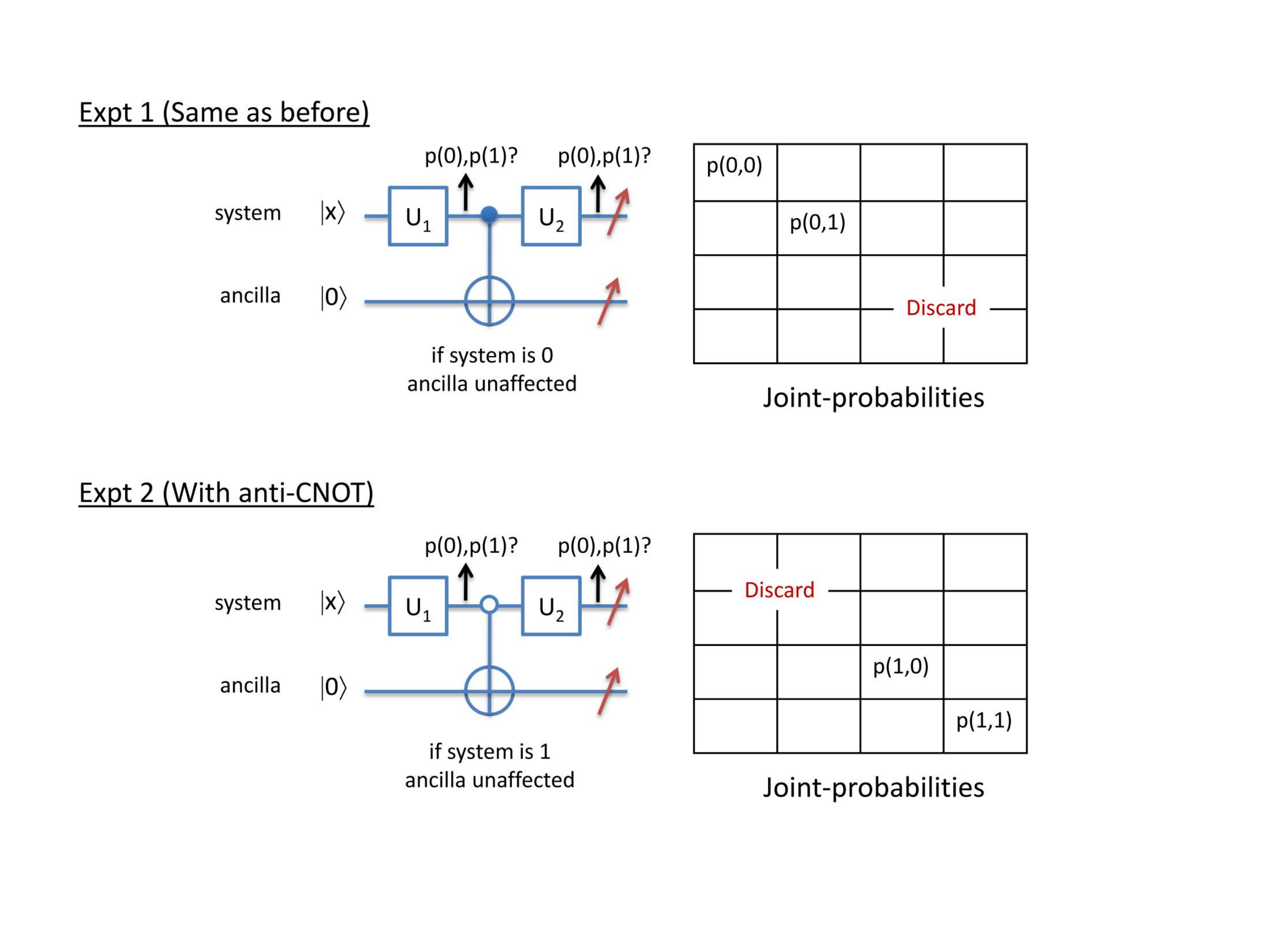}
\caption{INRM procedure for extracting probabilities $P_{0}$ (a) and $P_{1}$ (b) of a single qubit system.}
\label{inrm}
\end{center}
\end{figure}

In the earlier LGI experiments, NIM have been performed
by either 
(i) a weak measurement which has minimum effect on the post measurement quantum state 
\cite{palacios2010experimental,goggin2011violation,jordanPRL2011} or 
(ii) initializing the system qubit in the maximally mixed state so that
the system density matrix remains unchanged by the measurements \cite{elgimahesh,Oliveira2011timeinequality}.
Recently however, it was noted by Knee \textit{et al.} that a sceptical macrorealist
is not convinced by either of the above methods \cite{knee2012violation}.  Instead, they had proposed 
convincingly, a more non-invasive procedure, known as ideal negative result' measurements (INRM) \cite{knee2012violation}.  
The idea for INRM is as follows.  
The CNOT gate is able to flip the ancilla qubit
only if the system qubit is in state $\vert 1 \rangle$, and does nothing
if the system qubit is in state $\vert 0 \rangle$.
Therefore after the CNOT gate, if
we measure the probability of unflipped ancilla, this corresponds to an 'interaction-free'
or NIM of $P(q=0)$. 
Similarly, we can implement an anti-CNOT gate, which
flips the ancilla only if the system qubit is in state $\vert 0 \rangle$, and does nothing
otherwise, such that the probability of unflipped qubit now gives $P(q=1)$.  
Note that in both the cases, the probabilities
wherein the system interacted with the ancilla, resulting in its flip, are discarded Fig. \ref{inrm}.
To see this property consider a one qubit general state (for system) and an ancilla in the state
$ \outpr{0}{0} $, then the encoding of probability using C-NOT is as follows
\begin{eqnarray}
& \left( p_0 \outpr{0}{0} + p_1 \outpr{1}{1} + a \outpr{1}{0} + a^\dagger \outpr{0}{1} \right) _S  
\otimes \outpr{0}{0}_A & \nonumber \\
&\downarrow \mathrm{CNOT}& \nonumber \\
& \outpr{0}{0}_S \otimes p_0 \outpr{0}{0}_A 
+ \outpr{1}{1}_S \otimes p_1 \outpr{1}{1}_A & \nonumber \\
& + \outpr{1}{0}_S \otimes a \outpr{1}{0}_A
+ \outpr{0}{1}_S \otimes a^\dagger \outpr{0}{1}_A.& \nonumber
\end{eqnarray}
Now measuring the diagonal terms of the ancilla qubit, we can retrieve $ p_0 $ and $ p_1 $.

Since we are not concerned about any further evolution, the last measurement need not be NIM. 
In our experiments we combine the two methods, i.e., (i) first we prepare the system 
in a maximally mixed state i.e., $\rho_S = \mathbb{I}/2$, and 
(ii) we perform INRM.
%The experiment is performed twice one with CNOT and another with open-CNOT and the data when the ancilla qubit 
%interacted with the system qubit is discarded . For the CNOT gate the data corresponding to qubit state being in $ \ket{1}\bra{1} $
%and for open-CNOT being in $ \ket{0}\bra{0} $ is discarded. 
In this case, $P(0_i) = P(1_i) = 1/2$,
and JP are 
\begin{eqnarray}
&P(0_i,0_j) = |\cos(\theta_{ij}/2)|^2/2 = P(1_i,1_j),& ~~\mathrm{and}, \nonumber \\
&P(0_i,1_j) = |\sin(\theta_{ij}/2)|^2/2 = P(1_i,0_j),&
\label{pqiqj}
\end{eqnarray}
where $ \theta_{ij}=\omega(t_j-t_i)$ \cite{elgiUshadevi}.
The only SEP needed for the ELGI test is $H(Q_1)$,
since $H(Q_t)$ is constant for the maximally mixed system state.
Further, since $H(Q_1,Q_2) = H(Q_2,Q_3)$ in the case of uniform
time intervals, only two joint entropies $H(Q_1,Q_2)$ and $H(Q_1,Q_3)$
are needed to be measured for evaluating $D_3$.
In the following we describe the experimental implementation of these
circuits for the three-measurement LGI test.
%%%
\subsection{Experiment \label{414}}

\begin{figure}[h]
\begin{center}
\includegraphics[trim = 0.3cm 4.5cm 2.5cm 4cm,clip=true,width= 13cm]{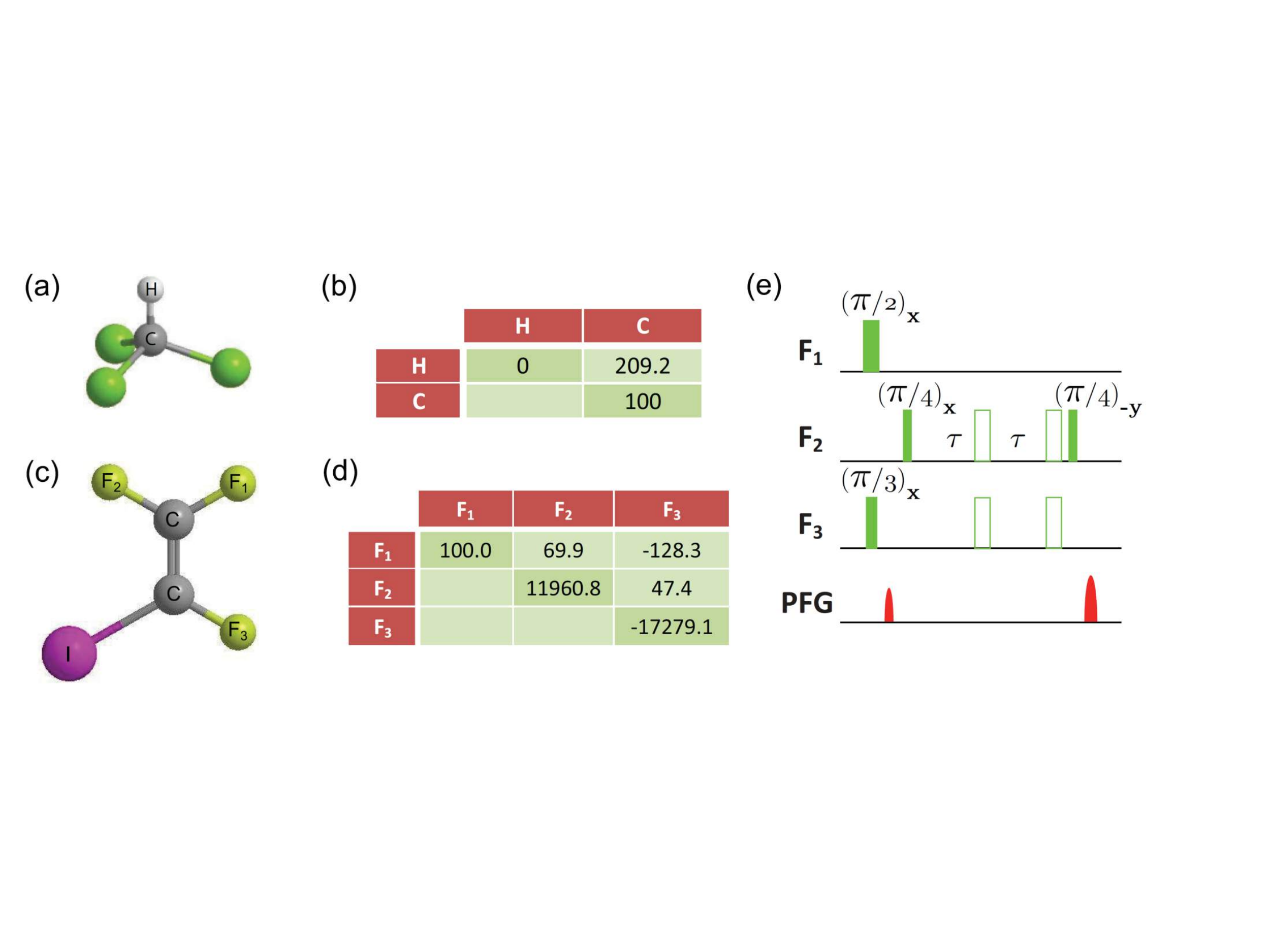}
\caption{The molecular structures of chloroform (a) and trifluoroiodoethylene (c) 
and the corresponding tables (b and d) of relative resonance frequencies 
(diagonal elements) and the J-coupling constants.  The pulse sequence for 
initializing trifluoroiodoethylene is shown in (e).
In (e) the open pulses are $\pi$ pulses and the delay $\tau = 1/(4J_{23})$.}
\label{exptscheme1}
\end{center}
\end{figure}
We have used $^{13}$CHCl$_3$ (dissolved in CDCl$_3$)
as the two qubit system and treat its $^{13}$C and 
$^1$H nuclear spins as the system and the ancilla qubits respectively
as shown in Fig. \ref{exptscheme1} a.
The resonance offset of $^{13}$C was set to 100 Hz and
that of $^1$H to 0 Hz (on resonant) Fig. \ref{exptscheme1} b.  The two spins
have an indirect spin-spin coupling constant $J=209.2$ Hz.
All the experiments were carried out at an ambient temperature
of 300 K on a 500 MHz Bruker UltraShield NMR spectrometer.

\begin{figure}[h]
\begin{center}
\includegraphics[trim = 0cm 0cm 0cm 0cm, clip=true, width=10cm]{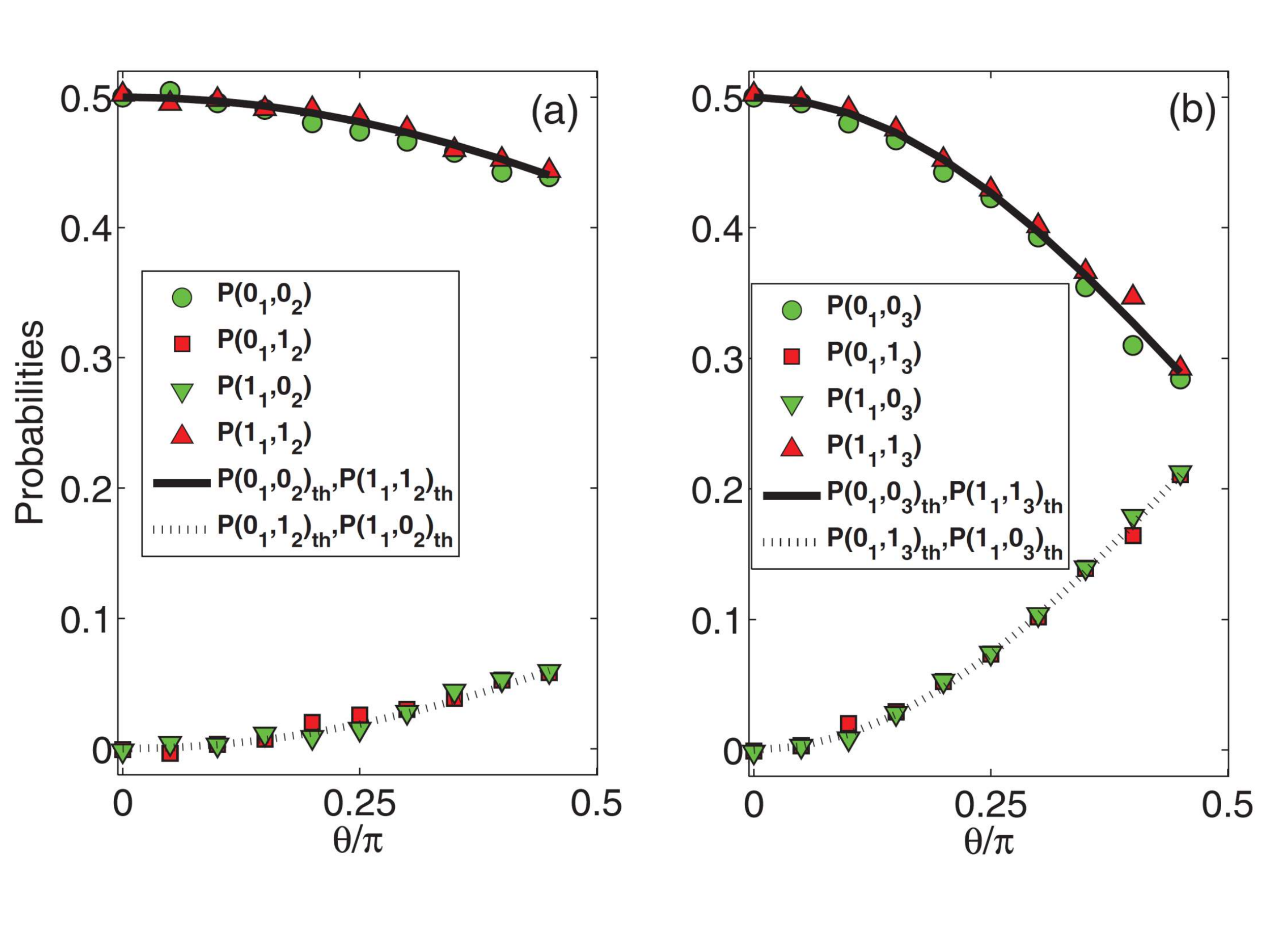}
\caption{
The lines indicate the theoretical joint
probabilities (a) $P(q_1,q_2)_{\mathrm th}$ and (b) $P(q_1,q_3)_{\mathrm th}$, and the symbols
indicate the mean experimental probabilities (a) $P(q_1,q_2)$ and
(b) $P(q_1,q_3)$ obtained by the INRM procedure.
}
\label{jp2}
\end{center}
\end{figure}

\begin{figure}[h]
\begin{center}
\includegraphics[trim= 1.5cm 1cm 0cm 0cm,clip=true,width=12cm]{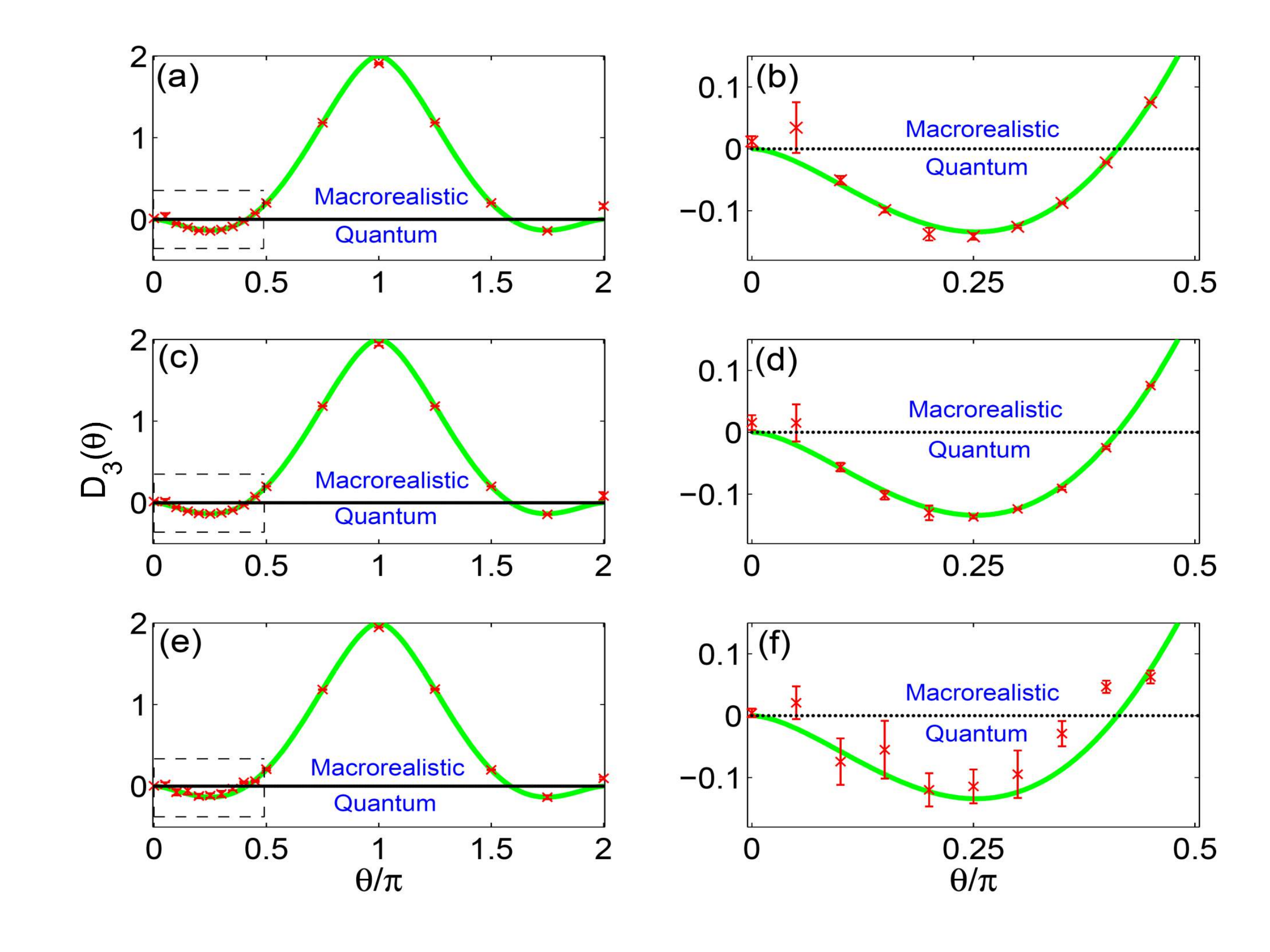}
\caption{Information deficit $D_3$ versus $\theta$ obtained
using (a), (b) CNOT; (c), (d) anti-CNOT; and (e), (f) INRM procedure.
The boxed areas in the left plots [(a), (c), (e)] are magnified in the
right plots [(b), (d), (f)], respectively. The mean experimental D3 (in
bits) values are shown as symbols. The curves indicate theoretical D3
(in bits). The horizontal lines at D3 = 0 indicate the lower bounds of
the macrorealism territories.}
\label{violation}
\end{center}
\end{figure}
The initialization involved preparing the maximally mixed state $\rho_{S} =\mathbb{I}/2$
 on the system qubit $^{13}$C. This is achieved by a $\pi/2$ pulse on $^{13}$C followed by a strong pulsed field gradient (PFG).
The evolution propagator $U_{j}^\dagger U_i = e^{-i S_x \omega (t_j-t_i)}$ 
is realized by the cascade ${H}U_d {H}$,
where ${H}$ is the Hadamard gate, and the delay propagator 
$U_d = e^{-i S_z \omega (t_j-t_i)}$ corresponds to the $\hat{z}$-precession
of the system qubit at $\omega = 200 \pi$ rad/s resonance
off-set.
The $J$-evolution during this delay is refocused by a $\pi$ pulse
on the ancilla qubit.
The CNOT, ${H}$, as well as the $\pi$ pulses are realized by
numerically optimized amplitude and phase modulated RF
pulses, and are robust against the RF inhomogeneity with a average
Hilbert-Schmidt fidelity better than 0.998 \cite{maheshsmp,coryrfijcp,khaneja2005optimal}.
The final measurement of probabilities are carried out by diagonal
tomography. It involved dephasing all the coherences using a strong 
pulsed field gradient followed by a $\pi/30$ detection pulse. 
The intensities of the resulting spectral lines yielded a traceless
diagonal density matrix $d_{ii}$, which was normalized by theoretical deviation density matrix 
 and a trace is introduced by adding the identity matrix to the normalized deviation matrix
 such that they both have the same root mean square value $\sqrt{\sum_{i} d_{ii}^{2}}$.
As described in Fig. \ref{circuit}b,
two sets of experiments were performed, one with CNOT and 
the other with anti-CNOT.
We extracted $P(0,q)$ ($q=\{0,1\}$)
from the CNOT set, and $P(1,q)$ from the anti-CNOT set.
The probabilities thus obtained by INRM 
procedure are plotted in Fig. \ref{jp2}.
These sets of experiments also allow us to compare the
results from (i) only CNOT, (ii) only anti-CNOT, and
(iii) INRM procedures.
The joint entropies were calculated in each case using the
experimental probabilities and 
the information deficit (in bits) was calculated using the expression
$D_3 = 2H(Q_2 \vert Q_1) - H(Q_3 \vert Q_1)$.
The theoretical and experimental values of $D_3$ for various rotation
angles $\theta$ are shown in Fig. \ref{violation}.  We find a general
agreement between the mean experimental $D_3$ values with that of the quantum 
theory.  The error bars indicate the standard deviations obtained by
a series of independent measurements.
According to quantum theory, a maximum violation of $D_3 = -0.134$ should occur 
at $\theta = \pi/4$.  The experimental values of $D_3(\pi/4)$ are
$-0.141 \pm 0.005$, $-0.136 \pm 0.002$, and $-0.114 \pm 0.027$ for
the CNOT, anti-CNOT, and INRM cases respectively.  Thus in all the
cases, we found a clear violation of ELGI.
\begin{figure}
\begin{center}
\includegraphics[trim= 0cm 0cm 0cm 2.83cm, clip=true, width=12cm] {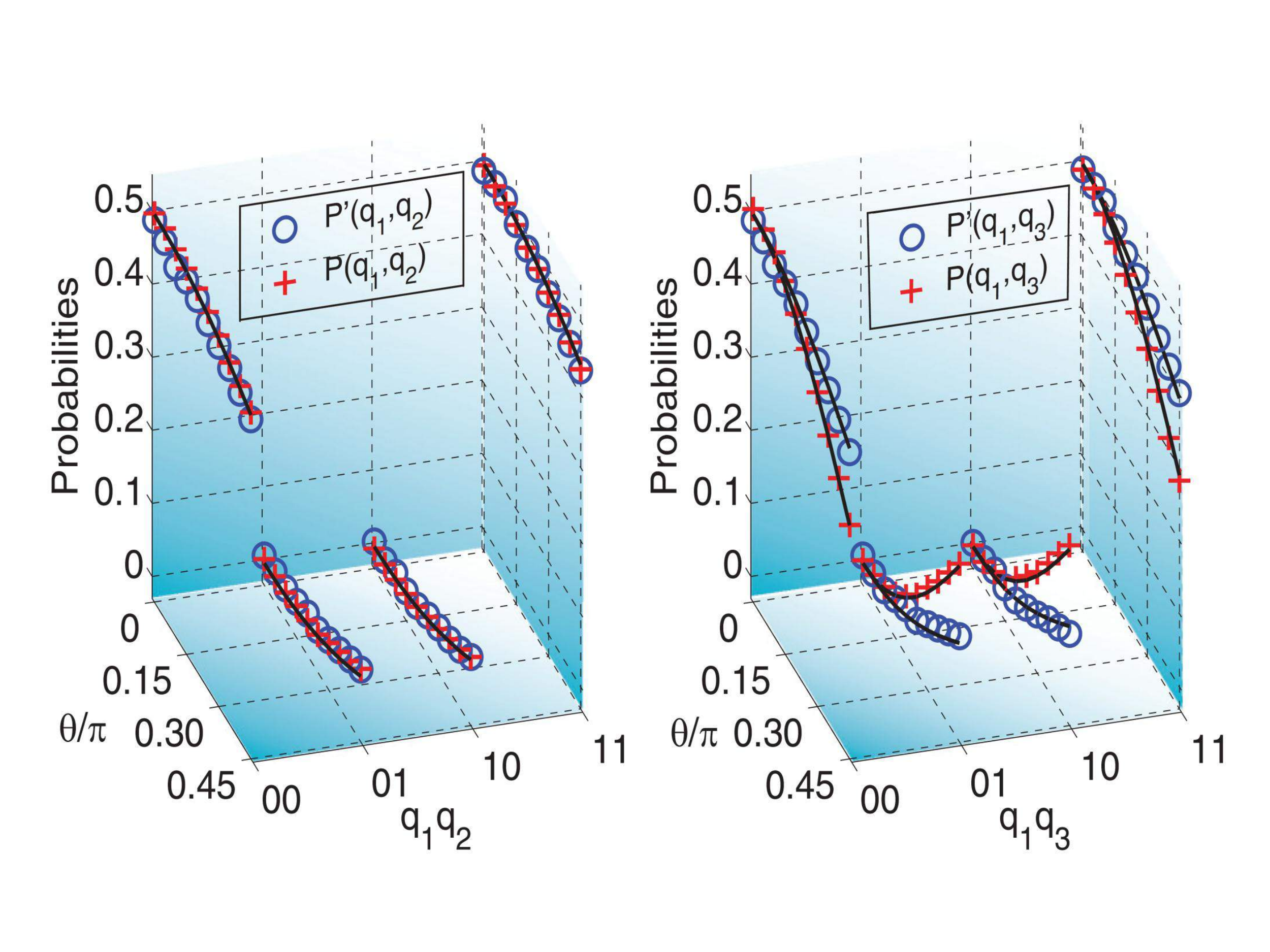}
\caption{
(a) Joint probabilities $P(q_1,q_2)$ and
marginal probabilities $P'(q_1,q_2)$, and (b) joint probabilities $P(q_1,q_3)$
and marginal probabilities $P
'(q_1,q_3)$. The lines correspond to theoretical
values and the symbols are mean experimental values.}
\label{jp3}
\end{center}
\end{figure}
\subsection{Three-time Joint Probabilities \label{415}}
In the above, we have described extracting the two-time JP
$P(q_i,q_j)$ directly. However, it should also be possible to generate them as 
marginals $P'(q_i,q_j)$ of three-time joint probabilities:
\begin{eqnarray}
P'(q_1,q_2) &=& \sum_{q_3} P(q_1,q_2,q_3), \nonumber \\
P'(q_2,q_3) &=& \sum_{q_1} P(q_1,q_2,q_3), \mathrm{and} \nonumber \\
P'(q_1,q_3) &=& \sum_{q_2} P(q_1,q_2,q_3).
\end{eqnarray}
Now $P(q_1,q_2,q_3)$ can reproduce $P(q_1,q_2)$ and $P(q_2,q_3)$, i.e., 
$P'(q_1,q_2) = P(q_1,q_2)$ and
$P'(q_2,q_3) = P(q_2,q_3)$.  However, 
$P(q_1,q_2,q_3)$ can not reproduce 
$P(q_1,q_3)$,
i.e., $P'(q_1,q_3) \ne P(q_1,q_3)$,
in general.  
While for a macrorealistic world $P'(q_i,q_j)= P(q_i,q_j)$
The above concept can be investigated experimentally
by measuring the three-time JP, as described in Fig. \ref{circuit}c.
Since this experiment requires measurements at three time instants,
 we need two ancilla qubits along with the system qubit.
Here the first spin (F$_1$) is used as the system qubit 
and the others (F$_2$ and F$_3$) are chosen as the ancilla qubits 
shown in Fig. \ref{exptscheme1}d.  The effective $^{19}$F transverse 
relaxation time constants $T_{2}^{*}$ were about 0.8 s and their longitudinal
 relaxation time constants were all longer than 6.3 s. The experiments 
 were carried out at an ambient temperature of 300 K.
  The initialization involved evolution of an equilibrium deviation density
matrix under the following sequence to prepare the state
$\rho_{in} = \frac{1-\epsilon}{8}\mathbb{I} + \epsilon \left\{ 
\frac{1}{2}\mathbb{I}_S \otimes \ket{00}\bra{00}_A \right\}$ as shown in 
Fig. \ref{exptscheme1}e.  Where
$\epsilon \sim 10^{-5}$ is the purity factor  \cite{corypnas1997}.
\begin{eqnarray}
&S_{1z} + S_{2z} + S_{3z}& \nonumber \\
&\downarrow& \hspace*{-1.5cm}(\pi/2)_{1x} (\pi/3)_{3x}, \mathrm{PFG} \nonumber \\
&S_z^2 + \frac{1}{2}S_{3z} & \nonumber \\
&\downarrow& \hspace*{-1.5cm}(\pi/4)_{2x} \nonumber \\
&\frac{1}{\sqrt{2}}S_{2z} - \frac{1}{\sqrt{2}}S_{2y} + \frac{1}{2}S_{3z}& \nonumber \\
&\downarrow& \hspace*{-1.5cm} 1/(2J_{23}) \nonumber \\
&\frac{1}{\sqrt{2}}S_{2z} + \sqrt{2}S_{2x} S_{3z} + \frac{1}{2}S_{3z}&  \nonumber \\
&\downarrow& \hspace*{-1.5cm}(\pi/4)_{-2y},\mathrm{PFG} \nonumber \\
&\frac{1}{2} (S_{2z} + 2S_{2z} S_{3z} + S_{3z} ). \nonumber&
\end{eqnarray}
First the experimental three-time JP $P(q_1,q_2,q_3)$ were obtained using the
circuit Fig. \ref{circuit}c.
Two-time MP $P'(q_i,q_j)$ were obtained using Eqs. (7).  
Note that the circuits measuring higher order JP can also
be used to retrieve lower order JP by selectively tracing
out qubits.  Therefore, two-time JP $P(q_i,q_j)$ were
measured directly with the same circuit Fig. \ref{circuit}c; here the JP
 are completely stored in the ancilla qubits, and were obtained
 by tracing out the system qubit. The experimental results of $P(q_1,q_2)$ and
$P'(q_2,q_3)$ are shown in Fig. \ref{jp3}a. It is evident that the marginals
agree quite well with the corresponding JP. Similarly 
experimental results of $P(q_1,q_3)$ and $P'(q_1,q_3)$ are shown in Fig. \ref{jp3}b
.  These results show, in contrary to the macrorealistic theory,
 that the grand probability $P(q_1,q_2,q_3)$ can not reproduce all the
  two-time joint probabilities as the marginals. Therefore the
   grand probability is not legitimate in the quantum case, 
   which is the fundamental reason for the violation of the 
   ELGI by quantum systems \cite{elgiUshadevi}.
It is interesting to note that even for those values of $\theta$ for
 which $D_3$ is positive, the three-time joint probability is illegitimate.  
 Therefore, while the violation of the ELGI indicates the quantumness of
  the system, its satisfaction does not rule out the quantumness.
%%%
\subsection{Conclusions \label{416}}
The entropic Leggett-Garg inequalities, imposes bounds on the statistical outcomes of temporal correlations of observables.  I described an experimental study of the entropic Leggett-Garg
inequality in nuclear spins using NMR techniques.  We employed the recently described `ideal negative result' 
procedure to noninvasively extract JP.
Our results indicate the macrorealistic
bound being violated by over four standard deviations, confirming the
non-macrorealistic nature of the spin-1/2 particles.  
Quantum systems do not have legitimate joint probability 
distribution, which results in the violation of bounds set-up for
macrorealistic systems.  We have experimentally measured the 
three-time JP and confirmed that it can not
reproduce all the two-time JP.

One distinct feature of the entropic LGI is that, the dichotomic
nature of observables assumed in the original formulation of LGI
can be relaxed, thus allowing one to study the quantum behavior 
of higher dimensional systems such as spin $>1/2$ systems.
This could be an interesting topic for future experimental investigations.
\section{\textbf{Retrieving joint probabilities by inversion of moments} \label{42}}
\subsection{Introduction \label{421}}
The issue of determining a probability distribution uniquely in terms of its moment sequence -- known as {\em classical moment problem} -- has been developed for more than 100 years~\cite{Tamarkin,Akhiezer}.  In the case of discrete distributions with the associated random variables taking finite values, moments faithfully capture the essence of the probabilities i.e., the probability distribution is moment determinate~\cite{Shi}. 
In the special case of classical random variables $X_i$ assuming dichotomic values $x_i=\pm 1$, it is easy to see that the sequence of  moments. 
\begin{equation}
\mu_{n_1,n_2,\cdots, n_k}=\langle X_1^{n_1}X_2^{n_2}\cdots X_k^{n_k}\rangle= \sum_{x_1,x_2,\cdots, x_k=\pm 1} \{x^{n_1}_1, x^{n_2}_2,\cdots,x^{n_k}_k\} P(x_1,x_2,\cdots , x_k)             
\end{equation}
where $n_1,n_2,\cdots, n_k=0,1,$  can be readily inverted to obtain the joint probabilities  $P(x_1,x_2,\cdots , x_k)$ uniquely.  More explicitly,  the joint probabilities $P(x_1,x_2,\cdots , x_k)$ are given in terms of the $2^k$ moments  $\mu_{n_1, n_2, \cdots, n_k}, n_1,n_2,\cdots n_k=0,1$ as, 
\begin{eqnarray}
\label{1}
 P(x_1,x_2,\cdots, x_k)&=&\frac{1}{2^k}\sum_{n_1,\cdots, n_k=0, 1} x_{1}^{n_1}x_{2}^{n_2} \cdots x_{k}^{n_k} \mu_{n_1, \cdots, n_k}\nonumber \\ 
& =&\frac{1}{2^k} \sum_{n_1,\cdots, n_k=0, 1} x_{1}^{n_1}x_{2}^{n_2} \cdots x_{k}^{n_k} \langle X_1^{n_1}X_2^{n_2}\cdots X_k^{n_k}\rangle. \nonumber \\
\end{eqnarray}
%%%
Does this feature prevail in the quantum scenario? This results in a negative answer as it is wellknown that the moments associated with measurement outcomes on spatially separated parties are not compatible with the joint probability distribution. This feature reflects itself in the violation of Bell inequalities. In this paper we investigate whether moment-indeterminacy persists when we focus on sequential measurements on a single quantum system. We show that the discrete joint probabilities originating in the sequential measurement of a  single qubit dichotomic observable $\hat{X}(t_i)=\hat{X}_i$ at  different time intervals are not consistent  with the ones reconstructed from the moments.  More explicitly, considering  sequential measurements of $\hat{X}_1$, $\hat{X}_2$, $\hat{X}_3$, we reconstruct the trivariate joint probabilties $P_{\mu}(x_1,x_2,x_3)$  based on the set of eight moments $\{\langle \hat{X}_1\rangle, \langle \hat{X}_2\rangle,  \langle \hat{X}_3\rangle, \langle \hat{X}_1, \hat{X}_2\rangle, \langle \hat{X}_2, \hat{X}_3\rangle, \langle \hat{X}_1, \hat{X}_3\rangle,\\ \langle \hat{X}_1, \hat{X}_2, \hat{X}_3\rangle \}$ and prove that they do not agree with the three-time  joint probabilities (TTJP) $P_d(x_1,x_2,x_3)$ evaluated directly based on the correlation outcomes in the sequential measurement of all the three observables. Interestingly, the moments and TTJP can be independently extracted experimentally in NMR system -- demonstrating the difference between moment inverted three time probabilities with the ones directly drawn from experiment, in agreement with theory. For obtaining TTJP directly we use the procedure of Ref. \cite{nmr_elgi} and for extracting moments we extend the Moussa protocol~\cite{moussa1qbitcontextuality2010} to a set of non-commutating observables. The specifics are given in the experimental section.
Disagreement between moment inverted joint probabilities with the ones based on measurement outcomes in turn  reflects the inherent inconsistency  that the family of all marginal probabilities do not arise from the grand joint probabilities. The non-existence of a legitimate grand joint probability distribution, consistent with the set of all pairwise marginals is attributed to be the common origin of a wide range of no-go theorems on non-contextuality, locality and macrorealism in the foundations of quantum theory~\cite{Fine, bellRevModPhys, KS, Peres, Mermin, elgimahesh, elgiUshadevi, TP2013}. Absence of a valid grand joint probability distribution in the sequential measurement on a single quantum system is brought out here in terms of its mismatch with moment sequence. 
I organize the chapter as follows.  In subsec. \ref{422}, I begin  with a discussion on moment inversion to obtain joint probabilities of three classical random variables assuming dichotomic values $\pm 1$.  In subsec. \ref{423}, I report our study for the quantum scenario with the help of a specific example of sequential measurements of dichotomic observable at three different times on a spin-1/2 system.  We have shown that the TTJP constructed from  eight moments do not agree with those originated from the measurement outcomes.  Subsec. \ref{424} is devoted to report experimental results with NMR implementation on an ensemble of spin-1/2 nuclei, demonstrating that moment constructed TTJP do not agree with those directly extracted.  Subsec. \ref{425} has concluding remarks.            
\subsection{Reconstruction of joint probability of classical dichotomic random variables from moments \label{422}} 
Let $X$ denote a dichotomic random variable with outcomes $x=\pm 1$. The moments associated with statistical outcomes involving the variable $X$  are given by  
$\mu_n=\langle X^n\rangle=\sum_{x=\pm 1} x^n P(x), n=0,1,2,3, \cdots$, where $0\leq P(x=\pm 1)\leq 1; \sum_{x=\pm 1}P(x)=1$ are the corresponding probabilities.  Given the moments $\mu_0$ and $\mu_1$ from a statistical trial, 
one can readily obtain the probability mass function: 
\begin{eqnarray}
 P(1)&=&\frac{1}{2}(\mu_0+\mu_1)=\frac{1}{2}(1+\mu_1),\\ 
 P(-1)&=& \frac{1}{2}(\mu_0-\mu_1)=\frac{1}{2}(1-\mu_1); 
\end{eqnarray}
i.e., moments  determine the probabilities uniquely. 
In the case of two dichotomic random variables $X_1$, $X_2$, the moments  
$\mu_{n_1,n_2}=\langle X_1^{n_1}X_2^{n_2}\rangle=\sum_{x_1=\pm 1, x_2=\pm 1} x_1^{n_1}x_2^{n_2} P(x_1,x_2)$ where $n_1,n_2=0,1\cdots$ encode the bivariate probabilities $P(x_1,x_2)$. Explicitly, 
%%\begin{widetext*}
\begin{eqnarray} 
\mu_{00}&=& \sum_{x_1, x_2=\pm 1} P(x_1,x_2)=P(1,1)+P(1,-1)+P(-1,1)+P(-1,-1)=1, \nonumber \\ 
\mu_{10}&=& \sum_{x_1, x_2=\pm 1} x_1 P(x_1,x_2)= \sum_{x_1=\pm 1} x_1 P(x_1) \nonumber \\
&=& P(1,1)+P(1,-1)-P(-1,1)-P(-1,-1), \nonumber \\ 
\mu_{01}&=& \sum_{x_1, x_2=\pm 1} x_2 P(x_1,x_2)= \sum_{x_2} x_2 P(x_2) \nonumber \\
&=& P(1,1)-P(1,-1)+P(-1,1)-P(-1,-1), \nonumber \\ 
\mu_{11}&=& \sum_{x_1, x_2=\pm 1} x_{1}x_{2} P(x_1,x_2)=P(1,1)-P(1,-1)-P(-1,1)+P(-1,-1). 
\end{eqnarray}  
%%%\end{widetext*}
Note that the  moments $\mu_{10}$, $\mu_{01}$  involve the MP $P(x_1)=\sum_{x_2=\pm 1} P(x_1,x_2)$, $P(x_2)=\sum_{x_1=\pm 1} P(x_1,x_2)$ respectively and they could be evaluated based on statistical trials  drawn independently from the two random variables $X_1$ and $X_2$.  
Given the moments $\mu_{00}, \mu_{10}, \mu_{01}, \mu_{11}$ the reconstruction of  the probabilities $P(x_1,x_2)$ is straightforward:   
 \begin{eqnarray}
 P(x_1,x_2)&=& \frac{1}{4}\sum_{n_1, n_2=0,1} x_1^{n_1}x_2^{n_2} \mu_{n_1, n_2} \nonumber \\ 
 &=& \frac{1}{4}\sum_{n_1, n_2=0,1} \langle X_1^{n_1}X_2^{n_2}\rangle.
\end{eqnarray}
Further, a reconstruction of trivariate JP $P(x_1,x_2,x_3)$ requires the following set of  eight moments: 
$\{\mu_{000}=1, \mu_{100}=\langle X_1\rangle, \mu_{010}=\langle X_2\rangle, \mu_{001}=\langle X_3\rangle, \mu_{110}=\langle X_1, X_2\rangle, \mu_{011}=\langle X_2, X_3\rangle, \mu_{101}=\langle X_1, X_3\rangle,\ \mu_{111}=\langle X_1, X_2, X_3\rangle\}$. The probabilities are retrieved faithfully in terms of the eight moments as,    
\begin{eqnarray}
\label{MI}
P(x_1,x_2,x_3)&=& \frac{1}{8}\sum_{n_1, n_2, n_3=0,1} x_1^{n_1}x_2^{n_2}x_3^{n_3} \mu_{n_1, n_2, n_3} \nonumber \\ 
&=&\frac{1}{8}\sum_{n_1, n_2, n_3=0,1} x_1^{n_1}x_2^{n_2}x_3^{n_3} \langle X_1^{n_1}X_2^{n_2}X^{n_3}_3\rangle. \nonumber \\
\end{eqnarray}
It is implicit that the moments  $\mu_{100}, \mu_{010}, \mu_{001}$ are determined through independent statistical trials involving the random variables $X_1, X_2, X_3$ separately;  $\mu_{110}, \mu_{011}, \mu_{101}$ are obtained based on the correlation outcomes of $(X_1, X_2)$, $(X_2, X_3)$, and $(X_1, X_3)$ respectively.  More specifically, in the classical probability setting there is a  tacit underlying assumption that the set of all MP $P(x_1), P(x_2), P(x_3), P(x_1,x_2), P(x_2,x_3), and P(x_1,x_3)$ are consistent with the trivariate JP $P(x_1,x_2,x_3)$. This underpinning does not get imprinted automatically in the quantum scenario.  Suppose the observables $\hat{X}_1, \hat{X}_2, \hat{X}_3$ are non-commuting and we consider their sequential measurement.  The moments $\mu_{100}=\langle \hat{X}_1\rangle, \mu_{010}=\langle \hat{X}_2\rangle, \ \mu_{001}=\langle \hat{X}_3\rangle$ may be evaluated from the measurement outcomes of dichotomic observables $\hat{X}_1, \hat{X}_2, \hat{X}_3$ independently;  the correlated statistical outcomes in the sequential measurements of $(\hat{X}_1,\ \hat{X}_2)$, $(\hat{X}_2, \hat{X}_3)$ and $(\hat{X}_1, \hat{X}_3)$ allow one to extract the set of moments $\mu_{110}=\langle \hat{X}_1\hat{X}_2\rangle, \mu_{011}=\langle \hat{X}_2\hat{X}_3\rangle, \mu_{101}=\langle \hat{X}_1\hat{X}_3 \rangle$; further the moment $\mu_{111}=\langle \hat{X}_1\hat{X}_2\hat{X}_3\rangle $ is evaluated based on the correlation outcomes when all the three observables are measured sequentially. The JP $P_\mu(x_1,x_2,x_3)$ retrieved from the moments as given in (\ref{MI}) differ from the ones evaluated directly in terms of the correlation outcomes in the sequential measurement of all the three observables.  We illustrate this inconsistency appearing in the quantum setting in the next section.
\subsection{Quantum three-time JP and moment inversion \label{423}}
 Let us consider a spin-1/2 system, dynamical evolution of which is governed by the Hamiltonian 
\begin{equation}
\hat{H}=\frac{1}{2} \hbar \omega \sigma_x.
\end{equation} 
We choose z-component of spin as our dynamical observable: 
\begin{eqnarray}
\hat{X}_i&=&\hat{X}(t_i)=\sigma_z'(t_i) \nonumber \\
&=& \hat{U}^\dag(t_i) \sigma_z \hat{U}(t_i) \nonumber \\
&=&\sigma_z \cos\omega t_i+\sigma_y \sin\omega t_i.
\end{eqnarray} 
where  $\hat{U}(t_i)=e^{-i\sigma_x\omega t_i/2}=\hat{U}_i$  
and consider sequential measurements of the observable $\hat{X}_i$ at three different times $t_1=0$, $t_2=\Delta t$, and $t_3=2\Delta t$: 
\begin{eqnarray}
\hat{X}_1&=&\sigma_z, \nonumber \\
 \hat{X}_2&=&\sigma_z'(\Delta t)=\sigma_z \cos(\omega\Delta t)+\sigma_y \sin(\omega\Delta t), \nonumber \\ 
\hat{X}_3&=& \sigma_z'(2\Delta t)=\sigma_z \cos(2\omega\Delta t)+\sigma_y \sin(2\omega\Delta t).
\end{eqnarray}
Note that these three operators are not commuting in general. 

The moments $\langle \hat{X}_1\rangle, \langle \hat{X}_2\rangle, \langle \hat{X}_3\rangle$ are readily evaluated to be  
\begin{eqnarray}
\label{1mom}
 \mu_{100}&=&\langle \hat{X}_1\rangle = {\rm Tr}[\hat{\rho}_{\rm in} \sigma_z]=0,  \\ 
\mu_{010}&=&\langle \hat{X}_2\rangle={\rm Tr}[\hat{\rho}_{\rm in} \sigma_z'(\Delta t)]=0,  \\ 
\mu_{001}&=& \langle \hat{X}_3\rangle={\rm Tr}[\hat{\rho}_{\rm in} \sigma_z' (2\Delta t)]=0.
\end{eqnarray}
when the system density matrix is prepared initially in a maximally mixed state $\hat{\rho}_{\rm in}=\mathbb{I}/2$. The probabilities of outcomes $x_i=\pm 1$ in the completely random initial state are given by $P(x_i=\pm 1)={\rm Tr}[\hat{\rho}_{\rm in} \hat{\Pi}_{x_i}]=\frac{1}{2}$, where  $\hat{\Pi}_{x_i}=\vert x_i\rangle\langle x_i\vert$ is the projection operator corresponding to measurement of the observable $\hat{X}_i$. 

The two-time JP arising in the sequential measurements of the observables $\hat{X}_i, \hat{X}_{j}, for j>i$ are evaluated as follows. 
The measurement of the observable $\hat{X}_i$ yielding the outcome $x_i=\pm 1$ projects the density operator to $\hat{\rho}_{x_i}=\frac{\hat{\Pi}_{x_i} \hat{\rho}_{\rm in} \hat{\Pi}_{x_i}}{{\rm Tr}[\hat{\rho}_{\rm in} \hat{\Pi}_{x_i}]}$. Further, a sequential measurement of  $\hat{X}_j$ leads to the two-time JP as, 
\begin{eqnarray}
\label{pxixj}
P(x_i,x_j)&=& P(x_i) P(x_j\vert x_i)\nonumber \\ 
&=& {\rm Tr}[\hat{\rho}_{\rm in} \hat{\Pi}_{x_i}] {\rm Tr}[\hat{\rho}_{x_i} \hat{\Pi}_{x_j}] \nonumber \\
&=&{\rm Tr}[\hat{\Pi}_{x_i} \hat{\rho}_{\rm in} \hat{\Pi}_{x_i} \hat{\Pi}_{x_j}] \nonumber \\
&=& \langle x_i\vert \hat{\rho}_{\rm in} \vert x_i\rangle \vert\langle x_i\vert x_j\rangle\vert^2.
\end{eqnarray}
We evaluate the two-time JP associated with the sequential measurements of 
$(\hat{X}_1,\hat{X}_2)$, $(\hat{X}_2,\hat{X}_3)$, and $(\hat{X}_1,\hat{X}_3)$ explicitly: 
\begin{eqnarray}
P(x_1, x_2)&=&\frac{1}{4} [1+ x_1 x_2 \cos(\omega\Delta t)], \\
P(x_2, x_3)&=&\frac{1}{4} [1+ x_2 x_3 \cos(\omega\Delta t)], \\
P(x_1, x_3)&=&\frac{1}{4} [1+ x_1 x_3 \cos(2\omega\Delta t)].
\end{eqnarray}
We then obtain two-time correlation moments as,
\begin{eqnarray}
\label{2mom}
\mu_{110} = \langle \hat{X}_1 \hat{X}_2\rangle &=& \sum_{x_1,x_2=\pm 1} x_1x_2 P(x_1, x_2) \nonumber \\
          &=&\frac{1}{2} \cos(\omega\Delta t)], \\
\mu_{011}=\langle \hat{X}_2\hat{X}_3\rangle&=&\sum_{x_2,x_3=\pm 1} x_2x_3 P(x_2, x_3) \nonumber \\
&=&\frac{1}{2} \cos(\omega\Delta t)], \\
\mu_{101}=\langle \hat{X}_1\hat{X}_3\rangle&=&\sum_{x_1,x_3=\pm 1} x_1x_3 P(x_1, x_3) \nonumber \\
&=&\frac{1}{2} \cos(2 \omega\Delta t)].
\end{eqnarray}
Further, the three-time JP $P(x_1,x_2,x_3)$ arising in the sequential measurements of $\hat{X}_1, \hat{X}_2$ followed by  $\hat{X}_3$ are given by
\begin{eqnarray}
P(x_1,x_2,x_3)&=& P(x_1) P(x_2\vert x_1) P(x_3\vert x_1,x_2)\nonumber \\ 
&=& {\rm Tr}[\hat{\rho}_{\rm in} \hat{\Pi}_{x_1}] {\rm Tr}[\hat{\rho}_{x_1} \hat{\Pi}_{x_2}] {\rm Tr}[\hat{\rho}_{x_2} \hat{\Pi}_{x_3}] \nonumber \\
\end{eqnarray} 
where $\hat{\rho}_{x_2}=\frac{\hat{\Pi}_{x_2}, \hat{\rho}_{x_1}, \hat{\Pi}_{x_2}}{{\rm Tr}[\hat{\rho}_{x_1}, \hat{\Pi}_{x_2}]}.$ 
We obtain, 
\begin{eqnarray}
\label{dirTTJP}
P(x_1,x_2,x_3)&=&{\rm Tr}[\hat{\Pi}_{x_2} \hat{\Pi}_{x_1} \hat{\rho}_{\rm in} \hat{\Pi}_{x_1} \hat{\Pi}_{x_2} \hat{\Pi}_{x_3}]\nonumber \\ 
 &=&  \langle x_1\vert \hat{\rho}_{\rm in} \vert x_1\rangle \vert\langle x_1\vert x_2\rangle\vert^2 \vert\langle x_2\vert x_3\rangle\vert^2 \nonumber \\
 &=& \frac{P(x_1, x_2) P(x_2, x_3)}{\langle x_2\vert \hat{\rho}_{\rm in} \vert x_2\rangle}\nonumber \\ 
 &=& \frac{P(x_1, x_2) P(x_2, x_3)}{P(x_2)}, 
 \end{eqnarray}
where in the third line of (\ref{dirTTJP}) we have used (\ref{pxixj}).   
The three-time correlation moment is evaluated to be, 
\begin{eqnarray}
\label{3mom}
\mu_{111}=\langle \hat{X}_1 \hat{X}_2 \hat{X}_3\rangle&=& \sum_{x_1,x_2,x_3=\pm 1} x_1 x_2 x_3 P(x_1, x_2, x_3)\nonumber \\
&=& 0.
\end{eqnarray}
From the set of eight moments (\ref{1mom}), (\ref{2mom}) and (\ref{3mom}), we construct the TTJP (see (\ref{MI}) as,  
%%\begin{widetext}
\begin{eqnarray}
\label{mip}
P_{\mu}(1,1,1)&=&\frac{1}{8} [1+2\cos(\omega\Delta t)+\cos(2\omega\Delta t)] = P_{\mu}(-1,-1,-1), \nonumber \\ 
P_{\mu}(-1,1,1)&=&\frac{1}{8} [1-\cos(2\omega\Delta t)]=P_{\mu}(-1,-1,1) = P_{\mu}(1,1,-1) = P_{\mu}(1,-1,-1), \nonumber \\
P_{\mu}(1,-1,1)&=&\frac{1}{8} [1-2\cos(\omega\Delta t)+\cos(2\omega\Delta t)] = P_{\mu}(-1,1,-1). 
\end{eqnarray}
%%\end{widetext}
On the other hand, the three dichotomic variable quantum probabilities $P(x_1,x_2, x_3)$ evaluated  directly are given by,  
\begin{eqnarray}
\label{pdirect}
P_{d}(1,1,1)&=& \frac{1}{8} [1+\cos(\omega \Delta t)]^2=P_{d}(-1,-1,-1), \nonumber \\
P_{d}(-1,1,1)&=& \frac{1}{8} [1-\cos^2(\omega \Delta t)] = P_{d}(-1,-1,1)=P_{d}(1,1,-1)=P_{d}(1,-1,-1),  \nonumber \\  
P_{d}(1,-1,1)&=& \frac{1}{8} [1-\cos(\omega \Delta t)]^2=P_{d}(-1,1,-1).
\end{eqnarray} 
%%%\end{widetext}
%%
Clearly, there is no agreement between the moment inverted TTJP (\ref{mip}) and the ones of (\ref{pdirect}) directly evaluated. In other words, the TTJP  realized in a sequential measurement are not invertible in terms of the moments which in turn reflects the incompatibility of the set of all MP with the grand JP  $P_d(x_1,x_2,x_3).$ In fact, it may be explicitly verified that $P(x_1,x_3)\neq \sum_{x_2=\pm 1} P_d(x_1,x_2,x_3)$. 
Moment-indeterminacy points towards the absence of a valid grand probability distribution consistent with all the marginals.   

The TTJP and moments can be independently extracted experimentally using NMR methods on an ensemble of spin-1/2 nuclei. The experimental approach and results are reported in the next section.   
\subsection{Experiment \label{424}}

The projection operators at time $t=0$ ($ \hat{X}_1 = \sigma_z $) are 
$\{\hat{\Pi}_{x^0_i} =  \outpr{x^0_i}{x^0_i} \}_{x^0_i= 0,1}$.
This measurement basis is rotating under the unitary $\hat{U}_i$, resulting in time dependent basis given by,
$ \hat{\Pi}_{x_i^t}= \hat{U}^\dagger_i \hat{\Pi}_{x^0_i} \hat{U}_i$.
While doing experiments it is convenient to perform the measurement in the computational basis as 
compared to the time dependent basis. This can be done as follows: We can expand the measurement on an 
instantaneous state $\rho(t_i)$ as,
$\hat{\Pi}_{x_i^t} \hat{\rho}(t_i) \hat{\Pi}_{x_i^t} = \hat{U}_i^\dagger \hat{\Pi}_{x_i^0} \left( \hat{U}_i \hat{\rho}(t_i) \hat{U}_i^\dagger \right) \Pi_{x_i^0} U_i$. Thus, measuring in time dependent basis is equivalent to  evolving the state under the unitary $\hat{U}_i$, followed by measuring in the computational basis and lastly evolving under the unitary $\hat{U}_i^\dagger$.  As explained before, the JP were measured using the circuit shown in Fig. \ref{circuit}c.  Circuit shown in Fig. \ref{circuit}c has two controlled gates for encoding the outcomes of first and second
measurements on to the first and second ancilla qubits respectively.  A set of four experiments are to be
performed, with following arrangement of first 
and second controlled gates for measurement of the TTJP:
{(i)} CNOT; CNOT,
{(ii)} anti-CNOT; CNOT,
{(iii)} CNOT; anti-CNOT, and
{(iv)} anti-CNOT; anti-CNOT.
The propagators $\hat{U}_i=e^{-i \sigma_x \omega t_i/2}$ 
is realized by the cascade ${H} \hat{U}_d {H}$, where ${H}$ is the Hadamard gate, and the delay propagator $\hat{U}_d = e^{-i \sigma_z \omega t_i/2}$ corresponds to the z-precession
of the system qubit at $\omega = 2 \pi100$ rad/s resonance off-set. The diagonal tomography was performed at the end to determine the probabilities \cite{nmr_elgi}.  The experimental profile shown in Fig. \ref{3d} was obtained by varying $\Delta t$ such that $\theta= \omega \Delta t \in [0,\pi]$. 
\begin{figure}[b]
\centering
\includegraphics[trim = 0cm 0cm 0cm 0cm,clip=true,width=8cm]{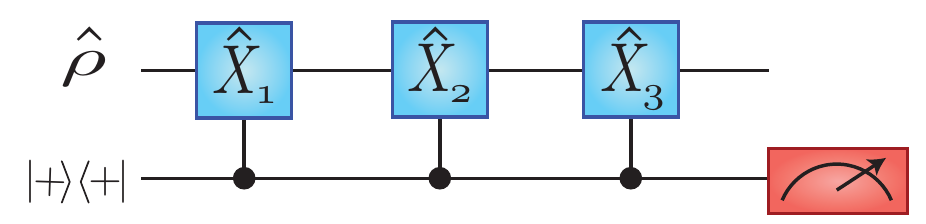}
\caption{Moussa Protocol for obtaining the $3$-time correlated moments. 
One and two time moments can be calculated using the appropriate number of controlled gates.}
\label{moments}
\end{figure}
\begin{figure}
\centering
\hspace{-.4cm}
\includegraphics[width=10cm]{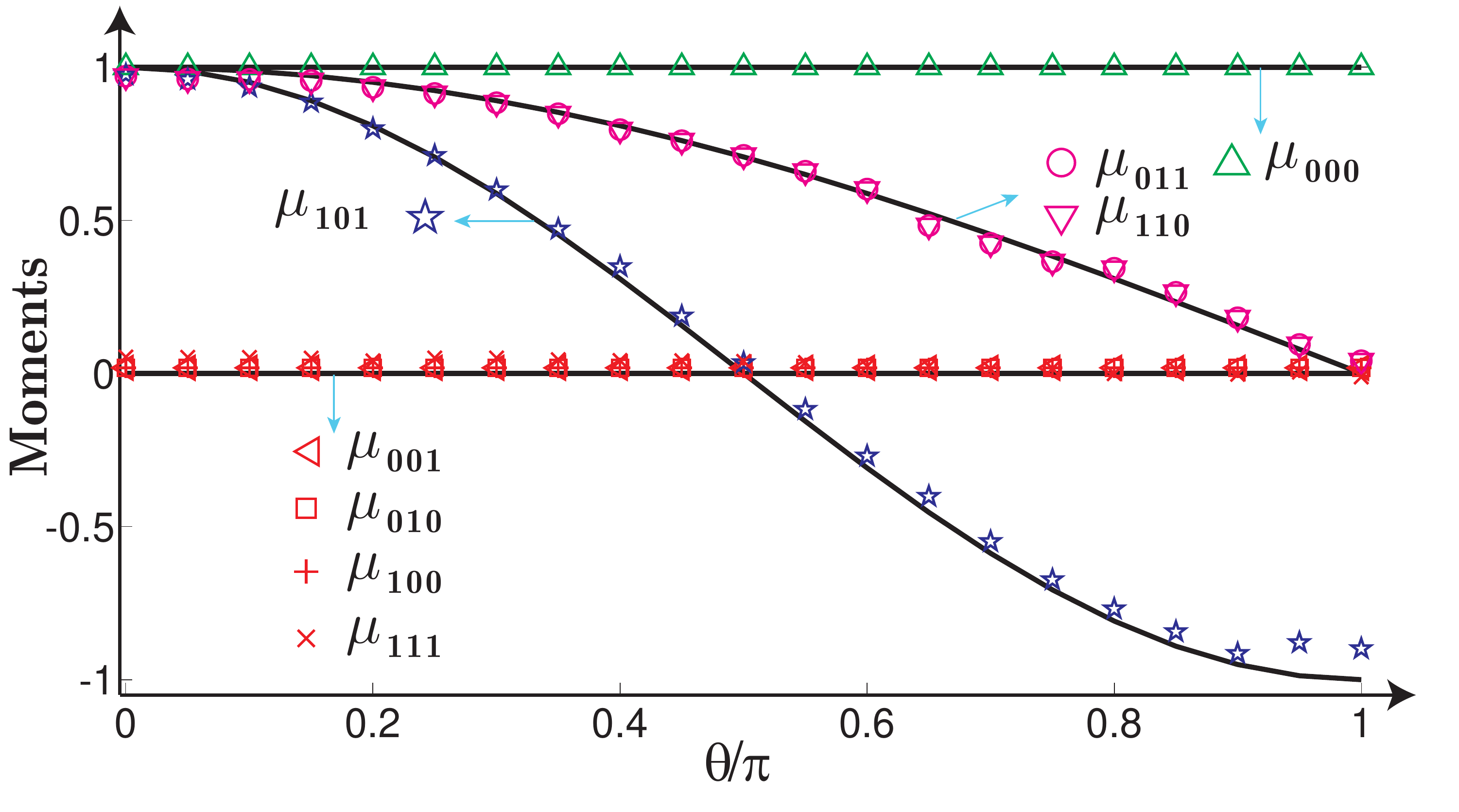}
\caption{Moments obtained experimentally from Moussa Protocol. The symbols represent experimentally obtained
values of the indicated moments with the solid lines showing the corresponding theoretical values.  Here $\theta = \omega \Delta t $.}
\label{moments_results}
\end{figure}
The three qubits were provided by the three $^{19}$F nuclear spins of 
trifluoroiodoethylene dissolved in acetone-D6. The structure of the molecule is shown in Fig. \ref{exptscheme1}c and the chemical shifts and the scalar coupling values (in Hz) in Fig. \ref{exptscheme1}d.
The effective $^{19}$F spin-lattice (T$ _2^* $) and spin-spin (T$ _1 $) relaxation time
constants  were about $ 0.8 $s and $ 6.3 $ s respectively. The
experiments were carried out at an ambient temperature
of 290 K on a 500 MHz Bruker UltraShield NMR spectrometer.
The first spin (F$_1$) is used as the system qubit and, other spins (F$_2$ and F$_3$) as the ancilla qubits. 
Initialization involved preparing the state,
$\frac{1-\epsilon}{8}\mathbb{I} + \epsilon \left\{\frac{I}{2}\mathbb{I}_S \otimes \outpr{00}{00}_A \right\}$ where
$\epsilon \sim 10^{-5}$ is the purity factor \cite{cory}. The pulse sequence to prepare this state from the equilibrium state is shown in Fig. \ref{exptscheme1}(e). All pulses were numerically optimized using the GRAPE 
technique \cite{khaneja2005optimal} and had fidelities better than $ 0.999 $.
With our choice of measurement model (Fig.~\ref{circuit}c) we find a striking agreement with theoretical results on TTJP (\ref{pdirect}) . 
%One might have also run the post measured state resulting after the first dashed block in Fig.~\ref{circuit}c through an arbitrary CP map before the next step. However such post processing CP map would have affected the results. In other words,
 Our measurement scheme provides an optimal procedure to preserve the state information, thus resulting in an excellent agreement of experimental results on TTJP with theoretical prediction (see Fig.~\ref{3d}).     

For calculating the moments we utilize the Moussa protocol \cite{moussa1qbitcontextuality2010}, 
which requires only two spins in our case. We utilize F$_1$ as the system 
and F$_2$ as the ancilla qubit. F$_3$ was decoupled using $ \pi $ pulses 
and the initialization involved preparing the state,
$\frac{1-\epsilon}{8}\mathbb{I} + \epsilon \left\{\frac{1}{2}\mathbb{I}_S \otimes \outpr{+}{+}_A 
\otimes \outpr{0}{0} \right\}$, which is obtained by applying the Hadamard gate to F$_2$ 
after the pulse sequence shown in Fig. \ref{exptscheme1}c.
The circuit for measuring moments by Moussa protocol is shown in Fig. \ref{moments}
and it proceeds as follows, 
\begin{eqnarray}
&\hat{\rho}\otimes\ket{+}\bra{+} & \nonumber \\
&\downarrow \mathrm{c\hat{X}_1} & \nonumber \\
&\hat{\rho}\mathrm{\hat{X}_1}^\dagger\otimes\outpr{0}{1}  + \mathrm{\hat{X}_1}\hat{\rho}\otimes\outpr{1}{0} + & \nonumber \\
&\hat{\rho}\otimes\outpr{0}{0} + \mathrm{\hat{X}_1}\hat{\rho}\mathrm{\hat{X}_1}^\dagger\otimes\outpr{1}{1}& \nonumber \\
& \downarrow \mathrm{c\hat{X}_2} & \nonumber \\
&\hat{\rho}\mathrm{\hat{X}_1}^\dagger\mathrm{\hat{X}_2}^\dagger\otimes\outpr{0}{1} + \mathrm{\hat{X}_2}\mathrm{\hat{X}_1}\hat{\rho}\otimes\outpr{1}{0} + & \nonumber \\
&\hat{\rho}\otimes\outpr{0}{0} + \mathrm{\hat{X}_2}\mathrm{\hat{X}_1}\rho\mathrm{\hat{X}_1}^\dagger \mathrm{\hat{X}_2}^\dagger\otimes\outpr{1}{1}& \nonumber \\
& \downarrow \mathrm{c\hat{X}_3} & \nonumber \\
&\hat{\rho}\mathrm{\hat{X}_1}^\dagger\mathrm{\hat{X}_2}^\dagger\mathrm{\hat{X}_3}^\dagger\otimes\outpr{0}{1} + \mathrm{\hat{X}_3}\mathrm{\hat{X}_2}\mathrm{\hat{X}_1}\hat{\rho}\otimes\outpr{1}{0} + & \nonumber \\
&\hat{\rho}\otimes\outpr{0}{0} + \mathrm{\hat{X}_3}\mathrm{\hat{X}_2}\mathrm{\hat{X}_1}\hat{\rho}\mathrm{\hat{X}_1}^\dagger \mathrm{\hat{X}_2}^\dagger\mathrm{\hat{X}_3}^\dagger\otimes\outpr{1}{1},& \nonumber
\end{eqnarray}
where, $ \mathrm{c\hat{X}_i} $ represents the controlled gates and $ \hat{\rho} $ is the initial state of the system.
The state of the ancilla qubit ($ \hat{\rho}_a $) at the end of the circuit is given by,
\begin{eqnarray}
\hat{\rho}_a  &=& \outpr{0}{1}{\rm Tr}(\hat{\rho}  \hat{X}_1^\dagger \hat{X}_2^\dagger \hat{X}_3^\dagger) + \outpr{1}{0}{\rm Tr}(\hat{X}_3 \hat{X}_2 \hat{X}_1 \hat{\rho}) \nonumber \\
&&+ \outpr{0}{0}{\rm Tr}(\hat{\rho}) + \outpr{1}{1}{\rm Tr}( \hat{X}_3 \hat{X}_2 \hat{X}_1 \hat{\rho} \hat{X}_1^\dagger \hat{X}_2^\dagger \hat{X}_3^\dagger). \nonumber
\end{eqnarray}
Moussa protocol was originally proposed for commuting observables, however, it can be easily extended to non-commuting
observables. The NMR measurements correspond to the expectation values of spin angular momentum operators $ I_x $ or $ I_y $\cite{cavanagh}.
The measurement of the $ I_x $ for ancilla qubit at the end of the circuit gives: 
\begin{equation}
{\rm Tr}[\hat{\rho}_a I_x] = {\rm Tr}[\hat{X}_3 \hat{X}_2 \hat{X}_1 \hat{\rho}]/2 + {\rm Tr}[\hat{\rho} \hat{X}_1^\dagger \hat{X}_2^\dagger \hat{X}_3^\dagger]/2. 
\label{exp_Ix}
\end{equation}
If, $ \hat{X}_1, \hat{X}_2, \hat{X}_3 $
commute, then the above expression gives $ \mathrm{Tr}[\hat{\rho} \hat{X}_1 \hat{X}_2 \hat{X}_3] $.
In case of non-commuting hermitian observables, we also measure expectation value of $ I_y $, which gives :
\begin{equation}
i{\rm Tr}[\rho_a I_y] = {\rm Tr}[\hat{X}_3 \hat{X}_2 \hat{X}_1 \hat{\rho}]/2 - {\rm Tr}[\hat{\rho} \hat{X}_1^\dagger \hat{X}_2^\dagger \hat{X}_3^\dagger]/2. 
\label{exp_Iy}
\end{equation}
From (\ref{exp_Ix}) and (\ref{exp_Iy}) we can calculate  
$ \mathrm{Tr}[\hat{\rho} \hat{X}_1 \hat{X}_2 \hat{X}_3] \equiv \langle \hat{X}_1 \hat{X}_2 \hat{X}_3 \rangle$
for the $3$-measurement case. Hence, by using the different number of controlled gates in appropriate order we can calculate all the moments.
The experimentally obtained moments for various values $\theta = \omega \Delta t \in [0,\pi]$  are shown in Fig. \ref{moments_results}.  
\begin{figure}
\centering
\vspace{.3cm}
\includegraphics[trim = 0cm 0cm 0cm 0cm,clip=true,width=11cm]{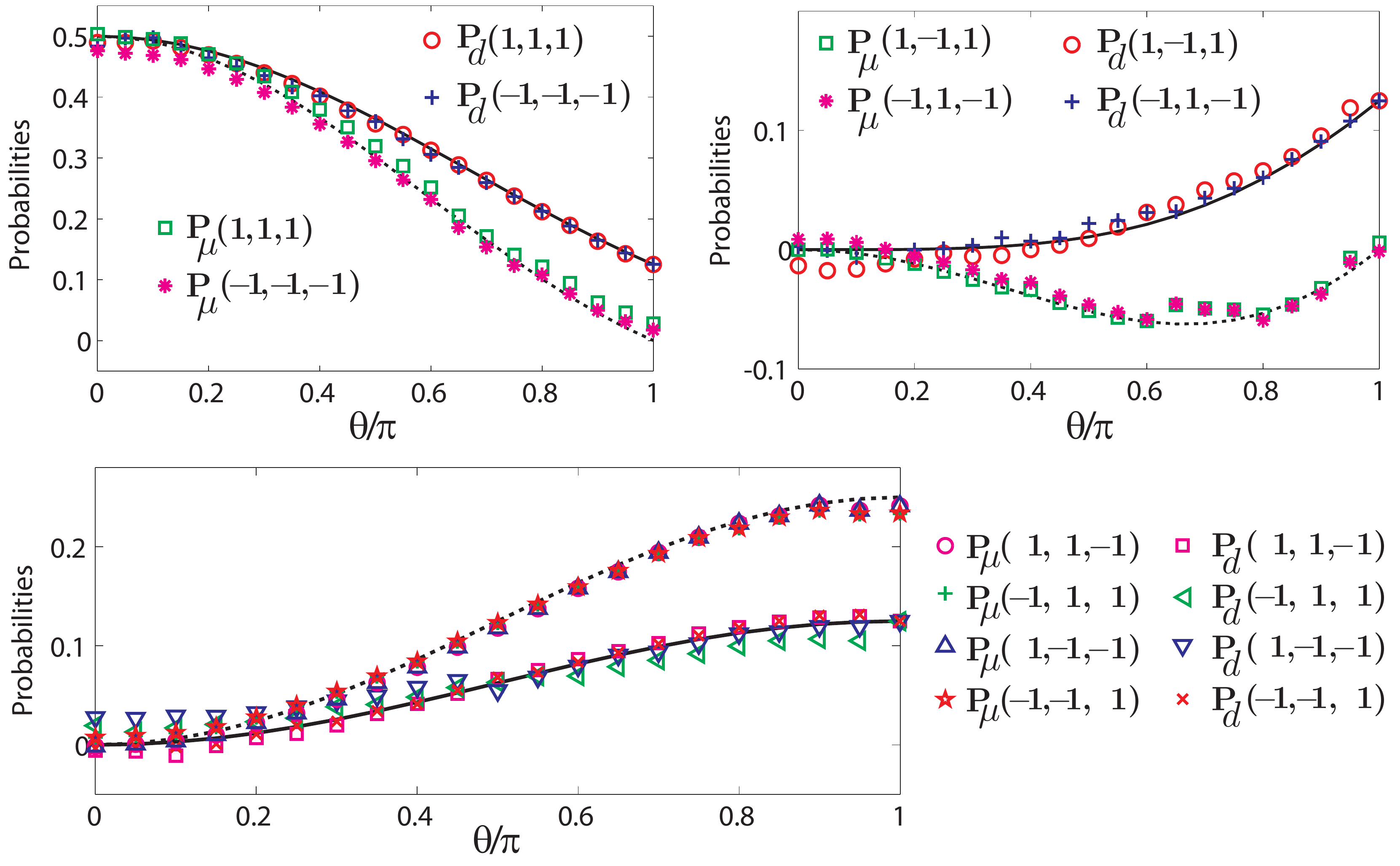}
\caption{Three-time JP (TTJP): The solid curves represent the probabilities measured
directly and the dashed curve the probabilities obtained by inverting the moments.
The symbols represents the experimental data. The mismatch between the directly measured and the inverted moments indicate the illegitimacy of the grand probability distribution.}
\label{3d}
\end{figure}
These experimentally obtained moments are inverted according to Eq. (\ref{MI}) to calculate
the TTJP and are plotted along with the directly obtained TTJP using circuit shown in Fig. \ref{exptscheme1} c as symbols in Fig. \ref{3d}. The theoretical values for TTJP from moments and the one directly obtained are plotted as solid and dashed lines respectively. The results agree with the predictions of Eqs. (\ref{mip}) and (\ref{pdirect}) that the TTJP obtained directly and the one obtained from the inversion of moments do not agree.

\subsection{Conclusion \label{425}}
In classical probability theory, statistical moments associated with dichotomic random variables determine the probabilities uniquely. When the same issue is explored in the quantum context -- with random variables replaced by Hermitian observables (which are in general non-commuting) and  the statistical outcomes of observables in sequential measurements are considered  -- it is shown that the JP do not agree with the ones inverted from the moments. This is explicitly illustrated by considering sequential measurements of a dynamical variable at three different times in the  specific example of a spin-1/2 system.An  experimental investigation based on NMR methods, where moments and the JP are extracted independently, demonstrates the moment indeterminacy of probabilities, concordant with theoretical observations.   

The failure to revert joint probability distribution from its moments points towards its inherent incompatibility with the family of all marginals. In turn, the moment indeterminacy reveals the absence of a legitimate joint probability distribution compatible with the set of all marginal distributions  %-- a common underpinning of various no-go theorems in the foundational aspects of quantum theory.  In other words .. 
is the root cause behind various no go theorems like Bell theorem and Kochen-Specker theorem which supports the underlying randomness in quantum mechanics and puts constraints on various local hidden variable theories.
%
%
%%\newpage

\thispagestyle{empty}

%%% % % % % % % % ch5
\titlespacing*{\chapter}{0pt}{-50pt}{20pt}
\titleformat{\chapter}[display]{\normalfont\Large\bfseries}{\chaptertitlename\ \thechapter}{20pt}{\Large}
\chapter{ Applications of NOON states \label{chp5}}

\section{\textbf{Introduction} \label{51}}
The use of multiple-quantum coherences has several applications 
in NMR \cite{bodenhausenMQ1980}.
The NOON state in an N-spin system is the highest multiple quantum coherence i.e. for an $N$-spin system it is $N^{th}$ quantum coherence which can be written in the
form $(\vert 00 \cdots 0 \rangle+\vert 11 \cdots 1 \rangle)/\sqrt{2}$.  Such states have 
found several applications in optics and other fields \cite{lee2002quantum,dowling2008quantum,chen2010heralded}.
Recently NMR NOON states have been used to sense ultra-weak magnetic fields \cite{jones2009magnetic}.
The NOON state can be easily prepared in star-topology systems using CNOT and Hadamard gates as described in the later section.
The NOON state has high sensitivity to phase encoding.  In NMR, phase encoding 
can be achieved easily either by static fields or by RF fields.  The former is used in
the characterization of translational diffusion and the latter is used to characterize
the RF inhomogeneities of the NMR probes.  Here we describe the advantages of
NOON state in both of these applications.

Driven by the internal thermal energy, the atoms or molecules in a bulk matter
may exhibit random translational motion, which is termed as translational
diffusion \cite{ghez2001diffusion,march2002introduction}.  
The diffusion constant ($D$) is described as the amount 
of a particular substance that diffuses across a unit area in unit time 
under the influence of a unit concentration gradient \cite{ghez2001diffusion}.  
Here we describe the application of NOON state for measuring diffusion constant.

The strength of NMR over other spectroscopy techniques is in the excellent control over
quantum dynamics \cite{levitt2001spin}.  
Coherent control of nuclear spins is achieved by a calibrated set of radio frequency pulses.  
Achieving high-fidelity quantum control in practice is however limited by 
radio frequency inhomogeneity (RFI) over the sample volume.
RFI characterization is important not only for conventional NMR experiments, but also for
designing robust and high-fidelity  quantum gates for quantum information studies \cite{coryrfijcp}.  
In MRI, RFI characterization can help to understand certain image distortions and to
correct them \cite{mri2}.  Here we describe a NOON state method to 
characterize RFI at high RF amplitudes.  In \S \ref{52}, I describe the preparation of NOON states in star topology spin systems.
\S \ref{53} contains, the description of the measurement of diffusion using NOON state method, and the experimental results obtained on a ten-spin star-topology system.  In \S \ref{54} I describe the characterization of RF inhomogeneity 
(RFI) using NOON states.  Finally I conclude in \S \ref{55}.

\section{\textbf{NOON state in a star-topology system} \label{52}}
\begin{figure}[t]
 \begin{center}
 \includegraphics[trim = 0cm 5.5cm 0cm 8cm,clip=true,width=11.5cm]{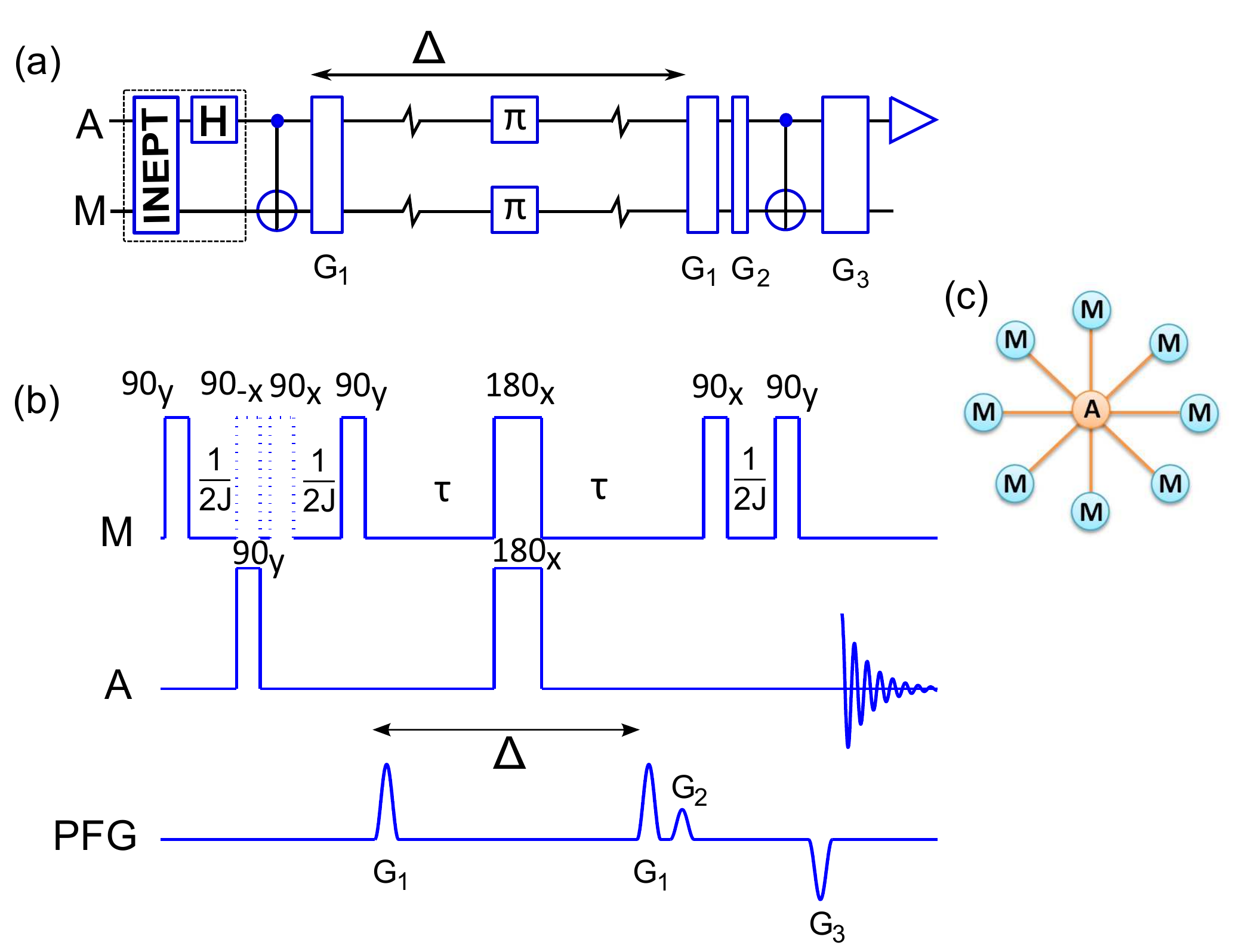}
 \caption{(a) Circuit for the measurement of diffusion constant
 using NOON state, (b) the corresponding NMR pulse sequence, and
 (c) a star-topology spin-system.
 Here $\Delta$ is the diffusion time, $G_1$ are diffusion gradients, and
 $G_2$ and $G_3$ are gradients which select out the NOON state. We have used
 INEPT sequence to transfer magnetization from the M spins to the A spin.  In this
 case, the pseudo-Hadamard gate is not necessary. In (b) second pulse of INEPT 
 cancels with the first pulse of CNOT as shown by dotted lines.}
 \label{nooncircuit}
\end{center}
\end{figure}

The NOON state of an $N$-qubit system is a superposition
\begin{eqnarray}
\vert N00N \rangle &=&(\vert N, 0 \rangle + \vert 0, N \rangle)/\sqrt{2} \nonumber \\
& = & (\vert 00 \cdots 0 \rangle + \vert 11 \cdots 1 \rangle)/\sqrt{2},
\end{eqnarray}
where $\vert N, 0 \rangle$ is the state with
$N$ qubits being $\vert 0 \rangle$ and 0 qubits being $\vert 1 \rangle$, and 
$\vert 0, N \rangle$ is the state with 0 qubits being 
$\vert 0 \rangle$ and $N$ qubits being $\vert 1 \rangle$
\cite{boto2000quantum}.  The circuit (Fig. \ref{nooncircuit}a) for preparing the NOON state consists of a Hadamard
gate followed by a C-NOT gate \cite{jones2009magnetic} on a quantum register
with a single `control' qubit and a set of $(N-1)$ `target'
qubits initialized in $\vert 00 \cdots 0 \rangle$  state:
\begin{eqnarray}
\vert 00 \cdots 0\rangle 
\stackrel{\mathrm{H}}{\longrightarrow}
\frac{1}{\sqrt{2}}(\vert 0 \rangle + \vert 1 \rangle) \vert 00 \cdots 0 \rangle
\nonumber
\stackrel{\mathrm{CNOT}^{\otimes (N-1)}}{\longrightarrow} \\
 \frac{1}{\sqrt{2}}(\vert 00 \cdots 0 \rangle + \vert 11 \cdots 1 \rangle)
= \vert N00N \rangle.
\end{eqnarray}

A star-topology spin system AM$_{N-1}$ consisting of a single spin A coupled to a 
set of $N-1$ magnetically equivalent spins M (see Fig. \ref{nooncircuit}c).
The star-topology allows parallel implementation of ${N-1}$ C-NOT gates.
For the diffusion experiments, all the CNOT gates can be realized in parallel by just two pulses:
$90_x^M - \frac{1}{2J} - 90_y^M$ as shown in Fig. \ref{nooncircuit}b. 
In our experiments, it is unnecessary to begin with a pure $\vert 00 \cdots 0 \rangle$ state.
We can start from the thermal equilibrium state at room temperature, 
and select the coherence pathway that passes through the NOON state.
The coherence selection can be achieved rather conveniently with two PFGs, $G_2$ and $G_3 = -g G_2$, where
$g=\gamma_\mathrm{eff}/\gamma_A$.

\section{\textbf{Measurement of Diffusion constant} \label{53}}
Diffusion constant of a liquid can be measured by NMR either 
with the help of relaxation studies or more conveniently using 
pulsed-field-gradients (PFGs) \cite{carr1954effects}.   The standard method 
for measuring diffusion constant consists of two identical PFGs (G$_1$)
separated by diffusion time $\Delta$ with a refocusing 
$\pi$ pulse in the middle \cite{stejskal1965spin,price2}.\\  
Due to the random molecular motion,  each nuclei in the NMR sample
acquires a different relative phase, leading to the decay of average Hahn-echo signal. 
Theoretically, the decay of this signal $S$ can be given as,
\begin{eqnarray}
S(G_1) / S_0 = \exp\left\{-\gamma^2 G_1^2 \delta^2 D (\Delta  - \delta/3 )\right\},
\label{sgz}
\end{eqnarray}
where $D$ is the diffusion constant, $S_0 = S(0)$ is the normalization
factor, and $-\delta/3$ is the correction to $\Delta$ 
due to the finite durations $\delta$ of the PFGs \cite{pricea}.
Thus the diffusion constant $D$ can be extracted  
by systematically varying the strength of the PFGs, $G_1$,
and fitting a Gaussian curve to the experimentally obtained 
echo intensities. 
Such PFG methods are widely used for the measurement of diffusion constants
and already numerous improved sequences are available \cite{pricea,priceb,johnson1999diffusion}.
For example, long-lived singlet states have been used to study slow diffusion \cite{Bodenhausen2005,yadav2010nmr},
and single-scan measurement of diffusion has been realized by
effectively z-coordinate dependent PFG strengths \cite{Keeler1dDOSY2003}.\\
 Although the NOON state method is a special form of the multiple-quantum diffusion experiments
\cite{kay1986application,chapman1993mqdiffusion,zheng2009mq}, the star-topology of the spin system, if available, 
provides a simple way to prepare the NOON states and to convert them back to single quantum coherences.

\subsection{Diffusion constant via the NOON state}
Under the $G_1-\pi-G_1$
sequence, a spin system in NOON state diffusing through a distance
$dz$ acquires a net relative phase:
\begin{eqnarray}
\frac{1}{\sqrt{2}}(\vert 00 \cdots 0 \rangle + \vert 11 \cdots 1\rangle) \stackrel{G_1-\pi-G_1}{\longrightarrow} \nonumber \\ 
\frac{1}{\sqrt{2}}(\vert 00 \cdots 0 \rangle + e^{id\phi} \vert 11 \cdots 1\rangle).
\end{eqnarray}
The relative phase acquired is $d\phi = \gamma_\mathrm{eff} dz G_1 \delta$,
where $\gamma_\mathrm{eff} = \{\gamma_A+(N-1)\gamma_M \} = g\gamma_A$. In the
standard diffusion experiments with uncoupled spin 1/2 particles, $N=0$ and 
$\gamma_\mathrm{eff}$ reduces to $\gamma_A$.  The larger the value of $g$, the
 more sensitive is the NOON state for phase encoding,  
and allows the study of diffusion with weaker PFGs and smaller durations ($\Delta$) 
between them.  This is the main advantage of using the NOON states.
Since NOON state is a multiple quantum coherence, it is necessary to convert
it back to single quantum coherence before detection.  This conversion can
efficiently be carried out using a second set of CNOT gates:
\begin{eqnarray}
\frac{1}{\sqrt{2}}(\vert 00 \cdots 0 \rangle + e^{id\phi} \vert 11 \cdots 1\rangle) \stackrel{\mathrm{CNOT}^{\otimes (N-1)}}{\longrightarrow} 
\nonumber \\
\frac{1}{\sqrt{2}} 
(\vert 0 \rangle + e^{id\phi}\vert 1 \rangle) \vert 00 \cdots 0 \rangle.
\end{eqnarray}
Thus the phase encoding due to the diffusion, i.e., $d\phi$ 
has been transfered to one transition of the control spin. 
In an ensemble of nuclei the above phase encoding results 
in the attenuation of the control transition.  
Diffusion constant can be measured by monitoring this attenuation as 
a function of the PFG strengths \cite{pricea}.

\begin{figure}[t]
\begin{center}
\includegraphics[trim= 0cm 4.5cm 0cm 0.5cm,clip=true,width= 15cm]{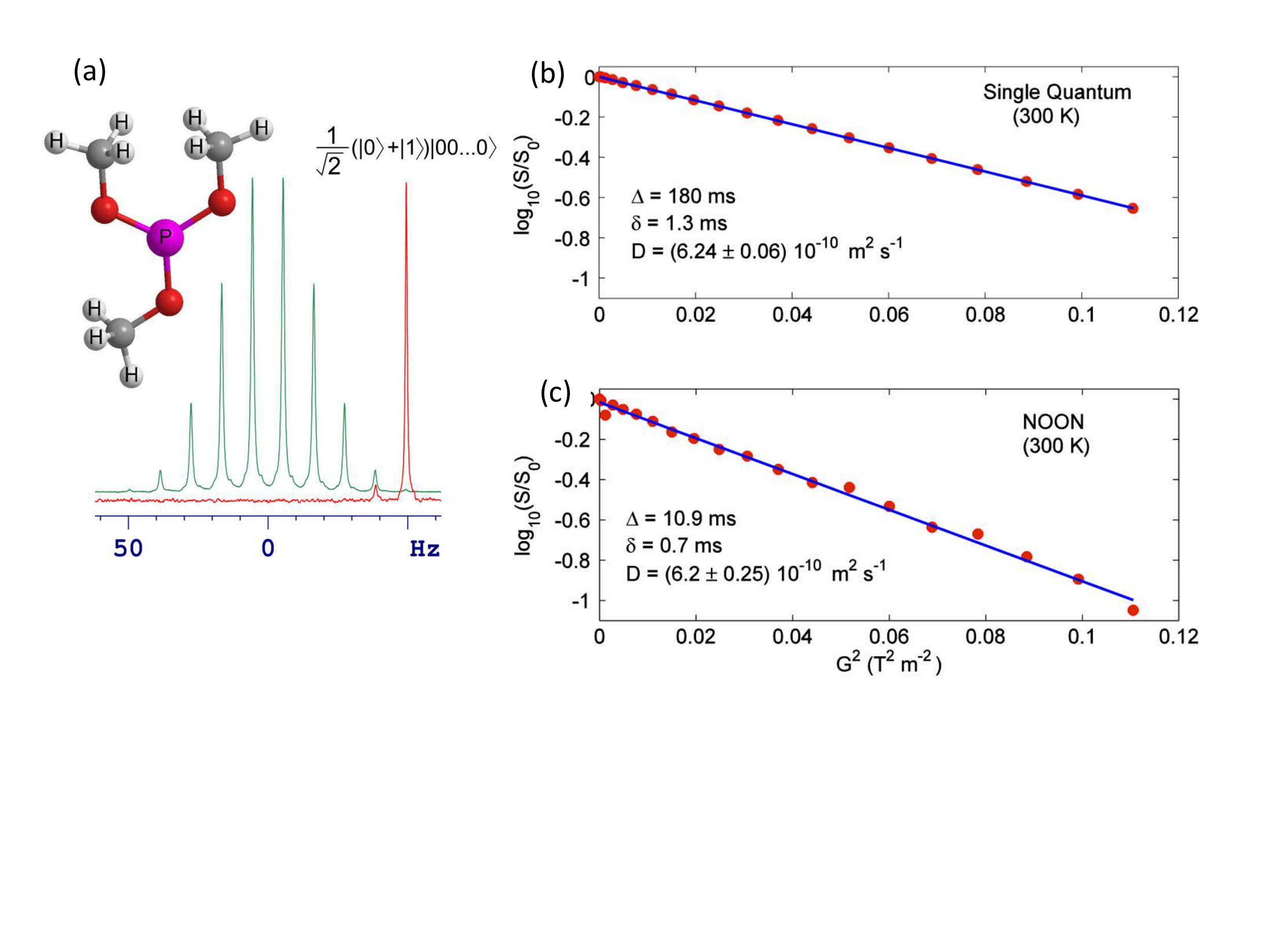}
\caption
{(a) $^{31}$P spectra of trimethylphosphite corresponding to single quantum excitation
from thermal equilibrium (upper trace) and corresponding to the NOON state (lower trace). 
The molecular structure of trimethylphosphite is shown in the inset.  Intensity of the
 echo signals as a function of the gradient strength G$_1$
with (b) standard method and (c) the NOON state method.  The dots represent the 
experimental data and the lines represent the linear fit.}
\label{difftmp}
\end{center}
\end{figure}
\begin{figure}[t]
\begin{center}
\includegraphics[trim= 0cm 0cm 0cm 1cm,clip=true,width= 13cm]{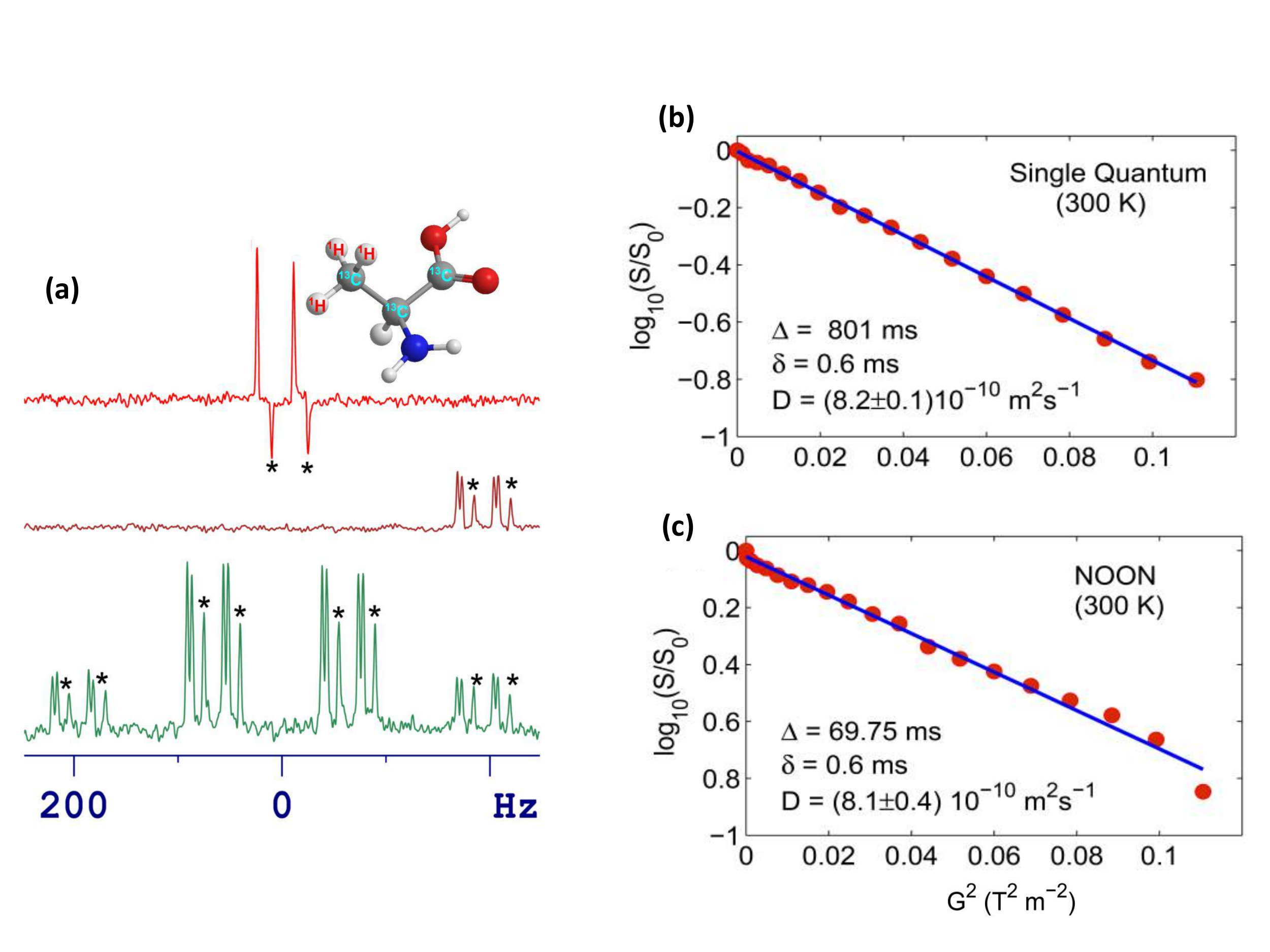} 
\caption{ 
(a) $^{13}$C spectra of methyl carbon of Alanine corresponding to single quantum excitation
from thermal equilibrium (lower trace), and after converting the NOON state into the single 
quantum coherence using CNOT gates without (middle trace) and with (top trace) 
subsequent $^1$H decoupling. The peaks marked with asterisks correspond to different
molecules and are to be ignored.  Intensity of the echo signals as a function of the gradient strength G$_1$
with (b) standard method and (c) the NOON state method.  The dots represent the
 experimental data and the lines represent the linear fit.}
\label{ala} 
\end{center}
\end{figure}

\subsection{Diffusion Experiments \label{diffexp}}
All the experiments are carried out at 300 K on a 
500 MHz Bruker NMR spectrometer.  We have studied the diffusion
 of two samples using the standard method as well as the NOON state
  method in each case.
  
First we used 100 $\mu$l of trimethylphosphite (P(OCH$_3$)$_3$ dissolved in 600 $\mu$l DMSO-D6) 
as the star-topology system (Fig. \ref{difftmp}a).  Here each of the nine magnetically 
equivalent $^1$H spins are coupled to the $^{31}$P spin via indirect spin-spin interaction with a coupling 
constant of $J=11$ Hz thus forming an AM$_9$ spin system.
The parameters and the results of single quantum and NOON state experiments are shown in
 Fig. \ref{difftmp}b, Fig. \ref{difftmp}c consecutively. 
The logarithms of normalized intensities obtained via diffusion experiments 
are plotted versus $G_1^2$ in Fig. \ref{difftmp}. 
In case of single quantum experiment 
 diffusion constant $D = (6.24 \pm 0.06) 10^{-10}$ m$^2$ s$^{-1}$
was obtained by applying a linear fit based on the expression \ref{sgz}.
The NOON state diffusion experiments were carried out 
as described by the pulse sequence in Fig. \ref{nooncircuit}b.
 After an initial INEPT transfer to
enhance the $^{31}$P polarization, the
NOON state was prepared and at the end of the diffusion delay, it was converted
back to a single quantum coherence. 
To select the 10-quantum coherence pathway, the PFGs $G_2$ and $G_3$ were
 adjusted such that $G_3/G_2 = -g = -(\gamma_P+9\gamma_H)/\gamma_P = -23.2$.

The results of the NOON state diffusion experiments are shown in Fig. \ref{difftmp}b. 
Again a linear fit based on expression \ref{sgz}, with an effective gyromagnetic ratio 
$\gamma_\mathrm{eff}= (\gamma_P+9\gamma_H)$ lead to the diffusion constant of 
$D = (6.2 \pm 0.25) 10^{-10}$ m$^2$ s$^{-1}$, which is close to the value obtained
from the single quantum method.
It can be seen that, in the NOON state method, the diffusion delay is reduced by a 
factor of 16 and the duration $\delta$ of diffusion PFG is reduced by a factor of 2.

The second sample consisted of 5 $mg$ of $^13$C-labeled alanine in 0.7 $ml$ of D$_{2}$O.  
The molecular spectra is shown in Fig. \ref{ala}a.  We prepared 4-spin NOON state using the three
 equivalent protons interacting with the methyl carbon with a coupling con-
 stant of $J = 130$ Hz.  The parameters and the results of single quantum and
  NOON state experiments are shown in Fig. \ref{ala}b, Fig. \ref{ala}c consecutively. 
The logarithms of normalized intensities obtained via diffusion experiments 
 are plotted versus $G_1^2$ in Fig. \ref{ala}.  In case of single quantum experiment 
diffusion constant $D = (8.2 \pm 0.1) 10^{-10}$ m$^2$ s$^{-1}$ (see Fig. \ref{ala})
 was obtained by applying a linear fit based on the expression \ref{sgz}.
 The NOON state diffusion experiments were carried out 
as described by the pulse sequence in Fig. \ref{nooncircuit}b.
  After an initial INEPT transfer to enhance the $^{31}$P polarization, the
NOON state was prepared and at the end of the diffusion delay, it was converted
back to a single quantum coherence.  To select the 4-quantum coherence pathway, the PFGs $G_2$ and $G_3$ were
 adjusted such that $G_3/G_2 = -g = -(\gamma_C+3\gamma_H)/\gamma_P = -12.92$.  During the INEPT
  transfer, CNOT gates, and the diffusion delays, the evolution of all un-necessary
   interactions were refocused by using $\pi$ pulses selectively on methyl carbon.  The spin-selective $\pi$ pulse 
   was realized using a strongly modulated
 RF sequence \cite{coryrfijcp}.  The results of the NOON state diffusion experiments are shown in Fig. \ref{ala}c. 
Again a linear fit based on expression \ref{sgz}, with an effective gyromagnetic ratio 
$\gamma_\mathrm{eff}= (\gamma_P+9\gamma_H)$ lead to the diffusion constant of 
$D = (8.1 \pm 0.4) 10^{-10}$ m$^2$ s$^{-1}$, which is close to the value obtained
from the single quantum method.  It can be seen that, in the NOON state method, the diffusion delay is reduced by a 
factor of 12.

\section{\textbf{Radio frequency inhomogeneity}\label{54}}
Consider an ensemble of spin-1/2 nuclei with long relaxation time constants.
The standard method for studying RFI involves a single, constant low-amplitude, on-resonant RF pulse of
variable duration.  The corresponding pulse sequence is shown in Fig. \ref{pprfi}a.
The intensity $s(\nu_0,t)$ of the obtained signal oscillates due to
the varying transverse magnetization and decays mainly due to RFI.  This oscillation is
known as  the `Torrey oscillation' \cite{torrey1949transient}. The Fourier transform $S(\nu_0,\nu)$ of the oscillation $s(\nu_0,t)$
leads to a distribution over the actual RF amplitudes $\nu$.  We can extract RFI distribution $p(\nu)$
by normalizing the positive real part of $S(\nu_0,\nu)$ to unit area.
In typical NMR probes, one obtains an asymmetric Lorentzian 
distribution with a higher weight towards the amplitudes lower than the nominal value
\cite{coryrfijcp}. Such a profile can be modelled by an asymmetric Lorentzian
\begin{eqnarray}
p(\nu) &=& \frac{a \lambda_-^2}{(1 - \frac{\nu}{\nu_0})^2+\lambda_-^2} ~~ \mathrm{if} ~~ \nu <   \nu_0, ~~ \mathrm{and,} \nonumber \\
p(\nu) &=& \frac{a \lambda_+^2}{(1 - \frac{\nu}{\nu_0})^2+\lambda_+^2} ~~ \mathrm{if} ~~ \nu \ge \nu_0
\label{rfimodel}
\end{eqnarray}
where $\lambda_\pm$ are the Lorentzian line-width parameters
and $a$ is the normalization constant.\\
 The main disadvantage of the single-quantum method is the requirement of 
a long RF pulse to capture artefact-free RFI profile.
The duty cycle limit of the probe introduces a limitation on the highest
amplitude at which RFI can be studied.  In the following we describe the NOON state
Torrey oscillation which is ultra-sensitive for incoherence and hence allows 
capturing RFI profile with a shorter RF pulse and at higher RF amplitudes.
\subsection{Measurement of RFI via NOON states}

In order to measure RFI while preserving the coherence order of NOON states, we utilize
Z-nutation.  The pulse sequence for the NOON state method is shown in Fig. \ref{pprfi}b.
A $\phi_z$ pulse acting on a NOON state introduces a large relative phase shift: 
\begin{eqnarray}
&&\frac{1}{\sqrt{2}}(\vert 00 \cdots 0 \rangle + \vert 11 \cdots 1 \rangle)  
\stackrel{\phi_z^M}{\longrightarrow} \\
&&\frac{1}{\sqrt{2}}(\vert 00 \cdots 0 \rangle + e^{(N-1)i\phi_z} \vert 11 \cdots 1 \rangle)
\nonumber 
\stackrel{CNOT}{\longrightarrow}  \\
&&\frac{1}{\sqrt{2}}(\vert 0 \rangle + e^{i(N-1)\phi_z} \vert 1 \rangle) \vert 00 \cdots 0\rangle.
\end{eqnarray}
Therefore the resultant phase-shift of the control spectral line is 
$\phi_z^M = (N-1)\phi_z$.
A $\phi_z^M$ pulse, on $M$-spins can be realized by $\left(\frac{\pi}{2}\right)^M_{-y} \phi_x^M \left(\frac{\pi}{2}\right)_{y}^M$, where
$\phi_x^M = 2\pi (N-1) \nu_0 \tau$ is the corresponding x-pulse of duration $\tau$
for a nominal RF amplitude $\nu_0$.
 Ideally, in the absence of RFI, one should see a regular oscillatory behavior of this coherence with $\tau$.
In practice we see Torrey oscillation, i.e., decaying oscillations, due to the RFI during $\phi_x^M$
pulse.  

\subsection{Experimental characterization of RFI}
The single-quantum Torrey oscillations were studied using a sample consisting of 
600 $\upmu$l of 99\% D$_2$O. All the RFI experiments are carried out at 300 K on a 
500 MHz NMR Bruker NMR spectrometer with QXI probe.
 The NOON Torrey oscillations were studied using the trimethylphosphite sample (Fig. \ref{difftmp}a).
 \begin{figure}[t]
 \begin{center}
 \includegraphics[trim=0cm 0cm 0cm 0cm, clip=true,width= 14cm,angle =0]{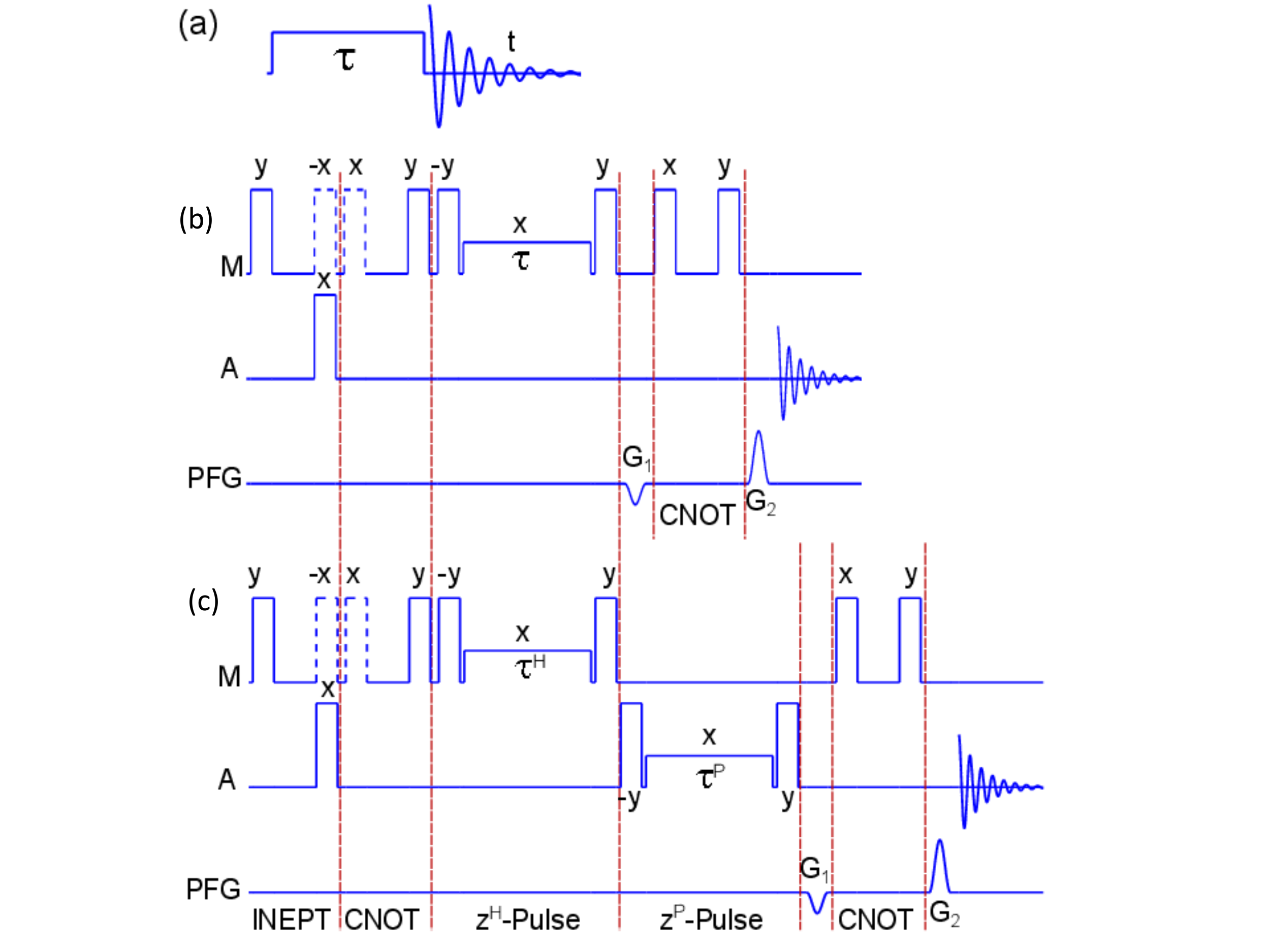}
 \caption{
  (a) Pulse sequence for single-quantum Torrey oscillation,
  (b) pulse sequence for NOON Torrey oscillation, and the RFI,
  (c) pulse sequence for measuring RFI correlation between $^{1}$H and $^{31}$P channels.  In (b) and (c)
 except the variable duration ($\tau$) pulses, all other pulses are 90$^{\circ}$
}
 \label{pprfi}
 \end{center}
 \end{figure}
 
\begin{figure}
\begin{center}
\includegraphics[trim=0.2cm 0.2cm 0cm 2cm, clip=true,width= 13.5cm,angle =0]{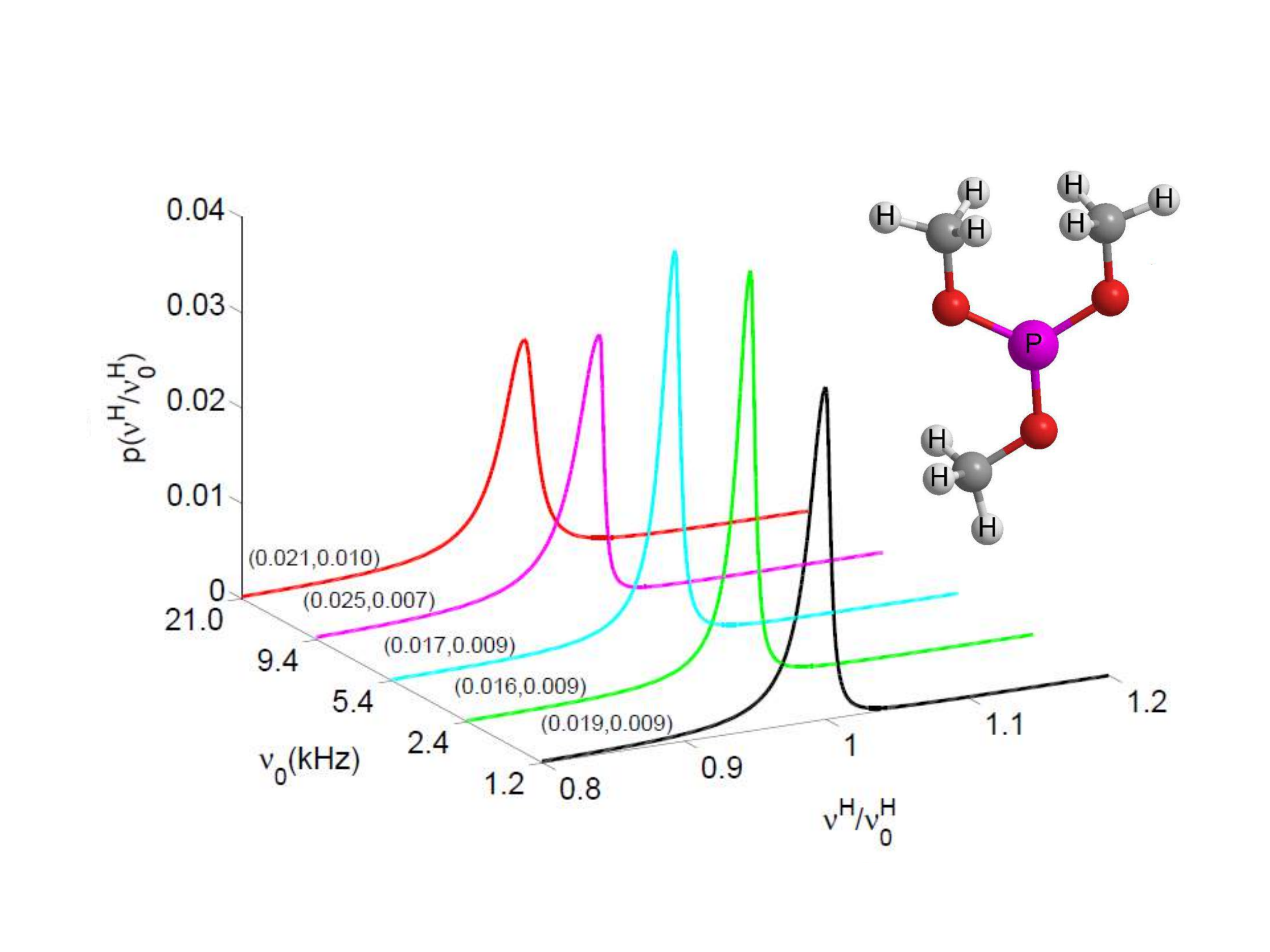}
\caption{The RFI profiles at various nominal amplitudes obtained from NOON state experiments and
corresponding values of asymmetry parameters $(\lambda_{-} ,\lambda_{+})$.
}
\label{qxiplotall}
\end{center}
\end{figure}
The single quantum Torrey oscillations were studied using the pulse sequence shown
in Fig. \ref{pprfi}a.  A series of experiments were recorded by incrementing the duration
of the on-resonant pulse with an amplitude of 1 kHz on $^1$H channel of QXI probe.
The $\tau$-increments are set to 250 $\upmu$s respectively, and a total of 256 transients were
recorded.  The asymmetry parameters $(\lambda_- = 0.018,~\lambda_+ =0.009)$ are obtained by fitting
 the profile in expression \ref{rfimodel} to the normalized real part $S(\nu_0,\nu)$ of the Fourier transform of Torrey oscillation.  
The asymmetry parameters $\lambda_\pm$ for NOON RFI experiments 
in the NOON state RFI experiments, can were extracted in similar way 
using the pulse-sequence Fig. \ref{pprfi}b and the results are 
displayed in Fig. \ref{qxiplotall}.
Its evident from values of $\lambda_\pm$ that the RFI is some what stronger at 
higher amplitudes.  For example,
the asymmetry parameters at 9.4 kHz and 21 kHz are significantly larger than those
at 5.4 kHz.

\subsection{Correlation between RFI of two channels}
In a two-channel probe, the regions of high RF intensity of first channel may
not correspond to regions of high RF intensity of the second.  In other words,
RFI profiles of the two channels may be spatially correlated.  Such a correlation
can easily be studied using a 
NOON state consisting of a collective coherence of heteronuclear species.
The pulse diagram for the RFI-correlation study is shown in Fig. \ref{pprfi}c.
A $\phi_z^{A}$ pulse is introduced after the $\phi_z^{M}$ pulse, and the two pulses are independently
incremented to obtain a 2D dataset $s(\nu^{M_9},\nu^A)$.  
The Fourier transform along the two dimensions results
in the frequency profile $S(\nu^{M_9},\nu^A)$.

\begin{figure}
\begin{center}
\includegraphics[trim=0cm 5.5cm 0cm 2.2cm, clip=true,width= 14cm,angle=0]{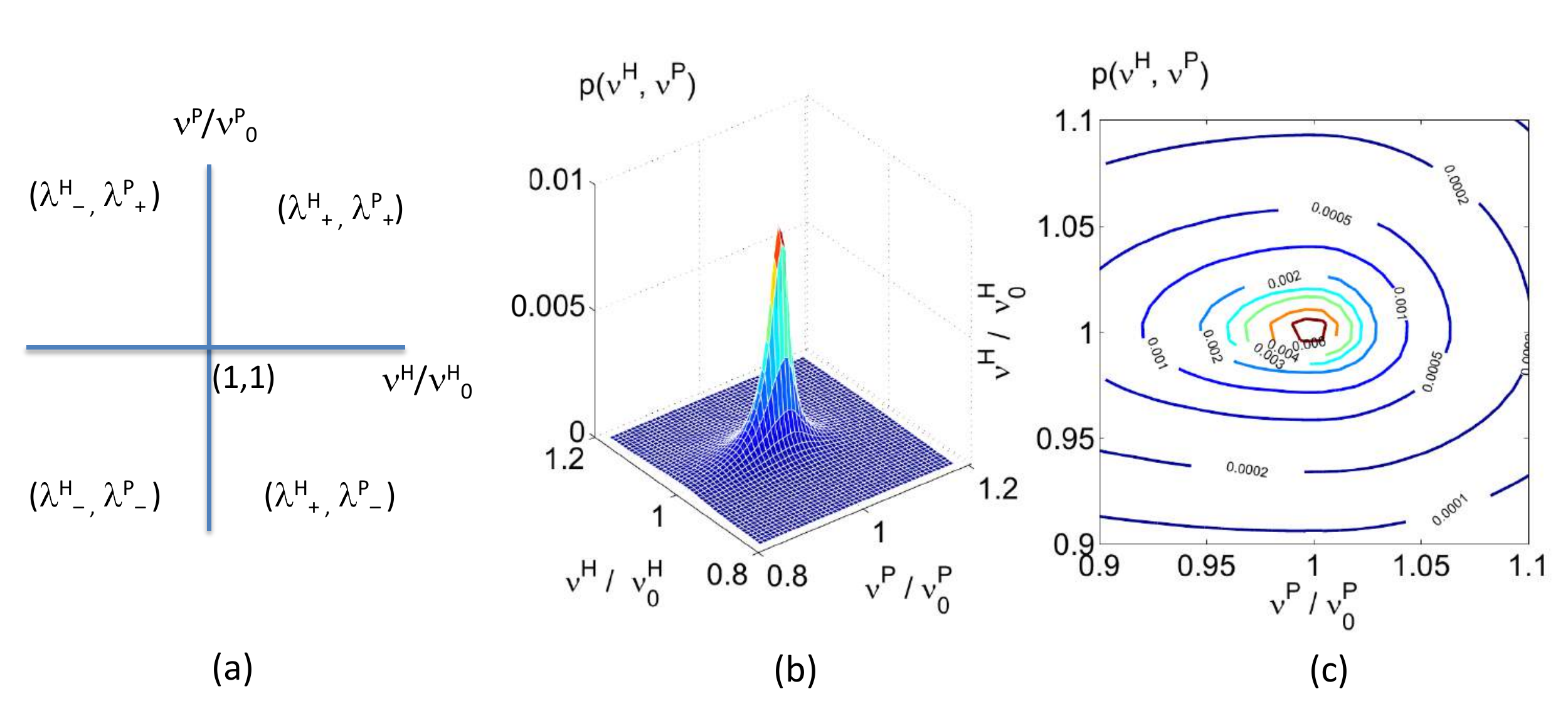}
\caption{Characterization of RFI correlations between $^1$H and $^{31}$P channels of
QXI probe.(a) The $\nu^H$ - $\nu^P$ plane, (b) the surface plot of the RFI
profile $p(\nu^H,\nu^P)$, and (b) the contour plot of $p(\nu^H,\nu^P)$}.
\label{rficorr}
\end{center}
\vspace{0.3cm}
\end{figure}

We now model the RFI correlation using a 3D asymmetric
Lorentzian:
\begin{eqnarray}
p(\nu^H/\nu^H_0,\nu^P/\nu^P_0) = 
\frac{\lambda_0^2}{d^2(\nu^H,\nu^P) + \lambda_0^2},
\label{eqrficorr}
\end{eqnarray}
where $d(\nu^H,\nu^P)$ is the scaled distance of 
$\left(\frac{\nu^H}{\nu_0^H},\frac{\nu^P}{\nu_0^P}\right)$ from the nominal point 
$(1,1)$:
\begin{eqnarray}
d^2(\nu^H,\nu^P) = \lambda_\pm^H \left( 1 - \frac{\nu^H}{\nu^H_0} \right)^2 
        + \lambda_\pm^P \left(1 - \frac{\nu^P}{\nu^P_0} \right)^2.
\end{eqnarray}
Here $\lambda_\pm^H$ and $\lambda_\pm^P$ are the four asymmetry parameters
on the four quadrants of $\nu^H$ - $\nu^P$ plane (Fig. \ref{rficorr}b).  These 
parameters together with
$\lambda_0$ completely characterize the RFI correlation.  

In the following, we describe experimental characterizations of 
the RFI correlation between $^1$H and $^{31}$P channels of an NMR probe.
  We used trimethylphosphite (P(OCH$_3$)$_3$ dissolved in DMSO-D6) 
  for correlation experiments.  The molecular properties of the sample are explained in subsec. \ref{diffexp}.  The 2D dataset was recorded by independently incrementing
the $\tau^H$ and $\tau^P$ pulses respectively by 11.1 $\upmu$s and 50.0 $\upmu$s.
The nominal RF amplitudes in $^1$H and $^{31}$P channels were 2.4 kHz and 2.5 kHz 
respectively.
A total of 128 data points in $^1$H dimension and 96 data points in $^{31}$P dimension were
collected.  The real positive part of the Fourier transform of the 2D Torrey oscillations 
$S(\nu^{H_9},\nu^P)$ was obtained after a zero-fill to 256 points in each dimension.
 The asymmetric Lorentzian parameters $(\lambda_0,\lambda_\pm^{H/P})$ were obtained by 
fitting a model profile obtained by expression \ref{eqrficorr} to 
to the experimental profile $S_{1,9}$. 
$\{\lambda_0 = 0.005, \lambda_+^H = 0.114, \lambda_-^H = 0.226, \lambda_+^P = 0.095, \lambda_-^P = 0.028\}$.
The corresponding RFI profile $p(\nu^H,\nu^P)$ is shown in Fig. \ref{rficorr}b and its 
contour plot is shown in Fig. \ref{rficorr}c.
It can be observed that $^{31}$P-channel has a wider RFI distribution than the $^1$H-channel.

\section{\textbf{Conclusions}\label{55}}
We described two applications of NOON states: (i) studying translational
diffusion in liquids, and (ii) characterizing RF inhomogeneity of
NMR probes.
  
 We first described the experimental measurement of diffusion constant in a model system.
Both single-quantum and the NOON-state 
experiments lead to identical values for the diffusion constant, but
the errors were slightly larger in the latter case due to  additional
complexities.  However, the ultra-sensitivity of a NOON state to incoherence allowed an order of magnitude shorter 
diffusion delay, indicating the possible applications in studying slow diffusion.  
It might also be possible to combine the NOON state technique with the
single scan 2D techniques to achieve ultra-fast diffusion measurements.  

In the second part, we observed that the 
NOON state Torrey oscillations decay much faster than the single quantum Torrey
oscillations allowing the characterization of RFI at higher RF amplitudes.
Using this method, we have studied RFI of an NMR probe at different RF amplitudes and 
compared the results.\\
   We then extended the NOON state method, using a 3D Lorentzian model, 
to characterize RFI correlations between two RF channels.
Although the basic principle is general, the methods are 
particularly convenient with spin-systems having star-topology allowing parallel
implementation of CNOT gates. 

Such spin-systems can be found in many organic compounds and in biomolecules.
The ultra-sensitivity of NOON states to phase encoding may also have
potential applications in MRI.

%\end{document}

\thispagestyle{empty}

% % % % % % % % ch6
\mathversion{normal2}
\titlespacing*{\chapter}{0pt}{-50pt}{20pt}
\titleformat{\chapter}[display]{\normalfont\Large\bfseries}{\chaptertitlename\ \thechapter}{20pt}{\Large}
\chapter{Dynamical Decoupling of Spin-Clusters using Solid State NMR \label{chp6}}

\section{\textbf{Introduction}\label{61}}
The study of dynamics and control of quantum many body systems has 
renewed interest in the field of quantum information.  While encoding
information onto a quantum channel can potentially speed up certain 
computations and allow secure data transmission, the practical realization
of these applications are hindered by the extreme sensitivity of the
quantum channel to environmental noises.
Systems based on nuclear spin-clusters is one among the various 
architectures being investigated to realize quantum channels.
Several experimental demonstrations of quantum information processing (QIP)
using solid-state nuclear magnetic resonance (SSNMR) have already been reported
\cite{ding2001quantum,nmrqipsinglecrystal,laflammessnmr,moussa1qbitcontextuality2010,simmons2011entanglement}.
By sophisticated control of spin-dynamics it is in principle possible to achieve 
a larger number of quantum bits (qubits) using SSNMR, because of the 
%The advantange in SSNMR is the 
availability of large spin-clusters coupled
mutually through long-range dipole-dipole interactions.  
However in such a spin-cluster, fluctuating local fields at the site of each
spin induced by its environment leads to the decoherence of the encoded quantum 
information.

Due to the availability of large spin-clusters it is possible to prepare coherences
of large quantum numbers by a widening network of correlated spins evolving under
two-quantum average Hamiltonian \cite{pines83,pinesmq85}.  These higher order coherences
are not directly observable as macroscopic magnetizations, but can be converted
into obervable single quantum coherence (SQC) using a time-reversed two-quantum
average Hamiltonian.  This method, often known as a `spin-counting experiment'
has been used to study the evolution of coherences of large quantum numbers exceeding 
4000 \cite{suter2006decoherence,suterDecoh2007,suter2010}.

Under the standard Zeeman Hamiltonian any spin-coherence is a non-equilibrium state 
and decays via various relaxation processes, ultimately leading to the equilibrium
longitudinal magnetization.  It has long been discovered that the decay process
of the spin coherence can be prolonged by applying a series of spin flips at regular
intervals of time.  This sequence known as `CPMG sequence', not only refocusses the
effect of spectrometer inhomogeneities, but also reverses the phase evolution of
the coherences under the random fluctuations, provided the spin flips are applied 
sufficiently frequent \cite{cp1954,cpmg1958}.  Such a dynamical method for the suppression 
of decoherence of a qubit due its interaction with environment is often termed 
as `dynamical decoupling' (DD) \cite{lloydDD1999}.
Recently Uhrig introduced a non-periodic spin-flip sequence which he proved
theoretically to proivde optimal decoupling performance for dephasing spin-bath
interactions \cite{uhrig2007keeping}.  
CPMG and other similar periodic spin-flip sequences suppress spin-environment 
interaction to $n$th order using up to $O(2^n)$ pulses, while Uhrig dynamical 
decoupling (UDD) suppresses the same using only $n$ pulses.  
In a high-frequency dominated bath with a sharp cutoff,  
UDD works well provided the frequency of the spin-flips exceeds the cut-off
frequency \cite{biercuk2009optimized,biercuktrappedions,biercuk2011dynamical}.
On the other hand when the spectral density of the
bath has a soft cutoff (such as a broad Gaussian or Lorentzian),
the CPMG sequence was found to outperform the UDD
sequence \cite{du2009preserving,alvarez2010performance,DDSQUIDS2008,de2010universal,barthel2010DDsinglettriplet,rDDdiamond2010,processtomoDD2010,ajoy2011optimal}.  The original sequence for UDD is based the assumption of instantaneous 
spin-flips, which requires infinite bandwidth.  
Later on, Uhrig provided an improved sequence - 'realistic UDD' (RUDD) for 
practical implimentations with finite bandwidth \cite{rudd2011}.   

Most of the theory and experiments of DD sequences are for single spin 
systems. Du et al have studied DD of electron spin coherence in solids \cite{du2011}
while Suter and co-workers have reported systematic experimental comparisons
of various DD schemes on an ensemble of single spins in SSNMR 
\cite{alvarez2010performance,ajoy2011optimal}.  A few studies of DD on two-qubit systems have also
been reported \cite{agarwal2010saving,mukhtar}.  Experimentally, Wang et. al. have
studied DD on electron-nuclear spin pairs in a solid-state
system \cite{du2011}, and Soumya et. al. have studied the performance
of UDD an a two-qubit liquid-state NMR system
\cite{roy2010initialization}.

We experimentally studied the performance of 
various DD schemes on an extended network of spin-1/2 nuclei forming a
large spin cluster.   In this chapter I report these studies.  This chapter is organized as
follows.  In the next \S I briefly describe the method of preparing
and detecting multiple quantum coherences (MQC) in SSNMR.  In \S \ref{63} I 
summarize the construction of various DD sequences.  The experimental
details are described in \S \ref{64}.  Finally I conclude in \S \ref{65}.

\section{\textbf{Multiple Quantum SSNMR}\label{62}} 
\label{mqddsec}
The SSNMR Hamiltonian for a spin cluster with $M$ spin-1/2 nuclei is
\begin{equation}
{\cal H}_{int}= {\cal H}_Z+ {\cal H}_D,
\end{equation}
where the Zeeman and the secular part of dipolar interaction are
respectively,
\begin{eqnarray}
{\cal H}_Z  = \sum_{i=1}^{M} \omega_{i} I_z^i, \nonumber 
%{\cal H}_D =  \left[3I_z^i I_z^j- \sum_{i < j}D_{ij}\left[3I_z^i I_z^j- I^i\cdotI^j\right]
{\cal H}_D = \sum_{i < j}D_{ij}\left[3I_z^i I_z^j-I^{i} \cdot I^{j} \right].
\end{eqnarray}
Here ${\bf I}^i$ and $I_z^i$ are spin angular momentum operator and its
z-component corresponding to the $i^{th}$ spin, and $w_i$ and $D_{ij}$ are
the chemical shift and the dipolar coupling constants.  
The equilibrium density matrix for the above Hamiltonian corresponds to
the longitudinal magnetization expressed as $\sum_i I_z^i$.
The density matrices for the longitudinal spin order can be expressed using
product of longitudinal spin operators, eg. $I_z^1 I_z^2 \cdots $.  
The coherences are described by the product of transverse 
(or of transverse and longitudinal) spin operators, eg. $I_z^1 I_x^2 I_x^3 \cdots $.
The transverse spin operators can also be expressed in terms of raising and lowering
operators:  $I_x = (I_+ + I_-)/2$ and $I_y = -i(I_+ - I_-)/2$.  The difference
between the total number of raising and lowering operators gives the quantum number
$n$ of a particular coherence.  For example, operators $I_+^j I_-^k$, $I_+^j$, and $I_+^j I_+^k$
describe zero, single, and two-quantum coherences respectively.
\begin{figure}
\begin{center}
\includegraphics[width=12cm]{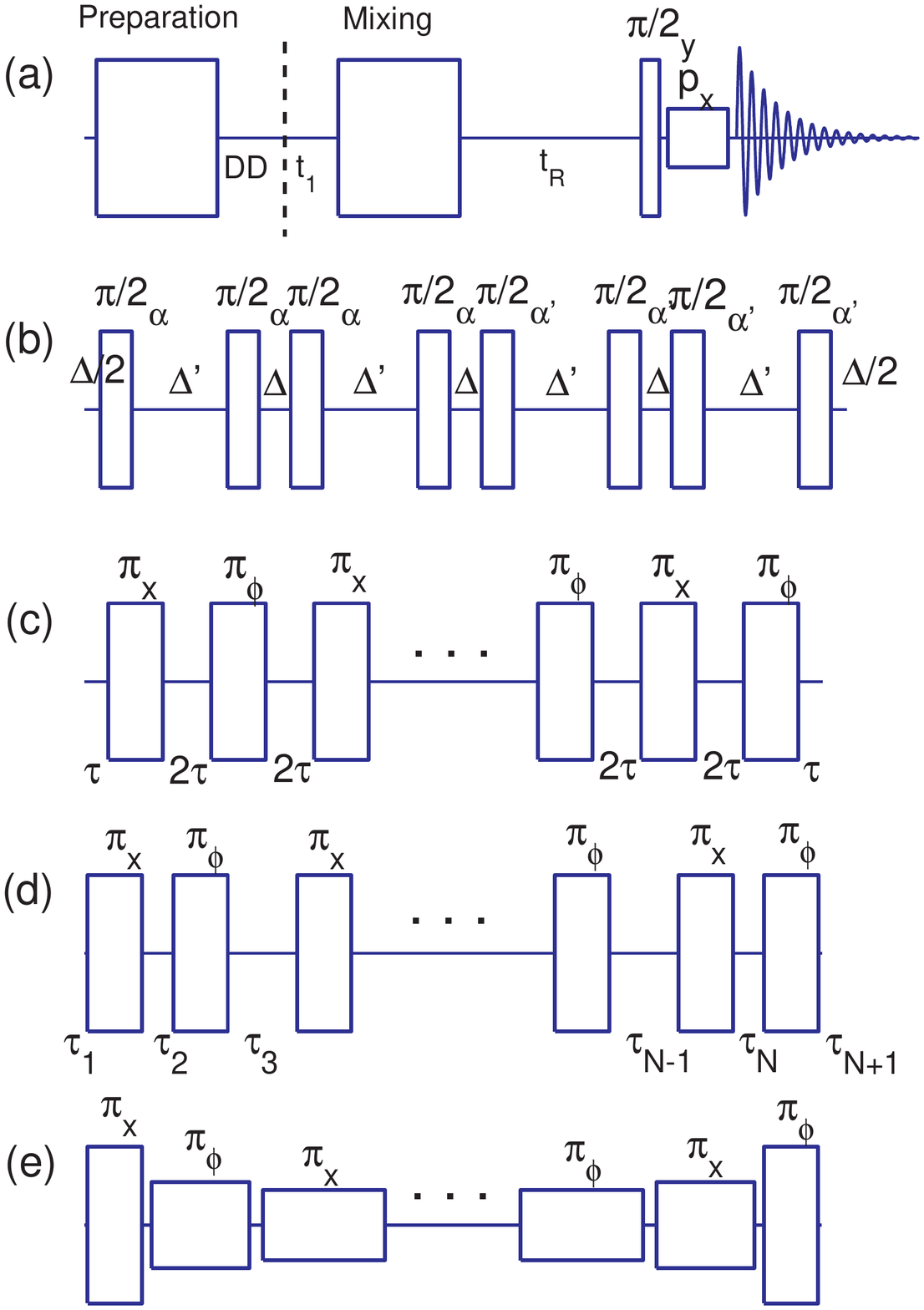}
\end{center}
\caption{\label{mqdd}
The experimental scheme (a) for studying the performance of DD on
large spin-clusters and the 8-pulse sequence (b) implementing ${\cal H}_1(\alpha)$.
In (b), $\Delta' = 2\Delta+\tau_{\pi/2}$, where $\tau_{\pi/2}$
is the duration of each $\pi/2$ pulse.  The DD schemes are described
in (c-e). The phase $\phi$ is set to $x$ for CPMG, UDD, and RUDD 
schemes, while it is alternated between $x$ and $-x$ for CPMGp, UDDp,
and RUDDp.}
\end{figure}
The pulse sequence for preparing and detecting MQC
is shown in Fig. \ref{mqdd}(a-b).  The sequence Fig. \ref{mqdd}a involves preparation
of MQC, application of DD schemes, free-evolution ($t_1$),
converting MQC into longitudinal spin order (mixing), 
destroying the residual coherences by transverse relaxation ($t_R$),  followed 
by detection after converting the logitudinal spin order into SQC.
The 8-pulse sequence in Fig. \ref{mqdd}b corresponds to the two-quantum average Hamiltonian
\begin{equation}
{\cal H}_1 = \frac{D_{ij}}{2} \left( I_+^i I_+^j + I_-^i I_-^j \right),
\end{equation}
for $\alpha=0$.
Preparation and mixing parts involve $m$-cycles of the 8-pulse sequence
${\cal H}_m(\alpha)$ and ${\cal H}_m(0)$ \cite{pinesmq85}.
The possible quantum numbers and the corresponding cluster size increases
with the number of cycles.  Only even quantum coherences are prepared
as shown in Fig. \ref{nVsM}.
To separate the MQC, the relative phase 
$\alpha$ between the prepation and mixing is incremented in proportion 
to the evolution time $t_1$ (Fig. \ref{mqdd}a).  Spurious transverse coherences are
suppressed by an extended delay $t_R$.  The desired signal, stored 
as population information along $z$-axis, is detected with a 
$\pi/2_y$ pulse and a final purge pulse $p_x$ is used to keep only the 
$x$-component.

\begin{figure}
\begin{center}
\includegraphics[width=7cm]{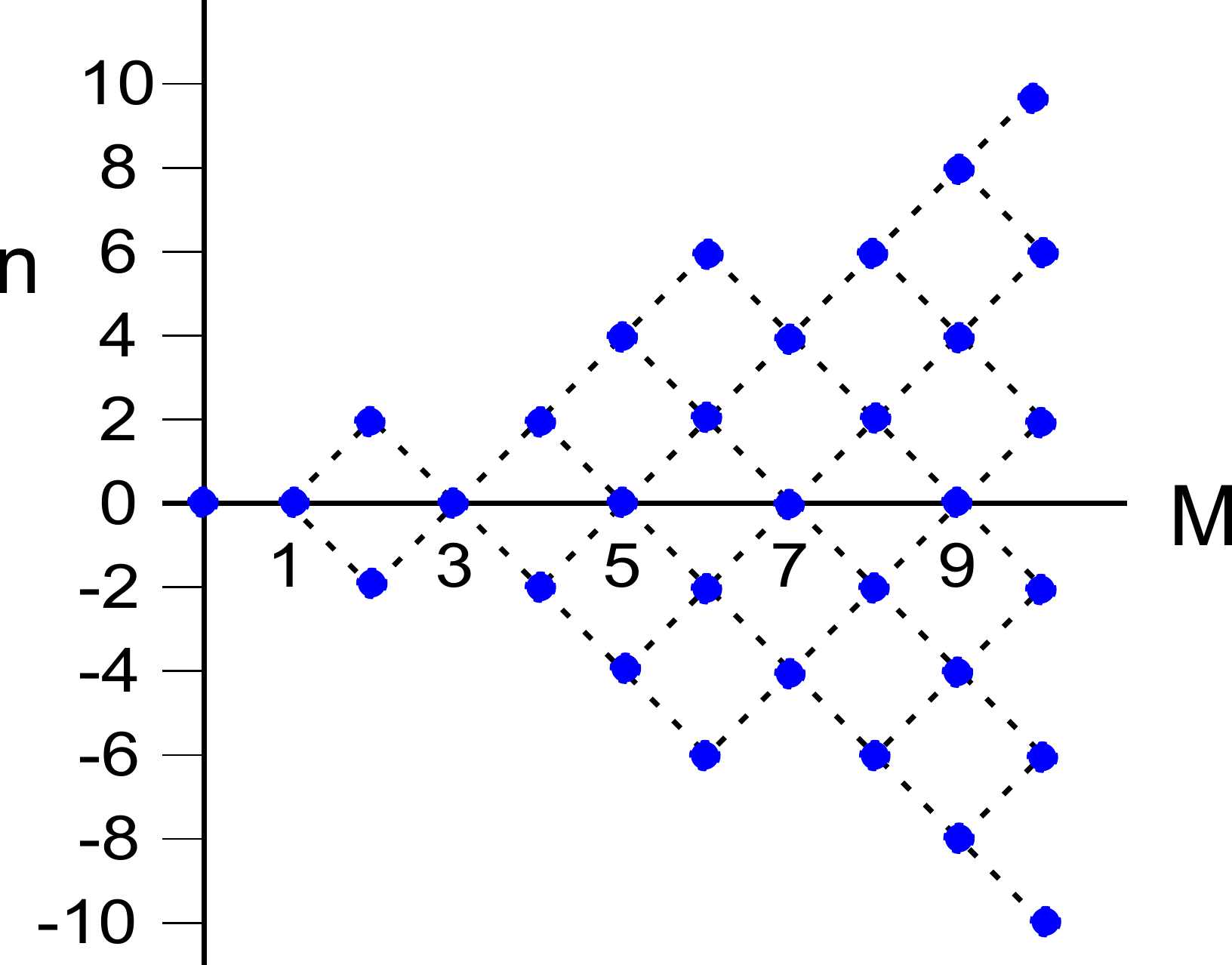}
\end{center}
\caption{\label{nVsM}
Possible quantum numbers ($n$) for $M$-spin cluster after excitation with 
several cycles of two-quantum average Hamiltonian $({\cal H}_m)$.
}
\end{figure}
 
\section{\textbf{DD schemes}\label{63}}
In the following three different DD sequences and their phase variants are described. The performance of a particular sequence depends on the noise spectrum of the system.  For example, CPMG generally performs better against a noise spectrum with a slow frequency cut-off, while UDD works better against the one with a sharp frequency cut-off.

\subsection{CPMG and CPMGp}
\label{cpmg}
CPMG and CPMGp schemes involve periodic spin flips as shown in Figure \ref{mqdd}c.
In CPMG phase $\phi$ set to $x$.  CPMGp scheme is obtained by alternating the phase $\phi$ 
between $x$ and $-x$.  CPMG and CPMGp have different performances depending on
the initial states \cite{suter2010}.  The total duration of the N-pulse 
CPMG is $T = N(2\tau+\tau_\pi)$, where $\tau_\pi$ is the duration of the $\pi$ pulse.
The same parameters $N$ and $T$ are used to compare
CPMG with the following schemes.

\subsection{UDD and UDDp}
The pulse distributions for UDD and UDDp schemes are shown in Figure \ref{mqdd}d.
Here the spin flips are symmetric but not periodic \cite{uhrig2007keeping}.  
The $j^{th}$ $\pi$ pulse is applied at the time instant
\begin{eqnarray}
t_j= T \sin^{2} \left[ \frac{\pi j}{2N+2} \right],
\label{uddtj}
\end{eqnarray}
where $T$ is the total duration of the sequence and $N$ is the total number of pulses.
For a finite bandwidth case, with a $\pi$ pulse
of duration $\tau_\pi$, the delays $\tau_j$ are given by
$\tau_1 = \tau_{N+1} = t_1-\tau_\pi/2$,
$\tau_j = t_{j+1}-t_j-\tau_\pi$, for $2 \le j \le N$.
Like in the previous scheme, UDD and UDDp are differed by the constant phase
and the phase alternation in $\phi$.
\subsection{RUDD}
\label{rudd}
In RUDD and RUDDp, both the delays and the pulse durations vary, but the
overall sequence remains symmetric.  The pulse durations are given by
\begin{equation}
\tau_\pi^j = T \left[ \sin \left( \frac{\pi j}{N+1} \right) \sin \theta_p \right],
\end{equation}
where $T$ is total duration of the sequence and $N$ is the number of pulses.
Here $\theta_p$ is a constant and can be determined by the allowed bandwidth.
We choosed $\tau_\pi^1 =\tau_\pi$, and calculated $\theta_p$ based on the minimum allowed pulse duration:
\begin{equation}
\mathcal
\sin\theta_p = \frac{\tau_\pi}{T \sin \left( \frac{\pi}{N+1} \right)}.
\end{equation}
The amplitude $a_j$ of $j^{th}$ pulse is calibrated such that $2\pi a_j\tau_\pi^j = \pi$.
Time instants of the center of each pulse is same as in equation (\ref{uddtj}).  Using these
time instants, the delays between the pulses can be calculated as
$\tau_{1} = \tau_{N+1} = t_1-\tau_\pi/2$, and
$\tau_{j} = t_j-t_{j-1}-\tau_\pi^j/2-\tau_\pi^{j-1}/2$ for $2 \le j \le N$.
Like in the previous schemes, RUDD and RUDDp are differed by the constant phase
and the phase alternation in $\phi$.

\section{\textbf{Experiment}\label{64}}
The sample consists of crystallites of powdered Hexamethylbenzene. At room temperature
the entire molecule undergoes six fold hopping about the $C_{6}$ axis of benzene ring.
Further the methyl group rapidly reorients about its $C_{3}$ axis.  Due to these motions,
the intramolecular dipolar interactions are averaged out.  Intermolecular dipolar 
coupling is retained and each molecule acts as a point dipole.  Under free precession (no DD),
this sample has a spin-spin relaxation constant of about 25 $\mu$s and a spin-lattice 
relaxation constant of 1.7s.  All the experiments are carried out on a Bruker 500 MHz
spectrometer.

\begin{figure}
\begin{center}

\includegraphics[trim = 1cm 0cm 0cm 0cm, clip=true,width=10cm,angle=-90]{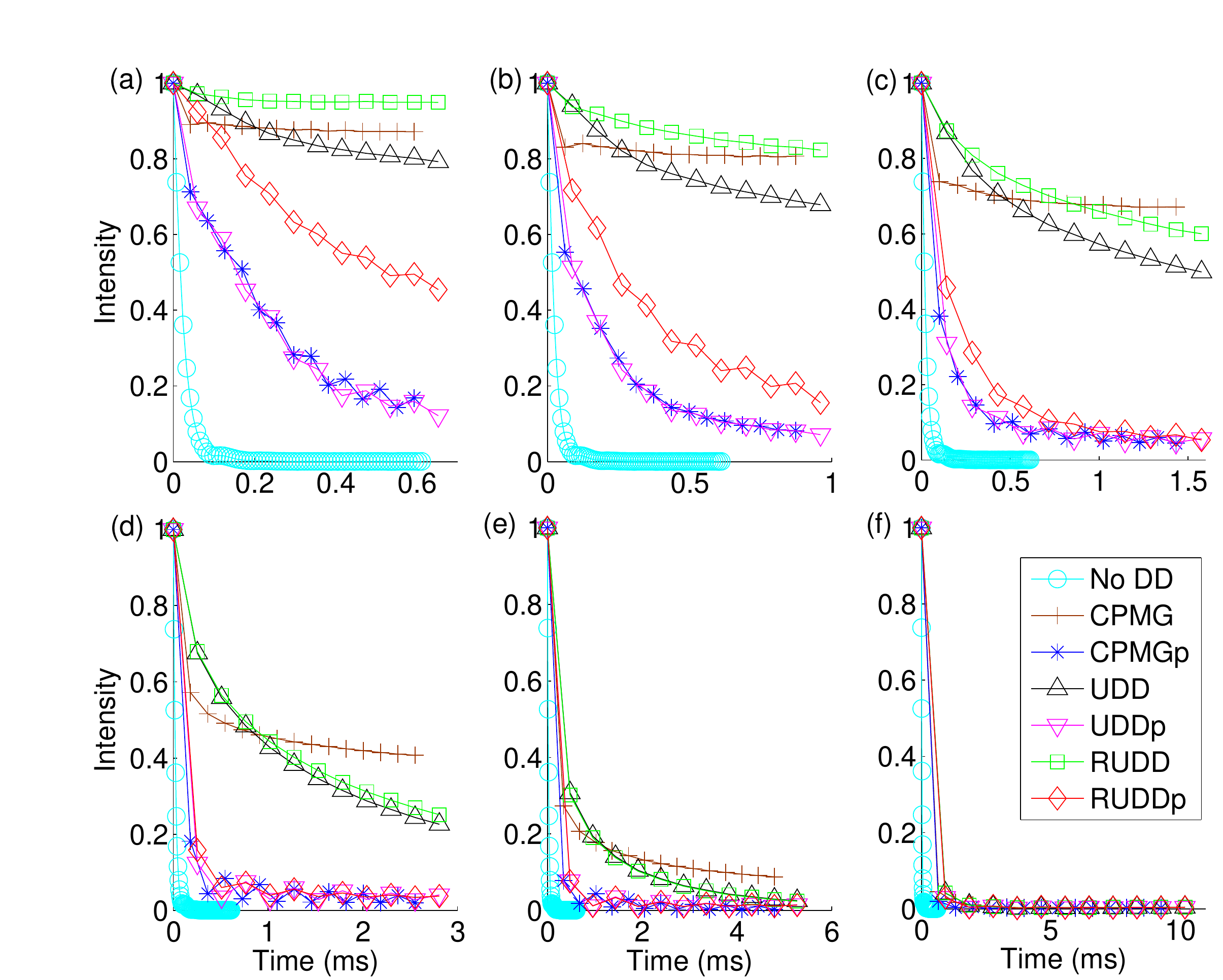}
\end{center}
\caption{\label{sqdddat} 
The performance of 7-pulse DD cycles to suppress decoherence measured as intensities of the preserved coherences.  Here initial state is $\sum_i\sigma_{ix}$, where $i$ is the proton spin index.    The sub-plots correspond to
various CPMG dealys: $\tau=2\;\mu$s (a), $\tau=4\;\mu$s (b), $\tau=8\;\mu$s (c), $\tau=16\;\mu$s (d),
$\tau=32\;\mu$s (e), and $\tau=64\;\mu$s (f). 
} 
\end{figure}

\subsection{DD on SQC}
First we describe the performance of various DD schemes on SQC.
SQC was prepared by using an initial $(\pi/2)_y$ pulse
on equilibrium longitudinal magnetization.
As described in \S \ref{cpmg} and Fig. \ref{mqdd}c, CPMG sequences were 
constructed by periodic distribution of $\pi$ pulses in $\tau-\pi-\tau$ fashion.
The minimum $\tau$ in our experiments was set to 2 $\mu$s owing to the duty cycle
limitation of the probe coil.  In our spectrometer the minimum duration of $\pi$ pulse 
was found to be $\tau_\pi=4.3 \;\mu$s.
Under these experimental conditions, the allowed values of $N$ for UDD and RUDD are 1 to 7.
For $N \ge 8$, one obtains negative delays between the pulses.  Therefore to study DD schemes for 
longer durations, we cycled these 7-pulse DD sequences.  
The 7-pulse CPMG has a total cycle time of $T(\tau) = 7(2\tau+\tau_\pi)$.  
The results of these experiments
are shown in Fig. \ref{sqdddat}.  
The graphs correspond 
to $\tau=2\;\mu$s (a), $\tau=4\;\mu$s (b), $\tau=8\;\mu$s (c), $\tau=16\;\mu$s (d),
$\tau=32\;\mu$s (e), and $\tau=64\;\mu$s (f).  The corresponding $T(\tau)$ values
are used to select the sampling points in no DD, as well as to construct other DD sequences. 
It is clear from these data that  RUDD displays superior
performance for shorter $\tau$ values, CPMG shows better performance while for longer $\tau$ values.
We can also notice from these plots that the performance of RUDDp is better than 
CPMGp and UDDp which have almost same performance.  UDD has better behavior than RUDDp, and 
for longer $\tau$ values UDD and RUDD have same behaviour.
However, these performances in general may also be dependent on initial states \cite{suter2010}.
\subsection{DD on MQC}
\begin{figure}
%\vspace{-3cm}
\begin{center}
\includegraphics[trim = 1.2cm 1.5cm 2cm 1.2cm,clip=true,width=12cm]{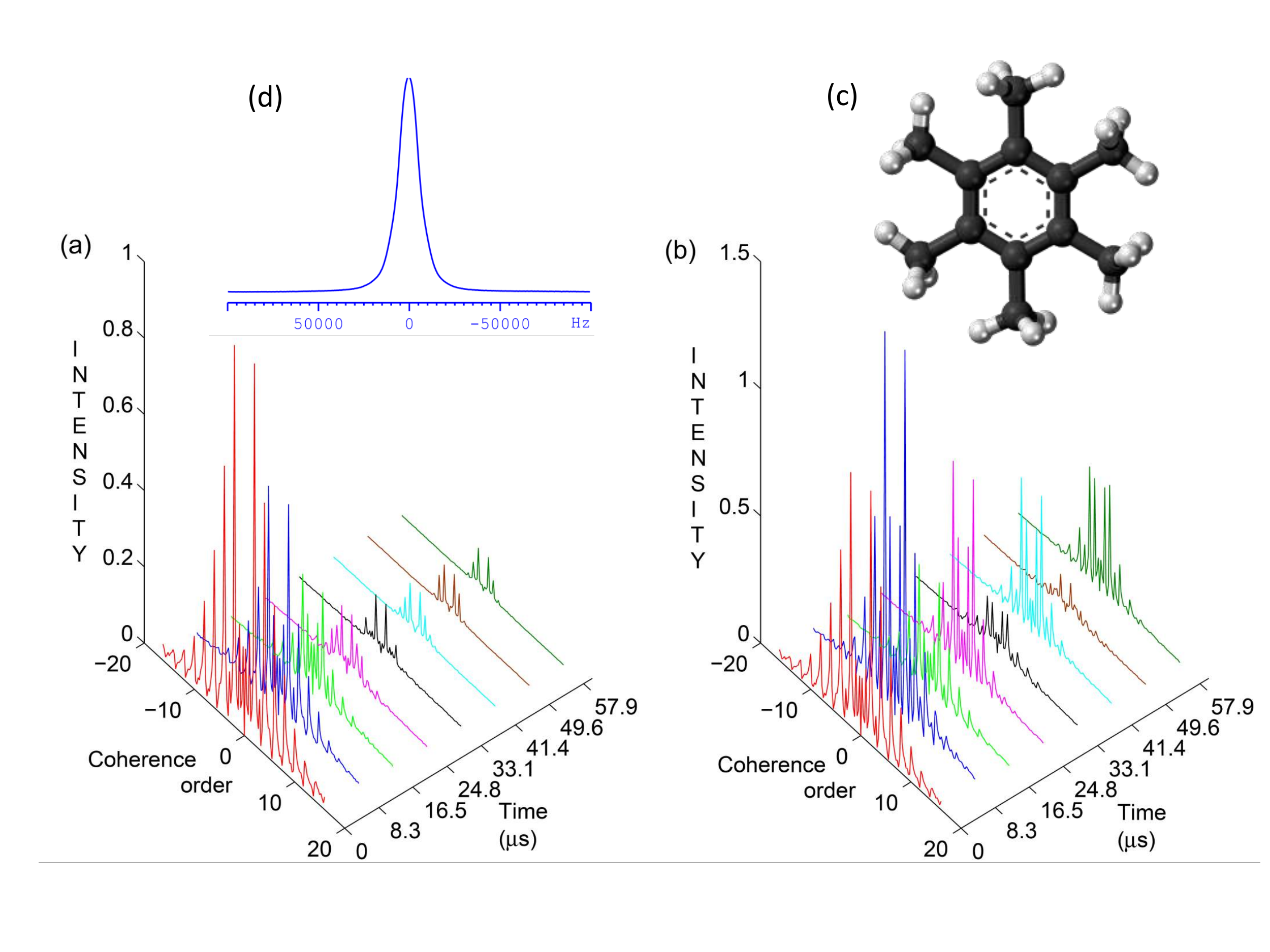}
\end{center}
\caption{\label{noddrudd}  Multiple quantum spectra showing different coherence orders
detected after inserting various delays (a) and RUDD sequences of same durations (b).  Structure of Hexamethylbenzene(c). $^{1}$H spectra of powder Hexamethylbenzene (d).  The line width of the resonance peak is 12 kHZ. 
}
\end{figure}
As described in \S \ref{mqddsec} and Fig. \ref{mqdd}(a-b), the scheme for 
studying DD on MQC involves preparation of MQC, evolution of MQC, followed by 
storing MQC onto longitudinal spin-orders.
A delay $t_R = 5$ ms was introduced to destroy the residual transverse magnetization.
The longitudinal spin order is then converted into SQC using a $(\pi/2)_y$
pulse, followed by a purge pulse $p_x$ of duration $50 \; \mu$s.  A 180 degree phase
alternation of the detection pulse, purge pulse, and the receiver is used to reduce
artifacts arising from receiver ringing.  For efficient generation of MQC
five cycles of 8-pulse sequence shown in Fig. \ref{mqdd}b was used in preparation and 
mixing periods, and the parameter $\Delta$ was optimized to $2 \;\mu$s.
In our experiments the coherences of successive quantum number are separated by 
$\Delta \omega = 2\pi \times 200$ kHz
In order to separate a maximum of $n_\mathrm{max}$
coherences, the relative phase $\alpha$ between the prepation and mixing is incremented by
$\Delta \alpha = \pi/n_\mathrm{max}$. 
We have choosen $n_\mathrm{max} = 64$.
The corresponding increment in the evolution period is
given by $\Delta t_1 = \Delta\alpha/\Delta\omega$.
The signal intensities of the spectrum corresponding to these increments after cosine transform
display strong peaks at even multiples of $\Delta \omega$.  
Mean value of the signal intensities is made to zero to suppress strong zero-quantum
peak.
Fig. \ref{noddrudd}a displays these even MQCs detected after inserting various delays,
and Fig. \ref{noddrudd}b displays those detected after applying RUDDp sequences of 
same durations. 
The first spectrum corresponding to no-delay is same in both cases, in which MQCs 
of order up to 24 can easily be observed.  Other spectra in (b)
were obtained by RUDDp sequences constructed with increasing number of pulses, i.e., 
$N={1,2,\cdots,7}$.
Under no DD (Fig. \ref{noddrudd}a), the intensities decay monotonically with delays 
while under RUDDp (Fig. \ref{noddrudd}b) the dependence of intensities is oscillatory w.r.t. 
$N$.  Similar behavior was earlier observed in a two-qubit liquid state
NMR system \cite{roy2010initialization}.  The spectra in (b) at odd $N$ clearly show better intensities 
compared to the corresponding spectra in (a). 
Comprasions of performance of different DD schemes for preserving MQCs of 
various orders are discribed in the following.
\begin{figure}
\begin{center}
\includegraphics[trim = 0cm 3cm 0cm 0cm,clip=true,width=9cm,angle=-90]{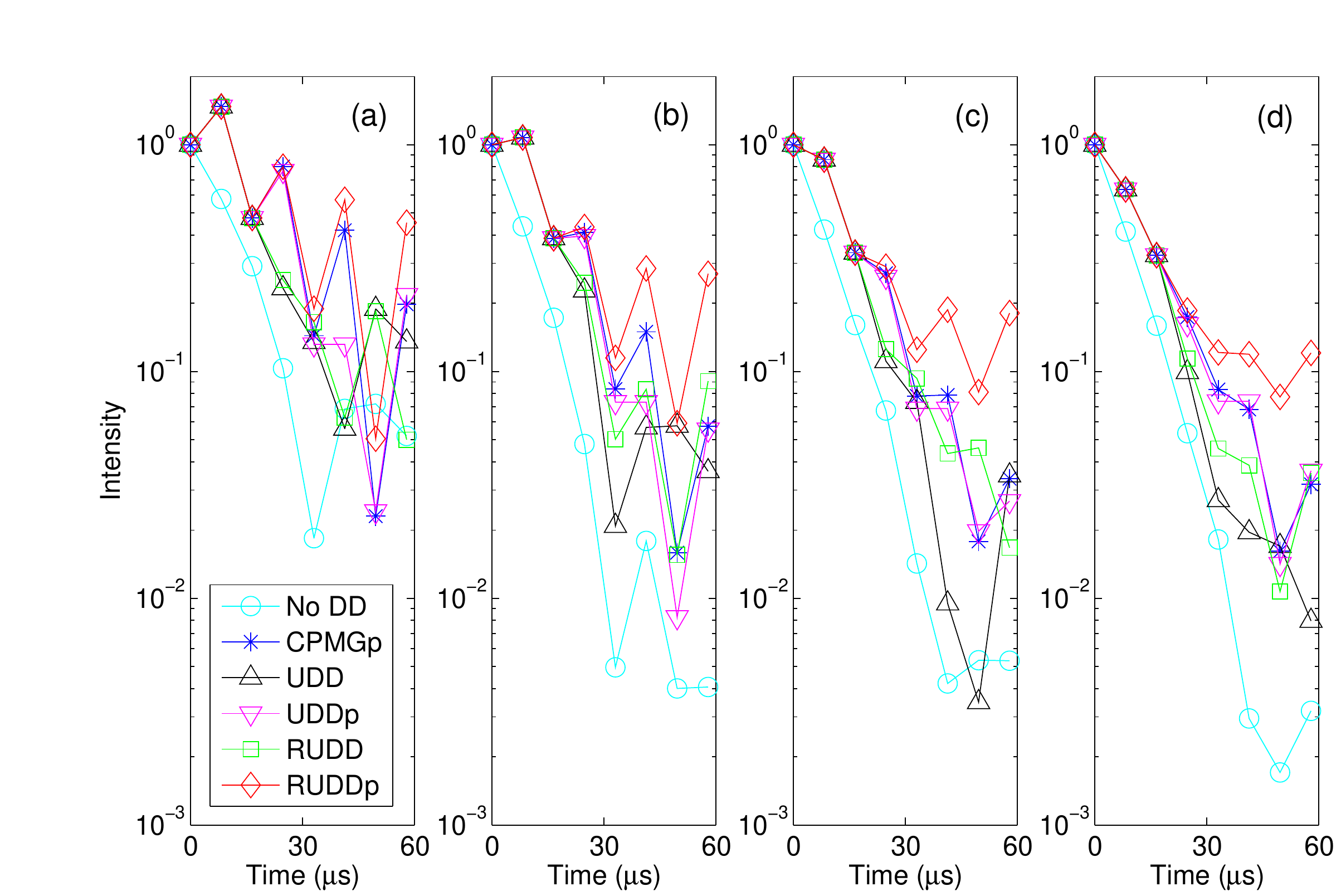}
%\end{center}
\caption{\label{btpspec}Performance of various DD schemes in preserving MQCs of
order 2 (a), 4 (b), 6 (c), and 8 (d).  Each data set has 8 points, in which the
first point corresponds to no DD, and the rest correspond to different size of
the DD sequence with $N={1,2,\cdots,7}$.
}
\label{mqdddat1}
\end{center}
\end{figure}
The intensities of MQCs of even orders between 2 and 8 w.r.t. size of various DD 
schemes are plotted in Fig. \ref{mqdddat1}.  The first data point in each data set 
corresponds to no DD, and the rest correspond to different size of the DD sequence 
with $N={1,2,\cdots,7}$. As observed in Fig. \ref{noddrudd}b, we see the oscillatory
behaviour of each MQC under various DD schemes.  But all the DD schemes display
an overall improvement w.r.t. no DD.  However it can be noticed that RUDDp has
significantly better performance than all other schemes, even for higher order 
coherences.  Surprisingly, unlike the single-quantum case, where in RUDD displayed
the best performance, in multiple-quantum case RUDDp is the best scheme.

\begin{figure}
\begin{center}
\includegraphics[trim = 0cm 3cm 0cm 0cm,clip=true,width=9cm,angle=-90]{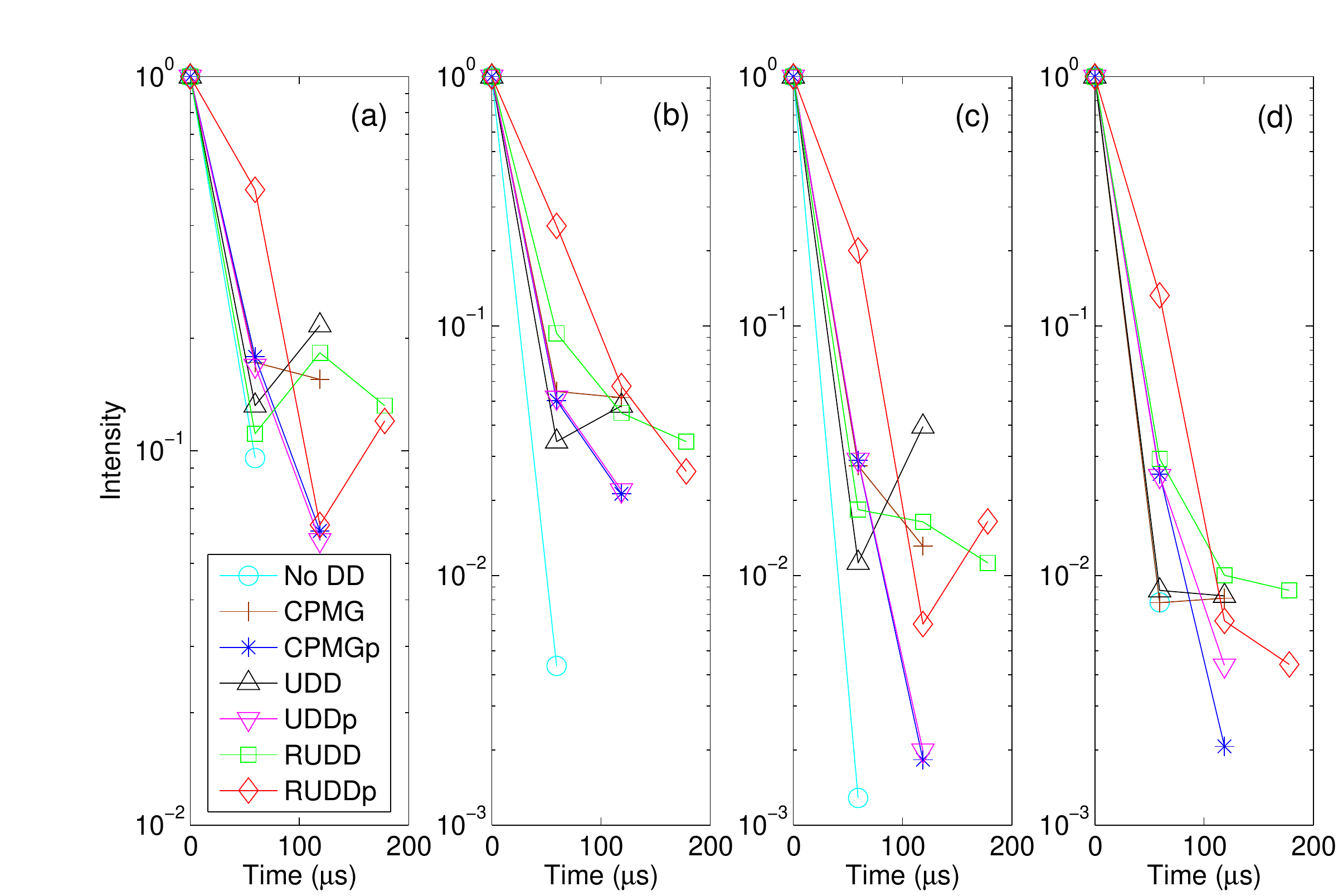}
\caption{\label{mqdddat2}
Performance of various DD schemes in preserving MQCs of
order 2 (a), 4 (b), 6 (c), and 8 (d).  The intensities were obtained
from spectra detected after applying up to a maximum of 3 cycles of 
7-pulse DD schemes. 
}
\end{center}
\end{figure}
The intensities of MQCs of even orders between 2 and 8 for different
cycles of 7-pulse DD schemes are plotted in Fig. \ref{mqdddat2}. 
 The first data point in each case corresponds to
no DD.  The fast decay of magnetization under no DD allowed to detect
intensities corresponding to a duration of only one cycle, while for 
RUDD and RUDDp, intensities up to 3 cycles could be detected.

\section{\textbf{Filter function analysis of various DD sequences}}
The most detrimental source of error in coherent evolution of quantum systems is the decoherence.  The fluctuations in environmental couplings lead to random qubit errors and hence to loss of coherence.  In following, I will revisit the effect of decoherence and dynamical decoupling quantitatively as described in \cite{biercuk2011dynamical}.
%e^{-i\frac{\phi}{2}}
Consider a qubit initially prepared in a superposition state,
\begin{eqnarray}
\ket{\psi} = c_1 \ket{0} + c_{2} \ket{1}.
\end{eqnarray}
Here $c_1$ and $c_2$ are probability amplitudes of state $\ket{0}$ and $\ket{1}$ respectively, such that ${\modulous{c_1}}^2+ {\modulous{c_2}}^2 =1$.  In a semiclassical picture, the average effect of the environment is captured by a random field $\beta(t)$, under which  the qubit experiences a dephasing  Hamiltonian $H = (\frac{\Omega}{2} + \beta(t)) \sigma_{z}$. The Rabi frequency $\Omega$ can be dropped by transforming into a rotating frame. The state after the evolution under the Hamiltonian at time t is
\begin{eqnarray}
\ket{\psi(t)} = c_1 e^{-i\int\limits_{0}^{t} \beta(t) dt} \ket{0} + c_2 e^{i\int\limits_{0}^{t} \beta(t) dt} \ket{1}.
\end{eqnarray}
Here $\int\limits_{0}^{t} \beta(t) dt$ is the accumulated random phase by the qubit in time t.  In the density matrix formulation the off diagonal terms in the density matrix 
\begin{eqnarray}
\rho(t) = {\modulous{c_1}}^2 \outpr{0}{0}+{\modulous{c_2}}^2 \outpr{1}{1}+ c_1c_2^* e^{-2i\int\limits_{0}^{t} \beta(t) dt} \outpr{0}{1} + c_2c_1^* e^{2i\int\limits_{0}^{t} \beta(t) dt} \outpr{1}{0}
\end{eqnarray}
represent coherence and decays due to ensemble average.  Typical source of dephasing involves error in experimental control, noise, and random parameters due to environment.  In particular, magnetic field fluctuations in atomic systems \cite{biercuk2009optimized}, charge fluctuations in solid state charge qubits \cite{hayashi2003coherent}, and effective Overhauser field due to nuclear spins in semiconductors systems \cite{reilly2008suppressing,johnson2005triplet}. 

Suppose we prepare our initial state $\ket{\psi(0)}$ along $x$ direction, then the coherence at time $t$ is the ensemble average of expectation value i.e.,
\begin{equation}
W(\tau) = \modulous{\overline{\expv{\sigma_{x}(\tau)}}} = e^{-\chi(\tau)},        
\label{coherence}
\end{equation}
where
 \begin{eqnarray}
\chi = \frac{2}{\pi}\int\limits_{0}^{\infty} \frac{S_{\beta}(\omega)}{{\omega}^2}F_{n}(\omega \tau)d \omega .
\label{chi}
\end{eqnarray}
In the above equation
\begin{eqnarray}
	S_{\beta}(\omega) = \int\limits_{-\infty}^{\infty} e^{-i\omega t}({\beta(t+\tau)\beta(t)})dt
\end{eqnarray}
is the power spectral density of environment and contains noise information in the frequency domain which is the Fourier transform of auto-correlation function of time domain noise term $\beta(t)$.
It is often convenient to characterize noise in frequency domain using $S_\beta(\omega)$.

\begin{figure}[h]
\begin{center}
\includegraphics[trim=3cm 0cm 3cm 0cm,width=8cm]{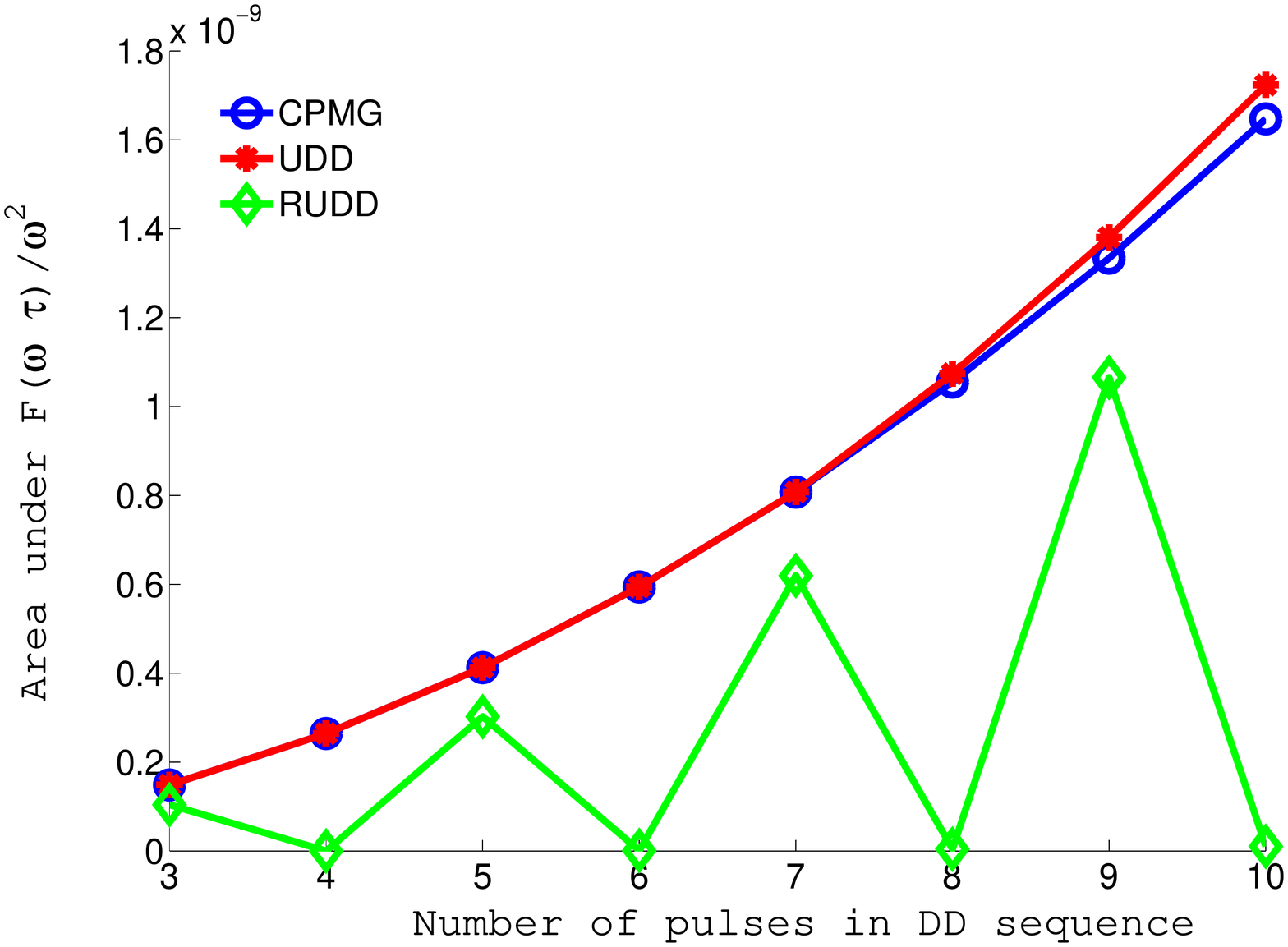}
\caption{Filter function analysis (FF) of various DD sequences.  Each point in the plot corresponds to the area under $\frac{F_{n}(\omega \tau)}{\omega^2}$ for respective sequence of given block size.}
\label{ff1}
\end{center}
\end{figure}

\begin{figure}[h]
\begin{center}
\includegraphics[trim=6cm 1.5cm 5cm 15cm,width=6.3cm]{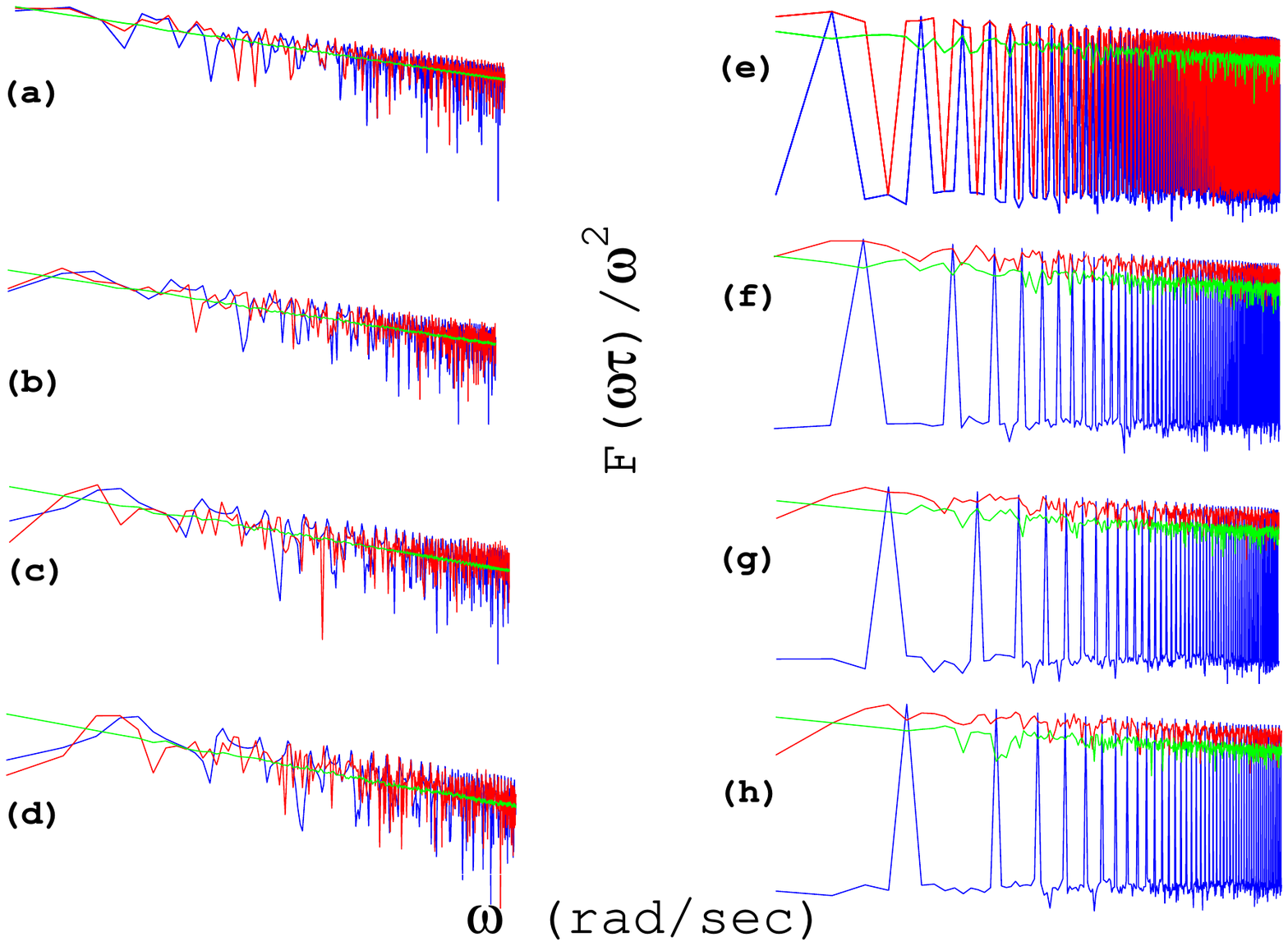}
\caption{Plot shows $\frac{F(\omega \tau)}{\omega^2}$ versus $\omega$  for various DD sequences (CPMG, UDD, RUDD).  The subplots (a) to (h) respectively correspond to the total number of $\pi$ pulses ranging from 3 to 10.}
\label{ff2}
\end{center}
\end{figure}

The term $F_{n}(\omega \tau)$ known as filter function captures the experimentally induced modulation to the coherence-decay $e^{-\chi(\tau)}$.  A 
DD sequence introduces modulations to accumulated random phase such that every $\pi$ pulse switches the phase between $\int\limits_{0}^{\infty} \beta(t) dt$ and $-\int\limits_{0}^{\infty} \beta(t) dt$.  The DD sequence and corresponding modulation function $y_{n}(t)$ is shown in figure 1 of \cite{biercuk2011dynamical}.  The convolution of $\beta(t)$ and  $y_{n}(t)$ provides desired noise suppression. 
The Fourier transform of this convolution  provides relevant spectral information.  Filter function form in Fourier domain is
\begin{eqnarray}
F_{n}(\omega \tau) = \left \vert 1+(-1)^{n+1}e^{i\omega \tau'}+ 2\sum_{j=1}^{n}(-1)^{j}e^{i\delta_{j}\omega \tau'}\cos\left( \frac{\omega\tau_{\pi}}{2}\right) \right \vert^2
\label{ffeq}
\end{eqnarray}
Here $\tau'$ is the total time of the pulse sequence, $\delta_{j}\tau'$ and $\tau_{\pi}$ are respectively the time instant and the duration of the j$^{th}$ $\pi$ pulse.  From the eqn. $\ref{coherence}$ it is evident that minimum value of $F_{n}(\omega \tau)$ leads to minimum decay $\chi$ and hence maximum coherence $W$.  The filter function takes values between $0$ and $1$.  For free evolution case, filter function $F_{n}(\omega \tau)$ is $1$ and for perfect refocusing it is zero.  Filter function analysis using eqn. \ref{chi} and \ref{ffeq} provides a way to examine performance of various DD sequences.  The minimum area under the filter function for a given DD leads to maximum coherence further one may also design an optimal DD sequence for a given power spectral density by minimizing $F_{n}(\omega \tau)$.
Here we compare performance of DD sequences used in above experiments via filter function analysis.  Results of the analysis are shown in Fig. \ref {ff1} and in Fig. \ref {ff2}.  Figure \ref{ff2}
 shows plots of $\frac{F_{n}(\omega \tau)}{\omega^{2}}$ versus $\omega$ for different number of pulses ranging from 3-10. In each subplot, the filter function for RUDD sequence has lesser area than for corresponding CPMG and UDD sequences.  The $\pi$ pulse durations and delays used for calculating F$_{n}(\omega \tau)$ are $\tau = 2\mu s$ and $\tau_{\pi}= 4.27\mu s$, same as in our experiments.  Fig. \ref{ff1} shows area under the function $\frac{F_{n}(\omega \tau)}{\omega^{2}}$ for CPMG, UDD, and RUDD sequences.  Clearly, in the case of RUDD sequence, the area under the curve is less than other sequences. One may also see typical even-odd behavior of RUDD sequence supporting our experimental results. 

\section{\textbf{Conclusions}\label{65}}
We studied the performance of various DD schemes on nuclear spins with
long-range interactions using a solid state NMR system.  First applied
these DD schemes on a single quantum coherence.  The experiments were
carried out for different number of $\pi$ pulses and for different 
delays between them.  The results clearly show that all the DD schemes
are able to preserve the single quantum coherence for longer durations 
of time compared to no DD.  However, for small delays between the $\pi$
pulses, RUDD showed the best performance.  For longer dealys between 
the $\pi$ pulses, CPMG was better. 

Then we prepared MQCs of even orders 
using multiple cycles of the well known 8-pulse sequence implementing a 
two-quantum average Hamiltonian.  The MQCs so prepared could be
detected using standard spin-counting experiment.  Various DD schemes
were inserted after the preparation of MQCs.  We studied the
performance of various DD sequences of different sizes.  The intensity
behaviour under all the DD sequences were oscillatory, but they showed
an overall improvement over no DD.  However, RUDDp sequence showed
the best performance over all other sequences. The superior performance of 
RUDD sequence over other DD sequences is attributed to lower filter function
 area of RUDD than others.

\mathversion{normal}
\thispagestyle{empty}

% % % % % % % % BIBLIOGRAPHY
\bibliographystyle{unsrt}
\bibliography{biblio}
\thispagestyle{empty}
\end{document}